\documentstyle[12pt,fleqn]{book}

\font\twlgot =eufm10 scaled \magstep1
\font\egtgot =eufm8
\font\sevgot =eufm7

\font\twlmsb =msbm10 scaled \magstep1
\font\egtmsb =msbm8
\font\sevmsb =msbm7

\newfam\gotfam
\def\pgot{\fam\gotfam\twlgot}
\textfont\gotfam\twlgot
\scriptfont\gotfam\egtgot
\scriptscriptfont\gotfam\sevgot
\def\got{\protect\pgot}

\newfam\msbfam
\textfont\msbfam\twlmsb
\scriptfont\msbfam\egtmsb
\scriptscriptfont\msbfam\sevmsb
\def\Bbb{\protect\pBbb}
\def\pBbb{\relax\ifmmode\expandafter\Bb\else\typeout{You cann't use
Bbb in text mode}\fi}
\def\Bb #1{{\fam\msbfam\relax#1}}

\def\op#1{\mathop{{\it\fam0} #1}\limits}

\newcommand{\id}{{\rm Id\,}}

\newcommand{\pr}{{\rm pr}}

\newcommand{\di}{{\rm dim\,}}

\newcommand{\Ker}{{\rm Ker\,}}
\newcommand{\im}{{\rm Im\, }}

\newcommand{\hm}{{\rm Hom\,}}
\newcommand{\dif}{{\rm Diff\,}}

\newcommand{\nm}[1]{|{#1}|}
\newcommand{\sU}{{\{U\}}}

\newcommand{\bll}{\bullet}

\newcommand{\beq}{\begin{equation}}
\newcommand{\eeq}{\end{equation}}
\newcommand{\ben}{\begin{eqnarray}}
\newcommand{\een}{\end{eqnarray}}
\newcommand{\be}{\begin{eqnarray*}}
\newcommand{\ee}{\end{eqnarray*}}

\newcommand{\nw}[1]{[{#1}]}

\newcommand{\gU}{{\got U}}
\newcommand{\gO}{{\got O}}
\newcommand{\gA}{{\got A}}

\newcommand{\cG}{{\got g}}
\newcommand{\gd}{{\got d}}

\newcommand{\gE}{{\got E}}

\newcommand{\gR}{{\got R}}

\newcommand{\gQ}{{\got Q}}

\newcommand{\gf}{{\got f}}

\newcommand{\cJ}{{\cal J}}
\newcommand{\cA}{{\cal A}}
\newcommand{\cO}{{\cal O}}
\newcommand{\cT}{{\cal T}}
\newcommand{\cP}{{\cal P}}
\newcommand{\cR}{{\cal R}}
\newcommand{\cL}{{\cal L}}
\newcommand{\cV}{{\cal V}}

\newcommand{\cQ}{{\cal Q}}
\newcommand{\cE}{{\cal E}}

\newcommand{\cF}{{\cal F}}
\newcommand{\cC}{{\cal C}}

\newcommand{\cI}{{\cal I}}

\newcommand{\cK}{{\cal K}}

\newcommand{\ccG}{{\cal G}}

\newcommand{\cS}{{\cal S}}

\newcommand{\bL}{{\bf L}}

\newcommand{\bb}{{\bf 1}}

\newcommand{\bE}{{\bf E}}

\newcommand{\bp}{{\bf p}}
\newcommand{\bu}{{\bf u}}

\newcommand{\rA}{{\rm Ann\,}}

\newcommand{\al}{\alpha}
\newcommand{\bt}{\beta}
\newcommand{\dl}{\delta}
\newcommand{\la}{\lambda}
\newcommand{\La}{\Lambda}
\newcommand{\f}{\phi}
\newcommand{\vf}{\varphi}
\newcommand{\F}{\Phi}
\newcommand{\p}{\pi}
\newcommand{\s}{\psi}

\newcommand{\om}{\omega}
\newcommand{\Om}{\Omega}
\newcommand{\m}{\mu}
\newcommand{\n}{\nu}
\newcommand{\g}{\gamma}
\newcommand{\G}{\Gamma}
\newcommand{\e}{\epsilon}
\newcommand{\ve}{\varepsilon}
\newcommand{\thh}{\theta}

\newcommand{\vr}{\varrho}
\newcommand{\up}{\upsilon}
\newcommand{\vt}{\vartheta}

\newcommand{\si}{\sigma}
\newcommand{\Si}{\Sigma}

\newcommand{\bT}{{\bf T}}

\newcommand{\Y}{Y\to X}
\newcommand{\w}{\wedge}

\newcommand{\wt}{\widetilde}
\newcommand{\wh}{\widehat}
\newcommand{\ol}{\overline}

\newcommand{\dr}{\partial}
\newcommand{\pdr}{\partial}

\newcommand{\rdr}{\stackrel{\leftarrow}{\dr}{}}
\newcommand{\llr}{\op\longleftarrow}

\newcommand{\mar}[1]{}

\newcommand{\lto}{{\leftarrow}}

\newcommand{\lla}{\op\longleftarrow}
\newcommand{\ar}{\op\longrightarrow}

\newcommand{\ot}{\otimes}
\newcommand{\ap}{\approx}

\let\ssection=\section
\renewcommand{\section}{\setcounter{equation}{0}\ssection}

\newcounter{eqalph}[section]
\newcounter{equationa}[section]
\newcounter{example}[section]
\newcounter{remark}[section]
\newcounter{theorem}[section]
\newcounter{proposition}[section]
\newcounter{lemma}[section]
\newcounter{corollary}[section]
\newcounter{definition}[section]

\def\theremark{\arabic{chapter}.\arabic{section}.\arabic{remark}}

\def\thedefinition{\arabic{chapter}.\arabic{section}.\arabic{definition}}

\newenvironment{example}{\refstepcounter{remark} {\bf Example
\theremark.}}{{\Large $\bullet$}  }
\newenvironment{remark}{\refstepcounter{remark} {\bf Remark
\theremark.}}{{\Large $\bullet$}  }
\newenvironment{theorem}{\refstepcounter{definition} {\sc
Theorem \thedefinition}.}{$\Box$ }

\newenvironment{lemma}{\refstepcounter{definition} {\sc Lemma
\thedefinition}.}{ $\Box$ }
\newenvironment{condition}{\refstepcounter{definition} {\sc
Condition \thedefinition}.}{ $\Box$ }
\newenvironment{notation}{\refstepcounter{definition} {\sc
Notation \thedefinition}.}{ $\Box$ }
\newenvironment{corollary}{\refstepcounter{definition} {\sc
Corollary \thedefinition}.}{ $\Box$}
\newenvironment{definition}{\refstepcounter{definition} {\sc
Definition \thedefinition}.}{$\Box$ }

\makeindex

\begin{document}

\hbox{}

\thispagestyle{empty}

\setcounter{page}{0}

\vskip 3cm

\begin{center}

{\large \bf Fibre Bundles, Jet Manifolds and Lagrangian Theory.

\bigskip Lectures for Theoreticians}

\bigskip
\bigskip
\bigskip

{\sc G. Sardanashvily}
\bigskip

Department of Theoretical Physics, Moscow State University,
Moscow, Russia

\bigskip
\bigskip
\bigskip
\bigskip
\bigskip
\bigskip

{\bf Abstract}
\bigskip
\end{center}

\noindent In contrast with QFT, classical field theory can be
formulated in a strict mathematical way by treating classical
fields as sections of smooth fibre bundles. Addressing to the
theoreticians, these Lectures aim to compile the relevant material
on fibre bundles, jet manifolds, connections, graded manifolds and
Lagrangian theory. They follow the perennial course of lectures on
geometric methods in field theory at the Department of Theoretical
Physics of Moscow State University.


\tableofcontents

\chapter{Geometry of fibre bundles}

Throughout the Lectures, all morphisms are smooth (i.e. of class
$C^\infty$) and manifolds are smooth real and finite-dimensional.
A smooth real {\sl manifold} \index{manifold} is \index{smooth
manifold} customarily assumed to be Hausdorff and second-countable
(i.e., it has a countable base for topology). Consequently, it is
a locally compact space which is a union of a countable number of
compact subsets, a separable space (i.e., it has a countable dense
subset), a paracompact and completely regular space. Being
paracompact, a smooth manifold admits a partition of unity by
smooth real functions. Unless otherwise stated, manifolds are
assumed to be connected (and, consequently, arcwise connected). We
follow the notion of a manifold without boundary.

\section{Fibre bundles}

Let $Z$ be a manifold.  \index{$TZ$} By \index{$T^*Z$}
\be
\pi_Z:TZ\to Z, \qquad \pi^*_Z:T^*Z\to Z
\ee
are denoted its tangent \index{bundle!tangent} and cotangent
bundles, \index{bundle!cotangent} respectively. \index{tangent
bundle} \index{cotangent bundle}  Given coordinates $(z^\al)$ on
$Z$, they are equipped with the \index{holonomic!coordinates} {\sl
holonomic coordinates}
\be
&& (z^\la,\dot z^\la), \qquad \dot z'^\la= \frac{\dr z'^\la}{\dr
z^\mu}\dot z^\m, \\
&&(z^\la,\dot z_\la), \qquad \dot z'_\la= \frac{\dr z'^\m}{\dr
z^\la}\dot z_\m,
\ee
with respect to the {\sl holonomic frames} \index{holonomic!frame}
$\{\dr_\la\}$ \index{frame!holonomic} and {\sl coframes}
\index{coframe} $\{dz^\la\}$ \index{holonomic!coframe} in the
tangent and cotangent spaces to $Z$, respectively. Any manifold
morphism $f:Z\to Z'$ yields the \index{tangent morphism} {\sl
tangent morphism} \index{$Tf$}
\be
Tf:TZ\to TZ', \qquad \dot z'^\la\circ Tf = \frac{\dr f^\la}{\dr
x^\m}\dot z^\m.
\ee

Let us consider manifold morphisms of maximal rank. They are
immersions (in particular, imbeddings) and submersions. An
injective immersion is a submanifold, and a surjective submersion
is a fibred manifold (in particular, a fibre bundle).

Given manifolds $M$ and $N$, by  the {\sl rank of a morphism}
\index{rank of a morphism} $f:M\to N$ at a point $p\in M$ is meant
the rank of the linear morphism
\be
T_pf:T_pM\to T_{f(p)}N.
\ee
For instance, if $f$ is of maximal rank at $p\in M$,  then $T_pf$
is injective when $\di M\leq \di N$ and surjective when $\di N\leq
\di M$. In this case, $f$ is called an {\sl immersion}
\index{immersion} and a {\sl submersion} \index{submersion} at a
point $p\in M$, respectively.

Since $p \to {\rm rank}_pf$ is a lower semicontinuous function,
then the morphism $T_pf$ is of maximal rank on an open
neighbourhood of $p$, too. It follows from the inverse function
theorem that:

$\bll$ if $f$ is an immersion at $p$, then it is locally injective
around $p$.

$\bll$ if $f$ is a submersion at $p$, it is locally surjective
around $p$.

\noindent If $f$ is both an immersion and a submersion, it is
called a {\sl local diffeomorphism} \index{local diffeomerphism}
at $p$. In this case, there exists an open neighbourhood $U$ of
$p$ such that $f: U\to f(U)$ is a diffeomorphism onto an open set
$f(U)\subset N$.

A manifold morphism  $f$ is called the immersion (resp.
submersion) if it is an immersion (resp. submersion) at all points
of $M$. A submersion is necessarily an {\sl open map}, \index{open
map} i.e., it sends open subsets of $M$ onto open subsets of $N$.
If an immersion $f$ is open (i.e., $f$ is a homeomorphism onto
$f(M)$ equipped with the relative topology from $N$), it is called
the {\sl imbedding}. \index{imbedding}

A pair $(M,f)$ is called a {\sl submanifold} \index{submanifold}
of $N$ if $f$ is an injective immersion. A submanifold $(M,f)$ is
an {\sl imbedded submanifold} \index{imbedded submanifold} if $f$
is an imbedding. For the sake of simplicity, we usually identify
$(M,f)$ with $f(M)$. If $M\subset N$, its natural injection is
denoted by $i_M:M\to N$.

There are the following criteria for a submanifold to be imbedded.

\begin{theorem}\label{subman3} \mar{subman3}
Let $(M,f)$ be a submanifold of $N$.

(i) The map $f$ is an imbedding iff, for each point $p\in M$,
there exists a (cubic) coordinate chart $(V,\psi)$ of $N$ centered
at $f(p)$ so that $f(M)\cap V$ consists of all points of $V$ with
coordinates $(x^1,\ldots,x^m,0,\ldots,0)$.

(ii) Suppose that $f:M\to N$ is a {\sl proper map}, \index{proper
map} i.e., the pre-images of compact sets are compact. Then
$(M,f)$ is a closed imbedded submanifold of $N$. In particular,
this occurs if $M$ is a compact manifold.

(iii) If $\di M =\di N$, then $(M,f)$ is an open imbedded
submanifold of $N$.
\end{theorem}

A triple
\mar{11f1}\beq
\p :Y\to X, \qquad \di X=n>0, \label{11f1}
\eeq
is called a {\sl fibred manifold} \index{fibred manifold} if
\index{manifold!fibred} a manifold morphism $\p$ is a surjective
submersion, i.e., the tangent morphism $T\pi:TY\to TX$ is a
surjection. One says that $Y$ is a {\sl total space} \index{total
space} of a fibred manifold (\ref{11f1}), $X$ is its {\sl base},
\index{base of a fibred manifold} $\p$ is a {\sl fibration},
\index{fibration} and $Y_x=\p^{-1}(x)$ is a {\sl fibre}
\index{fibre} over $x\in X$.

Any fibre is an imbedded submanifold of $Y$ of dimension $\di
Y-\di X$.

\begin{theorem} \mar{11t1} \label{11t1} A surjection (\ref{11f1})
is a fired manifold iff a manifold $Y$ admits an atlas of
coordinate charts $(U_Y; x^\la, y^i)$ such that $(x^\la)$ are
coordinates on $\p(U_Y)\subset X$ and coordinate transition
functions read
\be
x'^\la =f^\la(x^\m), \qquad y'^i=f^i(x^\m,y^j).
\ee
These coordinates are called {\sl fibred coordinates}
\index{fibred coordinates} compatible with a fibration $\p$.
\end{theorem}

By a {\sl local section} \index{section!local} of a surjection
(\ref{11f1}) is meant an injection $s:U\to Y$ of an open subset
$U\subset X$ such that $\p\circ s=\id U$, i.e., a section sends
any point $x\in X$ into the fibre $Y_x$ over this point. A local
section also is defined over any subset $N\in X$ as the
restriction to $N$ of a local section over an open set containing
$N$. If $U=X$, one calls $s$ the {\sl global section}.
\index{section!global} Hereafter, by a {\sl section}
\index{section} is meant both a global section and a local section
(over an open subset).

\begin{theorem}
A surjection $\p$ (\ref{11f1}) is a fibred manifold iff, for each
point $y\in Y$, there exists a local section $s$ of $\p :Y\to X$
passing through $y$.
\end{theorem}

The range $s(U)$ of a local section $s:U\to Y$ of a fibred
manifold $Y\to X$ is an imbedded submanifold of $Y$. It also is a
{\sl closed map}, \index{closed map} which sends closed subsets of
$U$ onto closed subsets of $Y$. If  $s$ is a global section, then
$s(X)$ is a closed imbedded submanifold of $Y$. Global sections of
a fibred manifold need not exist.

\begin{theorem} \label{mos9} \mar{mos9}
Let $Y\to X$ be a fibred manifold whose fibres are diffeomorphic
to $\Bbb R^m$.  Any its section over a closed imbedded submanifold
(e.g., a point) of $X$ is extended to a global section. In
particular, such a fibred manifold always has a global section.
\end{theorem}

Given fibred coordinates $(U_Y;x^\la,y^i)$, a section $s$ of a
fibred manifold $Y\to X$ is represented by collections of local
functions $\{s^i=y^i\ \circ s\}$ on $\p(U_Y)$.

A fibred manifold $Y\to X$ is called a {\sl fibre bundle}
\index{fibre bundle} if admits a fibred coordinate atlas
$\{(\pi^{-1}(U_\xi); x^\la, y^i)\}$ over a cover
$\{\pi^{-1}(U_\iota)\}$ of $Y$ which is the inverse image of a
cover $\gU=\{U_\xi\}$ is a cover of $X$. In this case, there
exists a manifold $V$, called a {\sl typical fibre},
\index{typical fibre} such that $Y$ is locally diffeomorphic to
the \index{$\psi_\xi$} splittings
\mar{mos02}\beq
\psi_\xi:\pi^{-1}(U_\xi) \to U_\xi\times V, \label{mos02}
\eeq
glued together by means of \index{transition functions} {\sl
transition functions} \index{$\vr_{\xi\zeta}$}
\mar{mos271}\beq
\vr_{\xi\zeta}=\psi_\xi\circ\psi_\zeta^{-1}: U_\xi\cap
U_\zeta\times V \to  U_\xi\cap U_\zeta\times V \label{mos271}
\eeq
on overlaps $U_\xi\cap U_\zeta$. Transition functions
$\vr_{\xi\zeta}$ fulfil the \index{cocycle condition} {\sl cocycle
condition}
\mar{+9}\beq
\vr_{\xi\zeta}\circ\vr_{\zeta\iota}=\vr_{\xi\iota} \label{+9}
\eeq
on all overlaps $U_\xi\cap U_\zeta\cap U_\iota$. Restricted to a
point $x\in X$, {\sl trivialization morphisms}
\index{trivialization morphism} $\psi_\xi$ (\ref{mos02}) and
transition functions $\vr_{\xi\zeta}$ (\ref{mos271}) define
diffeomorphisms of fibres
\mar{sp21,2}\ben
&&\psi_\xi(x): Y_x\to V, \qquad x\in U_\xi,\label{sp21}\\
&& \vr_{\xi\zeta}(x):V\to V, \qquad x\in U_\xi\cap U_\zeta. \label{sp22}
\een
{\sl Trivialization charts} \index{trivialization chart} $(U_\xi,
\psi_\xi)$ together with transition functions $\vr_{\xi\zeta}$
(\ref{mos271}) constitute a {\sl bundle atlas}
\index{bundle!atlas}
\mar{sp5}\beq
\Psi = \{(U_\xi, \psi_\xi), \vr_{\xi\zeta}\} \label{sp5}
\eeq
of a fibre bundle $Y\to X$. Two bundle atlases are said to be {\sl
equivalent} \index{equivalent bundle atlases} if their union also
is a bundle atlas, i.e., there exist transition functions between
trivialization charts of different atlases.

A fibre bundle $Y\to X$ is uniquely defined by a bundle atlas.
Given an atlas $\Psi$ (\ref{sp5}), there is a unique manifold
structure on $Y$ for which $\p:Y\to X$ is a fibre bundle with the
typical fibre $V$ and the bundle atlas $\Psi$. All atlases of a
fibre bundle are equivalent.

\begin{remark} \label{11r1} \mar{11r1}
The notion of a fibre bundle introduced above is the notion of a
{\sl smooth locally trivial fibre bundle}. \index{bundle!smooth}
\index{bundle!locally trivial} In a general setting, a {\sl
continuous fibre bundle} \index{bundle!continuous} is defined as a
continuous surjective submersion of topological spaces $Y\to X$. A
continuous map $\pi:Y\to X$ is called a {\sl submersion}
\index{submersion!continuous} if, for any point $y\in Y$, there
exists an open neighborhood $U$ of the point $\pi(y)$ and a right
inverse $\si:U\to Y$ of $\pi$ such that $\si\circ\pi(y)=y$, i.e.,
there exists a local section of $\pi$. The notion of a {\sl
locally trivial continuous fibre bundle}
\index{bundle!continuous!locally trivial} is a repetition of that
of a smooth fibre bundle, where trivialization morphisms
$\psi_\xi$ and transition functions $\vr_{\xi\zeta}$ are
continuous.
\end{remark}

We have the following useful criteria for a fibred manifold to be
a fibre bundle.

\begin{theorem} \label{1110} \mar{1110}
If a fibration $\p:Y\to X$ is a proper map, then $Y\to X$ is a
fibre bundle. In particular, a fibred manifold with a compact
total space is a fibre bundle.
\end{theorem}

\begin{theorem} \label{11t2} \mar{11t2} A fibred manifold whose
fibres are diffeomorphic either to a compact manifold or $\Bbb
R^r$ is a fibre bundle.
\end{theorem}

A comprehensive relation between fibred manifolds and fibre
bundles is given in Remark \ref{Ehresmann}. It involves the notion
of an Ehresmann connection.

Unless otherwise stated, we restrict our consideration to fibre
bundles. Without a loss of generality, we further assume that a
cover $\gU$ for a bundle atlas of $Y\to X$ also is a cover for a
manifold atlas of the base $X$. Then, given a bundle atlas $\Psi$
(\ref{sp5}), a fibre bundle $Y$ is provided with the associated
{\sl bundle coordinates} \index{bundle!coordinates}
\be
x^\la(y)=(x^\la\circ \pi)(y), \qquad y^i(y)=(y^i\circ\psi_\xi)(y),
\qquad y\in \pi^{-1}(U_\xi),
\ee
where $x^\la$ are coordinates on $U_\xi\subset X$ and $y^i$,
called {\sl fibre coordinates}, \index{fibre coordinates} are
coordinates on a typical fibre $V$.

The forthcoming Theorems \ref{11t3} -- \ref{sp2} describe the
particular covers which one can choose for a bundle atlas.
Throughout the Lectures, only {\sl proper covers} \index{proper
cover} of manifolds are considered, i.e., $U_\xi\neq U_\zeta$ if
$\zeta\neq \xi$. Recall that a cover $\gU'$ is a refinement of a
cover $\gU$ if, for each $U'\in\gU'$, there exists $U\in\gU$ such
that $U'\subset U$. If a fibre bundle $Y\to X$ has a bundle atlas
over a cover $\gU$ of $X$, it admits a bundle atlas over any
refinement of $\gU$.

A fibred manifold $Y\to X$ is called trivial if $Y$ is
diffeomorphic to the product $X\times V$. Different
trivializations of $Y\to X$ differ from each other in surjections
$Y\to V$.

\begin{theorem} \label{11t3} \mar{11t3} Any fibre
bundle over a contractible base is trivial.
\end{theorem}

However, a fibred manifold over a contractible base need not be
trivial, even its fibres are mutually diffeomorphic.

It follows from Theorem \ref{11t3} that any cover of a base $X$
consisting of {\sl domains} \index{domain} (i.e., contractible
open subsets) is a bundle cover.

\begin{theorem} \label{sp1} \mar{sp1}
Every fibre bundle $Y\to X$ admits a bundle atlas over a countable
cover $\gU$ of $X$ where each member $U_\xi$ of $\gU$ is a domain
whose closure $\ol U_\xi$ is compact.
\end{theorem}

If a base $X$ is compact, there is a bundle atlas of $Y$ over a
finite cover of $X$ which obeys the condition of Theorem
\ref{sp1}.

\begin{theorem} \label{sp2} \mar{sp2}
Every fibre bundle $Y\to X$ admits a bundle atlas over a finite
cover $\gU$ of $X$, but its members need not be contractible and
connected.
\end{theorem}

Morphisms of fibre bundles, by definition, are {\sl fibrewise
morphisms}, \index{fibrewise morphism} sending a fibre to a fibre.
Namely, a {\sl bundle morphism} \index{bundle!morphism}
\index{morphism!of fibre bundles} of a fibre bundle $\pi:Y\to X$
to a fibre bundle $\pi': Y'\to X'$ is defined as a pair $(\Phi,f)$
of manifold morphisms which form a commutative diagram
\be
\begin{array}{rcccl}
& Y &\ar^\Phi & Y'&\\
_\pi& \put(0,10){\vector(0,-1){20}} & & \put(0,10){\vector(0,-1){20}}&_{\pi'}\\
& X &\ar^f & X'&
\end{array}, \qquad \pi'\circ\Phi=f\circ\pi.
\ee

Bundle injections and surjections are called {\sl bundle
monomorphisms and epimorphisms}, \index{bundle!monomorphism}
\index{bundle!epimorphism} respectively. A bundle diffeomorphism
is called a {\sl bundle isomorphism}, \index{bundle!isomorphism}
or a {\sl bundle automorphism} \index{bundle!automorphism} if it
is an isomorphism to itself. For the sake of brevity, a bundle
morphism over $f=\id X$ is often said to be a bundle morphism over
$X$, and is denoted by $Y\ar_XY'$. In particular, a bundle
automorphism over $X$ is called a {\sl vertical automorphism}.
\index{vertical automorphism} \index{bundle!automorphism!vertical}

A bundle monomorphism $\Phi:Y\to Y'$ over $X$ is called a {\sl
subbundle} \index{subbundle} of a fibre bundle $Y'\to X$ if
$\Phi(Y)$ is a submanifold of $Y'$. There is the following useful
criterion  for an image and an inverse image of a bundle morphism
to be subbundles.

\begin{theorem}\label{pomm} \mar{pomm}
Let $\Phi: Y\to Y'$ be a bundle morphism over $X$. Given a global
section $s$ of the fibre bundle $Y'\to X$ such that $s(X)\subset
\F(Y)$, by the {\sl kernel of a bundle morphism} \index{kernel!of
a bundle morphism} $\F$ with respect to a section $s$ is meant the
inverse image
\be
\Ker_s\F = \F^{-1}(s(X))
\ee
of $s(X)$ by $\F$. If $\Phi: Y\to Y'$ is a bundle morphism of
constant rank over $X$, then $\Phi(Y)$ and $\Ker_s\F$ are
subbundles of $Y'$ and $Y$, respectively.
\end{theorem}

Let us describe the following standard constructions of new fibre
bundles from the old ones.

$\bullet$ Given a fibre bundle $\pi:Y\to X$ and a manifold
morphism $f: X'\to X$, the {\sl pull-back}
\index{pull-back!bundle} of $Y$ by $f$ is called the manifold
\index{$f^*Y$}
\mar{mos106}\beq
f^*Y =\{(x',y)\in X'\times Y \,: \,\, \pi(y) =f(x')\}
\label{mos106}
\eeq
together with the natural projection $(x',y)\to x'$. It is a fibre
bundle over $X'$ such that the fibre of $f^*Y$ over a point $x'\in
X'$ is that of $Y$ over the point $f(x')\in X$. There is the
canonical bundle morphism
\mar{mos81}\beq
f_Y:f^*Y\ni (x',y)|_{\pi(y) =f(x')} \op\to_f y\in Y. \label{mos81}
\eeq
Any section $s$ of a fibre bundle $Y\to X$ yields the {\sl
pull-back section} \index{pull-back!section}
\be
f^*s(x')=(x',s(f(x'))
\ee
of $f^*Y\to X'$.

$\bullet$ If $X'\subset X$ is a submanifold of $X$ and $i_{X'}$ is
the corresponding natural injection, then the pull-back bundle
\be
i_{X'}^*Y=Y|_{X'}
\ee
is called the {\sl restriction} \index{restriction of a bundle} of
a fibre bundle $Y$ to the submanifold $X'\subset X$. If $X'$ is an
imbedded submanifold, any section of the pull-back bundle
\be
Y|_{X'}\to X'
\ee
is the restriction to $X'$ of some section of $Y\to X$.

$\bullet$ Let $\pi:Y\to X$ and $\pi':Y'\to X$ be fibre bundles
over the same base $X$. Their {\sl bundle product}
\index{bundle!product} $Y\times_X Y'$ over $X$ is defined as the
pull-back
\be
Y\op\times_X Y'=\pi^*Y'\quad {\rm or} \quad Y\op\times_X
Y'={\pi'}^*Y
\ee
together with its natural surjection onto $X$.  Fibres of the
bundle product $Y\times Y'$ are the Cartesian products $Y_x\times
Y'_x$ of fibres of fibre bundles $Y$ and $Y'$.

Let us consider the composition
\mar{1.34}\beq
\pi: Y\to \Si\to X \label{1.34}
\eeq
of fibre \index{$\pi_{\Si X}$} bundles \index{$\pi_{Y\Si}$}
\mar{z275,6}\ben
&& \pi_{Y\Si}: Y\to\Si, \label{z275}\\
&& \pi_{\Si X}: \Si\to X. \label{z276}
\een
One can show that it is a fibre bundle, called the
\index{bundle!composite} {\sl composite bundle}. \index{composite
bundle} It is provided with bundle coordinates
$(x^\la,\si^m,y^i)$, where $(x^\la,\si^m)$ are bundle coordinates
on the fibre bundle (\ref{z276}), i.e., transition functions of
coordinates $\si^m$ are independent of coordinates $y^i$.

\begin{theorem}\label{comp10} \mar{comp10}
Given a composite bundle (\ref{1.34}), let $h$ be a global section
of the fibre bundle $\Si\to X$. Then the restriction \index{$Y^h$}
\mar{S10}\beq
Y^h=h^*Y \label{S10}
\eeq
of the fibre bundle $Y\to\Si$ to $h(X)\subset \Si$ is a subbundle
of the fibre bundle $Y\to X$.
\end{theorem}

\begin{theorem} \label{mos61} \mar{mos61}  Given
a section $h$ of the fibre bundle $\Si\to X$ and a section $s_\Si$
of the fibre bundle $Y\to\Si$, their composition $s=s_\Si\circ h$
is a section of the composite bundle $Y\to X$ (\ref{1.34}).
Conversely, every section $s$ of the fibre bundle $Y\to X$ is a
composition of the section $h=\pi_{Y\Si}\circ s$ of the fibre
bundle $\Si\to X$ and some section $s_\Si$ of the fibre bundle
$Y\to \Si$ over the closed imbedded submanifold $h(X)\subset \Si$.
\end{theorem}

\section{Vector and affine bundles}

A {\sl vector bundle} \index{vector bundle} is a fibre bundle
$Y\to X$ such that:

$\bullet$ its typical fibre $V$ and all the fibres
$Y_x=\pi^{-1}(x)$, $x\in X$, are real finite-dimensional vector
spaces;

$\bullet$ there is a bundle atlas $\Psi$ (\ref{sp5}) of $Y\to X$
whose trivialization morphisms $\psi_\xi$ (\ref{sp21}) and
transition functions $\vr_{\xi\zeta}$ (\ref{sp22}) are linear
isomorphisms.

\noindent Accordingly, a vector bundle is provided with {\sl
linear bundle coordinates} \index{bundle!coordinates!linear}
$(y^i)$ possessing linear transition functions $y'^i=A^i_j(x)y^j$.
We have
\mar{trt}\beq
y=y^ie_i(\pi(y))=y^i \psi_\xi(\pi(y))^{-1}(e_i), \qquad \pi(y)\in
U_\xi, \label{trt}
\eeq
where $\{e_i\}$ is a fixed basis for the typical fibre $V$ of $Y$,
and $\{e_i(x)\}$ are the fibre bases (or the {\sl frames})
\index{frame} for the fibres $Y_x$ of $Y$ associated to the bundle
atlas $\Psi$.

By virtue of Theorem \ref{mos9}, any vector bundle has a global
section, e.g., the canonical global {\sl zero-valued section}
\index{section!zero-valued} $\wh 0(x)=0$. \index{$\wh 0$} Global
sections of a vector bundle $Y\to X$ constitute a projective
$C^\infty(X)$-module $Y(X)$ of \index{$Y(X)$} finite rank. It is
called the {\sl structure module of a vector bundle}.
\index{structure module!of a vector bundle}

\begin{theorem} \label{12t10} \mar{12t10} Let a vector bundle $Y\to
X$ admit $m=\di V$ nowhere vanishing global sections $s_i$ which
are linearly independent, i.e., $\op\w^m s_i\neq 0$. Then $Y$ is
trivial.
\end{theorem}

\begin{remark}
Theorem \ref{sp60} state the categorial equivalence between the
vector bundles over a smooth manifold $X$ and projective
$C^\infty(X)$-modules of finite rank. Therefore, the differential
calculus (including linear differential operators, linear
connections) on vector bundles can be algebraically formulated as
the differential calculus on these modules.
\end{remark}

By a morphism of vector bundles is meant a {\sl linear bundle
morphism}, \index{bundle!morphism!linear} which is a fibrewise map
whose restriction to each fibre is a linear map.

Given a linear bundle morphism $\Phi: Y'\to Y$ of vector bundles
over $X$, its {\sl kernel} \index{kernel!of a vector bundle
morphism} Ker$\,\Phi$ is defined as the inverse image
$\Phi^{-1}(\wh 0(X))$ of the canonical zero-valued section $\wh
0(X)$ of $Y$. By virtue of Theorem \ref{pomm}, if $\Phi$ is of
constant rank, its kernel and its range are vector subbundles of
the vector bundles $Y'$ and $Y$, respectively. For instance,
monomorphisms and epimorphisms of vector bundles fulfil this
condition.

There are the following particular constructions of new vector
bundles from the old ones.

$\bullet$ Let $Y\to X$ be a vector bundle with a typical fibre
$V$. By $Y^*\to X$ is denoted the {\sl dual vector bundle}
\index{dual vector bundle} \index{vector bundle!dual} with the
typical fibre $V^*$ dual of $V$. The {\sl interior product}
\index{interior product!of vector bundles} of $Y$ and $Y^*$ is
defined as a fibred morphism
\be
\rfloor: Y\otimes Y^*\ar_X X\times \Bbb R.
\ee

$\bullet$ Let $Y\to X$ and $Y'\to X$ be vector bundles with
typical fibres $V$ and $V'$, respectively. Their {\sl Whitney sum}
\index{Whitney sum!of vector bundles} $Y\op\oplus_X Y'$ is a
vector bundle over $X$ with the typical fibre $V\oplus V'$.

$\bullet$ Let $Y\to X$ and $Y'\to X$ be vector bundles with
typical fibres $V$ and $V'$, respectively. Their {\sl tensor
product} \index{tensor product!of vector bundles} $Y\op\ot_X Y'$
is a vector bundle over $X$ with the typical fibre $V\ot V'$.
Similarly, the {\sl exterior product} of vector bundles
\index{exterior product!of vector bundles} $Y\op\w_X Y'$  is
defined. The exterior product
\mar{ss12f11}\beq
\w Y=X\times \Bbb R \op\oplus_X Y \op\oplus_X \op\w^2
Y\op\oplus_X\cdots \op\w^k Y, \qquad k=\di Y-\di X,
\label{ss12f11}
\eeq
is \index{$\w Y$} called the {\sl exterior bundle}.
\index{exterior bundle} \index{bundle!exterior}

\begin{remark}\label{mos30} \mar{mos30}
Given vector bundles $Y$ and $Y'$ over the same base $X$, every
linear bundle morphism
\be
\Phi: Y_x\ni \{e_i(x)\}\to \{\Phi^k_i(x)e'_k(x)\}\in Y'_x
\ee
over $X$ defines a global section
\be
\Phi: x\to \Phi^k_i(x)e^i(x)\ot e'_k(x)
\ee
of the tensor product $Y\ot Y'^*$, and {\it vice versa}.
\end{remark}

A sequence
\be
Y'\ar^i Y\ar^j Y''
\ee
of vector bundles over the same base $X$ is called {\sl exact}
\index{exact sequence!of vector bundles} at $Y$ if Ker$\,j=\im i$.
A sequence of vector bundles
\mar{sp10}\beq
0\to Y'\ar^i Y\ar^j Y'' \to 0 \label{sp10}
\eeq
over $X$ is said to be a {\sl short exact sequence} \index{exact
sequence!of vector bundles!short} if it is exact at all terms
$Y'$, $Y$, and $Y''$. This means that $i$ is a bundle
monomorphism, $j$ is a bundle epimorphism, and Ker$\,j=\im i$.
Then $Y''$ is the {\sl factor bundle} $Y/Y'$ \index{factor!bundle}
whose structure module is the quotient $Y(X)/Y'(X)$ of the
structure modules of $Y$ and $Y'$. Given an exact sequence of
vector bundles (\ref{sp10}), there is the exact sequence of their
duals
\be
0\to Y''^*\ar^{j^*} Y^*\ar^{i^*} Y'^* \to 0.
\ee
One says that an exact sequence (\ref{sp10}) is {\sl split}
\index{exact sequence!of vector bundles!split} if there exists a
bundle monomorphism $s:Y''\to Y$ such that $j\circ s=\id Y''$ or,
equivalently,
\be
Y=i(Y')\oplus s(Y'')= Y'\oplus Y''.
\ee

\begin{theorem} \label{sp11} \mar{sp11}
Every exact sequence of vector bundles (\ref{sp10}) is split.
\end{theorem}

The tangent bundle $TZ$ and the cotangent bundle $T^*Z$ of a
manifold $Z$ exemplify vector bundles.

\begin{remark}
Given an atlas $\Psi_Z =\{(U_\iota,\phi_\iota)\}$ of a manifold
$Z$, the tangent bundle is provided with the
\index{holonomic!atlas} {\sl holonomic bundle atlas}
\index{bundle!atlas!holonomic}
\mar{mos150}\beq
\Psi_T =\{(U_\iota, \psi_\iota = T\phi_\iota)\}, \label{mos150}
\eeq
where $T\phi_\iota$ is the tangent morphism to $\f_\iota$. The
associated linear bundle coordinates are holonomic (or {\sl
induced}) \index{induced coordinates} coordinates $(\dot z^\la)$
with respect to the holonomic frames $\{\dr_\la\}$ in tangent
spaces $T_zZ$.
\end{remark}

The tensor product of tangent and cotangent bundles
\mar{sp20}\beq
T=(\op\ot^mTZ)\ot(\op\ot^kT^*Z), \qquad m,k\in \Bbb N,
\label{sp20}
\eeq
is called a {\sl tensor bundle}, \index{tensor bundle} provided
with holonomic bundle coordinates $\dot
x^{\al_1\cdots\al_m}_{\bt_1\cdots\bt_k}$ possessing transition
functions
\be
\dot x'^{\al_1\cdots\al_m}_{\bt_1\cdots\bt_k}=\frac{\dr
x'^{\al_1}}{\dr x^{\m_1}}\cdots\frac{\dr x'^{\al_m}}{\dr
x^{\m_m}}\frac{\dr x^{\nu_1}}{\dr x'^{\bt_1}}\cdots\frac{\dr
x^{\nu_k}}{\dr x'^{\bt_k}} \dot
x^{\m_1\cdots\m_m}_{\nu_1\cdots\nu_k}.
\ee

Let $\pi_Y:TY\to Y$ be the tangent bundle of a fibre bundle $\pi:
Y\to X$. Given bundle coordinates $(x^\la,y^i)$ on $Y$, it is
equipped with the holonomic coordinates $(x^\la,y^i,\dot x^\la,
\dot y^i)$. The tangent bundle $TY\to Y$ has the subbundle $VY =
\Ker (T\pi)$,  \index{$VY$} which consists of the vectors tangent
to fibres of $Y$. It is called the {\sl vertical tangent bundle}
\index{bundle!tangent!vertical} of $Y$ \index{tangent
bundle!vertical} and is provided with the holonomic coordinates
$(x^\la,y^i,\dot y^i)$ with respect to the vertical frames
$\{\dr_i\}$. Every bundle morphism $\Phi: Y\to Y'$ yields the
linear bundle morphism over $\Phi$ of the vertical tangent
\index{$V\Phi$} bundles
\mar{ws538}\beq
V\Phi: VY\to VY', \qquad \dot y'^i\circ V\Phi=\frac{\dr
\Phi^i}{\dr y^j}\dot y^j. \label{ws538}
\eeq
It is called the {\sl vertical tangent morphism}. \index{tangent
morphism!vertical}

In many important cases, the vertical tangent bundle $VY\to Y$ of
a fibre bundle $Y\to X$ is trivial, and is isomorphic to the
bundle product
\mar{48'}\beq
VY= Y\op\times_X\ol Y \label{48'}
\eeq
where $\ol Y\to X$ is some vector bundle. It follows that $VY$ can
be provided with bundle coordinates $(x^\la,y^i,\ol y^i)$ such
that transition functions of coordinates $\ol y^i$ are independent
of coordinates $y^i$. One calls (\ref{48'}) the {\sl vertical
splitting}. \index{vertical splitting} For instance, every vector
bundle $Y\to X$ admits the \index{vertical splitting!of a vector
bundle} {\sl canonical vertical splitting}
\mar{12f10}\beq
VY= Y\op\oplus_X Y. \label{12f10}
\eeq

The {\sl vertical cotangent bundle} $V^*Y\to Y$ \index{$V^*Y$} of
a fibre bundle $Y\to X$ \index{cotangent bundle!vertical} is
defined as the dual of the vertical tangent bundle $VY\to Y$. It
is not a subbundle of the cotangent bundle $T^*Y$, but there is
the canonical surjection
\mar{z11}\beq
\zeta: T^*Y\ni \dot x_\la dx^\la +\dot y_i dy^i \to \dot y_i \ol
dy^i\in V^*Y, \label{z11}
\eeq
where $\{\ol dy^i\}$, \index{$\ol dy^i$} possessing transition
functions
\be
\ol dy'^i=\frac{\dr y'^i}{\dr y^j}\ol dy^j,
\ee
are the duals of the holonomic frames $\{\dr_i\}$ of $VY$.

For any fibre bundle $Y$, there exist the exact sequences of
vector bundles
\mar{1.8a,b}\ben
&& 0\to VY\ar TY\ar^{\pi_T} Y\op\times_X TX\to 0,
\label{1.8a} \\
&& 0\to Y\op\times_X T^*X\to T^*Y\to V^*Y\to 0.
\label{1.8b}
\een
Their splitting, by definition, is a connection on $Y\to X$.

For the sake of simplicity, we agree to denote the pull-backs
\be
Y\op\times_X TX, \qquad Y\op\times_X T^*X
\ee
by $TX$ and $T^*X$, respectively.

Let $\ol\pi:\ol Y\to X$ be a vector bundle with a typical fibre
$\ol V$. An {\sl affine bundle} \index{affine bundle} modelled
\index{bundle!affine} over the vector bundle $\ol Y\to X$ is a
fibre bundle $\pi:Y\to X$ whose typical fibre $V$ is an affine
space modelled over $\ol V$ such that the following conditions
hold.

$\bullet$ All the fibres $Y_x$ of $Y$ are affine spaces modelled
over the corresponding fibres $\ol Y_x$ of the vector bundle $\ol
Y$.

$\bullet$ There is an affine bundle atlas
\be
\Psi=\{(U_\al,\psi_\chi),\vr_{\chi\zeta}\}
\ee
of $Y\to X$ whose local trivializations morphisms $\psi_\chi$
(\ref{sp21}) and transition functions $\vr_{\chi\zeta}$
(\ref{sp22}) are affine isomorphisms.

Dealing with affine bundles, we use only {\sl affine bundle
coordinates} \index{bundle!coordinates!affine} $(y^i)$ associated
to an affine bundle atlas $\Psi$. There are the bundle morphisms
\be
&&Y\op\times_X\ol Y\ar_X Y,\qquad (y^i, \ol y^i)\to  y^i +\ol y^i,\\
&&Y\op\times_X Y\ar_X \ol Y,\qquad (y^i, y'^i)\to  y^i - y'^i,
\ee
where $(\ol y^i)$ are linear coordinates on the vector bundle $\ol
Y$.

By virtue of Theorem \ref{mos9}, affine bundles  have  global
sections, but in contrast with vector bundles, there is no
canonical global section of an affine bundle. Let $\pi:Y\to X$ be
an affine bundle. Every global section $s$ of an affine bundle
$Y\to X$ modelled over a vector bundle $\ol Y\to X$ yields the
bundle morphisms
\mar{mos31,'}\ben
&& Y\ni y\to y-s(\pi(y))\in \ol Y, \label{mos31}\\
&& \ol Y\ni \ol y\to s(\pi(y))+\ol y\in Y. \label{mos31'}
\een
In particular, every vector bundle $Y$ has a natural structure of
an affine bundle due to the morphisms (\ref{mos31'}) where $s=\wh
0$ is the canonical zero-valued section of $Y$. For instance, the
tangent bundle $TX$ of a manifold $X$ is naturally an affine
bundle $ATX$ called the \index{bundle!tangent!affine} {\sl affine
tangent bundle}. \index{tangent bundle!affine}

\begin{theorem} \label{11t60} \mar{11t60}
Any affine bundle $Y\to X$ admits bundle coordinates $(x^\la, \wt
y^i)$ with linear transition functions $\wt y'^i=A^i_j(x)\wt y^j$
(see Example \ref{exe1}).
\end{theorem}

By a morphism of affine bundles is meant a bundle morphism
$\Phi:Y\to Y'$ whose restriction to each fibre of $Y$ is an affine
map. It is called \index{bundle!morphism!affine} an {\sl affine
bundle morphism}. \index{affine bundle!morphism} Every affine
bundle morphism $\Phi:Y\to Y'$ of an affine bundle $Y$ modelled
over a vector bundle $\ol Y$ to an affine bundle $Y'$ modelled
over a vector bundle $\ol Y'$ yields an unique linear bundle
morphism
\mar{1355'}\beq
\ol \Phi: \ol Y\to \ol Y', \qquad \ol y'^i\circ \ol\Phi=
\frac{\dr\Phi^i}{\dr y^j}\ol y^j, \label{1355'}
\eeq
called the {\sl linear derivative} \index{linear derivative of an
affine morphism} of $\Phi$.

Similarly to vector bundles, if $\Phi:Y\to Y'$ is an affine
morphism of affine bundles of constant rank, then $\Phi(Y)$ and
Ker$\,\Phi$ are  affine subbundles of $Y'$ and $Y$, respectively.

Every affine bundle $Y\to X$ modelled over a vector bundle $\ol
Y\to X$ admits the \index{vertical splitting!of an affine bundle}
{\sl canonical vertical splitting}
\mar{48}\beq
VY= Y\op\times_X\ol Y. \label{48}
\eeq

Note that Theorems \ref{sp1} and \ref{sp2} on a particular cover
for bundle atlases remain true in the case of linear and affine
atlases of vector and affine bundles.

\section{Vector fields}

{\sl Vector fields} \index{vector field} on a manifold $Z$ are
global sections of the tangent bundle $TZ\to Z$.

The set $\cT(Z)$ of \index{$\cT(Z)$} vector fields on $Z$ is both
a $C^\infty(Z)$-module and a real Lie algebra with respect to the
\index{Lie bracket} {\sl Lie bracket}
\be
&& u=u^\la\dr_\la, \qquad v=v^\la\dr_\la,\\
&& [v,u] = (v^\la\dr_\la u^\m - u^\la\dr_\la v^\m)\dr_\m.
\ee

Given a vector field $u$ on $X$, a {\sl curve}
\be
c:\Bbb R\supset (,)\to Z
\ee
in $Z$ is said to be an {\sl integral curve} \index{integral
curve} of \index{curve integral} $u$ if $Tc=u(c)$. Every vector
field $u$ on a manifold $Z$ can be seen as an {\sl infinitesimal
generator} \index{infinitesimal generator} of a local
one-parameter group of diffeomorphisms (a {\sl flow}),
\index{flow} and {\it vice versa}. One-dimensional orbits of this
group are integral curves of $u$. A vector field is called {\sl
complete} \index{vector field!complete} if its flow is a
one-parameter group of diffeomorphisms of $Z$. For instance, every
vector field on a compact manifold is complete.

A vector field $u$ on a fibre bundle $Y\to X$ is called {\sl
projectable} \index{vector field!projectable} if it projects onto
a vector field on $X$, i.e., there exists a vector field $\tau$ on
$X$ such that
\be
\tau\circ\pi= T\pi\circ u.
\ee
A projectable vector field takes the coordinate form
\mar{11f30}\beq
u=u^\la(x^\m) \dr_\la + u^i(x^\m,y^j) \dr_i, \qquad
\tau=u^\la\dr_\la. \label{11f30}
\eeq
Its flow is a local one-parameter group of automorphisms of $Y\to
X$ over a local one-parameter group of diffeomorphisms of $X$
whose generator is $\tau$. A projectable vector field is called
{\sl vertical} \index{vector field!vertical} if its projection
onto $X$ vanishes, i.e., if it lives in the vertical tangent
bundle $VY$.

A vector field $\tau=\tau^\la\dr_\la$ on a base $X$ of a fibre
bundle $Y\to X$ gives rise to a vector field on $Y$ by means of a
connection on this fibre bundle (see the formula (\ref{b1.85})).
Nevertheless, every tensor bundle (\ref{sp20}) admits the {\sl
canonical lift} \index{lift of a vector field!canonical} of vector
fields \index{$\wt\tau$}
\mar{l28}\beq
\wt\tau = \tau^\m\dr_\m + [\dr_\nu\tau^{\al_1}\dot
x^{\nu\al_2\cdots\al_m}_{\bt_1\cdots\bt_k} + \ldots
-\dr_{\bt_1}\tau^\nu \dot
x^{\al_1\cdots\al_m}_{\nu\bt_2\cdots\bt_k} -\ldots]\dot \dr
_{\al_1\cdots\al_m}^{\bt_1\cdots\bt_k}, \label{l28}
\eeq
where we employ the compact notation  \index{$\dot\dr_\la$}
\mar{vvv}\beq
\dot\dr_\la = \frac{\dr}{\dr\dot x^\la}. \label{vvv}
\eeq
This lift is functorial, i.e., it is an $\Bbb R$-linear
monomorphism of the Lie algebra $\cT(X)$ of vector fields on $X$
to the Lie algebra $\cT(Y)$ of vector fields on $Y$ (see Section
5.1). In particular, we have the functorial lift
\mar{l27}\beq
\wt\tau = \tau^\m\dr_\m +\dr_\nu\tau^\al\dot
x^\nu\frac{\dr}{\dr\dot x^\al} \label{l27}
\eeq
of vector fields on $X$ onto the tangent bundle $TX$ and their
functorial lift
\mar{l27'}\beq
\wt\tau = \tau^\m\dr_\m -\dr_\bt\tau^\nu\dot
x_\nu\frac{\dr}{\dr\dot x_\bt} \label{l27'}
\eeq
onto the cotangent bundle $T^*X$.

A fibre  bundle admitting functorial lift of vector fields on its
base is called the natural bundle (see Chapter 5).

A subbundle $\bT$ of the tangent bundle $TZ$ of a manifold $Z$ is
called a regular {\sl distribution} \index{distribution} (or,
simply, a distribution). A vector field $u$ on $Z$ is said to be
{\sl subordinate} \index{vector field!subordinate to a
distribution} to a distribution $\bT$ if it lives in $\bT$. A
distribution $\bT$ is called {\sl involutive}
\index{distribution!involutive} if the Lie bracket of
$\bT$-subordinate vector fields also is subordinate to $\bT$.

A subbundle of the cotangent bundle $T^*Z$ of $Z$ is called a {\sl
codistribution} $\bT^*$ \index{codistribution} on a manifold $Z$.
For instance, the {\sl annihilator} \index{annihilator of a
distribution} $\rA\bT$ of a distribution $\bT$ is a codistribution
\index{codistribution} whose fibre over $z\in Z$ consists of
covectors $w\in T^*_z$ such that $v\rfloor w=0$ for all $v\in
\bT_z$.

\begin{theorem} \label{warr} \mar{warr} Let $\bT$ be a distribution and $\rA\bT$ its
annihilator. Let $\w\rA\bT(Z)$ be the ideal of the exterior
algebra $\cO^*(Z)$ which is generated by sections of $\rA\bT\to
Z$. A distribution $\bT$ is involutive iff the ideal $\w\rA\bT(Z)$
is a {\sl differential ideal}, \index{differential ideal} i.e.,
\be
d(\w\rA\bT(Z))\subset\w\rA\bT(Z).
\ee
\end{theorem}

The following local coordinates can be associated to an involutive
distribution.

\begin{theorem}\label{c11.0} \mar{c11.0} Let $\bT$ be an involutive
$r$-dimensional distribution on a manifold $Z$, $\di Z=k$. Every
point $z\in Z$ has an open neighborhood $U$ which is a domain of
an adapted coordinate chart $(z^1,\dots,z^k)$ such that,
restricted to $U$, the distribution $\bT$ and its annihilator
$\rA\bT$ are spanned by the local vector fields $\dr/\dr z^1,
\cdots,\dr/\dr z^r$ and the one-forms $dz^{r+1},\dots, dz^k$,
respectively.
\end{theorem}

A connected submanifold $N$ of a manifold $Z$ is called an {\sl
integral manifold} \index{integral manifold} of a distribution
$\bT$ on $Z$ if $TN\subset \bT$. Unless otherwise stated, by an
integral manifold is meant an integral manifold of dimension of
$\bT$. An integral manifold is called {\sl maximal}
\index{integral manifold!maximal} if no other integral manifold
contains it. The following is the classical theorem of Frobenius.

\begin{theorem}\label{to.1}  \mar{to.1} Let $\bT$ be an
involutive distribution on a manifold $Z$. For any $z\in Z$, there
exists a unique maximal integral manifold of $\bT$ through $z$,
and any integral manifold through $z$ is its open subset.
\end{theorem}

Maximal integral manifolds of an involutive distribution on a
manifold $Z$ are assembled into a regular foliation $\cF$ of $Z$.
A regular $r$-dimensional {\sl foliation} (or, simply, a
foliation) \index{foliation} $\cF$ of a $k$-dimensional manifold
$Z$ is defined as a partition of $Z$ into connected
$r$-dimensional submanifolds (the {\sl leaves} of a foliation)
\index{leaf} $F_\iota$, $\iota\in I$, which possesses the
following properties. A foliated manifold $(Z,\cF)$ admits an
adapted coordinate atlas
\mar{spr850}\beq
\{(U_\xi;z^\la; z^i)\},\quad \la=1,\ldots,n-r, \qquad
i=1,\ldots,r, \label{spr850}
\eeq
such that transition functions of coordinates $z^\la$ are
independent of the remaining coordinates $z^i$ and, for each leaf
$F$ of a foliation $\cF$, the connected components of $F\cap
U_\xi$ are given by the equations $z^\la=$const. These connected
components and coordinates $(z^i)$ on them make up a coordinate
atlas of a leaf $F$.

It should be emphasized that leaves of a foliation need not be
closed or imbedded submanifolds. Every leaf has an open {\sl
tubular neighborhood} \index{tubular neighborhood} $U$, i.e., if
$z\in U$, then a leaf through $z$ also belongs to $U$.

A pair $(Z,\cF)$ where $\cF$ is a foliation of $Z$ is called a
\index{foliated manifold} {\sl foliated manifold}.  For instance,
any submersion $f:Z\to M$ yields a foliation
\be
\cF=\{F_p=f^{-1}(p)\}_{p\in f(Z)}
\ee
of $Z$ indexed by elements of $f(Z)$, which is an open submanifold
of $M$, i.e., $Z\to f(Z)$ is a fibred manifold. Leaves of this
foliation are closed imbedded submanifolds. Such a foliation is
called {\sl simple}. \index{foliation!simple} It is a fibred
manifold over $f(Z)$. Any (regular) foliation is locally simple.

\section{Exterior and tangent-valued forms}

An {\sl exterior $r$-form} \index{exterior form} on a manifold $Z$
is a section
\be
\f =\frac{1}{r!}\f_{\la_1\dots\la_r} dz^{\la_1}\w\cdots\w
dz^{\la_r}
\ee
of the exterior product $\op\w^r T^*Z\to Z$, where
\be
&& dz^{\la_1}\w\cdots\w dz^{\la_r}=
\frac{1}{r!}\e^{\la_1\ldots\la_r}{}_{\m_1\ldots\m_r}dx^{\m_1}\ot\cdots\ot
dx^{\m_r},\\
&& \e^{\ldots \la_i\ldots\la_j\ldots}{}_{\ldots
\m_p\ldots\m_k\ldots}= -\e^{\ldots
\la_j\ldots\la_i\ldots}{}_{\ldots \m_p\ldots\m_k\ldots} = -
\e^{\ldots \la_i\ldots\la_j\ldots}{}_{\ldots
\m_k\ldots\m_p\ldots}, \\
&& \e^{\la_1\ldots\la_r}{}_{\la_1\ldots\la_r}=1.
\ee

Let $\cO^r(Z)$ denote \index{$\cO^*(Z)$} the vector space of
exterior $r$-forms on a manifold $Z$. By definition,
$\cO^0(Z)=C^\infty(Z)$ is the ring of smooth real functions on
$Z$. All exterior forms on $Z$ constitute the ${\Bbb N}$-graded
commutative algebra $\cO^*(Z)$ of global sections of the exterior
bundle $\w T^*Z$ (\ref{ss12f11}) endowed with the {\sl exterior
product} \index{exterior product}
\be
&&\f=\frac{1}{r!}\f_{\la_1\dots\la_r} dz^{\la_1}\w\cdots\w
dz^{\la_r}, \qquad \si= \frac{1}{s!}\si_{\m_1\dots\m_s}
dz^{\m_1}\w\cdots\w dz^{\m_s},\\
&& \f\w\si=\frac{1}{r!s!}\f_{\nu_1\ldots\nu_r}\si_{\nu_{r+1}\ldots\nu_{r+s}}
dz^{\nu_1}\w\cdots\w
dz^{\nu_{r+s}}=\\
&& \qquad
\frac{1}{r!s!(r+s)!}\e^{\nu_1\ldots\nu_{r+s}}{}_{\al_1\ldots\al_{r+s}}
\f_{\nu_1\ldots\nu_r}\si_{\nu_{r+1}\ldots\nu_{r+s}}dz^{\al_1}\w\cdots\w
dz^{\al_{r+s}},
\ee
such  that
\be
\f\w\si=(-1)^{|\f||\si|}\si\w\f,
\ee
where the symbol $|\f|$ stands for the form degree. The algebra
$\cO^*(Z)$  also is provided with \index{exterior differential}
the {\sl exterior differential} \index{differential!exterior}
\be
d\f= dz^\m\w \dr_\m\f=\frac{1}{r!} \dr_\m\f_{\la_1\ldots\la_r}
dz^\m\w dz^{\la_1}\w\cdots \w dz^{\la_r}
\ee
which obeys the relations
\be
d\circ d=0, \qquad d(\f\w\si)= d(\f)\w \si +(-1)^{|\f|}\f\w
d(\si).
\ee
The exterior differential $d$ makes $\cO^*(Z)$ into a differential
graded algebra (henceforth DGA) which is the minimal
Chevalley--Eilenberg differential calculus $\cO^*\cA$ over the
real ring $\cA=C^\infty(Z)$. Its de Rham complex is (\ref{t37}).

Given a manifold morphism $f:Z\to Z'$, any exterior $k$-form $\f$
on $Z'$ yields the {\sl pull-back exterior form}
\index{pull-back!form} $f^*\f$ on $Z$ \index{$f^*\f$} given by the
condition
\be
f^*\f(v^1,\ldots,v^k)(z) = \f(Tf(v^1),\ldots,Tf(v^k))(f(z))
\ee
for an arbitrary collection of tangent vectors $v^1,\cdots, v^k\in
T_zZ$. We have the relations
\be
f^*(\f\w\si) =f^*\f\w f^*\si, \qquad df^*\f =f^*(d\f).
\ee

In particular, given a fibre bundle $\pi:Y\to X$, the pull-back
onto $Y$ of exterior forms on $X$ by $\pi$ provides the
monomorphism of graded commutative algebras $\cO^*(X)\to
\cO^*(Y)$. Elements of its range $\pi^*\cO^*(X)$ are called
\index{exterior form!basic} {\sl basic forms}. \index{basic form}
Exterior forms
\be
\phi : Y\to\op\w^r T^*X, \qquad \phi
=\frac{1}{r!}\phi_{\la_1\ldots\la_r}dx^{\la_1}\w\cdots\w
dx^{\la_r},
\ee
on $Y$ such that $u\rfloor\f=0$ for an arbitrary vertical vector
field $u$ on $Y$ are said to \index{exterior form!horizontal} be
{\sl horizontal forms}. \index{horizontal!form} Horizontal forms
of degree $n=\dim X$ are called {\sl densities}. \index{density}
We use for them the compact \index{$\om$} notation
\mar{gm141}\ben
&& L=\frac1{n!}L_{\m_1\ldots\m_n}dx^{\m_1}\w\cdots \w dx^{\m_n} =
\cL\om, \qquad \cL=L_{1\ldots n},\nonumber\\
&&\om=dx^1\w\cdots\w dx^n=
\frac1{n!}\e_{\m_1\ldots\m_n}dx^{\m_1}\w\cdots \w dx^{\m_n},
\label{gm141} \\
&& \om_\la=\dr_\la\rfloor\om, \qquad
\om_{\m\la}=\dr_\m\rfloor\dr_\la\rfloor\om, \nonumber
\een
where \index{$\om_\la$} $\e$ is the skew-symmetric {\sl
Levi--Civita symbol} \index{Levi--Civita symbol} with the
component $\e_{\m_1\ldots\m_n}=1$.

The {\sl interior product} \index{interior product} (or {\sl
contraction}) \index{contraction} of a vector field $u$ and an
exterior $r$-form $\f$ on a manifold $Z$ is given by the
coordinate expression \index{$u\rfloor\f$}
\be
&& u\rfloor\f = \op\sum_{k=1}^r \frac{(-1)^{k-1}}{r!} u^{\la_k}
\f_{\la_1\ldots\la_k\ldots\la_r} dz^{\la_1}\w\cdots\w\wh
{dz}^{\la_k}\w\cdots \w dz^{\la_r}= \\
&& \qquad \frac{1}{(r-1)!}u^\m\f_{\m\al_2\ldots\al_r} dz^{\al_2}\w\cdots\w
dz^{\al_r},
\ee
where the caret $\,\wh{}\,$ denotes omission. It obeys the
relations
\be
&& \f(u_1,\ldots,u_r)=u_r\rfloor\cdots u_1\rfloor\f,\nonumber\\
&& u\rfloor(\f\w\si)= u\rfloor\f\w\si +(-1)^{|\f|}\f\w
u\rfloor\si.
\ee

The {\sl Lie derivative} \index{Lie derivative!of an exterior
form} of an exterior form $\f$ along a vector \index{$\bL_u$}
field $u$ is
\be
&& \bL_u\f = u\rfloor d\f +d(u\rfloor\f), \\
&& \bL_u(\f\w\si)= \bL_u\f\w\si +\f\w\bL_u\si.
\ee
It is a derivation of the graded algebra $\cO^*(Z)$ such that
\be
\bL_u\circ \bL_{u'}-\bL_{u'}\circ\bL_u=\bL_{[u,u']}.
\ee
In particular, if $f$ is a function, then
\be
\bL_u f =u(f)=u\rfloor d f.
\ee

An exterior form $\f$ is invariant under a local one-parameter
group of diffeomorphisms $G(t)$ of $Z$ (i.e., $G(t)^*\f=\f$) iff
its Lie derivative along the infinitesimal generator $u$ of this
group vanishes, i.e., $\bL_u\f=0$.

A {\sl tangent-valued $r$-form} \index{tangent-valued form} on a
manifold $Z$ is a section
\mar{spr611}\beq
\phi = \frac{1}{r!}\phi_{\la_1\ldots\la_r}^\m dz^{\la_1}\w\cdots\w
dz^{\la_r}\ot\dr_\m \label{spr611}
\eeq
of the tensor bundle
\be
\op\w^r T^*Z\ot TZ\to Z.
\ee

\begin{remark}
There is one-to-one correspondence between the tangent-valued
one-forms $\f$ on a manifold $Z$ and the linear bundle
endomorphisms
\mar{29b,b'}\ben
&& \wh\f:TZ\to TZ,\quad
\wh\f: T_zZ\ni v\to v\rfloor\f(z)\in T_zZ, \label{29b} \\
&&\wh\f^*:T^*Z\to T^*Z,\quad \wh\f^*: T_z^*Z\ni v^*\to
\f(z)\rfloor v^*\in T_z^*Z, \label{29b'}
\een
over $Z$ (see Remark \ref{mos30}). For instance, the
\index{tangent-valued form!canonical} {\sl canonical
tangent-valued one-form}
\mar{b1.51}\beq
\thh_Z= dz^\la\ot \dr_\la \label{b1.51}
\eeq
on \index{$\thh_Z$} $Z$ corresponds to the identity morphisms
(\ref{29b}) and (\ref{29b'}).
\end{remark}

\begin{remark}
Let $Z=TX$, and let $TTX$ be the tangent bundle of $TX$. There is
the bundle endomorphism
\mar{z117}\beq
J(\dr_\la)= \dot\dr_\la, \qquad J(\dot\dr_\la)=0 \label{z117}
\eeq
of $TTX$ over $X$. It  corresponds to the canonical tangent-valued
form
\mar{z117'}\beq
\thh_J=dx^\la\ot\dot\dr_\la \label{z117'}
\eeq
on the tangent bundle $TX$. It is readily observed that $J\circ
J=0$.
\end{remark}

The space $\cO^*(Z)\ot \cT(Z)$ of tangent-valued forms is provided
with the \index{Fr\"olicher--Nijenhuis bracket} {\sl
Fr\"olicher--Nijenhuis bracket}
\mar{1149}\ben
&& [\; ,\; ]_{\rm FN}:\cO^r(Z)\ot \cT(Z)\times \cO^s(Z)\ot \cT(Z)
\to\cO^{r+s}(Z)\ot \cT(Z), \nonumber \\
&& [\al\ot u,\, \bt\ot v]_{\rm FN} = (\al\w\bt)\ot [u, v] +
(\al\w \bL_u \bt)\ot v - \label{1149} \\
&& \qquad(\bL_v \al\w\bt)\ot u +  (-1)^r (d\al\w u\rfloor\bt)\ot v
+ (-1)^r(v\rfloor\al\w d\bt)\ot u, \nonumber \\
&&  \al\in\cO^r(Z), \qquad \bt\in \cO^s(Z), \qquad u,v\in\cT(Z). \nonumber
\een
Its coordinate expression is
\be
&& [\phi,\si]_{\rm FN} = \frac{1}{r!s!}(\phi_{\la_1
\dots\la_r}^\nu\dr_\n\si_{\la_{r+1}\dots\la_{r+s}}^\m -
\si_{\la_{r+1}
\dots\la_{r+s}}^\nu\dr_\nu\phi_{\la_1\dots\la_r}^\m -\\
&& \qquad r\phi_{\la_1\ldots\la_{r-1}\nu}^\m\dr_{\la_r}\si_{\la_{r+1}
\dots\la_{r+s}}^\nu + s \si_{\nu\la_{r+2}\ldots\la_{r+s}}^\m
\dr_{\la_{r+1}}\phi_{\la_1\ldots\la_r}^\nu)\\
&& \qquad dz^{\la_1}\wedge\cdots \wedge dz^{\la_{r+s}}\otimes\dr_\m,\\
&&
\f\in \cO^r(Z)\ot \cT(Z), \qquad \si\in \cO^s(Z)\ot \cT(Z).
\ee
There are the relations
\mar{1150,'}\ben
&& [\f,\si]_{\rm FN}=(-1)^{|\f||\s|+1}[\si,\f]_{\rm FN}, \label{1150} \\
&& [\f, [\si, \thh]_{\rm FN}]_{\rm FN} = [[\f, \si]_{\rm FN}, \thh]_{\rm
FN} +(-1)^{|\f||\si|}  [\si, [\f,\thh]_{\rm FN}]_{\rm FN}, \label{1150'}\\
&& \f,\si,\thh\in \cO^*(Z)\ot \cT(Z). \nonumber
\een

Given a tangent-valued form  $\thh$, the {\sl Nijenhuis
differential} on $\cO^*(Z)\ot\cT(Z)$ \index{Nijenhuis
differential} is defined as the morphism
\be
d_\thh : \psi\to d_\thh\psi = [\thh,\psi]_{\rm FN}, \qquad
\psi\in\cO^*(Z)\ot\cT(Z).
\ee
By virtue of (\ref{1150'}), it has the property
\be
d_\f[\psi,\thh]_{\rm FN} = [d_\phi\psi,\thh]_{\rm FN}+
(-1)^{|\f||\psi|} [\psi,d_\f\thh]_{\rm FN}.
\ee
In particular, if $\thh=u$ is a vector field, the Nijenhuis
differential is \index{Lie derivative!of a tangent-valued form}
the {\sl Lie derivative of tangent-valued forms}
\be
&& \bL_u\si= d_u\si=[u,\si]_{\rm FN} =\frac{1}{s!}(u^\n\dr_\n\si_{\la_1\ldots\la_s}^\m -
\si_{\la_1\ldots\la_s}^\n\dr_\n u^\m +\\
&& \qquad s\si^\m_{\nu\la_2\ldots\la_s}\dr_{\la_1}u^\nu)dx^{\la_1}
\w\cdots\w dx^{\la_s}\ot\dr_\m, \qquad \si\in\cO^s(Z)\ot\cT(Z).
\ee

Let $\Y$ be a fibre bundle. We consider the following subspaces of
the space $\cO^*(Y)\ot \cT(Y)$ of tangent-valued forms on $Y$:

$\bullet$ {\sl horizontal tangent-valued forms}
\index{tangent-valued form!horizontal}
\be
&& \phi : Y\to\op\w^r T^*X\op\otimes_Y TY,\\
&& \phi =dx^{\la_1}\wedge\cdots\wedge dx^{\la_r}\otimes
\frac{1}{r!}[\phi_{\la_1\ldots\la_r}^\m(y) \dr_\m
+\phi_{\la_1\ldots\la_r}^i(y) \dr_i],
\ee

$\bullet$ {\sl projectable horizontal tangent-valued forms}
\index{tangent-valued form!horizontal!projectable}
\be
\phi =dx^{\la_1}\wedge\cdots\wedge dx^{\la_r}\otimes
\frac{1}{r!}[\phi_{\la_1\ldots\la_r}^\m(x)\dr_\m
+\phi_{\la_1\ldots\la_r}^i(y) \dr_i],
\ee

$\bullet$ {\sl vertical-valued form} \index{vertical-valued form}
\be
&& \phi : Y\to\op\w^r T^*X\op\otimes_Y VY,\\
&& \phi =\frac{1}{r!}\phi_{\la_1\ldots\la_r}^i(y)dx^{\la_1}\wedge\cdots
\wedge dx^{\la_r}\otimes\dr_i,
\ee

$\bullet$ vertical-valued one-forms, called {\sl soldering forms},
\index{soldering form}
\be
\si = \si_\la^i(y) dx^\la\otimes\dr_i,
\ee

$\bullet$ {\sl basic soldering forms} \index{soldering form!basic}
\be
\si = \si_\la^i(x) dx^\la\otimes\dr_i.
\ee

\begin{remark} \label{mos161} \mar{mos161}
The tangent bundle $TX$ is provided with the canonical soldering
form $\thh_J$ (\ref{z117'}). Due to the canonical vertical
splitting
\mar{mos163}\beq
VTX=TX\op\times_X TX, \label{mos163}
\eeq
the canonical soldering form (\ref{z117'}) on $TX$ defines the
canonical tangent-valued form $\thh_X$ (\ref{b1.51}) on $X$. By
this reason, tangent-valued one-forms on a manifold $X$ also are
called soldering forms.
\end{remark}

\begin{remark} \label{mos80} \mar{mos80}
Let $Y\to X$ be a fibre bundle, $f:X'\to X$ a manifold morphism,
$f^*Y\to X'$ the pull-back of $Y$ by $f$, and
\be
f_Y:f^*Y\to Y
\ee
the corresponding bundle morphism (\ref{mos81}). Since
\be
Vf^*Y=f^*VY=f^*_YVY, \qquad V_{y'}Y'=V_{f_Y(y')}Y,
\ee
one can define the {\sl pull-back}
\index{pull-back!vertical-valued form} $f^*\f$ onto $f^*Y$ of any
vertical-valued form $\f$ on $Y$ in accordance with the relation
\be
f^*\f(v^1,\ldots,v^r)(y')=\f(Tf_Y(v^1),\ldots,Tf_Y(v^r))(f_Y(y')).
\ee
\end{remark}

We also mention the  $TX$-valued forms
\mar{1.11}\ben
&&\f:Y\to \op\w^r T^*X\op\ot_Y TX, \label{1.11}\\
&&\phi =\frac{1}{r!}\phi_{\la_1\ldots\la_r}^\m  dx^{\la_1}\w\cdots\w dx^{\la_r}
\otimes \dr_\m,\nonumber
\een
and  $V^*Y$-valued forms
\mar{87}\ben
&& \f :Y\to \op\w^r T^*X\op\ot_Y V^*Y, \label{87}\\
&& \phi =\frac{1}{r!}\phi_{\la_1\ldots\la_ri}dx^{\la_1}\w\cdots\w dx^{\la_r}\ot
\ol dy^i. \nonumber
\een
It should be emphasized that (\ref{1.11}) are not tangent-valued
forms, while (\ref{87}) are not exterior forms. They exemplify
{\sl vector-valued forms}. \index{vector-valued form} Given a
vector bundle $E\to X$, by a $E$-valued $k$-form on $X$, is meant
a section of the fibre bundle
\be
(\op\w^k T^*X)\op\ot_X E^*\to X.
\ee

\chapter{Jet manifolds}

There are different notions of jets. Here we are concerned with
jets of sections of fibre bundles. They are the particular jets of
maps and the jets of submanifolds. Let us also mention the jets of
modules over a commutative ring. In particular, given a smooth
manifold $X$, the jets of a projective $C^\infty(X)$-module $P$ of
finite rank are exactly the jets of sections of the vector bundle
over $X$ whose module of sections is $P$ in accordance with the
Serre--Swan Theorem \ref{sp60}. The notion of jets is extended to
modules over graded commutative rings. However, the jets of
modules over a noncommutative ring can not be defined.

\section{First order jet manifolds}

Given a fibre bundle $Y\to X$ with bundle coordinates
$(x^\la,y^i)$, let us consider the equivalence classes $j^1_xs$ of
its sections $s$, which are identified by their values $s^i(x)$
and the values of their partial derivatives $\dr_\mu s^i(x)$ at a
point $x\in X$. They are called the {\sl first order jets}
\index{jet!first order} of sections at $x$. One can justify that
the definition of jets is coordinate-independent.  The key point
is that the set $J^1Y$ \index{$J^1Y$} of first order jets
$j^1_xs$, $x\in X$, is a smooth manifold with respect to the
adapted coordinates $(x^\la,y^i,y_\la^i)$ such that
\mar{50}\beq
y_\la^i(j^1_xs)=\dr_\la s^i(x),\qquad {y'}^i_\la = \frac{\dr
x^\m}{\dr{x'}^\la}(\dr_\m +y^j_\m\dr_j)y'^i.\label{50}
\eeq
It is called the first order {\sl jet manifold} \index{jet
manifold} of a fibre bundle $Y\to X$. We call $(y_\la^i)$ the {\sl
jet coordinates}. \index{jet coordinates}

The jet manifold $J^1Y$ admits the natural \index{$\pi^1$}
fibrations \index{$\pi^1_0$}
\mar{1.14,5}\ben
&&\pi^1:J^1Y\ni j^1_xs\to x\in X, \label{1.14}\\
&&\pi^1_0:J^1Y\ni j^1_xs\to s(x)\in Y. \label{1.15}
\een
A glance at the transformation law (\ref{50}) shows that $\pi^1_0$
is an affine bundle modelled over the vector bundle
\mar{cc9}\beq
T^*X \op\otimes_Y VY\to Y.\label{cc9}
\eeq
It is convenient to call $\pi^1$ (\ref{1.14}) the {\sl jet
bundle}, \index{jet bundle} while $\pi^1_0$ (\ref{1.15}) is said
to be the {\sl affine jet bundle}. \index{jet bundle!affine}

Let us note that, if $Y\to X$ is a vector or an affine bundle, the
jet bundle $\pi_1$ (\ref{1.14}) is so.

Jets can be expressed in terms of familiar tangent-valued forms as
follows. There are the canonical imbeddings
\mar{18,24}\ben
&&\la_{(1)}:J^1Y\op\to_Y
T^*X \op\otimes_Y TY,\nonumber\\
&& \la_{(1)}=dx^\la
\otimes (\dr_\la + y^i_\la \dr_i)=dx^\la\otimes d_\la, \label{18}\\
&&\thh_{(1)}:J^1Y \op\to_Y T^*Y\op\otimes_Y VY,\nonumber\\
&&\thh_{(1)}=(dy^i- y^i_\la dx^\la)\otimes \dr_i=\thh^i \otimes
\dr_i,\label{24}
\een
where $d_\la$ \index{$d_\la$} are called {\sl total derivatives},
\index{total derivative}and  $\thh^i$ \index{$\thh^i$} are {\sl
local contact forms}. \index{contact form!local}

\begin{remark}
We further identify the jet manifold $J^1Y$ with its images under
the canonical morphisms (\ref{18}) and (\ref{24}), and represent
the jets $j^1_xs=(x^\la,y^i,y^i_\m)$ by the tangent-valued forms
$\la_{(1)}$ (\ref{18}) and $\thh_{(1)}$ (\ref{24}).
\end{remark}

Any section $s$ of $Y\to X$ has the {\sl jet prolongation}
\index{jet prolongation!of a section} to the \index{$J^1s$}
section
\be
(J^1s)(x)= j_x^1s, \qquad y_\la^i\circ J^1s= \dr_\la s^i(x),
\ee
of the jet bundle $J^1Y\to X$. A {\sl section of the jet bundle}
\index{section!of a jet bundle} $J^1Y\to X$ is called  {\sl
integrable} \index{section!of a jet bundle!integrable} if it is
the jet prolongation of some section of $Y\to X$.

\begin{remark} \label{jj1} \mar{jj1}
By virtue of Theorem \ref{mos9}, the affine jet bundle $J^1Y\to Y$
admits global sections. If $Y$ is trivial, there is the canonical
zero section $\wh 0(Y)$ of $J^1Y\to Y$ taking its values into
centers of its affine fibres.
\end{remark}

Any bundle morphism $\Phi:Y\to Y'$ over a diffeomorphism $f$
admits a  {\sl jet prolongation} \index{jet prolongation!of a
morphism} to a bundle morphism of affine jet \index{$J^1\Phi$}
bundles
\mar{1.21a}\beq
J^1\Phi : J^1Y \ar_\Phi J^1Y', \qquad {y'}^i_\la\circ
J^1\Phi=\frac{\dr(f^{-1})^\m}{\dr x'^\la}d_\m\Phi^i. \label{1.21a}
\eeq

Any projectable vector field $u$ (\ref{11f30}) on a fibre bundle
$Y\to X$ has a {\sl jet prolongation} \index{jet prolongation!of a
vector field} to the projectable vector field \index{$J^1u$}
\mar{1.21}\ben
&&J^1u =r_1\circ J^1u: J^1Y\to J^1TY\to TJ^1Y,\nonumber \\
&& J^1u =u^\la\dr_\la + u^i\dr_i + (d_\la u^i
- y_\m^i\dr_\la u^\m)\dr_i^\la, \label{1.21}
\een
on the jet manifold $J^1Y$. To obtain (\ref{1.21}), the canonical
bundle morphism
\be
  r_1: J^1TY\to TJ^1Y,\qquad
\dot y^i_\la\circ r_1 = (\dot y^i)_\la-y^i_\m\dot x^\m_\la
\ee
is used. In particular, there is the canonical isomorphism
\mar{d020}\beq
VJ^1Y=J^1VY, \qquad \dot y^i_\la=(\dot y^i)_\la.\label{d020}
\eeq

\section{Higher order jet manifolds}

The notion of first jets of sections of a fibre bundle is
naturally extended to higher order jets.

Let $Y\to X$ be a fibre bundle. Given its bundle coordinates
$(x^\la,y^i)$, a {\sl multi-index} \index{multi-index} $\La$ of
the length $|\La|=k$ throughout denotes a collection of indices
$(\la_1...\la_k)$ modulo permutations. By $\La+\Si$ is meant a
multi-index $(\la_1\ldots\la_k\si_1\ldots\si_r)$. For instance
$\la+\La=(\la\la_1...\la_r)$. By $\La\Si$ is denoted the union of
collections $(\la_1\ldots\la_k;\si_1\ldots\si_r)$ where the
indices $\la_i$ and $\si_j$ are not permitted. Summation over a
multi-index $\La$ means separate summation over each its index
$\la_i$. We use the compact notation
\be
\dr_\La=\dr_{\la_k}\circ\cdots\circ\dr_{\la_1}, \qquad
\La=(\la_1...\la_k).
\ee

The {\sl $r$-order jet manifold}  $J^rY$ \index{$J^rY$}  of
sections of a bundle $Y\to X$ \index{jet manifold!higher order} is
defined as the disjoint union of equivalence classes $j^r_xs$ of
sections $s$ of $Y\to X$ such that sections $s$ and $s'$ belong to
the same equivalence class $j^r_xs$ iff
\be
s^i(x)={s'}^i(x), \qquad\dr_\La s^i(x)=\dr_\La {s'}^i(x), \qquad
0< |\La| \leq r.
\ee
In brief, one can say that sections of $Y\to X$ are identified by
the $r+1$ terms of their Taylor series at points of $X$. The
particular choice of coordinates does not matter for this
definition. The equivalence classes $j^r_xs$ are called the {\sl
$r$-order jets} of sections. \index{jet!higher order} Their set
$J^rY$ is endowed with an atlas of the adapted coordinates
\mar{55.1,21}\ben
&& (x^\la, y^i_\La),\qquad   y^i_\La\circ s= \dr_\La s^i(x), \qquad
0\leq|\La| \leq r, \label{55.1}\\
&& {y'}^i_{\la+\La}=\frac{\dr x^\m}{\dr'x^\la}d_\m y'^i_\La,
\label{55.21}
\een
where the symbol $d_\la$ stands for \index{total derivative!higher
order} the {\sl higher order total derivative}
\mar{5.32}\beq
d_\la = \dr_\la + \op\sum_{0\leq|\La|\leq r-1}
y^i_{\La+\la}\dr_i^\La, \qquad d'_\la =\frac{\dr x^\m}{\dr
{x'}^\la}d_\m. \label{5.32}
\eeq
These derivatives act on exterior forms on $J^rY$ and obey the
relations
\be
&& [d_\la,d_\m]=0, \qquad d_\la\circ d=d\circ d_\la,\\
&& d_\la(\f\w\si)=d_\la(\f)\w\si +\f\w d_\la(\si),\qquad
d_\la(d\f)=d(d_\la(\f)),\\
&& d_\la(dx^\m)=0, \qquad d_\la(dy^i_\La)= dy^i_{\la+\La}.
\ee
We use the compact notation
\be
 d_\La=d_{\la_r}\circ\cdots\circ d_{\la_1}, \qquad
\La=(\la_r...\la_1).
\ee

The coordinates (\ref{55.1}) bring the set $J^rY$  into a smooth
finite-dimensional manifold. The coordinates (\ref{55.1}) are
compatible with the natural surjections \index{$\pi_k^r$}
\be
\pi_k^r: J^rY\to J^kY, \quad r>k,
\ee
which form the composite bundle
\be
\pi^r: J^rY\op\ar^{\pi^r_{r-1}} J^{r-1}Y\op\ar^{\pi^{r-1}_{r-2}}
\cdots \op\ar^{\pi^1_0} Y\op\ar^\pi X
\ee
with the properties
\be
\pi^k_s\circ\pi^r_k=\pi^r_s, \qquad \pi^s\circ\pi^r_s=\pi^r.
\ee
A glance at the transition functions (\ref{55.21}) shows that the
fibration
\be
\pi_{r-1}^r: J^rY\to J^{r-1}Y
\ee
is an affine bundle modelled over the vector bundle
\mar{5.117}\beq
\op\vee^rT^*X\op\ot_{J^{r-1}Y}VY \to J^{r-1}Y. \label{5.117}
\eeq

\begin{remark} \label{15r3} \mar{15r3}
Let us recall that a base of any affine bundle is a strong
deformation retract of its total space. Consequently, $Y$ is a
strong deformation retract of $J^1Y$, which in turn is a strong
deformation retract of $J^2Y$, and so on. It follows that a fibre
bundle $Y$ is a strong deformation retract of any finite order jet
manifold $J^rY$. Therefore, by virtue of the Vietoris--Begle
theorem, there is an isomorphism
\mar{j19}\beq
H^*(J^rY;\Bbb R)=H^*(Y;\Bbb R) \label{j19}
\eeq
of cohomology of $J^rY$ and $Y$ with coefficients in the constant
sheaf $\Bbb R$.
\end{remark}

\begin{remark}
To introduce higher order jet manifolds, one can use the
construction of repeated jet manifolds. Let us consider the
$r$-order jet manifold $J^rJ^kY$ of a jet bundle $J^kY\to X$. It
is coordinated by $(x^\m, y^i_{\Si\La})$, $|\La| \leq k$, $|\Si|
\leq r$. There is a canonical monomorphism
\be
\si_{rk}: J^{r+k}Y\to J^rJ^kY, \qquad y^i_{\Si\La}\circ \si_{rk}=
y^i_{\Si+\La}.
\ee
\end{remark}

In the calculus in higher order jets, we have the $r$-order {\sl
jet prolongation functor} \index{jet prolongation!functor} such
that, given fibre bundles $Y$ and $Y'$ over $X$, every bundle
morphism $\Phi:Y\to Y'$ over a diffeomorphism $f$ of $X$ admits
the $r$-order {\sl jet prolongation} \index{jet prolongation!of a
morphism!higher order}to a morphism \index{$J^r\Phi$} of $r$-order
jet manifolds
\mar{5.152}\beq
J^r\Phi: J^rY\ni j^r_xs\to j^r_{f(x)}(\Phi\circ s\circ f^{-1}) \in
J^rY'. \label{5.152}
\eeq
The jet prolongation functor is exact. If $\Phi$ is an injection
or a surjection, so is  $J^r\Phi$. It also preserves an algebraic
structure. In particular, if $Y\to X$ is a vector bundle, $J^rY\to
X$ is well. If $Y\to X$ is an affine bundle modelled over the
vector bundle $\ol Y\to X$, then $J^rY\to X$ is an affine bundle
modelled over the vector bundle $J^r\ol Y\to X$.

Every section $s$ of a fibre bundle $Y\to X$ admits the $r$-order
{\sl jet prolongation} \index{jet prolongation!of a section!higher
order} to the  \index{section!of a jet bundle!integrable!higher
order} {\sl integrable section} \index{$J^rs$} $(J^rs)(x)= j^r_xs$
 of the jet bundle $J^rY\to X$.

Let $\cO_k^*=\cO^*(J^kY)$ \index{$\cO_k^*$} be the DGA of exterior
forms on a jet manifold $J^kY$. Every exterior form $\f$ on a jet
manifold $J^kY$ gives rise to the pull-back form
$\pi^{k+i}_k{}^*\f$ on a jet manifold $J^{k+i}Y$. We  have the
direct sequence of $C^\infty(X)$-algebras
\be
\cO^*(X)\op\longrightarrow^{\pi^*} \cO^*(Y)
\op\longrightarrow^{\pi^1_0{}^*} \cO_1^*
\op\longrightarrow^{\pi^2_1{}^*} \cdots
\op\longrightarrow^{\pi^r_{r-1}{}^*}
 \cO_r^*.
\ee

\begin{remark} \label{15r4} \mar{15r4}
By virtue of de Rham Theorem \ref{t60}, the cohomology of the de
Rham complex of $\cO_k^*$ equals the cohomology $H^*(J^kY;\Bbb R)$
of $J^kY$ with coefficients in the constant sheaf $\Bbb R$. The
latter in turn coincides with the sheaf cohomology $H^*(Y;\Bbb R)$
of $Y$ (see Remark \ref{15r3}) and, thus, it equals the de Rham
cohomology $H^*_{\rm DR}(Y)$ of $Y$.
\end{remark}

Given a $k$-order jet manifold $J^kY$ of $Y\to X$, there exists
the canonical bundle morphism
\be
r_{(k)}: J^kTY\to TJ^kY
\ee
over a surjection
\be
J^kY\op\times_XJ^kTX\to J^kY\op\times_XTX
\ee
whose coordinate expression is
\be
\dot y^i_\La\circ r_{(k)}= (\dot y^i)_\La -\sum (\dot
y^i)_{\m+\Si}(\dot x^\m)_\Xi, \qquad 0\leq|\La|\leq k,
\ee
where the sum is taken over all partitions $\Si +\Xi=\La$ and
$0<|\Xi|$. In particular,  we have the canonical isomorphism over
$J^kY$
\mar{5.30}\beq
r_{(k)}:J^kVY\to VJ^kY, \qquad (\dot y^i)_\La = \dot y^i_\La\circ
r_{(k)}. \label{5.30}
\eeq
As a consequence, every projectable vector field $u$ (\ref{11f30})
on a fibre bundle $Y\to X$ has the following $k$-order {\sl jet
prolongation} to \index{jet prolongation!of a vector field!higher
order} a vector field on $J^kY$: \index{$J^k u$}
\mar{55.5}\ben
&& J^k u =r_{(k)}\circ J^ku: J^kY\to TJ^kY, \nonumber\\
&&  J^k u =
u^\la\dr_\la + u^i\dr_i + \op\sum_{0<|\La|\leq k}(d_\La(u^i-y^i_\m
u^\m) +y^i_{\m+\La}u^\m)\dr_i^\La, \label{55.5}
\een
(cf. (\ref{1.21}) for $k=1$). In particular, the $k$-order jet
prolongation (\ref{55.5}) of a vertical vector field $u=u^i\dr_i$
on $Y\to X$ is a vertical vector field
\mar{16f1}\beq
J^k u = u^i\dr_i + \op\sum_{0<|\La|\leq k}d_\La u^i\dr_i^\La
\label{16f1}
\eeq
on $J^kY\to X$ due to the isomorphism (\ref{5.30}).

A vector field $u_r$ on an $r$-order jet manifold  $J^rY$ is
called projectable if, for any $k< r$, there exists a projectable
vector field $u_k$ on $J^kY$ such that
\be
u_k\circ \pi^r_k=T\pi^r_k\circ u_r.
\ee
A {\sl projectable vector field} \index{vector
field!projectable!on a jet manifold} $u_k$ on $J^kY$ has the
coordinate expression
\be
u_k= u^\la \dr_\la + \op\sum_{0\leq |\La|\leq k} u^i_\La \dr^\La_i
\ee
such that $u_\la$ depends only on coordinates $x^\m$ and every
component $u^i_\La$ is independent of coordinates $y^i_\Xi$,
$|\Xi|>|\La|$. In particular, the $k$-order jet prolongation
$J^ku$ (\ref{55.5}) of a projectable vector field on $Y$ is a
projectable vector field on $J^kY$. It is called an {\sl
integrable vector field}. \index{vector field!integrable}

Let $\cP^k$ denote  a vector space of  projectable vector fields
on a jet manifold $J^kY$. It is easily seen that $\cP^r$ is a real
Lie algebra and that the morphisms $T\pi^r_k$, $k<r$, constitute
the inverse system
\mar{55.6}\beq
\cP^0 \op\longleftarrow^{T\pi^1_0}
\cP^1\op\longleftarrow^{T\pi^2_1}\cdots
\op\longleftarrow^{T\pi^{r-1}_{r-2}}\cP^{r-1}
\op\longleftarrow^{T\pi^r_{r-1}} \cP^r \label{55.6}
\eeq
of these Lie algebras. One can show the following.

\begin{theorem}\label{ch532} \mar{ch532}
The  $k$-order jet prolongation (\ref{55.5}) is a Lie algebra
monomorphism of the Lie algebra $\cP^0$ of projectable vector
fields on $Y\to X$ to the Lie algebra $\cP^k$ of projectable
vector fields on $J^kY$ such that
\mar{5.33}\beq
T\pi^r_k(J^r u)=J^k u\circ \pi^r_k. \label{5.33}
\eeq
\end{theorem}

Every projectable vector field $u_k$ on $J^kY$ is decomposed into
the sum
\mar{+404}\beq
u_k= J^k(T\pi^k_0(u_k)) +v_k \label{+404}
\eeq
of the integrable vector field $J^k(T\pi^k_0(u_k))$ and a
projectable vector field $v_k$ which is vertical  with respect to
a fibration $J^kY\to Y$.

Similarly to the canonical monomorphisms (\ref{18}) -- (\ref{24}),
there are the canonical bundle monomorphisms over $J^kY$:
\mar{5.16}\ben
&&\la_{(k)}: J^{k+1}Y\ar T^*X\op\otimes_{J^kY}
TJ^kY,\nonumber\\
 &&\la_{(k)} =dx^\la\otimes d_\la, \label{5.16}
\een
\mar{5.15'}\ben
&&\thh_{(k)}: J^{k+1}Y\ar T^*J^kY\op\otimes_{J^kY}
VJ^kY,\nonumber\\
&&\thh_{(k)} =\op\sum_{|\La|\leq k}(dy^i_\La-
y^i_{\la+\La}dx^\la)\otimes\dr_i^\La. \label{5.15'}
\een
The one-forms \index{$\thh^i_\La$}
\mar{55.15}\beq
\thh^i_\La=dy^i_\La- y^i_{\la+\La}dx^\la \label{55.15}
\eeq
are called the {\sl local contact forms}. \index{contact
form!local!of higher jet order} The monomorphisms (\ref{5.16}) --
(\ref{5.15'}) yield the bundle monomorphisms over $J^{k+1}Y$:
\be
&& \wh\la_{(k)}: TX\op\times_XJ^{k+1}Y\ar
TJ^kY\op\times_{J^kY}J^{k+1}Y,  \\
&&\wh\thh_{(k)}: V^*J^kY\op\times_{J^kY}
\op\ar T^*J^kY\op\times_{J^kY}J^{k+1}Y
\ee
(cf.  (\ref{z20}) -- (\ref{z21}) for $k=1$). These monomorphisms
in turn define the {\sl canonical horizontal splittings}
\index{horizontal!splitting!canonical!of higher  jet order} of the
pull-back bundles
\mar{5.72}\ben
&&\pi^{k+1*}_kTJ^kY =
\wh\la_{(k)}(TX\op\times_XJ^{k+1}Y) \op\oplus_{J^{k+1}Y}VJ^kY,
\label{5.72}\\
&& \dot x^\la\dr_\la +\op\sum_{|\La|\leq k} \dot y^i_\La\dr^\La_i =\dot
x^\la d_\la +\op\sum_{|\La|\leq k} (\dot y^i_\La -\dot x^\la
y^i_{\la+\La}) \dr_i^\La,\nonumber
\een
\mar{55.19}\ben
&& \pi^{k+1*}_kT^*J^kY =T^*X
\op\oplus_{J^{k+1}Y}
\wh\thh_{(k)}(V^*J^kY\op\times_{J^kY}J^{k+1}Y),
\label{55.19} \\
&& \dot x_\la dx^\la +\op\sum_{|\La|\leq k} \dot y_i^\La dy_\La^i =
(\dot x_\la +\op\sum_{|\La|\leq k} \dot y_i^\La
y_{\la+\La}^i)dx^\la +\sum \dot y_i^\La \thh_\La^i.\nonumber
\een

For instance, it follows from the canonical horizontal splitting
(\ref{5.72}) that any vector field $u_k$ on $J^kY$ admits the
canonical decomposition
\mar{+402}\ben
&& u_k = u_H +u_V= (u^\la \dr_\la + \op\sum_{|\La|\leq k}
y^i_{\la+\La}\dr_i^\La) +\label{+402}\\
&& \qquad \op\sum_{|\La|\leq k} (u^i_\La - u^\la y^i_{\la+\La})
\dr_i^\La\nonumber
\een
over $J^{k+1}Y$ into the horizontal and vertical parts.

By virtue of the canonical horizontal splitting (\ref{55.19}),
every exterior one-form $\f$ on $J^kY$ admits the canonical
splitting of its pull-back onto $J^{k+1}Y$ into the horizontal and
vertical parts:
\mar{+403}\beq
\pi^{k+1*}_k\f =\f_H+\f_V=h_0\f +(\f- h_0(\f)), \label{+403}
\eeq
where $h_0$ is the \index{horizontal!projection} {\sl horizontal
projection} \index{$h_0$}
\be
h_0(dx^\la)=dx^\la, \qquad h_0(dy^i_{\la_1\cdots\la_k})=
y^i_{\m\la_1\ldots\la_k}dx^\m.
\ee
The vertical part of the splitting is called a {\sl contact
one-form} \index{contact form} on $J^{k+1}Y$.

Let us consider an ideal of the  algebra $\cO_k^*$ of exterior
forms on $J^kY$ which is generated by the contact one-forms on
$J^kY$. This ideal, called the {\sl ideal of contact forms}, is
locally generated by the contact forms $\thh^i_\La$ (\ref{55.15}).
One can show that an exterior form $\f$ on the a manifold $J^kY$
is a contact form iff its pull-back $\ol s^*\f$ onto a base $X$ by
means of any integrable section $\ol s$ of $J^kY\to X$ vanishes.

\section{Differential operators and equations}

Jet manifolds provides the conventional language of theory of
differential equations and differential operators if they need not
be linear.

\begin{definition}  \label{16d1} \mar{16d1}
A system of $k$-order partial differential equations on a fibre
bundle $Y\to X$ is defined as a closed subbundle $\gE$ of a jet
bundle $J^kY\to X$. For the sake of brevity, we agree to call
$\gE$ a  \index{differential equation} {\sl differential
equation}.
\end{definition}

By a {\sl classical solution} \index{classical solution} of a
differential equation $\gE$ on $Y\to X$ is meant a section $s$ of
$Y\to X$ such that its $k$-order jet prolongation $J^ks$ lives in
$\gE$.

Let $J^kY$ be provided with the adapted coordinates $(x^\la,
y^i_\La)$. There exists a local coordinate system $(z^A)$,
$A=1,\ldots,{\rm codim}\gE,$ on $J^kY$ such that $\gE$ is locally
given (in the sense of item (i) of Theorem \ref{subman3}) by
equations
\beq
\cE^A(x^\la, y^i_\La) =0,
 \qquad A=1,\ldots, {\rm codim}\gE. \label{5.88}
\eeq

Differential equations are often associated to differential
operators. There are several equivalent definitions of
differential operators.

\begin{definition}\label{ch538} \mar{ch538}
Let $Y\to X$ and $E\to X$ be fibre bundles, which are assumed to
have global sections. A {\sl $k$-order $E$-valued differential
operator} \index{differential operator!as a section} on a fibre
bundle $Y\to X$ is defined as a section $\cE$ of the pull-back
bundle
\mar{5.113}\beq
{\rm pr}_1:E^k_Y= J^kY\op\times_X E \to J^kY. \label{5.113}
\eeq
\end{definition}

Given bundle coordinates $(x^\la, y^i)$ on $Y$ and $(x^\la,
\chi^a)$ on $E$, the pull-back (\ref{5.113}) is provided with
coordinates $(x^\la, y^j_\Si,\chi^a)$, $0\leq|\Si|\leq k$. With
respect to these coordinates, a differential operator $\cE$ seen
as a closed imbedded submanifold $\cE\subset E^k_Y$ is given by
the equalities
\mar{5.132}\beq
\chi^a = \cE^a(x^\la, y^j_\Si). \label{5.132}
\eeq

There is obvious one-to-one correspondence between the sections
$\cE$ (\ref{5.132}) of the fibre bundle (\ref{5.113}) and the
bundle morphisms
\mar{5.115}\ben
&&\Phi: J^kY\ar_X E, \label{5.115}\\
&& \Phi= {\rm pr}_2\circ \cE \, \Longleftrightarrow \, \cE
=(\id J^kY, \Phi). \nonumber
\een
Therefore, we come to the following equivalent definition of
differential operators on $Y\to X$.

\begin{definition}\label{ch539} \mar{ch539}
Let $Y\to X$ and $E\to X$ be fibre bundles. A bundle morphism
$J^kY\to E$ over $X$ is called a {\sl $E$-valued $k$-order
differential operator} \index{differential operator!as a morphism}
on $Y\to X$.
\end{definition}

It is readily observed that the differential operator $\Phi$
(\ref{5.115}) sends each section $s$ of $Y\to X$ onto the section
$\Phi\circ J^ks$ of $E\to X$. The mapping
\be
&& \Delta_\F: \cS(Y) \to \cS(E),\\
&& \Delta_\F: s \to \Phi\circ J^ks, \qquad \chi^a(x) =
\cE^a(x^\la, \dr_\Si s^j(x)),
\ee
is called the {\sl standard form of a differential operator}.
\index{differential operator!of standard form}

Let $e$ be a global section of a fibre bundle $E\to X$, the {\sl
kernel of a $E$-valued differential operator} \index{kernel!of a
differential operator} $\Phi$ is defined as the kernel
\mar{16f2}\beq
\Ker_e\Phi= \Phi^{-1}(e(X)) \label{16f2}
\eeq
of the bundle morphism $\Phi$ (\ref{5.115}). If it is a closed
subbundle of the jet bundle $J^kY\to X$, one says that
$\Ker_e\Phi$ (\ref{16f2}) is a {\sl differential equation
associated to the differential operator} \index{differential
equation!associated to a differential operator} $\Phi$. By virtue
of Theorem \ref{pomm}, this condition holds if $\Phi$ is a bundle
morphism of constant rank.

If $E\to X$ is a vector bundle, by the kernel of a $E$-valued
differential operator is usually meant its kernel with respect to
the canonical zero-valued section $\wh 0$ of $E\to X$.

In the framework of Lagrangian formalism, we deal with
differential operators of the following type. Let
\be
F\to Y\to X, \qquad E\to Y\to X
\ee
be composite bundles where $E\to Y$ is a vector bundle. By a
$k$-order differential operator on $F\to X$ taking its values into
$E\to X$ is meant a bundle morphism
\mar{k5}\beq
\Phi: J^kF\ar_Y E, \label{k5}
\eeq
which certainly is a bundle morphism over $X$ in accordance with
Definition \ref{ch539}. Its kernel $\Ker \Phi$ is defined as the
inverse image of the canonical zero-valued section of $E\to Y$. In
an equivalent way, the differential operator (\ref{k5}) is
represented  by a section $\cE_\Phi$ of the vector bundle
\be
J^kF\op\times_Y E\to J^kF.
\ee
Given bundle coordinates $(x^\la,y^i,w^r)$ on $F$ and $(x^\la,y^i,
c^A)$ on $E$ with respect to the fibre basis $\{e_A\}$ for $E\to
Y$, this section reads
\mar{k6}\beq
\cE_\Phi=\cE^A(x^\la,y^i_\La, w_\La^r)e_A, \qquad 0\leq|\La|\leq
k. \label{k6}
\eeq
Then the differential operator (\ref{k5}) also is represented by a
function
\mar{k7}\beq
\cE_\Phi=\cE^A(x^\la,y^i_\La, w_\La^r)c_A \in
C^\infty(F\op\times_Y E^*) \label{k7}
\eeq
on the product $F\times_Y E^*$, where $E^*\to Y$ is the dual of
$E\to Y$ coordinated by $(x^\la,y^i, c_A)$.

If $F\to Y$ is a vector bundle, a differential operator $\Phi$
(\ref{k5}) on the composite bundle $F\to Y\to X$ is called linear
if it is linear on the fibres of the vector bundle $J^kF\to J^kY$.
In this case, its representations (\ref{k6}) and (\ref{k7}) take
the form
\mar{0608,'}\ben
&& \cE_\Phi=\op\sum_{0\leq|\Xi|\leq k} \cE^{A,\Xi}_r(x^\la,
y^i_\La) w^r_\Xi e_A, \qquad 0\leq |\La|\leq k, \label{0608}\\
&& \cE_\Phi=\op\sum_{0\leq|\Xi|\leq k} \cE^{A,\Xi}_r(x^\la,
y^i_\La) w^r_\Xi c_A, \qquad 0\leq |\La|\leq k. \label{0608'}
\een

\section{Infinite order jet formalism}

The finite order jet manifolds $J^kY$ of a fibre bundle $Y\to X$
form the inverse sequence
\mar{j1}\beq
Y\op\longleftarrow^\pi J^1Y \longleftarrow \cdots J^{r-1}Y
\op\longleftarrow^{\pi^r_{r-1}} J^rY\longleftarrow\cdots,
\label{j1}
\eeq
where $\pi^r_{r-1}$ are affine bundles modelled over the vector
bundles (\ref{5.117}). Its inductive limit $J^\infty Y$
\index{$J^\infty Y$} is defined as a minimal set such that there
exist surjections
\mar{5.74}\beq
\pi^\infty: J^\infty Y\to X, \quad \pi^\infty_0: J^\infty Y\to Y,
\quad \quad \pi^\infty_k: J^\infty Y\to J^kY, \label{5.74}
\eeq
obeying the relations \index{$\pi^\infty_r$}
$\pi^\infty_r=\pi^k_r\circ\pi^\infty_k$ for all admissible $k$ and
$r<k$. A projective limit of the inverse system (\ref{j1}) always
exists. It consists of those elements
\be
(\ldots,z_r,\ldots,z_k, \ldots),\qquad z_r\in J^rY,\qquad z_k\in
J^kY,
\ee
of the Cartesian product $\op\prod_k J^kY$ which satisfy the
relations $z_r=\pi^k_r(z_k)$ for all $k>r$. One can think of
elements of $J^\infty Y$ as being {\sl infinite order jets}
\index{jet!infinite order} of sections of $Y\to X$ identified by
their Taylor series at points of $X$.

The set $J^\infty Y$ is provided with the projective limit
topology. This is the coarsest topology such that the surjections
$\pi^\infty_r$ (\ref{5.74}) are continuous. Its base consists of
inverse images of open subsets of $J^rY$, $r=0,\ldots$, under the
maps $\pi^\infty_r$. With this topology, $J^\infty Y$ is a
paracompact Fr\'echet (complete metrizable) manifold modelled over
a locally convex vector space of number series
$\{a^\la,a^i,a^i_\la,\cdots\}$. It is called the {\sl infinite
order jet manifold}. \index{jet manifold!infinite order} One can
show that the surjections $\pi^\infty_r$ are open maps admitting
local sections, i.e., $J^\infty Y\to J^rY$ are continuous bundles.
A bundle coordinate atlas $\{U_Y,(x^\la,y^i)\}$ of $Y\to X$
provides $J^\infty Y$ with the manifold coordinate atlas
\mar{j3}\beq
\{(\pi^\infty_0)^{-1}(U_Y), (x^\la, y^i_\La)\}_{0\leq|\La|},
\qquad {y'}^i_{\la+\La}=\frac{\dr x^\m}{\dr x'^\la}d_\m y'^i_\La.
\label{j3}
\eeq

\begin{theorem} \label{17t1} \mar{17t1}
A fibre bundle $Y$ is a strong deformation retract of the infinite
order jet manifold $J^\infty Y$.
\end{theorem}

\begin{corollary} \label{17c1} \mar{17c1}
There is an isomorphism
\mar{j19'}\beq
H^*(J^\infty Y;\Bbb R)=H^*(Y;\Bbb R) \label{j19'}
\eeq
between cohomology of $J^\infty Y$ and $Y$ with coefficients in
the sheaf $\Bbb R$.
\end{corollary}

The inverse sequence (\ref{j1}) of jet manifolds yields the direct
sequence of DGAs $\cO_r^*$ of exterior forms on finite order jet
manifolds
\mar{5.7}\beq
\cO^*(X)\op\longrightarrow^{\pi^*} \cO^*(Y)
\op\longrightarrow^{\pi^1_0{}^*} \cO_1^* \longrightarrow \cdots
\cO_{r-1}^*\op\longrightarrow^{\pi^r_{r-1}{}^*}
 \cO_r^* \longrightarrow\cdots, \label{5.7}
\eeq
where $\pi^r_{r-1}{}^*$ are the pull-back monomorphisms. Its
direct limit \index{$\cO^*_\infty Y$}
\mar{5.77}\beq
\cO^*_\infty Y=\op\lim^\to \cO_r^* \label{5.77}
\eeq
exists \index{algebra!$\cO^*_\infty Y$} and consists of all
exterior forms on finite order jet manifolds modulo the pull-back
identification. In accordance with Theorem \ref{spr170'},
$\cO^*_\infty Y$ is a DGA which inherits the operations of the
exterior differential $d$ and exterior product $\w$ of exterior
algebras $\cO^*_r$. If there is no danger of confusion, we denote
\index{$\cO^*_\infty$} $\cO^*_\infty =\cO^*_\infty Y$.
\index{algebra!$\cO^*_\infty$}

\begin{theorem} \label{j4} \mar{j4} The cohomology $H^*(\cO_\infty^*)$ of
the de Rham complex
\mar{5.13} \beq
0\longrightarrow \Bbb R\longrightarrow \cO^0_\infty
\op\longrightarrow^d\cO^1_\infty \op\longrightarrow^d \cdots
\label{5.13}
\eeq
of the DGA $\cO^*_\infty$ equals the de Rham cohomology $H^*_{\rm
DR}(Y)$ of $Y$.
\end{theorem}

\begin{corollary} \label{j21} \mar{j21}
Any closed form $\f\in \cO^*_\infty$ is decomposed into the sum
$\f=\si +d\xi$, where $\si$ is a closed form on $Y$.
\end{corollary}

One can think of elements of $\cO_\infty^*$ as being {\sl
differential forms} \index{differential form} on the infinite
order jet manifold $J^\infty Y$ as follows. Let $\gO^*_r$ be a
sheaf of germs of exterior forms on $J^rY$ and $\ol\gO^*_r$ the
canonical presheaf of local sections of $\gO^*_r$. Since
$\pi^r_{r-1}$ are open maps, there is the direct sequence of
presheaves
\be
\ol\gO^*_0 \op\longrightarrow^{\pi^1_0{}^*} \ol\gO_1^* \cdots
\op\longrightarrow^{\pi^r_{r-1}{}^*}
 \ol\gO_r^* \longrightarrow\cdots.
\ee
Its direct limit $\ol\gO^*_\infty$ is a presheaf of DGAs on
$J^\infty Y$. Let $\gQ^*_\infty$ be the sheaf of DGAs of germs of
$\ol\gO^*_\infty$ on $J^\infty Y$. The structure module
\index{$\cQ^*_\infty$}
\mar{17t4}\beq
\cQ^*_\infty=\G(\gQ^*_\infty) \label{17t4}
\eeq
of global sections of $\gQ^*_\infty$ is a DGA such that, given an
element $\f\in \cQ^*_\infty$ and a point $z\in J^\infty Y$, there
exist an open neighbourhood $U$ of $z$ and an exterior form
$\f^{(k)}$ on some finite order jet manifold $J^kY$ so that
\be
\f|_U= \pi^{\infty*}_k\f^{(k)}|_U.
\ee
Therefore, one can regard $\cQ^*_\infty$ as an algebra of locally
exterior forms on finite order jet manifolds. There is a
monomorphism $\cO^*_\infty \to\cQ^*_\infty$.

\begin{theorem} \label{17t5} \mar{17t5}
The paracompact space $J^\infty Y$ admits a partition of unity by
elements of the ring $\cQ^0_\infty$.
\end{theorem}

Since elements of the DGA $\cQ^*_\infty$ are locally exterior
forms on finite order jet manifolds, the following Poincar\'e
lemma holds.

\begin{lemma} \label{j8} \mar{j8}
Given a closed element $\f\in \cQ^*_\infty$, there exists a
neighbourhood $U$ of each point $z\in J^\infty Y$ such that
$\f|_U$ is exact.
\end{lemma}

\begin{theorem} \label{j6} \mar{j6} The cohomology $H^*(\cQ_\infty^*)$ of
the de Rham complex
\mar{5.13'} \beq
0\longrightarrow \Bbb R\longrightarrow \cQ^0_\infty
\op\longrightarrow^d\cQ^1_\infty \op\longrightarrow^d \cdots\,.
\label{5.13'}
\eeq
of the DGA $\cQ^*_\infty$ equals the de Rham cohomology of a fibre
bundle $Y$.
\end{theorem}

Due to a monomorphism $\cO^*_\infty  \to\cQ^*_\infty$, one can
restrict $\cO^*_\infty$ to the coordinate chart (\ref{j3}) where
horizontal forms $dx^\la$ and contact one-forms
\be
\thh^i_\La=dy^i_\La -y^i_{\la+\La}dx^\la
\ee
make up a local basis for the $\cO^0_\infty$-algebra
$\cO^*_\infty$. Though $J^\infty Y$ is not a smooth manifold,
elements of $\cO^*_\infty$ are exterior forms on finite order jet
manifolds and, therefore, their coordinate transformations are
smooth. Moreover, there is the canonical decomposition
\be
\cO^*_\infty=\oplus\cO^{k,m}_\infty
\ee
of $\cO^*_\infty$ into $\cO^0_\infty$-modules $\cO^{k,m}_\infty$
\index{$\cO^{k,m}_\infty$} of $k$-contact and $m$-horizontal forms
together with the corresponding projectors
\be
h_k:\cO^*_\infty\to \cO^{k,*}_\infty, \qquad h^m:\cO^*_\infty\to
\cO^{*,m}_\infty.
\ee
Accordingly, the exterior differential on $\cO_\infty^*$ is
decomposed into the sum
\be
d=d_V+d_H
\ee
of the \index{differential!vertical} {\sl vertical differential}
\index{$d_V$}
\be
d_V \circ h^m=h^m\circ d\circ h^m, \qquad d_V(\f)=\thh^i_\La \w
\dr^\La_i\f, \qquad \f\in\cO^*_\infty,
\ee
and the {\sl total differential} \index{differential!total}
\be
d_H\circ h_k=h_k\circ d\circ h_k, \qquad d_H\circ h_0=h_0\circ d,
\qquad d_H(\f)= dx^\la\w d_\la(\f),
\ee
where \index{$d_H$}
\mar{17f10}\beq
d_\la= \dr_\la + y^i_\la \dr_i +
\op\sum_{0<|\La|}y^i_{\la+\La}\dr^\La_i \label{17f10}
\eeq
are the {\sl infinite order total derivatives}. \index{total
derivative!infinite order} They obey the nilpotent conditions
\mar{mbm}\beq
d_H\circ d_H=0, \qquad d_V\circ d_V=0, \qquad d_H\circ
d_V+d_V\circ d_H=0, \label{mbm}
\eeq
and make $\cO^{*,*}_\infty$ into a bicomplex.

Let us consider the $\cO^0_\infty$-module $\gd\cO^0_\infty$ of
derivations of the real ring $\cO^0_\infty$.

\begin{theorem} \label{g62} \mar{g62}
The derivation module $\gd\cO^0_\infty$ is isomorphic to the
$\cO^0_\infty$-dual $(\cO^1_\infty)^*$ of the module of one-forms
$\cO^1_\infty$.
\end{theorem}

One can say something more. The DGA $\cO^*_\infty$ is a minimal
Chevalley--Eilenberg differential calculus $\cO^*\cA$ over the
real ring $\cA=\cO^0_\infty$ of smooth real functions on finite
order jet manifolds of $Y\to X$. Let $\vt\rfloor\f$, $\vt\in
\gd\cO^0_\infty$, $\f\in \cO^1_\infty$, denote the interior
product.  Extended to the DGA $\cO^*_\infty$, the interior product
$\rfloor$ obeys the rule
\be
\vt\rfloor(\f\w\si)=(\vt\rfloor \f)\w\si
+(-1)^{|\f|}\f\w(\vt\rfloor\si).
\ee

Restricted to a coordinate chart (\ref{j3}), $\cO^1_\infty$ is a
free $\cO^0_\infty$-module generated by one-forms $dx^\la$,
$\thh^i_\La$. Since $\gd\cO^0_\infty=(\cO^1_\infty)^*$, any
derivation of the real ring $\cO^0_\infty$ takes the coordinate
form
\mar{g3,71}\ben
&&\vt=\vt^\la \dr_\la + \vt^i\dr_i + \op\sum_{0<|\La|}\vt^i_\La
\dr^\La_i, \label{g3}\\
&& \dr^\La_i(y_\Si^j)=\dr^\La_i\rfloor
dy_\Si^j=\dl_i^j\dl^\La_\Si, \nonumber\\
&& \vt'^\la=\frac{\dr x'^\la}{\dr x^\m}\vt^\m, \qquad
\vt'^i=\frac{\dr y'^i}{\dr y^j}\vt^j + \frac{\dr y'^i}{\dr
x^\m}\vt^\m, \nonumber\\
&& \vt'^i_\La=\op\sum_{|\Si|\leq|\La|}\frac{\dr y'^i_\La}{\dr
y^j_\Si}\vt^j_\Si + \frac{\dr y'^i_\La}{\dr x^\m}\vt^\m.
\label{g71}
\een

Any derivation $\vt$ (\ref{g3}) of the ring $\cO^0_\infty$ yields
a derivation (called the  Lie derivative) $\bL_\vt$ of the DGA
$\cO^*_\infty$ given by the relations
\be
&& \bL_\vt\f=\vt\rfloor d\f+ d(\vt\rfloor\f), \\
&& \bL_\vt(\f\w\f')=\bL_\vt(\f)\w\f' +\f\w\bL_\vt(\f').
\ee

\begin{remark}
In particular, the total derivatives (\ref{17f10}) are defined as
the local derivations of $\cO^0_\infty$ and the corresponding Lie
derivatives
\be
d_\la\f=\bL_{d_\la}\f
\ee
of $\cO^*_\infty$. Moreover, the $C^\infty(X)$-ring $\cO^0_\infty$
possesses the canonical connection
\mar{17f20}\beq
\nabla=dx^\la\ot d_\la \label{17f20}
\eeq
in the sense of Definition \ref{mos088}.
\end{remark}

\chapter{Connections on fibre bundles}

There are several equivalent definitions of a connection on a
fibre bundle. We start with the traditional notion of a connection
as a splitting of the exact sequences (\ref{1.8a}) --
(\ref{1.8b}), but then follow its definition as a global section
of an affine jet bundle. In the case of vector bundles, there is
an equivalent definition (\ref{t56}) of a linear connection on
their structure modules.

\section{Connections as tangent-valued forms}

A {\sl connection} \index{connection} on a fibre bundle $Y\to X$
is defined traditionally as a linear bundle monomorphism
\mar{150}\ben
&&\G: Y\op\times_X TX\to TY, \label{150}\\
&&\G: \dot x^\la\dr_\la \to \dot
x^\la(\dr_\la+\G^i_\la\dr_i),\nonumber
\een
over $Y$ which splits the exact sequence (\ref{1.8a}), i.e.,
\be
\pi_T\circ \G=\id (Y\op\times_X TX).
\ee
This is a definition of connections on fibred manifolds (see
Remark \ref{Ehresmann}).

By virtue of Theorem \ref{sp11}, a connection always exists. The
local functions $\G^i_\la(y)$ in (\ref{150}) are said to be {\sl
components} \index{component of a connection} of the connection
$\G$  with respect to the bundle coordinates $(x^\la,y^i)$ on
$Y\to X$.

The image of $Y\times TX$ by the connection $\G$ defines the {\sl
horizontal distribution} \index{horizontal!distribution}
$HY\subset TY$ \index{distribution!horizontal} which \index{$HY$}
splits the tangent bundle $TY$ as follows:
\mar{152}\ben
&& TY=HY\op\oplus_Y VY, \label{152} \\
&& \dot x^\la \dr_\la + \dot y^i \dr_i = \dot
x^\la(\dr_\la + \G_\la^i \dr_i) + (\dot y^i - \dot
x^\la\G_\la^i)\dr_i.  \nonumber
\een
Its annihilator is locally generated by the one-forms
$dy^i-\G^i_\la dx^\la$.

Given the {\sl horizontal splitting} \index{horizontal!splitting}
(\ref{152}), the surjection
\mar{mos35}\beq
\G: TY\op\to_Y VY,  \qquad \dot y^i\circ \G= \dot y^i -
\G_\la^i\dot x^\la, \label{mos35}
\eeq
defines a connection on $Y\to X$ in an equivalent way.

The linear morphism $\G$ over $Y$ (\ref{150}) yields uniquely  the
horizontal tangent-valued one-form
\mar{154}\beq
\G = dx^\la\otimes (\dr_\la + \G_\la^i\dr_i) \label{154}
\eeq
on $Y$ which projects onto the canonical tangent-valued form
$\thh_X$ (\ref{b1.51}) on $X$. With this form called the {\sl
connection form}, \index{connection form} the morphism (\ref{150})
reads
\be
\G:\dr_\la\to \dr_\la\rfloor\G=\dr_\la +\G^i_\la\dr_i.
\ee

Given a connection $\G$ and  the corresponding horizontal
distribution (\ref{152}), a vector field $u$ on a fibre bundle
$Y\to X$ is called {\sl horizontal} \index{horizontal!vector
field} if \index{vector field!horizontal} it lives in $HY$.  A
horizontal vector field takes the form
\mar{mos167}\beq
u=u^\la(y)(\dr_\la +\G^i_\la\dr_i). \label{mos167}
\eeq
In particular, let $\tau$ be a vector field on the base $X$. By
means of the connection form $\G$ (\ref{154}), we obtain the
projectable horizontal vector field \index{$\G\tau$}
\mar{b1.85}\beq
\G \tau=\tau\rfloor\G=\tau^\la(\dr_\la +\G^i_\la\dr_i)
\label{b1.85}
\eeq
on $Y$, called the {\sl horizontal lift} \index{lift of a vector
field!horizontal} of $\tau$ \index{horizontal!lift!of a vector
field} by means of a connection $\G$. Conversely, any projectable
horizontal vector field $u$ on $Y$ is the horizontal lift $\G\tau$
of its projection $\tau$ on $X$. Moreover, the horizontal
distribution $HY$ is generated by the horizontal lifts $\G\tau$
(\ref{b1.85}) of vector fields $\tau$ on $X$. The horizontal lift
\mar{13f50}\beq
\cT(X)\ni\tau\to \G\tau\in \cT(Y) \label{13f50}
\eeq
is a $C^\infty(X)$-linear module morphism.

Given the splitting (\ref{150}), the dual splitting of the exact
sequence (\ref{1.8b}) is
\mar{cc3}\beq
\G: V^*Y\to T^*Y,\qquad \G: \ol dy^i\to dy^i-\G^i_\la dx^\la.
\label{cc3}
\eeq
Hence, a connection $\G$ on $Y\to X$ is represented by the
vertical-valued form
\mar{b1.223}\beq
\G= (dy^i -\G^i_\la dx^\la)\ot\dr_i \label{b1.223}
\eeq
such that the morphism (\ref{cc3}) reads
\be
\G:\ol dy^i\to \G\rfloor\ol dy^i=dy^i-\G^i_\la dx^\la.
\ee
We call $\G$ (\ref{b1.223}) the {\sl vertical connection form}.
\index{connection form!vertical} The corresponding horizontal
splitting of the cotangent bundle $T^*Y$ takes the form
\mar{cc7}\ben
&&T^*Y=T^*X\op\oplus_Y \G(V^*Y),\label{cc7}\\
&&\dot x_\la dx^\la + \dot y_i dy^i= (\dot x_\la +\dot
y_i\G^i_\la)dx^\la +\dot y_i(dy^i-\G^i_\la dx^\la). \nonumber
\een
Then we have the surjection
\mar{000}\beq
\G=\pr_1: T^*Y\to T^*X,\qquad \dot x_\la \circ\G=\dot x_\la + \dot
y_i\G^i_\la, \label{000}
\eeq
which also defines a connection on a fibre bundle $Y\to X$.

\begin{remark} \label{mos251} \mar{mos251}
Treating a connection as the vertical-valued form (\ref{b1.223}),
we come to the following important construction. Given a fibre
bundle $Y\to X$, let $f:X'\to X$ be a morphism and $f^*Y\to X'$
the pull-back of $Y$ by $f$. Any connection $\G$ (\ref{b1.223}) on
$Y\to X$ induces the \index{pull-back!connection} {\sl pull-back
connection} \index{$f^*\G$}
\mar{mos82}\beq
f^*\G=\left(dy^i-(\G\circ f_Y)^i_\la\frac{\dr f^\la}{\dr
x'^\m}dx'^\m\right)\ot\dr_i \label{mos82}
\eeq
on $f^*Y\to X'$ (see Remark \ref{mos80}).
\end{remark}

\begin{remark} \label{Ehresmann}  Let $\pi:Y\to X$ be
a fibred manifold. Any connection $\G$ on $Y\to X$ yields a
horizontal lift of a vector field on $X$ onto $Y$, but need not
defines the similar lift of a {\sl path} \index{path} in $X$ into
$Y$. Let
\be
\Bbb R\supset[,]\ni t\to x(t)\in X, \qquad \Bbb R\ni t\to y(t)\in
Y,
\ee
be smooth paths in $X$ and $Y$, respectively. Then $t\to y(t)$ is
called a {\sl horizontal lift} \index{horizontal!lift!of a path}
of $x(t)$ if
\be
\pi(y(t))= x(t), \qquad \dot y(t)\in H_{y(t)}Y, \qquad t\in\Bbb R,
\ee
where $HY\subset TY$ is the horizontal subbundle associated to the
connection $\G$. If, for each path $x(t)$ $(t_0\leq t\leq t_1)$
and for any $y_0\in\pi^{-1}(x(t_0))$, there exists a horizontal
lift $y(t)$ $(t_0\leq t\leq t_1)$ such that $y(t_0)=y_0$, then
$\G$ is called the {\sl Ehresmann connection}. \index{Ehresmann
connection} A fibred manifold is a fibre bundle iff it admits an
Ehresmann connection.
\end{remark}

\section{Connections as jet bundle sections}

Throughout the Lectures, we follow the equivalent definition of
connections on a fibre bundle $Y\to X$ as sections of the affine
jet bundle $J^1Y\to Y$.

Let $Y\to X$ be a fibre bundle, and $J^1Y$ its first order jet
manifold. Given the canonical morphisms (\ref{18}) and (\ref{24}),
we have the corresponding morphisms
\mar{z20,1}\ben
&& \wh\la_{(1)}: J^1Y\op\times_X TX\ni\dr_\la\to d_\la = \dr_\la\rfloor
\la_{(1)}\in J^1Y\op\times_Y TY, \label{z20}\\
&&\wh\thh_{(1)}: J^1Y\op\times_Y V^*Y\ni
\ol dy^i\to\thh^i=\thh_{(1)} \rfloor dy^i\in J^1Y\op\times_Y T^*Y
\label{z21}
\een
(see Remark \ref{mos30}). These morphisms yield the {\sl canonical
horizontal splittings} \index{horizontal!splitting!canonical} of
the pull-back bundles
\mar{1.20,34}\ben
&&J^1Y\op\times_Y TY=\wh\la_{(1)}(TX)\op\oplus_{J^1Y} VY,\label{1.20}\\
&& \dot x^\la\dr_\la
+\dot y^i\dr_i =\dot x^\la(\dr_\la +y^i_\la\dr_i) + (\dot y^i-\dot
x^\la
y^i_\la)\dr_i, \nonumber\\
&& J^1Y\op\times_Y T^*Y=T^*X\op\oplus_{J^1Y} \wh\thh_{(1)}(V^*Y),\label{34}\\
&& \dot x_\la dx^\la
+\dot y_i dy^i =(\dot x_\la + \dot y_iy^i_\la)dx^\la + \dot
y_i(dy^i- y^i_\la dx^\la).\nonumber
\een
Let $\G$ be a global section of $J^1Y\to Y$. Substituting the
tangent-valued form
\be
\la_{(1)}\circ \G= dx^\la\ot(\dr_\la +\G^i_\la\dr_i)
\ee
in the canonical splitting (\ref{1.20}), we obtain the familiar
horizontal splitting (\ref{152}) of $TY$ by means of a connection
$\G$ on $Y\to X$. Accordingly, substitution of the tangent-valued
form
\be
\thh_{(1)}\circ \G=(dy^i-\G^i_\la dx^\la)\ot \dr_i
\ee
in the canonical splitting (\ref{34}) leads to the dual splitting
(\ref{cc7}) of $T^*Y$  by means of a connection $\G$.

\begin{theorem} \label{mnh} \mar{mnh}
There is one-to-one correspondence between the connections $\G$ on
a fibre bundle $\Y$ and the global sections
\mar{2115}\beq
\G :Y\to J^1Y, \qquad (x^\la,y^i,y_\la^i)\circ\G
=(x^\la,y^i,\G_\la^i), \label{2115}
\eeq
of the affine jet bundle $J^1Y\to Y$.
\end{theorem}

There are the following corollaries  of this theorem.

$\bullet$ Since $J^1Y\to Y$ is affine, a connection on a fibre
bundle $Y\to X$ exists in accordance with Theorem \ref{mos9}.

$\bullet$ Connections on a fibre bundle $Y\to X$ make up an affine
space modelled over the vector space of soldering forms on $Y\to
X$, i.e., sections of the vector bundle (\ref{cc9}).

$\bullet$ Connection components possess the coordinate
transformation law
\be
\G'^i_\la = \frac{\dr x^\m}{\dr{x'}^\la}(\dr_\m
+\G^j_\m\dr_j)y'^i.
\ee

$\bullet$ Every connection $\G$ (\ref{2115}) on a fibre bundle
$Y\to X$ yields the first order differential operator
\mar{2116}\ben
&& D_\G:J^1Y\op\to_Y T^*X\op\otimes_Y VY, \label{2116}\\
&& D_\G=\la_{(1)}- \G\circ \pi^1_0 =(y^i_\la -\G^i_\la)dx^\la\otimes\dr_i,
\nonumber
\een
on \index{$D_\G$} $Y$ called the {\sl covariant differential}
relative to the connection \index{covariant differential}  $\G$.
\index{differential!covariant} If $s:X\to Y$ is a section, from
(\ref{2116}) we obtain its covariant differential
\mar{+190}\ben
&&\nabla^\G s  =  D_\G\circ J^1s:X\to T^*X\ot VY, \label{+190}\\
&& \nabla^\G s  =  (\dr_\la s^i - \G_\la^i\circ s) dx^\la\ot \dr_i,
\nonumber
\een
and \index{$\nabla^\G$} the \index{covariant derivative}{\sl
covariant derivative}
\be
\nabla_\tau^\G =\tau\rfloor\nabla^\G
\ee
along a vector field $\tau$ on $X$. A section $s$ is said to be an
{\sl integral section} \index{section!integral} of a connection
$\G$ \index{integral section of a connection} if it belongs to the
kernel of the covariant differential $D_\G$ (\ref{2116}), i.e.,
\mar{b1.86}\beq
\nabla^\G s=0 \quad {\rm or} \quad J^1s=\G\circ s. \label{b1.86}
\eeq

\begin{theorem} \label{ints} \mar{ints}
For any global section $s:X\to Y$, there always exists a
connection $\G$ such that $s$ is an integral section of $\G$.
\end{theorem}

Treating connections as jet bundle sections, one comes to the
following two constructions.

(i) Let $Y$ and $Y'$ be fibre bundles over the same base $X$.
Given a connection $\G$ on $\Y$ and a connection $\G'$ on $Y'\to
X$, the bundle product $Y\times Y'$ is provided with the {\sl
product connection} \index{product connection}
\mar{b1.96}\ben
&&\G\times\G': Y\op\times_XY'\to
J^1(Y\op\times_XY')=J^1Y\op\times_XJ^1Y', \nonumber\\
&&\G\times\G' = dx^\la\ot\left(\dr_\la +\G^i_\la\frac{\dr}{\dr y^i} +
\G'^j_\la\frac{\dr}{\dr y'^j}\right). \label{b1.96}
\een

(ii) Let $i_Y:Y\to Y'$ be a subbundle of a fibre bundle  $Y'\to X$
and $\G'$ a connection on $Y'\to X$. If there exists a connection
$\G$ on $\Y$ such that the diagram
\be
\begin{array}{rcccl}
 & Y' & \op\longrightarrow^{\G'} & J^1Y &  \\
 _{i_Y} & \put(0,-10){\vector(0,1){20}} & & \put(0,-10){\vector(0,1){20}} &
_{J^1i_Y} \\
 & Y & \op\longrightarrow^\G & {J^1Y'} &
 \end{array}
 \ee
is commutative, we say that $\G'$ is {\sl reducible}
\index{reducible connection} \index{connection!reducible} to a
connection $\G$. The following conditions are equivalent:

$\bullet$  $\G'$ is reducible to $\G$;

$\bullet$ $Ti_Y(HY)=HY'\vert_{i_Y(Y)}$, where $HY\subset TY$ and
$HY'\subset TY'$ are the horizontal subbundles determined by $\G$
and $\G'$, respectively;

$\bullet$ for every vector field $\tau$ on $X$, the vector fields
$\G\tau$ and $\G'\tau$ are related as follows:
\mar{b1.117}\beq
Ti_Y\circ\G \tau=\G' \tau\circ i_Y.  \label{b1.117}
\eeq

\section{Curvature and torsion}

Let $\G$ be a connection on a fibre bundle $Y\to X$. Its {\sl
curvature} \index{curvature} is defined as the Nijenhuis
differential
\mar{1178a,'}\ben
&& R=\frac{1}{2} d_\G\G=\frac{1}{2} [\G,\G]_{\rm FN}:Y\to
\op\w^2T^*X\ot VY, \label{1178a}\\
&& R = \frac12 R_{\la\m}^i dx^\la\wedge dx^\m\otimes\dr_i,\label{1178a'} \\
&& R_{\la\m}^i = \dr_\la\G_\m^i - \dr_\m\G_\la^i + \G_\la^j\dr_j
\G_\m^i - \G_\m^j\dr_j \G_\la^i. \nonumber
\een
This is a $VY$-valued horizontal two-form on $Y$. Given vector
fields $\tau$, $\tau'$ on $X$ and their horizontal lifts $\G\tau$
and $\G\tau'$ (\ref{b1.85}) on $Y$, we have the relation
\mar{13f30}\beq
R(\tau,\tau')=-\G [\tau,\tau'] + [\G \tau, \G \tau']= \tau^\la
\tau'^\m R_{\la\m}^i\dr_i. \label{13f30}
\eeq

The curvature (\ref{1178a}) obeys the identities
\mar{1180,a}\ben
&& [R,R]_{\rm FN}=0, \label{1180}\\
&& d_\G R= [\G, R]_{\rm FN}=0. \label{1180a}
\een
They result from the identity (\ref{1150}) and the graded Jacobi
identity (\ref{1150'}), respectively. The identity (\ref{1180a})
is called the {\sl second Bianchi identity}. \index{Bianchi
identity!second} It takes the coordinate form
\mar{1182}\beq
\sum_{(\la\m\n)} (\dr_\la R_{\m\n}^i + \G_\la^j\dr_j R_{\m\n}^i -
\dr_j\G_\la^iR_{\m\n}^j)=0, \label{1182}
\eeq
where the sum is cyclic over the indices $\la$, $\m$ and $\n$.

Given a soldering form $\si$, \index{curvature!soldered} one
defines the {\sl soldered curvature} \index{soldered curvature}
\mar{1186}\ben
&& \rho =\frac{1}{2} d_\si \si = \frac{1}{2} [\si,\si]_{\rm FN}:Y\to
\op\w^2T^*X\ot VY, \label{1186} \\
&& \rho =\frac{1}{2} \rho_{\la\m}^i dx^\la\w dx^\m\ot \dr_i,
\qquad \rho_{\la\m}^i =\si_\la^j\dr_j\si_\m^i -
\si_\m^j\dr_j\si_\la^i. \nonumber
\een
It fulfills the identities
\be
[\rho,\rho]_{\rm FN}=0, \qquad d_\si \rho =[\si,\rho]_{\rm FN}=0,
\ee
similar to (\ref{1180}) -- (\ref{1180a}).

Given a connection $\G$ and a soldering form $\si$, the {\sl
torsion form} of $\G$ \index{torsion form} with respect to $\si$
is defined as
\mar{1190}\ben
&& T = d_\G \si = d_\si \G :Y\to \op\w^2 T^*X\ot VY, \nonumber\\
&& T = (\dr_\la\si_\m^i + \G_\la^j\dr_j\si_\m^i -
\dr_j\G_\la^i\si_\m^j) dx^\la\w dx^\m\ot \dr_i. \label{1190}
\een
It obeys the \index{Bianchi identity!first} {\sl first Bianchi
identity}
\mar{1191}\beq
d_\G T = d_\G^2 \si = [R, \si]_{\rm FN} = -d_\si R. \label{1191}
\eeq

If $\G' =\G + \si$, we have the relations
\mar{a,1193}\ben
&& T' =T + 2\rho, \label{1193a} \\
&& R' = R + \rho +T. \label{1193}
\een

\section{Linear and affine connections}

A connection $\G$ on a vector bundle $Y\to X$ is called the {\sl
linear connection} \index{connection!linear} if the section
\mar{167}\beq
\G: Y\to J^1Y, \qquad \G =dx^\la\ot(\dr_\la + \G_\la{}^i{}_j(x)
y^j\dr_i), \label{167}
\eeq
is a linear bundle morphism over $X$. Note that linear connections
are principal connections, and they always exist (see Theorem
\ref{53a1}).

The curvature $R$ (\ref{1178a'}) of a linear connection $\G$
(\ref{167}) reads
\mar{mos4}\ben
&&R=\frac12 R_{\la\m}{}^i{}_j(x)y^j dx^\la\w dx^\m\ot\dr_i,\nonumber\\
&& R_{\la\m}{}^i{}_j = \dr_\la \G_\m{}^i{}_j - \dr_\m
\G_\la{}^i{}_j + \G_\la{}^h{}_j \G_\m{}^i{}_h - \G_\m{}^h{}_j
\G_\la{}^i{}_h. \label{mos4}
\een
Due to the vertical splitting (\ref{12f10}), we have the linear
morphism
\mar{+102}\beq
R: Y\ni y^ie_i\to \frac12 R_{\la\m}{}^i{}_jy^j dx^\la\w dx^\m\ot
e_i\in \cO^2(X)\ot Y. \label{+102}
\eeq

There are the following standard constructions of new linear
connections from the old ones.

$\bullet$ Let $Y\to X$ be a vector bundle, coordinated by
$(x^\la,y^i)$, and $Y^*\to X$ its dual, coordinated by
$(x^\la,y_i)$. Any linear connection $\G$ (\ref{167}) on a vector
bundle $Y\to X$ defines the \index{connection!dual} {\sl dual
linear connection}
\mar{spr300}\beq
\G^*=dx^\la\ot(\dr_\la - \G_\la{}^j{}_i(x) y_j\dr^i)
\label{spr300}
\eeq
on $Y^*\to X$.

$\bullet$ Let $\G$ and $\G'$ be linear connections on vector
bundles $Y\to X$ and $Y'\to X$, respectively. The direct sum
connection $\G\oplus\G'$ on their Whitney sum $Y\oplus Y'$ is
defined as the product connection (\ref{b1.96}).

$\bullet$ Let $Y$ coordinated by $(x^\la,y^i)$ and $Y'$
coordinated by $(x^\la,y^a)$ be vector bundles over the same base
$X$. Their tensor product $Y\otimes Y'$ is endowed with the bundle
coordinates $(x^\la,y^{ia})$. Linear connections $\G$ and $\G'$ on
$Y\to X$ and $Y'\to X$ define the linear {\sl tensor product
connection} \index{tensor product connection}
\mar{b1.92}\beq
\G\otimes\G'=dx^\la\ot\left[\dr_\la +(\G_\la{}^i{}_j
y^{ja}+\G'_\la{}^a{}_b y^{ib}) \frac{\dr}{\dr y^{ia}}\right]
\label{b1.92}
\eeq
on $Y\ot Y'$.

An important example of linear connections is a linear connection
\mar{B}\beq
\G= dx^\la\otimes (\dr_\la +\G_\la{}^\m{}_\n \dot x^\n \dot\dr_\m)
\label{B}
\eeq
on the tangent bundle $TX$ of a manifold $X$. We agree to call it
a {\sl world connection} \index{world connection} on a manifold
$X$. The {\sl dual world connection} \index{world connection!on
the cotangent bundle} (\ref{spr300}) on the cotangent bundle
$T^*X$ is \index{$\G^*$}
\mar{C}\beq
\G^*= dx^\la\otimes (\dr_\la -\G_\la{}^\m{}_\n\dot x_\m
\dot\dr^\n). \label{C}
\eeq
Then, using the construction of the tensor product connection
(\ref{b1.92}), one can introduce the corresponding linear {\sl
world connection on an arbitrary tensor bundle} $T$ (\ref{sp20}).
\index{world connection!on a tensor bundle}

\begin{remark} \label{mos7} \mar{mos7}
It should be emphasized that the expressions (\ref{B}) and
(\ref{C}) for a world connection differ in a minus sign from those
usually used in the physical literature.
\end{remark}

The {\sl curvature of a world connection} \index{curvature!of a
world connection} is defined as the curvature $R$ (\ref{mos4}) of
the connection $\G$ (\ref{B}) on the tangent bundle $TX$. It reads
\mar{1203}\ben
&& R=\frac12R_{\la\m}{}^\al{}_\bt\dot x^\bt dx^\la\w dx^\m\ot\dot\dr_\al,
\label{1203}\\
&& R_{\la\m}{}^\al{}_\bt = \dr_\la \G_\m{}^\al{}_\bt - \dr_\m
\G_\la{}^\al{}_\bt + \G_\la{}^\g{}_\bt \G_\m{}^\al{}_\g -
\G_\m{}^\g{}_\bt \G_\la{}^\al{}_\g. \nonumber
\een

By the {\sl torsion of a world connection} \index{torsion form!of
a world connection} is meant the torsion (\ref{1190}) of the
connection $\G$ (\ref{B}) on the tangent bundle $TX$ with respect
to the canonical soldering form $\thh_J$ (\ref{z117'}):
\mar{191}\beq
T =\frac12 T_\m{}^\n{}_\la  dx^\la\w dx^\m\ot \dot\dr_\n, \qquad
T_\m{}^\n{}_\la  = \G_\m{}^\n{}_\la - \G_\la{}^\n{}_\m.
\label{191}
\eeq
A world connection is said to be {\sl symmetric} \index{world
connection!symmetric} if its torsion (\ref{191}) vanishes, i.e.,
$\G_\m{}^\n{}_\la = \G_\la{}^\n{}_\m$.

\begin{remark} \label{mos44} \mar{mos44}
For any vector field $\tau$ on a manifold $X$, there exists a
connection $\G$ on the tangent bundle $TX\to X$ such that $\tau$
is an integral section of $\G$, but this connection is not
necessarily linear. If a vector field $\tau$ is non-vanishing at a
point $x\in X$, then there exists a local symmetric world
connection $\G$ (\ref{B}) around $x$ for which $\tau$ is an
integral section
\mar{mos45}\beq
\dr_\nu\tau^\al=\G_\nu{}^\al{}_\bt\tau^\bt. \label{mos45}
\eeq
Then the canonical lift $\wt\tau$ (\ref{l27}) of $\tau$ onto $TX$
can be seen locally as the horizontal lift $\G\tau$ (\ref{b1.85})
of $\tau$ by means of this connection.
\end{remark}

\begin{remark} \label{mos51} \mar{mos51}
Every manifold $X$ can be provided with a non-degenerate fibre
metric
\be
g\in\op\vee^2\cO^1(X), \qquad g=g_{\la\m}dx^\la\ot dx^\m,
\ee
in the tangent bundle $TX$, and with the corresponding metric
\be
g\in\op\vee^2\cT^1(X), \qquad g=g^{\la\m}\dr_\la\ot \dr_\m,
\ee
in the cotangent bundle $T^*X$. We call it a {\sl world metric}
\index{world metric} on $X$. For any world metric $g$, there
exists a unique symmetric world connection $\G$ (\ref{B}) with the
components \index{$\{_\la{}^\nu{}_\m\}$}
\mar{b1.400}\beq
\G_\la{}^\nu{}_\m =
\{_\la{}^\nu{}_\m\}=-\frac{1}{2}g^{\n\rho}(\dr_\la g_{\rho\m} +
\dr_\m g_{\rho\la}-\dr_\rho g_{\la\m}), \label{b1.400}
\eeq
called \index{Christoffel symbols} the {\sl Christoffel symbols},
such that $g$ is an integral section of $\G$, i.e.
\be
\dr_\la\, g^{\al \bt} = g^{\al\g}\{_\la{}^\bt{}_\g\} +
g^{\bt\g}\{_\la{}^\al{}_\g\}.
\ee
It is called the {\sl Levi--Civita connection} \index{Levi--Civita
connection} associated to $g$.
\end{remark}

Let $Y\to X$ be an affine bundle modelled over a vector bundle
$\ol Y\to X$. A connection $\G$ on $Y\to X$ is called an {\sl
affine connection} \index{connection!affine} if the section
$\G:Y\to J^1Y$ (\ref{2115}) is an affine bundle morphism over $X$.
Associated to principal connections, affine connections always
exist (see Theorem \ref{53a1}).

For any affine connection $\G:Y\to J^1Y$, the corresponding linear
derivative $\ol \G:\ol Y\to J^1\ol Y$ (\ref{1355'}) defines a
unique linear connection on the vector bundle $\ol Y\to X$. Since
every vector bundle is an affine bundle, any linear connection on
a vector bundle also is an affine connection.

With respect to affine bundle coordinates $(x^\la,y^i)$ on $Y$, an
affine connection $\G$ on $Y\to X$ reads
\mar{184}\beq
\G_\la^i=\G_\la{}^i{}_j(x) y^j + \si_\la ^i(x). \label{184}
\eeq
The coordinate expression of the associated linear connection is
\mar{mos032}\beq
\ol\G_\la^i=\G_\la{}^i{}_j(x) \ol y^j, \label{mos032}
\eeq
where $(x^\la,\ol y^i)$ are the associated linear bundle
coordinates on $\ol Y$.

Affine connections on an affine bundle $Y\to X$ constitute an
affine space modelled over the soldering forms on $Y\to X$. In
view of the vertical splitting (\ref{48}), these soldering forms
can be seen as global sections of the vector bundle
\be
T^*X\op\ot_X\ol Y\to X.
\ee
If $Y\to X$ is a vector bundle, both the affine connection $\G$
(\ref{184}) and the associated linear connection $\ol\G$ are
connections on the same vector bundle $Y\to X$, and their
difference is a basic soldering form on $Y$. Thus, every affine
connection on a vector bundle $Y\to X$ is the sum of a linear
connection and a basic soldering form on $Y\to X$.

Given an affine connection $\G$ on a vector bundle $Y\to X$, let
$R$ and $\ol R$ be the curvatures of a connection $\G$ and the
associated linear connection $\ol \G$, respectively.  It is
readily observed that $R = \ol R + T$, where the $VY$-valued
two-form
\mar{mos036}\ben
&& T=d_\G\si=d_\si\G :X\to \op\wedge^2 T^*X\op\otimes_X VY, \label{mos036} \\
&& T =\frac12 T_{\la
\m}^i dx^\la\wedge dx^\m\otimes \dr_i, \nonumber\\
&&  T_{\la \m}^i = \dr_\la\si_\m^i - \dr_\m\si_\la^i + \si_\la^h
\G_\m{}^i{}_h - \si_\m^h \G_\la{}^i{}_h, \nonumber
\een
is the torsion (\ref{1190}) of $\G$ with respect to the basic
soldering form $\si$.

In particular, let us consider the tangent bundle $TX$ of a
manifold $X$. We have the canonical soldering form
$\si=\thh_J=\thh_X$ (\ref{z117'}) on $TX$. Given an arbitrary
world connection $\G$ (\ref{B}) on $TX$, the corresponding affine
connection
\mar{b1.97}\beq
 A=\G +\thh_X, \qquad
A_\la^\m=\G_\la{}^\m{}_\nu \dot x^\nu +\dl^\m_\la, \label{b1.97}
\eeq
on $TX$ is called the {\sl Cartan connection}. \index{Cartan
connection} Since the soldered curvature $\rho$ (\ref{1186}) of
$\thh_J$ equals zero, the torsion (\ref{1193a}) of the Cartan
connection coincides with the torsion $T$ (\ref{191}) of the world
connection $\G$, while its curvature (\ref{1193}) is the sum $R+T$
of the curvature and the torsion of $\G$.

\section{Flat connections}

By a  {\sl flat} or {\sl curvature-free} \index{connection!flat}
\index{curvature-free connection} connection is meant a connection
which satisfies the following equivalent conditions.

\begin{theorem} \label{flat} \mar{flat}
Let $\G$ be a connection on a fibre bundle $Y\to X$. The following
assertions are equivalent.

(i) The curvature $R$ of a connection $\G$ vanishes identically,
i.e., $R\equiv 0$.

(ii) The horizontal lift (\ref{13f50}) of vector fields on $X$
onto $Y$ is an $\Bbb R$-linear Lie algebra morphism (in accordance
with the formula (\ref{13f30})).

(iii) The horizontal distribution is involutive.

(iv) There exists a local integral section for $\G$ through any
point $y\in Y$.
\end{theorem}

By virtue of Theorem \ref{to.1} and item (iii) of Theorem
\ref{flat}, a flat connection $\G$ on a fibre bundle $Y\to X$
yields a {\sl horizontal foliation} \index{horizontal!foliation}
on \index{foliation!horizontal} $Y$, transversal to the fibration
$Y\to X$. The leaf of this foliation through a point $y\in Y$ is
defined locally by an integral section $s_y$ for the connection
$\G$ through $y$. Conversely, let a fibre bundle $Y\to X$ admit a
transversal foliation such that, for each point $y\in Y$, the leaf
of this foliation through $y$ is locally defined by a section
$s_y$ of $Y\to X$ through $y$. Then the map
\be
\G:Y\to J^1Y, \qquad \G(y)=j^1_xs_y, \qquad \pi(y)=x,
\ee
introduces a flat connection on $Y\to X$. Thus, there is
one-to-one correspondence between the flat connections and the
transversal foliations of a fibre bundle $Y\to X$.

Given a transversal foliation on a fibre bundle $Y\to X$, there
exists the associated atlas of bundle coordinates $(x^\la, y^i)$
of $Y$ such that every leaf of this foliation is locally generated
by the equations $y^i=$const., and the transition functions
$y^i\to {y'}^i(y^j)$ are independent of the base coordinates
$x^\la$. This is called the {\sl atlas of constant local
trivializations}. \index{bundle!atlas!of constant local
trivializations} Two such atlases are said to be equivalent if
their union also is an atlas of constant local trivializations.
They are associated to the same horizontal foliation. Thus, we
come to the following assertion.

\begin{theorem} \label{gena113} \mar{gena113}
There is one-to-one correspondence between the flat connections
$\G$ on a fibre bundle $Y\to X$ and the equivalence classes of
atlases of constant local trivializations of $Y$ such that
\be
\G=dx^\la\ot\dr_\la
\ee
relative to these atlases.
\end{theorem}

In particular, if $Y\to X$ is a trivial bundle, one associates to
each its trivialization a flat connection represented by the
global zero section $\wh 0(Y)$ of $J^1Y\to Y$ with respect to this
trivialization (see Remark \ref{jj1}).

\section{Connections on composite bundles}

Let $Y\to \Si\to X$ be a composite bundle (\ref{1.34}). Let us
consider the jet manifolds $J^1\Si$, $J^1_\Si Y$, \index{$J^1_\Si
Y$} and $J^1Y$ of the fibre bundles
\be
\Si\to X,\qquad Y\to \Si, \qquad Y\to X,
\ee
respectively. They are provided with the adapted coordinates
\be
( x^\la ,\si^m, \si^m_\la),\quad ( x^\la ,\si^m, y^i, \wt y^i_\la,
y^i_m),\quad ( x^\la ,\si^m, y^i, \si^m_\la ,y^i_\la).
\ee
One can show the following.

\begin{theorem}
There is the canonical map
\mar{1.38}\beq
\vr : J^1\Si\op\times_\Si J^1_\Si Y\ar_Y J^1Y, \qquad
y^i_\la\circ\vr=y^i_m{\si}^m_{\la} +\wt y^i_{\la}. \label{1.38}
\eeq
\end{theorem}

Using the canonical map (\ref{1.38}), we can get the relations
between connections on the fibre bundles $Y\to X$, $Y\to\Si$ and
$\Si\to X$. These connections are given by the corresponding
connection forms
\mar{spr290-2}\ben
&& \g=dx^\la\ot (\dr_\la +\g_\la^m\dr_m + \g_\la^i\dr_i), \label{spr290}\\
&&  A_\Si=dx^\la\ot (\dr_\la + A_\la^i\dr_i) +d\si^m\ot (\dr_m + A_m^i\dr_i),
\label{spr291}\\
&& \G=dx^\la\ot (\dr_\la + \G_\la^m\dr_m). \label{spr292}
\een

A connection $\g$ (\ref{spr290}) on the fibre bundle $Y\to X$ is
called {\sl projectable} \index{connection!projectable} onto a
connection $\G$ (\ref{spr292}) on the fibre bundle $\Si\to X$ if,
for any vector field $\tau$ on $X$, its horizontal lift $\g\tau$
on $Y$ by means of the connection $\g$ is a projectable vector
field over the horizontal lift $\G\tau$ of $\tau$ on $\Si$ by
means of the connection $\G$. This property holds iff
$\g_\la^m=\G_\la^m$, i.e., components $\g_\la^m$ of the connection
$\g$ (\ref{spr290}) must be independent of the fibre coordinates
$y^i$.

A connection $A_\Si$ (\ref{spr291}) on the fibre bundle $Y\to\Si$
and a connection $\G$ (\ref{spr292}) on the fibre bundle $\Si\to
X$ define a connection  on the composite bundle $Y\to X$ as the
composition of bundle morphisms
\be
\g:Y\op\times_XTX\ar^{(\id,\G)} Y\op\times_\Si T\Si\ar^{A_\Si} TY.
\ee
It is called the {\sl composite connection}. \index{composite
connection} This \index{connection!composite} composite connection
reads
\mar{b1.114}\beq
\g=dx^\la\ot (\dr_\la +\G_\la^m\dr_m + (A_\la^i +
A_m^i\G_\la^m)\dr_i). \label{b1.114}
\eeq
It is projectable onto $\G$.  Moreover, this is a unique
connection such that the horizontal lift $\g\tau$ on $Y$ of a
vector field $\tau$ on $X$ by means of the composite connection
$\g$ (\ref{b1.114}) coincides with the composition $A_\Si(\G\tau)$
of horizontal lifts of $\tau$ on $\Si$ by means of the connection
$\G$ and then on $Y$ by means of the connection $A_\Si$. For the
sake of brevity, let us write $\g=A_\Si\circ\G$.

Given a composite bundle $Y$ (\ref{1.34}), there are the exact
sequences of vector bundles over $Y$:
\mar{63}\ben
&& 0\to V_\Si Y\ar VY\to Y\op\times_\Si V\Si\to 0, \label{63a}\\
&& 0\to Y\op\times_\Si V^*\Si \ar V^*Y\to V^*_\Si Y\to 0,
\label{63b}
\een
where $V_\Si Y$ \index{$V_\Si Y$} and $V_\Si^*Y$
\index{$V_\Si^*Y$} are the vertical tangent and the vertical
cotangent bundles of $Y\to\Si$ which are coordinated by
$(x^\la,\si^m,y^i, \dot y^i)$ and $(x^\la,\si^m,y^i, \dot y_i)$,
respectively. Let us consider a splitting of these exact sequences
\mar{63c,d}\ben
&& B: VY\ni \dot y^i\dr_i +\dot \si^m\dr_m \to (\dot y^i\dr_i +\dot
\si^m\dr_m)\rfloor B= \label{63c}\\
&&\qquad (\dot y^i -\dot \si^m B^i_m)\dr_i\in V_\Si Y,
\nonumber\\
&& B: V_\Si^*Y\ni \ol dy^i\to B\rfloor \ol dy^i= \ol dy^i- B^i_m\ol
d\si^m\in V^*Y, \label{63d}
\een
given by the form
\mar{mmm}\beq
B=(\ol dy^i-B^i_m\ol d\si^m)\ot\dr_i. \label{mmm}
\eeq
Then the connection $\g$ (\ref{spr290}) on $Y\to X$ and the
splitting $B$ (\ref{63c}) define the connection
\mar{nnn}\ben
&& A_\Si=B\circ \g: TY\to VY \to V_\Si Y, \label{nnn}\\
&& A_\Si=dx^\la\ot(\dr_\la +(\g^i_\la - B^i_m\g^m_\la)\dr_i) +
d\si^m\ot (\dr_m +B^i_m\dr_i),\nonumber
\een
on the fibre bundle $Y\to\Si$.

Conversely, every connection $A_\Si$ (\ref{spr291}) on the fibre
bundle $Y\to\Si$ yields the splitting
\mar{46a}\beq
A_\Si: TY\supset VY \ni \dot y^i\dr_i + \dot\si^m\dr_m \to (\dot
y^i -A^i_m\dot\si^m)\dr_i \label{46a}
\eeq
of the exact sequence (\ref{63a}). Using this splitting, one can
construct a first order differential operator \index{$\wt D$}
\mar{7.10}\ben
&& \wt D: J^1Y\to T^*X\op\otimes_Y V_\Si Y, \label{7.10}\\
&& \wt D= dx^\la\otimes(y^i_\la- A^i_\la -A^i_m\si^m_\la)\dr_i,
\nonumber
\een
on the composite bundle $Y\to X$. It is called \index{covariant
differential!vertical} the {\sl vertical covariant differential}.
\index{differential!covariant!vertical} This operator also can be
defined as the composition
\be
\wt D=\pr_1\circ D^\g: J^1Y\to T^*X\op\ot_YVY\to T^*X\op\otimes_Y
VY_\Si,
\ee
where $D^\g$ is the covariant differential (\ref{2116}) relative
to some composite connection $A_\Si\circ\G$ (\ref{b1.114}), but
$\wt D$ does not depend on the choice of a connection $\G$ on the
fibre bundle $\Si\to X$.

The vertical covariant differential (\ref{7.10}) possesses the
following important property. Let $h$ be a section of the fibre
bundle $\Si\to X$, and let $Y^h\to X$ be the restriction
(\ref{S10}) of the fibre bundle $Y\to\Si$ to $h(X)\subset \Si$.
This is a subbundle
\be
i_h:Y^h\to Y
\ee
of the fibre bundle $Y\to X$. Every connection $A_\Si$
(\ref{spr291}) induces the pull-back connection
\mar{mos83}\beq
A_h=i_h^*A_\Si=dx^\la\ot[\dr_\la+((A^i_m\circ h)\dr_\la h^m
+(A\circ h)^i_\la)\dr_i] \label{mos83}
\eeq
on $Y^h\to X$. Then the restriction of $\wt D$ (\ref{7.10}) to
\be
J^1i_h(J^1Y^h)\subset J^1Y
\ee
coincides with the familiar covariant differential $D^{A_h}$
(\ref{2116}) on $Y^h$ relative to the pull-back connection $A_h$
(\ref{mos83}).

\begin{remark}
Let $\G :Y\to J^1Y$ be a connection on a fibre bundle $Y\to X$. In
accordance with the canonical isomorphism $VJ^1Y= J^1VY$
(\ref{d020}), the vertical tangent map
\be
V\G :VY\to VJ^1Y
\ee
to $\G$ defines the connection \index{$V\G$}
\mar{43}\ben
&& V\G : VY\to J^1VY, \nonumber\\
&& V\G = dx^\la\otimes(\dr_\la +\G^i_\la\dr_i+\dr_j\G^i_\la\dot y^j
\dot\dr_i), \label{43}
\een
on the composite vertical tangent bundle
\be
VY\to Y\to X.
\ee
This is called the {\sl vertical connection}
\index{connection!vertical} to $\G$. Of course, the connection
$V\G$ projects onto $\G$. Moreover, $V\G$ is linear over $\G$.
Then the dual connection of $V\G$ on the composite vertical
cotangent bundle
\be
V^*Y\to Y\to X
\ee
reads \index{$V^*\G$}
\mar{44}\ben
&& V^*\G :V^*Y\to J^1V^*Y, \nonumber \\
&& V^*\G =dx^\la\otimes(\dr_\la +\G^i_\la\dr_i-\dr_j\G^i_\la \dot y_i
\dot\dr^j). \label{44}
\een
It is called the {\sl covertical connection}
\index{connection!covertical} to $\G$. If $Y\to X$ is an affine
bundle, the connection $V\G$ (\ref{43}) can be seen as the
composite connection generated by the connection $\G$ on $Y\to X$
and the linear connection
\mar{gm420}\beq
\wt\G= dx^\la\ot(\dr_\la +\dr_j\G^i_\la\dot y^j\dot\dr_i) +
dy^i\ot\dr_i \label{gm420}
\eeq
on the vertical tangent bundle $VY\to Y$.
\end{remark}

\chapter{Geometry of principal bundles}

Classical gauge theory is adequately formulated as Lagrangian
field theory on principal and associated bundles where gauge
potentials are identified with principal connections. The main
ingredient in this formulation is the bundle of principal
connections $C=J^1P/G$ whose sections are principal connections on
a principal bundle $P$ with a structure group $G$.

\section{Geometry of Lie groups}

Let $G$ be a topological group which is not reduced to the unit
$\bb$. \index{$\bb$} Let $V$ be a topological space. By a {\sl
continuous action} \index{action of a group} of $G$ on $V$ {\sl on
the left} \index{action of a group!on the left} is meant a
continuous map
\mar{51f11}\beq
\zeta: G\times V \to V, \qquad \zeta(g'g,v)=\zeta(g',\zeta(g,v)),
\label{51f11}
\eeq
If there is no danger of confusion, we denote $\zeta(g,v)=gv$. One
says that a group $G$ acts on $V$ {\sl on the right} \index{action
of a group!on the right} if the map (\ref{51f11}) obeys the
relations
\be
\zeta(g'g,v)=\zeta(g,\zeta(g',v)).
\ee
In this case, we agree to write $\zeta(g,v)=vg$.

\begin{remark} \mar{51r50} \label{51r50}  Strictly speaking, by an
action of a group $G$ on $V$ is meant a class of morphisms $\zeta$
(\ref{51f11}) which differ from each other in inner automorphisms
of $G$, that is,
\be
\zeta'(g,v) = \zeta(g'^{-1}gg',v)
\ee
for some element $g'\in G$.
\end{remark}

An action of $G$ on $V$ is called:

$\bullet$ {\sl effective}  \index{action of a group!effective} if
there is no $g\neq \bb$ such that $\zeta(g,v)=v$ for all $v\in V$,

$\bullet$ {\sl free} \index{action of a group!free} if, for any
two elements $v,v\in V$, there exists an element $g\in G$ such
that $\zeta(g,v)=v'$.

$\bullet$ {\sl transitive} \index{action of a group!transitive} if
there is no element $v\in V$ such that $\zeta(g,v)=v$ for all
$g\in G$.

\noindent Unless otherwise stated, an action of a group is assumed
to be effective. If an action $\zeta$ (\ref{51f11}) of $G$ on $V$
is transitive, then $V$ is called the {\sl homogeneous space},
\index{homogeneous space} homeomorphic to the quotient $V=G/H$ of
$G$ with respect to some subgroup $H\subset G$. If an action
$\zeta$ is both free and transitive, then $V$ is homeomorphic to
the group space of $G$. For instance, this is the case of action
of $G$ on itself by left ($\zeta=L_G$) \index{$L_G$} and right
($\zeta=R_G$) \index{$R_G$} multiplications.

Let $G$ be a connected real Lie group of finite dimension $\di G
>0$. A vector field $\xi$ on $G$ is called {\sl left-invariant}
\index{vector field!left-invariant} if
\be
\xi(g)=TL_g(\xi(\bb)), \qquad g\in G,
\ee
where $TL_g$ denotes the tangent morphism to the \index{$L_g$} map
\be
L_g:G\to gG.
\ee
Accordingly, {\sl right-invariant} vector fields $\xi$
\index{vector field!right-invariant} on $G$ obey the condition
\be
\xi(g)=TR_g(\xi(\bb)),
\ee
where $TR_g$ is the tangent morphism to the \index{$R_g$} map
\be
T_g:G\to gG.
\ee
Let $\cG_l$ \index{$\cG_l$} (resp. $\cG_r$) \index{$\cG_r$} denote
the Lie algebra of left-invariant  (resp. right-invariant) vector
fields on $G$. They are called the {\sl left} \index{Lie
algebra!left} and {\sl right} \index{Lie algebra!right} Lie
algebras of $G$, respectively. Every left-invariant vector field
$\xi_l(g)$ (resp. a right-invariant vector field $\xi_r(g)$) can
be associated to the element $v=\xi_l(\bb)$ (resp. $v=\xi_r(\bb)$)
of the tangent space $T_\bb G$ at the unit $\bb$ of $G$.
Accordingly, this tangent space is provided both with left and
right Lie algebra structures. Given $v\in T_\bb G$, let $v_l(g)$
and $v_r(g)$ be the corresponding left-invariant and
right-invariant vector fields on $G$, respectively. There is the
relation
\be
v_l(g)=(TL_g\circ TR^{-1}_g)(v_r(g)).
\ee
Let $\{\e_m=\e_m(\bb)\}$ \index{$\e_m$} (resp.
$\{\ve_m=\ve_m(\bb)\}$) \index{$\ve_m$} denote the basis for the
left (resp. right) Lie algebra, and let $c^k_{mn}$ be the {\sl
right structure constants}: \index{right structure constants}
\be
[\ve_m,\ve_n]=c^k_{mn}\ve_k.
\ee
The map $g\to g^{-1}$ yields an isomorphism
\be
\cG_l\ni \e_m \to\ve_m= -\e_m\in \cG_r
\ee
of left and right Lie algebras.

The tangent bundle
\mar{50f1}\beq
\pi_G:TG\to G \label{50f1}
\eeq
of a Lie group $G$ is trivial because of the isomorphisms
\be
&& \vr_l: TG\ni q\to (g=\pi_G(q), TL^{-1}_g(q))\in G\times \cG_l,\\
&& \vr_r: TG\ni q\to (g=\pi_G(q), TR^{-1}_g(q))\in G\times \cG_r.
\ee
Let $\zeta$ (\ref{51f11}) be a smooth action of a Lie group $G$ on
a smooth manifold $V$. Let us consider the tangent morphism
\mar{51f22}\beq
T\zeta: TG\times TV\to TV \label{51f22}
\eeq
to this action. Given an element $g\in G$, the restriction of
$T\zeta$ (\ref{51f22}) to $(g,0)\times TV$ is the tangent morphism
$T\zeta_g$ to the map
\be
\zeta_g:g\times V\to V.
\ee
Therefore, the restriction
\mar{51f33} \beq
T\zeta_G:\wh 0(G)\times TV \to TV \label{51f33}
\eeq
of the tangent morphism $T\zeta$ (\ref{51f22}) to  $\wh 0(G)\times
TV$ (where $\wh 0$ is the canonical zero section of $TG\to G$) is
called the {\sl tangent prolongation} of a smooth action of $G$ on
$V$. \index{tangent prolongation!of a group action}

In particular, the above mentioned morphisms
\be
TL_g=TL_G|_{(g,0)\times TG}, \qquad TR_g=TR_G|_{(g,0)\times TG}
\ee
are of this type. For instance, the morphism $TL_G$ (resp. $TR_G$)
(\ref{51f33}) defines the adjoint representation $g\to {\rm Ad}_g$
\index{adjoint representation!of a Lie group} (resp. $g\to {\rm
Ad}_{g^{-1}}$) of \index{${\rm Ad}_g$} a group $G$ in its right
Lie algebra $\cG_r$ (resp. left Lie algebra $\cG_l$) and the
identity representation in its left (resp. right) one.

Restricting $T\zeta$ (\ref{51f22}) to $T_\bb G\times \wh 0(V)$,
one obtains a homomorphism of the right (resp. left) Lie algebra
of $G$ to the Lie algebra $\cT(V)$ of vector field on $V$ if
$\zeta$ is a left (resp. right) action. We call this homomorphism
a {\sl representation} of the Lie algebra of $G$ in $V$.
\index{representation of a Lie algebra} For instance, a vector
field on a manifold $V$ associated to a local one-parameter group
$G$ of diffeomorphisms of $V$ (see Section 1.3) is exactly an
image of such a homomorphism of the one-dimensional Lie algebra of
$G$ to $\cT(V)$.

In particular, the adjoint representation ${\rm Ad}_g$ of a Lie
group $G$ in its right Lie algebra $\cG_r$ yields the
corresponding \index{adjoint representation!of a Lie algebra} {\sl
adjoint representation}
\mar{51f34}\beq
\ve': \ve\to {\rm ad}_{\ve'} (\ve)=[\ve',\ve],\qquad {\rm
ad}_{\ve_m}(\ve_n)=c^k_{mn}\ve_k, \label{51f34}
\eeq
of the right Lie algebra $\cG_r$ in itself. Accordingly, the
adjoint representation of the left Lie algebra $\cG_l$ in itself
reads
\be
{\rm ad}_{\e_m}(\e_n)=-c^k_{mn}\e_k,
\ee
where $c^k_{mn}$ are the right structure constants (\ref{51f34}).

\begin{remark} \label{51r22} \mar{51f22}
Let $G$ be a {\sl matrix group}, \index{matrix group} i.e., a
subgroup of the algebra $M(V)$ of endomorphisms of some
finite-dimensional vector space $V$. Then its Lie algebras are Lie
subalgebras of $M(V)$. In this case, the adjoint representation
${\rm Ad}_g$ of $G$ reads
\mar{51f35}\beq
{\rm Ad}_g(e)= geg^{-1}, \qquad e\in \cG. \label{51f35}
\eeq
\end{remark}

An exterior form $\f$ on a Lie group $G$ is said to be {\sl
left-invariant} \index{left-invariant form} (resp. {\sl
right-invariant}) \index{right-invariant form} if
$\f(\bb)=L^*_g(\f(g))$ (resp. $\f(\bb)=R_g^*(\f(g))$). The
exterior differential of a left-invariant (resp right-invariant)
form is left-invariant (resp. right-invariant). In particular, the
left-invariant one-forms satisfy the \index{Mourer--Cartan
equation} {\sl Maurer--Cartan equation}
\mar{50f2}\beq
d\f(\e,\e')=-\frac12\f([\e,\e']), \qquad \e,\e'\in \cG_l.
\label{50f2}
\eeq
There is the {\sl canonical $\cG_l$-valued left-invariant
one-form} \index{left-invariant form!canonical}
\mar{mos270}\beq
\thh_l:T_\bb G\ni\e \to \e\in \cG_l \label{mos270}
\eeq
on a Lie group $G$. The components $\thh_l^m$ of its decomposition
$\thh_l=\thh_l^m\e_m$ with respect to the basis for the left Lie
algebra $\cG_l$ make up the basis for the space of left-invariant
exterior one-forms on $G$:
\be
\e_m\rfloor\thh^n_l=\dl^n_m.
\ee
The Maurer--Cartan equation (\ref{50f2}), written with respect to
this basis, reads
\be
d\thh_l^m=\frac12c^m_{nk}\thh_l^n\w \thh_l^k.
\ee

\section{Bundles with structure groups}

Principal bundles are particular bundles with a structure group.
Since equivalence classes of these bundles are topological
invariants (see Theorem \ref{51t4}), we consider continuous
bundles with a structure topological group.

Let $G$ be a topological group. Let $\pi: Y\to X$ be a locally
trivial continuous bundle (see Remark \ref{11r1}) whose typical
fibre $V$ is provided with a certain left action (\ref{51f11}) of
a topological group $G$ (see Remark \ref{51r50}). Moreover, let
$Y$ admit an atlas
\mar{51f12}\beq
\Psi=\{(U_\al,\psi_\al), \vr_{\al\bt}\}, \qquad
\psi_\al=\vr_{\al\bt}\psi_\bt,  \label{51f12}
\eeq
whose transition functions $\vr_{\al\bt}$ (\ref{mos271}) factorize
as
\mar{51f15}\beq
\vr_{\al\bt}: U_\al\cap U_\bt \times V \ar U_\al\cap U_\bt
\times(G\times V) \ar^{\id \times\zeta} U_\al\cap U_\bt \times V
\label{51f15}
\eeq
through local continuous $G$-valued functions
\mar{51f14}\beq
\vr_{\al\bt}^G: U_\al\cap U_\bt \to G \label{51f14}
\eeq
on $X$. This means that transition morphisms $\vr_{\al\bt}(x)$
(\ref{sp22}) are elements of $G$ acting on $V$. Transition
functions (\ref{51f15}) are called {\sl $G$-valued}.
\index{transition functions!$G$-valued}

Provided with an atlas (\ref{51f12}) with $G$-valued transition
functions, a locally trivial continuous bundle $Y$ is called the
{\sl bundle with a structure group} \index{bundle!with a structure
group} $G$ \index{structure group} or, in brief, a {\sl
$G$-bundle}. \index{$G$-bundle} Two $G$-bundles $(Y,\Psi)$ and
$(Y,\Psi')$ are called equivalent \index{equivalent $G$-bundles}
if their atlases $\Psi$ and $\Psi'$ are equivalent. Atlases $\Psi$
and $\Psi'$ with $G$-valued transition functions are said to be
{\sl equivalent} \index{equivalent $G$-bundle atlases} iff, given
a common cover $\{U_i\}$ of $X$ for the union of these atlases,
there exists a continuous $G$-valued function $g_i$ on each $U_i$
such that
\mar{51f16}\beq
\psi'_i(x)=g_i(x)\psi_i(x), \qquad x\in U_i. \label{51f16}
\eeq

Let $h(X,G,V)$ denote the set of equivalence classes of continuous
bundles over $X$ with a structure group $G$ and a typical fibre
$V$. In order to characterize this set, let us consider the
presheaf $G^0_\sU$ of continuous $G$-valued functions on a
topological space $X$. Let $G^0_X$ be the sheaf of germs of these
functions generated by the canonical presheaf $G^0_\sU$, and let
$H^1(X;G^0_X)$ \index{$H^1(X;G^0_X)$} be the first cohomology of
$X$ with coefficients in $G^0_X$ (see Remark \ref{rrr1}). The
group functions $\vr_{\al\bt}^G$ (\ref{51f14}) obey the cocycle
condition
\be
\vr_{\al\bt}^G \vr_{\bt\g}^G= \vr_{\al\g}^G
\ee
on overlaps $U_\al\cap U_\bt \cap U_\g$ (cf. (\ref{spr192'})) and,
consequently, they form a one-cocycle $\{\vr_{\al\bt}^G\}$ of the
presheaf $G^0_\sU$. This cocycle is a representative of some
element of the first cohomology $H^1(X;G^0_X)$ of $X$ with
coefficients in the sheaf $G^0_X$.

Thus, any atlas of a $G$-bundle over $X$ defines an element of the
cohomology set $H^1(X;G^0_X)$. Moreover, it follows at once from
the condition (\ref{51f16}) that equivalent atlases define the
same element of $H^1(X;G^0_X)$. Thus, there is an injection
\mar{51f17}\beq
h(X,G,V)\to H^1(X;G^0_X) \label{51f17}
\eeq
of the set of equivalence classes of $G$-bundles over $X$ with a
typical fibre $V$ to the first cohomology $H^1(X;G^0_X)$ of $X$
with coefficients in the sheaf $G^0_X$. Moreover, the injection
(\ref{51f17}) is a bijection as follows.

\begin{theorem} \label{51t1} \mar{51t1}
There is one-to-one correspondence between the equivalence classes
of $G$-bundles over $X$ with a typical fibre $V$ and the elements
of the cohomology set $H^1(X;G^0_X)$.
\end{theorem}

The bijection (\ref{51f17}) holds for $G$-bundles with any typical
fibre $V$. Two $G$-bundles $(Y,\Psi)$ and $(Y',\Psi')$ over $X$
with different typical fibres are called {\sl associated}
\index{associated bundles} if the cocycles of transition functions
of their atlases $\Psi$ and $\Psi'$ are representatives of the
same element of the cohomology set $H^1(X;G^0_X)$. Then Theorem
\ref{51t1} can be reformulated as follows.

\begin{theorem} \label{51t2} \mar{51t2}
There is one-to-one correspondence between the classes of
associated $G$-bundles over $X$ and the elements of the cohomology
set $H^1(X;G^0_X)$.
\end{theorem}

Let $f:X'\to X$ be a continuous map.  Every continuous $G$-bundle
$Y\to X$ yields the pull-back bundle $f^*Y\to X'$ (\ref{mos106})
with the same structure group $G$. Therefore, $f$ induces the map
\be
[f]:H^1(X;G_X^0)\to H^1(X';G_{X'}^0).
\ee

\begin{theorem} \label{mos65} \mar{mos65}
Given a continuous $G$-bundle $Y$ over a paracompact base $X$, let
$f_1$ and $f_2$ be two continuous maps of $X'$ to $X$. If these
maps are homotopic, the pull-back $G$-bundles $f_1^*Y$ and
$f_2^*Y$ over $X'$ are equivalent.
\end{theorem}

Let us return to smooth fibre bundles. Let $G$, $\di G>0$, be a
real Lie group which acts on a smooth manifold $V$ on the left. A
smooth fibre bundle $\pi:Y\to X$ is called a {\sl bundle with a
structure group} $G$ \index{bundle!with a structure group!smooth}
if \index{$G$-bundle!smooth} it is a continuous $G$-bundle
possessing a smooth atlas $\Psi$ (\ref{51f12}) whose transition
functions factorize  as those (\ref{51f12}) through smooth
$G$-valued functions (\ref{51f14}).

\begin{example} \label{gal} \mar{gal}
Any vector (resp. affine) bundle of fibre dimension $\di V=m$ is a
bundle with a structure group which is the general linear group
$GL(m,\Bbb R)$ (resp. the general affine group $GA(m,\Bbb R)$).
\end{example}

Let $G^\infty_X$ be the sheaf of germs of smooth $G$-valued
functions on $X$ and $H^1(X;G^\infty_X)$ the first cohomology of a
manifold $X$ with coefficients in the sheaf $G^\infty_X$. The
following theorem is analogous to Theorem \ref{51t2}.

\begin{theorem} \label{51t3} \mar{51t3}
There is one-to-one correspondence between the classes of
associated smooth $G$-bundles over $X$ and the elements of the
cohomology set $H^1(X;G^\infty_X)$.
\end{theorem}

Since a smooth manifold is paracompact, one can show the
following.

\begin{theorem} \label{51t4} \mar{51t4}
There is a bijection
\mar{51f20}\beq
H^1(X;G^\infty_X)=H^1(X;G^0_X), \label{51f20}
\eeq
where a Lie group $G$ is treated as a topological group.
\end{theorem}

The bijection (\ref{51f20}) enables one to classify smooth
$G$-bundles as the continuous ones by means of topological
invariants.

\section{Principal bundles}

We restrict our consideration to smooth bundles with a structure
Lie group of non-zero dimension.

Given a real Lie group $G$, let
\mar{51f1}\beq
\pi_P :P\to X \label{51f1}
\eeq
be a  $G$-bundle whose typical fibre is the group space of $G$,
which a group $G$ acts on by left multiplications.  It is called a
{\sl principal bundle} \index{bundle!with a structure
group!principal} with \index{$G$-bundle!principal} a structure
group \index{bundle!principal} $G$. \index{principal!bundle}
Equivalently, a principal $G$-bundle is defined as a fibre bundle
$P$ (\ref{51f1}) which admits an {\sl action of $G$ on $P$}
\index{structure group!action} on \index{action of a structure
group!on $P$} the right by a fibrewise morphism \index{$R_{gP}$}
\mar{1}\ben
&& R_{GP}: G\op\times_X P \ar_X P,  \label{1}\\
&& R_{gP}: p\to pg, \qquad \pi_P(p)=\pi_P(pg), \qquad p\in P, \nonumber
\een
which \index{$R_{GP}$} is free and transitive on each fibre of
$P$. As a consequence, the quotient of $P$ with respect to the
action (\ref{1}) of $G$ is diffeomorphic to a base $X$, i.e.,
$P/G=X$.

\begin{remark} \mar{spb} \label{spb}
The definition of a {\sl continuous principal bundle}
\index{principal!bundle!continuous} is a repetition of that of a
smooth one, but all morphisms are continuous.
\end{remark}

A principal $G$-bundle $P$ is equipped with a bundle atlas
\mar{51f2}\beq
\Psi_P=\{(U_\al,\psi^P_\al),\vr_{\al\bt}\} \label{51f2}
\eeq
whose trivialization morphisms
\be
\psi_\al^P: \pi_P^{-1}(U_\al)\to U_\al\times G
\ee
obey the condition
\be
\psi_\al^P(pg)=g\psi_\al^P(p), \qquad g\in G.
\ee
Due to this property, every trivialization morphism $\psi^P_\al$
determines a unique local section $z_\al:U_\al\to P$ such that
\be
(\psi^P_\al\circ z_\al)(x)=\bb, \qquad x\in U_\al.
\ee
The transformation law for $z_\al$ reads
\mar{b1.202}\beq
z_\bt(x)=z_\al(x)\vr_{\al\bt}(x),\qquad x\in U_\al\cap
U_\bt.\label{b1.202}
\eeq
Conversely, the family
\mar{vcv}\beq
\{(U_\al,z_\al),\vr_{\al\bt}\} \label{vcv}
\eeq
of local sections of $P$ which obey the transformation law
(\ref{b1.202}) uniquely determines a bundle atlas $\Psi_P$ of a
principal bundle $P$.

\begin{theorem} \label{52a3} \mar{52a3}
A principal bundle admits a global section iff it is trivial.
\end{theorem}

\begin{example} \mar{52e1} \label{52e1} Let $H$ be a closed
subgroup of a real Lie group $G$. Then $H$ is a Lie group. Let
$G/H$ be the quotient of $G$ with respect to an action of $H$ on
$G$ by right multiplications. Then
\mar{ggh}\beq
\pi_{GH}:G\to G/H \label{ggh}
\eeq
is a principal $H$-bundle. If $H$ is a maximal compact subgroup of
$G$, then $G/H$ is diffeomorphic to $\Bbb R^m$ and, by virtue of
Theorem \ref{11t3}, $G\to G/H$ is a trivial bundle, i.e., $G$ is
diffeomorphic to the product $\Bbb R^m\times H$.
\end{example}

\begin{remark} \mar{52r1} \label{52r1}
The pull-back $f^*P$ (\ref{mos106}) of a principal bundle also is
a principal bundle with the same structure group.
\end{remark}

\begin{remark}  \mar{52r1'} \label{52r1'}
Let $P\to X$ and $P'\to X'$ be principal $G$- and $G'$-bundles,
respectively. A bundle morphism $\Phi:P\to P'$ is a {\sl morphism
of principal bundles} \index{bundle!morphism!of principal bundles}
if there exists a Lie group homomorphism $\g:G\to G'$ such that
\be
\Phi(pg)=\Phi(p)\g(g).
\ee
In particular, equivalent principal bundles are isomorphic.
\end{remark}

Any class of associated smooth bundles on $X$ with a structure Lie
group $G$ contains a principal bundle. In other words, any smooth
bundle with a structure Lie group $G$ is associated with some
principal bundle.

Let us consider the tangent morphism
\mar{1aa}\beq
TR_{GP}: (G\times \cG_l)\op\times_X TP \ar_X TP  \label{1aa}
\eeq
to the right action $R_{GP}$ (\ref{1}) of $G$ on $P$. Its
restriction to $T_\bb G\op\times_X TP$ provides a homomorphism
\mar{52f44}\beq
\cG_l\ni \e\to \xi_\e\in \cT(P) \label{52f44}
\eeq
of the left Lie algebra $\cG_l$ of $G$ to the Lie algebra $\cT(P)$
of vector fields on $P$. Vector fields $\xi_\e$ (\ref{52f44}) are
obviously  vertical. They are called {\sl fundamental vector
fields}. \index{vector field!fundamental} Given a basis $\{\e_r\}$
for $\cG_l$, the corresponding fundamental vector fields
$\xi_r=\xi_{\e_r}$ form a family of $m=\di\cG_l$ nowhere vanishing
and linearly independent sections of the vertical tangent bundle
$VP$ of $P\to X$. Consequently, this bundle is trivial
\mar{ttt91}\beq
VP = P\times\cG_l \label{ttt91}
\eeq
by virtue of Theorem \ref{12t10}.

Restricting the tangent morphism $TR_{GP}$ (\ref{1aa}) to
\mar{2aa}\beq
TR_{GP}: \wh 0(G)\op\times_X TP \ar_X TP,  \label{2aa}
\eeq
we obtain the {\sl tangent prolongation} of the structure group
action $R_{GP}$ (\ref{1}). \index{tangent prolongation!of a
structure group action} If there is no danger of confusion, it is
simply called the {\sl action of $G$ on $TP$}. \index{action of a
structure group!on $TP$} Since the action of $G$ (\ref{1}) on $P$
is fibrewise, its action (\ref{2aa})  is restricted to the
vertical tangent bundle $VP$ of $P$.

Taking the quotient of the tangent bundle $TP\to P$ and the
vertical tangent bundle $VP$ of $P$ by $G$ (\ref{2aa}), we obtain
the vector bundles \index{$T_GP$}
\mar{b1.205}\beq
 T_GP=TP/G,\qquad  V_GP=VP/G \label{b1.205}
\eeq
over $X$. \index{$V_GP$} Sections of $T_GP\to X$ are $G$-invariant
vector fields on $P$. Accordingly, sections of $V_GP\to X$ are
$G$-invariant vertical vector fields on $P$. Hence, a typical
fibre of $V_GP\to X$ is the right Lie algebra $\cG_r$ of $G$
subject to the adjoint representation of a structure group $G$.
Therefore, $V_GP$ (\ref{b1.205}) is called the {\sl Lie algebra
bundle}. \index{Lie algebra bundle}

Given a bundle atlas $\Psi_P$ (\ref{51f2}) of $P$, there is the
corresponding atlas
\mar{52f57}\beq
\Psi=\{(U_\al,\psi_\al), {\rm Ad}_{\vr_{\al\bt}}\} \label{52f57}
\eeq
of the Lie algebra bundle $V_GP$, which is provided with bundle
coordinates $(U_\al; x^\m,\chi^m)$ with respect to the fibre
frames
\be
\{e_m=\psi_\al^{-1}(x)(\ve_m)\},
\ee
where $\{\ve_m\}$ is a basis for the Lie algebra $\cG_r$. These
coordinates obey the transformation rule
\mar{52f40}\beq
\vr(\chi^m) \ve_m=\chi^m {\rm Ad}_{\vr^{-1}}(\ve_m). \label{52f40}
\eeq
A glance at this transformation rule shows that $V_GP$ is a bundle
with a structure group $G$. Moreover, it is associated with a
principal $G$-bundle $P$ (see Example \ref{56e1}).

Accordingly, the vector bundle $T_GP$ (\ref{b1.205}) is endowed
with bundle coordinates $(x^\m,\dot x^\m,\chi^m)$ with respect to
the fibre frames $\{\dr_\m,e_m\}$. Their transformation rule is
\mar{52f40a}\beq
\vr(\chi^m)\ve_m=\chi^m{\rm Ad}_{\vr^{-1}}(\ve_m) +\dot x^\m
R^m_\m\ve_m. \label{52f40a}
\eeq
If $G$ is a matrix group (see Remark \ref{51r22}), this
transformation rule reads
\mar{52f40b}\beq
\vr(\chi^m)\ve_m = \chi^m\vr^{-1}\ve_m\vr - \dot
x^\m\dr_\m(\vr^{-1})\vr. \label{52f40b}
\eeq
Since the second term in the right-hand sides of expressions
(\ref{52f40a}) -- (\ref{52f40b}) depend on derivatives of a
$G$-valued function $\vr$ on $X$, the vector bundle $T_GP$
(\ref{b1.205}) fails to be a $G$-bundle.

The Lie bracket of $G$-invariant vector fields on $P$ goes to the
quotient by $G$ and defines the Lie bracket of sections of the
vector bundle $T_GP\to X$. This bracket reads
\mar{1128,9}\ben
&& \xi =\xi^\la\dr_\la + \xi^p e_p,\qquad \eta = \eta^\m \dr_\m +
\eta^q e_q, \label{1128}\\
&& [\xi,\eta ]=(\xi^\m\dr_\m\eta^\la -
\eta^\m\dr_\m\xi^\la)\dr_\la
 + \label{1129} \\
&& \qquad (\xi^\la\dr_\la\eta^r - \eta^\la\dr_\la\xi^r +
c_{pq}^r\xi^p\eta^q) e_r. \nonumber
\een
Putting $\xi^\la=0$ and $\eta^\m=0$ in the formulas (\ref{1128})
-- (\ref{1129}), we obtain the Lie bracket
\mar{1129'}\beq
[\xi,\eta]= c_{pq}^r\xi^p\eta^q e_r \label{1129'}
\eeq
of sections of the Lie algebra bundle $V_GP$. A glance at the
expression (\ref{1129'}) shows that sections of $V_GP$ form a
finite-dimensional Lie $C^\infty(X)$-algebra, called the {\sl
gauge algebra}. \index{gauge!algebra} Therefore, $V_GP$ also is
called the {\sl gauge algebra bundle}. \index{gauge!algebra
bundle}

\section{Principal connections}

Principal connections on a principal bundle $P$ (\ref{51f1}) are
connections on $P$ which are equivariant with respect to the right
action (\ref{1}) of a structure group $G$ on $P$. In order to
describe them, we follow the definition of connections on a fibre
bundle $Y\to X$ as global sections of the affine jet bundle
$J^1Y\to X$ (Theorem \ref{mnh}).

Let $J^1P$ be the first order jet manifold of a principal
$G$-bundle $P\to X$ (\ref{51f1}). Then connections on a principal
bundle $P\to X$ are global sections
\mar{c4c}\beq
A: P\to J^1P \label{c4c}
\eeq
of the affine jet bundle $J^1P\to P$ modelled over the vector
bundle
\be
 T^*X\op\ot_P VP=(T^*X\op\ot_P \cG_l).
\ee
In order to define principal connections on $P\to X$, let us
consider the jet prolongation
\be
J^1R_G: J^1(X\times G)\op\times_X J^1P \to J^1P
\ee
of the morphism $R_{GP}$ (\ref{1}). Restricting this morphism to
\be
J^1R_G: \wh 0(G)\op\times_X J^1P \to J^1P,
\ee
we obtain the {\sl jet prolongation} of the structure group action
$R_{GP}$ (\ref{1}) \index{jet prolongation!of a structure group
action} called, simply, the {\sl action of $G$ on $J^1P$}.
\index{action of a structure group!on $J^1P$} It reads
\mar{53f1}\beq
G\ni g: j^1_xp\to (j^1_xp)g =j^1_x(pg). \label{53f1}
\eeq
Taking the quotient of the affine jet bundle $J^1P$ by $G$
(\ref{53f1}), we obtain the affine bundle
\mar{B1}\beq
C=J^1P/G\to X\label{B1}
\eeq
modelled over the vector bundle
\be
\ol C=T^*X\op\ot_X V_GP\to X.
\ee
Hence, there is the vertical splitting
\be
VC= C\op\ot_X \ol C
\ee
of the vertical tangent bundle $VC$ of $C\to X$.

\begin{remark} \mar{53r1} \label{53r1} A glance at the expression
(\ref{53f1}) shows that the fibre bundle $J^1P\to C$ is a
principal bundle with the structure group $G$. It is canonically
isomorphic to the pull-back
\mar{b1.251}\beq
J^1P= P_C=C\op\times_X P\to C. \label{b1.251}
\eeq
\end{remark}

Taking the quotient with respect to the action of a structure
group $G$, one can reduce the canonical imbedding (\ref{18})
(where $Y=P$) to the bundle monomorphism
\mar{za1}\ben
&& \la_C: C\ar_X T^*X\op\ot_X T_GP, \nonumber\\
&& \la_C: dx^\m\ot(\dr_\m + \chi_\m^m e_m). \label{za1}
\een
It follows that, given atlases $\Psi_P$ (\ref{51f2}) of $P$ and
$\Psi$ (\ref{52f57}) of $T_GP$, the bundle of principal
connections $C$ is provided with bundle coordinates
$(x^\la,a^m_\m)$ possessing the transformation rule
\mar{53f10}\beq
\vr(a^m_\m)\ve_m=(a^m_\nu{\rm Ad}_{\vr^{-1}}(\ve_m) +
R^m_\nu\ve_m)\frac{\dr x^\nu}{\dr x'^\m}. \label{53f10}
\eeq
If $G$ is a matrix group, this transformation rule reads
\mar{53f11}\beq
\vr(a^m_\m)\ve_m = (a^m_\nu\vr^{-1}(\ve_m)\vr -
\dr_\m(\vr^{-1})\vr)\frac{\dr x^\nu}{\dr x'^\m}. \label{53f11}
\eeq
A glance at this expression shows that the bundle of principal
connections $C$ as like as the vector bundle $T_GP$ (\ref{b1.205})
fails to be a bundle with a structure group $G$.

As was mentioned above, a connection $A$ (\ref{c4c}) on a
principal bundle $P\to X$ is called a {\sl principal connection}
\index{connection!principal} if \index{principal!connection} it is
{\sl equivariant} \index{equivariant!connection} under the action
(\ref{53f1}) of a structure group $G$, i.e.,
\mar{b1.210}\beq
A(pg)= A(p)g \qquad  g\in G. \label{b1.210}
\eeq
There is obvious one-to-one correspondence between the principal
connections on a principal $G$-bundle $P$ and global sections
\mar{BB1}\beq
A:X\to C \label{BB1}
\eeq
of the bundle $C\to X$ (\ref{B1}), called the {\sl bundle of
principal connections}. \index{bundle!of principal connections}

\begin{theorem} \label{53a1} \mar{53a1}
Since the bundle of principal connections $C\to X$ is affine,
principal connections on a principal bundle always exist.
\end{theorem}

Due to the bundle monomorphism (\ref{za1}), any principal
connection $A$ (\ref{BB1}) is represented by a $T_GP$-valued form
\mar{1131}\beq
A=dx^\la\ot (\pdr_\la + A_\la^q e_q).  \label{1131}
\eeq
Taking the quotient with respect to the action of a structure
group $G$, one can reduce the exact sequence (\ref{1.8a}) (where
$Y=P$) to the exact sequence
\mar{1.33}\beq
0\to V_GP\ar_X T_GP\ar TX\to 0. \label{1.33}
\eeq
A principal connection $A$ (\ref{1131}) defines a splitting of
this exact sequence.

\begin{remark} \label{53rr} \mar{53rr}
A principal connection $A$ (\ref{c4c}) on a principal bundle $P\to
X$ can be represented by the vertical-valued form $A$
(\ref{b1.223}) on $P$ which is a $\cG_l$-valued form due to the
trivialization (\ref{ttt91}). It is the familiar $\cG_l$-valued
{\sl connection form} \index{connection form!of a principal
connection} on a principal bundle $P$. Given a local bundle
splitting $(U_\al,z_\al)$ of $P$, this form reads
\mar{mos166}\beq
\ol A=\psi_\al^*(\thh_l - \ol A^q_\la dx^\la\ot\e_q),
\label{mos166}
\eeq
where $\thh_l$ is the canonical $\cG_l$-valued one-form
(\ref{mos270}) on $G$ and $A^p_\la$ are local functions on $P$
such that
\be
\ol A^q_\la(pg)\e_q =\ol A^q_\la(p){\rm Ad}_{g^{-1}}(\e_q).
\ee
The pull-back $z^*_\al\ol A$ of the connection form $\ol A$
(\ref{mos166}) onto $U_\al$ is the well-known  \index{connection
form!of a principal connection!local} {\sl local connection
one-form}
\mar{b1.225}\beq
A_\al= -A^q_\la dx^\la\ot\e_q=A^q_\la dx^\la\ot\ve_q,
\label{b1.225}
\eeq
where $A^q_\la=\ol A^q_\la\circ z_\al$ are local functions on $X$.
It is readily observed that the coefficients $A^q_\la$ of this
form are exactly the coefficients of the form (\ref{1131}).
\end{remark}

There are both pull-back and push-forward operations of principal
connections.

\begin{theorem} \label{mos252} \mar{mos252}
Let $P$ be a principal bundle and $f^*P$ (\ref{mos106}) the
pull-back principal bundle with the same structure group. Let
$f_P$ be the canonical morphism (\ref{mos81}) of $f^*P$ to $P$. If
$A$ is a principal connection on $P$, then the pull-back
connection $f^*A$ (\ref{mos82}) on $f^*P$ is a principal
connection.
\end{theorem}

\begin{theorem} \label{mos253} \mar{mos253}
Let $P'\to X$ and $P\to X$ be principle bundles with structure
groups $G'$ and $G$, respectively. Let $\Phi: P'\to P$ be a
principal bundle morphism over $X$ with the corresponding
homomorphism $G'\to G$ (see Remark \ref{52r1'}). For every
principal connection $A'$ on $P'$, there exists a unique principal
connection $A$ on $P$ such that $T\Phi$ sends the horizontal
subspaces of $TP'$ $A'$ onto the horizontal subspaces of $TP$ with
respect to $A$.
\end{theorem}

Let $P\to X$ be a principal $G$-bundle. The Fr\"olicher--Nijenhuis
bracket (\ref{1149}) on the space $\cO^*(P)\ot\cT(P)$ of
tangent-valued forms on $P$ is compatible with the right action
$R_{GP}$ (\ref{1}). Therefore, it induces the
Fr\"olicher--Nijenhuis bracket on the space $\cO^*(X)\ot T_GP(X)$
of $T_GP$-valued forms on $X$, where $T_GP(X)$ is the vector space
of sections of the vector bundle $T_GP\to X$. Note that, as it
follows from the exact sequence (\ref{1.33}), there is an
epimorphism
\be
T_GP(X)\to \cT(X).
\ee

Let $A\in \cO^1(X)\ot T_GP(X)$ be a principal connection
(\ref{1131}). The associated Nijenhuis differential is
\mar{1159b}\ben
&& d_A : \cO^r(X)\ot T_GP(X)\to \cO^{r+1}(X)\ot
V_GP(X),
\nonumber \\
&& d_A\f = [A,\f]_{\rm FN}, \quad \f\in \cO^r(X)\ot T_GP(X).
\label{1159b}
\een
The {\sl strength} \index{strength} of a principal connection $A$
(\ref{1131}) is defined as the $V_GP$-valued two-form
\mar{mos36}\beq
F_A = \frac{1}{2} d_AA = \frac{1}{2} [A, A]_{\rm FN} \in
\cO^2(X)\ot V_GP(X). \label{mos36}
\eeq
Its coordinated expression
\mar{1136b}\ben
&& F_A
=\frac12 F^r_{\la\m} dx^\la\w dx^\m\ot e_r, \nonumber \\
&& F_{\la\m}^r = [\dr_\la +A^p_\la e_p, \dr_\m +A^q_\m
e_q]^r= \label{1136b}\\
&& \qquad \pdr_\la A_\m^r - \pdr_\m A_\la^r + c_{pq}^rA_\la^p
A_\m^q,\nonumber
\een
results from the bracket (\ref{1129}).

\begin{remark} \label{nnx} \mar{nnx}
It should be emphasized that the strength $F_A$ (\ref{mos36}) is
not the standard curvature (\ref{1178a}) of a principal connection
because $A$ (\ref{1131}) is not a tangent-valued form. The {\sl
curvature of a principal connection} \index{curvature!of a
principal connection} $A$ (\ref{c4c}) on $P$ is the $VP$-valued
two-form $R$ (\ref{1178a}) on $P$, which is brought into the
$\cG_l$-valued form owing to the canonical isomorphism
(\ref{ttt91}).
\end{remark}

\begin{remark} \label{conj} \mar{conj} Given a principal
connection $A$ (\ref{BB1}), let $\Phi_C$ be a vertical principal
automorphism of the bundle of principal connections $C$. The
connection $A'=\Phi_C \circ A$ is called {\sl conjugate}
\index{principal!connection!conjugate} to a principal connection
$A$. The strength forms (\ref{mos36}) of conjugate principal
connections $A$ and $A'$ coincide with each other, i.e.,
$F_A=F_{A'}$.
\end{remark}

\section{Canonical principal connection}

Given a principal $G$-bundle $P\to X$ and its jet manifold $J^1P$,
let us consider the canonical morphism $\thh_{(1)}$ (\ref{18})
where $Y=P$. By virtue of Remark \ref{mos30}, this morphism
defines the morphism
\be
\thh:J^1P\op\times_P TP\to VP.
\ee
Taking its quotient with respect to $G$, we obtain the morphism
\mar{264}\ben
&& C\op\times_X T_GP \ar^\thh V_GP,   \label{264} \\
&&  \thh(\dr_\la)= -a_\la^p e_p, \qquad \thh(e_p) = e_p. \nonumber
\een
Consequently, the exact sequence (\ref{1.33}) admits the canonical
splitting over $C$.

In view of this fact, let us consider the pull-back principal
$G$-bundle $P_C$ (\ref{b1.251}). Since
\mar{265}\beq
V_G(C\op\times_X P) = C\op\times_X V_GP, \qquad T_G(C\op\times_X
P)=TC\op\times_X T_GP, \label{265}
\eeq
the exact sequence (\ref{1.33}) for the principal bundle $P_C$
reads
\mar{mos34}\beq
0\to C\op\times_X V_GP\ar_C TC\op\times_XT_GP\ar TC\to 0.
\label{mos34}
\eeq
The morphism (\ref{264}) yields the horizontal splitting
(\ref{mos35}):
\be
TC\op\times_XT_GP \ar C\op\times_XT_GP\ar C\op\times_X V_GP,
\ee
of the exact sequence (\ref{mos34}). Thus, it defines the
principal connection
\mar{266}\ben
&& \cA: TC\to TC\op\times_XT_GP, \nonumber\\
&& \cA =dx^\la\ot(\dr_\la +a_\la^p e_p) + da^r_\la\ot\dr^\la_r,
\label{266}\\
&& \cA\in \cO^1(C)\ot T_G(C\op\times_X P)(X), \nonumber
\een
on the principal bundle
\mar{bng}\beq
P_C= C\op\times_X P\to C. \label{bng}
\eeq
It follows that the principal bundle $P_C$ admits  the {\sl
canonical principal connection} \index{canonical principal
connection} \index{principal!connection!canonical} (\ref{266}).
\index{connection!principal!canonical}

Following the expression (\ref{mos36}), let us define the strength
\mar{267}\ben
&& F_\cA =\frac{1}{2} d_\cA \cA = \frac{1}{2} [\cA,\cA]
\in \cO^2(C)\ot V_GP(X), \nonumber \\
&& F_\cA =(da_\m^r\w dx^\m + \frac{1}{2} c_{pq}^r a_\la^p
a_\m^q dx^\la\w dx^\m)\ot e_r, \label{267}
\een
of the canonical principal connection $\cA$ (\ref{266}). It is
called the {\sl canonical strength} \index{strength!canonical}
because, given a principal connection $A$ (\ref{BB1}) on a
principal bundle $P\to X$, the pull-back
\mar{268}\beq
A^*F_\cA = F_A \label{268}
\eeq
is the strength (\ref{1136b}) of $A$.

With the $V_GP$-valued two-form $F_\cA$ (\ref{267}) on $C$, let us
define the $V_GP$-valued horizontal two-form \index{$\cF$}
\mar{295}\ben
&& \cF=h_0(F_\cA)=\frac{1}{2} \cF_{\la\m}^r dx^\la\w dx^\m\ot
\ve_r, \nonumber \\
&& \cF_{\la\m}^r = a_{\la\m}^r -
a_{\m\la}^r +c_{pq}^r a_\la^p a_\m^q, \label{295}
\een
on $J^1C$. It is called the {\sl strength form}.
\index{strength!form} For each principal connection $A$
(\ref{BB1}) on $P$, the pull-back
\mar{268a}\beq
J^1A^*\cF = F_A \label{268a}
\eeq
is the strength (\ref{1136b}) of $A$.

The strength form (\ref{295}) yields an affine surjection
\mar{294}\beq
\cF/2:J^1C\ar_C C\op\times_X(\op\w^2T^*X\ot V_GP) \label{294}
\eeq
over $C$ of the affine jet bundle $J^1C\to C$ to the vector (and,
consequently, affine) bundle
\be
C\op\times_X(\op\w^2T^*X\ot V_GP) \to C.
\ee
By virtue of Theorem \ref{pomm}, its kernel $C_+=\Ker\, \cF/2$ is
an affine subbundle of $J^1C\to C$. Thus, we have the canonical
splitting of the affine jet bundle
\mar{296,'}\ben
&& J^1C =C_+\op\oplus_C C_-=C_+\op\oplus_C (C\op\times_X\op\w^2T^*X\ot V_GP),
\label{296}\\
&&a_{\la\m}^r = \frac12(\cF_{\la\m}^r + \cS_{\la\m}^r)= \frac{1}{2}(a_{\la\m}^r + a_{\m\la}^r
 - c_{pq}^r a_\la^p a_\m^q) +
\label{296'}\\
&& \qquad   \frac{1}{2}
(a_{\la\m}^r - a_{\m\la}^r + c_{pq}^r a_\la^p a_\m^q). \nonumber
\een

The jet manifold $J^1C$ plays a role of the configuration space of
classical gauge theory on principal bundles.

\section{Gauge transformations}

In classical gauge theory, {\sl gauge transformations}
\index{gauge!transformation} are defined as principal
automorphisms of a principal bundle $P$. In accordance with Remark
\ref{52r1'}, an automorphism $\Phi_P$ of a principal $G$-bundle
$P$ is called {\sl principal} \index{automorphism!principal} if
\index{principal!automorphism} it is {\sl equivariant}
\index{equivariant!automorphism} under the right action (\ref{1})
of a structure group $G$ on $P$, i.e.,
\mar{55ff1}\beq
\Phi_P(pg)=\Phi_P(p)g, \qquad g\in G, \qquad p\in P. \label{55ff1}
\eeq

In particular, every vertical principal automorphism of a
principal bundle $P$ is represented as
\mar{b3111}\beq
\Phi_P(p)=pf(p), \qquad p\in P, \label{b3111}
\eeq
where $f$ is a $G$-valued {\sl equivariant function}
\index{equivariant!function} on $P$, i.e.,
\mar{b3115}\beq
f(pg)=g^{-1}f(p)g, \qquad g\in G. \label{b3115}
\eeq
There is one-to-one correspondence between the equivariant
functions $f$ (\ref{b3115}) and the global sections $s$ of the
associated {\sl group bundle} \index{group bundle}
\mar{55f1}\beq
\pi_{P^G}: P^G\to X \label{55f1}
\eeq
whose \index{$P^G$} fibres are groups isomorphic to $G$ and whose
typical fibre is the group $G$ which acts on itself by the adjoint
representation. This one-to-one correspondence is defined by the
relation
\mar{56f2}\beq
s(\pi_P(p))p = pf(p), \qquad p\in P, \label{56f2}
\eeq
(see Example \ref{56e2}). The group of vertical principal
automorphisms of a principal $G$-bundle is called the {\sl gauge
group}. \index{gauge!group} It is isomorphic to the group $P^G(X)$
of global sections of the group bundle (\ref{55f1}). Its unit
element is the canonical global section $\wh \bb$ of $P^G\to X$
whose values are unit elements of fibres of $P^G$.

\begin{remark} \label{pol} \mar{pol}
Note that transition functions of atlases of a principle bundle
$P$ also are represented by local sections of the associated group
bundle $P^G$ (\ref{55f1}).
\end{remark}

Let us consider (local) one-parameter groups of principal
automorphisms of $P$. Their infinitesimal generators are
$G$-invariant projectable vector fields $\xi$ on $P$, and {\it
vice versa}. We call $\xi$ \index{vector field!principal} the {\sl
principal vector fields} \index{principal!vector field} or the
{\sl infinitesimal gauge transformations}.
\index{gauge!transformation!infinitesimal} They are represented by
sections $\xi$ (\ref{1128}) of the vector bundle $T_GP$
(\ref{b1.205}). Principal vector fields constitute a real Lie
algebra $T_GP(X)$ with respect to the Lie bracket (\ref{1129}).
{\sl Vertical principal vector fields} \index{principal!vector
field!vertical} are the sections
\mar{b3106}\beq
\xi=\xi^p e_p \label{b3106}
\eeq
of the gauge algebra bundle $V_GP\to X$ (\ref{b1.205}). They form
a finite-dimensional Lie $C^\infty(X)$-algebra $\ccG(X)=V_GP(X)$
(\ref{1129'}) \index{$\ccG(X)$} that has been called the gauge
algebra.

Any (local) one-parameter group of principal automorphism $\Phi_P$
(\ref{55ff1}) of a principal bundle $P$ admits the jet
prolongation $J^1\Phi_P$ (\ref{1.21a}) to a one-parameter group of
$G$-equivariant automorphism of the jet manifold $J^1P$ which, in
turn, yields a one-parameter group of {\sl principal
automorphisms} \index{principal!automorphism!of a connection
bundle} $\Phi_C$ of the bundle of principal connections $C$
(\ref{B1}). Its infinitesimal generator is a vector field on $C$,
called the principal vector field on $C$ and regarded as an
infinitesimal gauge transformation of $C$. Thus, any principal
vector field $\xi$ (\ref{1128}) on $P$ yields a principal vector
field $u_\xi$ on $C$, which can be defined as follows.

Using the morphism (\ref{264}), we obtain the morphism
\be
\xi\rfloor\thh :C\ar_X V_GP,
\ee
which is a section of of the Lie algebra bundle
\be
V_G(C\op\times_X P)\to C
\ee
in accordance with the first formula (\ref{265}). Then the
equation
\be
u_\xi\rfloor F_\cA = d_\cA(\xi\rfloor\thh)
\ee
uniquely determines a desired vector field $u_\xi$ on $C$. A
direct computation leads to \index{$u_\xi$}
\mar{277}\beq
u_\xi = \xi^\m\dr_\m + (\dr_\m\xi^r + c_{pq}^r a_\m^p\xi^q -
a_\nu^r\dr_\m \xi^\nu) \dr_r^\m. \label{277}
\eeq
In particular, if $\xi$ is a vertical principal field
(\ref{b3106}), we obtain the vertical vector field
\mar{279}\beq
u_\xi = (\dr_\m\xi^r + c_{pq}^r a_\m^p\xi^q)\dr_r^\m. \label{279}
\eeq

\begin{remark}
The jet prolongation (\ref{1.21}) of the vector field $u_\xi$
(\ref{277}) onto $J^1C$ reads
\mar{281}\ben
&&J^1u_\xi =u_\xi +(\dr_{\la\m}\xi^r + c_{pq}^r a_\m^p\dr_\la\xi^q
+c_{pq}^r a_{\la\m}^p\xi^q - \label{281}\\
&& \qquad
a_\nu^r\dr_{\la\m} \xi^\nu  - a_{\la\nu}^r\dr_\m \xi^\nu
 -a_{\nu\m}^r\dr_\la \xi^\nu)\dr_
r^{\la\m}. \nonumber
\een
\end{remark}

\begin{example}
Let $A$ (\ref{1131}) be a principal connection on $P$. For any
vector field $\tau$ on $X$, this connection yields a section
\be
\tau\rfloor A= \tau^\la\dr_\la + A_\la^p\tau^\la e_p
\ee
of the vector bundle $T_GP\to X$. It, in turn, defines a principal
vector field (\ref{277}) on the bundle of principal connection $C$
which reads
\mar{mos39}\ben
&& \tau_A =\tau^\la\dr_\la +(\dr_\m (A_\nu^r \tau^\nu)
+c_{pq}^ra_\m^p A_\nu^q \tau^\nu -a_\nu^r\dr_\m \tau^\nu)
\dr_r^\m, \label{mos39}\\
&& \xi^\la=\tau^\la, \qquad \xi^p=A^p_\nu\tau^\nu. \nonumber
\een
\end{example}

It is readily justified that the monomorphism
\mar{55f5}\beq
T_GP(X)\ni\xi \to u_\xi\in \cT(C) \label{55f5}
\eeq
obeys the equality
\mar{55f5'}\beq
u_{[\xi,\eta]}=[u_\xi, u_\eta], \label{55f5'}
\eeq
i.e., it is a monomorphism of the real Lie algebra $T_GP(X)$ to
the real Lie algebra $\cT(C)$. In particular, the image of the
gauge algebra $\ccG(X)$ in $\cT(C)$ also is a real Lie algebra,
but not the $C^\infty(X)$-one because
\be
u_{f\xi}\neq fu_\xi, \qquad f\in C^\infty(X).
\ee

\begin{remark} \label{55rrr} \mar{55rrr}
A glance at the expression (\ref{277}) shows that the monomorphism
(\ref{55f5}) is a linear first order differential operator which
sends sections of the pull-back bundle
\be
C\op\times_X T_GP\to C
\ee
onto sections of the tangent bundle $TC\to C$. Refereing to
Definition \ref{s7}, we therefore can treat principal vector
fields (\ref{277}) as infinitesimal gauge transformations
depending on gauge parameters $\xi\in T_GP(X)$.
\end{remark}

\section{Geometry of associated bundles}

Given a principal $G$-bundle $P$ (\ref{51f1}), any associated
$G$-bundle over $X$ with a typical fibre $V$ is equivalent to the
following one.

Let us consider the quotient
\mar{b1.230}\beq
Y=(P\times V)/G \label{b1.230}
\eeq
of the product $P\times V$ by identification of elements $(p,v)$
and $(pg,g^{-1}v)$ for all $g\in G$. Let $[p]$ denote the
restriction of the canonical surjection
\mar{mos75}\beq
P\times V\to (P\times V)/G \label{mos75}
\eeq
to the subset $\{p\}\times V$ so that
\be
[p](v)=[pg](g^{-1}v).
\ee
Then the map
\be
Y\ni[p](V)\to \pi_P(p)\in X
\ee
makes the quotient $Y$ (\ref{b1.230}) into a fibre bundle over
$X$. This is a smooth $G$-bundle with the typical fibre $V$ which
is associated with the principal $G$-bundle $P$. For short, we
call it the {\sl $P$-associated bundle}.
\index{bundle!$P$-associated}

\begin{remark}
The tangent morphism to the morphism (\ref{mos75}) and the jet
prolongation of the morphism (\ref{mos75}) lead to the bundle
isomorphisms
\mar{56f3,4}\ben
&& TY=(TP\times TV)/G, \label{56f3}\\
&& J^1Y=(J^1P\times V)/G. \label{56f4}
\een
\end{remark}

The peculiarity of the $P$-associated bundle $Y$ (\ref{b1.230}) is
the following.

(i) Every bundle atlas $\Psi_P=\{(U_\al, z_\al)\}$ (\ref{vcv}) of
$P$ defines a unique \index{bundle!atlas!associated} {\sl
associated bundle atlas}
\mar{aaq1}\beq
\Psi=\{(U_\al,\psi_\al(x)=[z_\al(x)]^{-1})\} \label{aaq1}
\eeq
of the quotient $Y$ (\ref{b1.230}).

\begin{example} \label{56e1} \mar{56e1} Because of the splitting (\ref{ttt91}),
the Lie algebra bundle
\be
V_GP=(P\times \cG_l)/G,
\ee
by definition, is of the form (\ref{b1.230}). Therefore, it is a
$P$-associated bundle.
\end{example}

\begin{example} \label{56e2} \mar{56e2}
The group bundle $\ol P$ (\ref{55f1}) is defined as the quotient
\mar{b3130}\beq
P^G =(P\times G)/G, \label{b3130}
\eeq
where the group $G$ which acts on itself by the adjoint
representation. There is the following fibre-to-fibre action of
the group bundle $P^G$ on any $P$-associated bundle $Y$
(\ref{b1.230}):
\be
&& P^G\op\times_X Y\ar_X Y, \\
&& ((p, g)/G, (p, v)/G) \to (p, gv)/ G, \qquad g\in G,
\qquad  v\in V.
\ee
For instance, the action of $P^G$ on $P$ in the formula
(\ref{56f2}) is of this type.
\end{example}

(ii) Any principal automorphism $\Phi_P$ (\ref{55ff1}) of $P$
yields a unique \index{principal!automorphism!of an associated
bundle} {\sl principal automorphism}
\mar{024}\beq
\Phi_Y: (p,v)/G\to  (\Phi_P(p),v)/G, \qquad p\in P, \qquad v\in V,
\label{024}
\eeq
of \index{automorphism!associated} the $P$-associated bundle $Y$
(\ref{b1.230}). For the sake of brevity, we agree to write
\be
\Phi_Y: (P\times V)/G\to (\Phi_P(P)\times V)/G.
\ee
Accordingly, any (local) one-parameter group of principal
automorphisms of $P$ induces a (local) one-parameter group of
automorphisms of the $P$-associated bundle $Y$ (\ref{b1.230}).
Passing to infinitesimal generators of these groups, we obtain
that any principal vector field $\xi$ (\ref{1128}) yields a vector
field $\up_\xi$ on $Y$ regarded as an infinitesimal gauge
transformation of $Y$. Owing to the bundle isomorphism
(\ref{56f3}), we have
\mar{56f5}\ben
&& \up_\xi: X\to (\xi(P)\times TV)/G\subset TY, \nonumber\\
&& \up_\xi= \xi^\la\dr_\la + \xi^p I_p^i\dr_i, \label{56f5}
\een
where $\{I_p\}$ is a representation of the Lie algebra $\cG_r$ of
$G$ in $V$.

(iii) Any principal connection on $P\to X$ defines a unique
connection on the $P$-associated fibre bundle $Y$ (\ref{b1.230})
as follows. Given a principal connection $A$ (\ref{b1.210}) on $P$
and the corresponding horizontal distribution $HP\subset TP$, the
tangent map to the canonical morphism (\ref{mos75}) defines the
horizontal splitting of the tangent bundle $TY$ of $Y$
(\ref{b1.230}) and the corresponding connection on $Y\to X$. Owing
to the bundle isomorphism (\ref{56f4}), we have
\mar{A}\ben
&& A: (P\times V)/G\to (A(P)\times V)/G\subset J^1Y, \nonumber\\
&& A=dx^\la\ot(\dr_\la +A^p_\la I^i_p\dr_i), \label{A}
\een
where $\{I_p\}$ is a representation of the Lie algebra $\cG_r$ of
$G$ in $V$. The connection $A$ (\ref{A}) on $Y$ is called the {\sl
associated principal connection}
\index{principal!connection!associated} or,
\index{connection!principal!associated} simply, a principal
connection on $Y\to X$. The {\sl curvature} (\ref{1178a'})
\index{curvature!of an associated principal connection} of this
connection takes the form
\mar{mos118}\beq
R=\frac12 F_{\la\m}^pI_p^i dx^\la\w dx^\m\ot\dr_i. \label{mos118}
\eeq

\begin{example} \mar{56e5} \label{56e5} A principal connection $A$
on $P$ yields the associated connection (\ref{A}) on the
associated Lie algebra bundle $V_GP$ which reads
\mar{56f20}\beq
A= dx^\la\ot(\dr_\la - c^m_{pq}A^p_\la\xi^q e_m). \label{56f20}
\eeq
\end{example}

\begin{remark} \mar{56e5a} \label{56e5a}
If an associated principal connection $A$ is linear, one can
define its \index{strength!of a linear connection} {\sl strength}
\mar{lkp}\beq
F_A=\frac12 F_{\la\m}^pI_p dx^\la\w dx^\m, \label{lkp}
\eeq
where $I_p$ are matrices of a representation of the Lie algebra
$\cG_r$ in fibres of $Y$ with respect to the fibre bases
$\{e_i(x)\}$. They coincide with the matrices of a representation
of $\cG_r$ in the typical fibre $V$ of $Y$ with respect to its
fixed basis $\{e_i\}$ (see the relation (\ref{trt})). It follows
that $G$-valued transition functions act on $I_p$ by the adjoint
representation. Note that, because of the canonical splitting
(\ref{12f10}), one can identify $e_i(x)=\dr_i$.
\end{remark}

In view of the above mentioned properties, the $P$-associated
bundle $Y$ (\ref{b1.230}) is called {\sl canonically associated}
\index{bundle!associated!canonically} to a principal bundle $P$.
Unless otherwise stated, only canonically associated bundles are
considered, and we simply call $Y$ (\ref{b1.230}) an {\sl
associated bundle}. \index{bundle!associated}

\section{Reduced structure}

Let $H$ and $G$ be Lie groups and $\f:H\to G$ a Lie group
homomorphism. If $P_H\to X$ is a principal $H$-bundle, there
always exists a principal $G$-bundle $P_G\to X$ together with the
principal bundle morphism
\mar{510f1}\beq
\Phi:P_H\ar_X P_G \label{510f1}
\eeq
over $X$ (see Remark \ref{52r1'}). It is the $P_H$-associated
bundle
\be
P_G=(P_H\times G)/H
\ee
with the typical fibre $G$ on which $H$ acts on the left by the
rule $h(g)=\f(h)g$, while $G$ acts on $P_G$ as
\be
G\ni g': (p,g)/ H \to (p,gg')/ H.
\ee
Conversely, if $P_G\to X$ is a principal $G$-bundle, a problem is
to find a principal $H$-bundle $P_H\to X$ together with a
principal bundle morphism (\ref{510f1}). If $H\to G$ is a closed
subgroup, we have the {\sl structure group reduction}.
\index{structure group!reduction} If $H\to G$ is a group
epimorphism, one says that $P_G$ {\sl lifts} \index{bundle!lift}
to \index{lift of a bundle} $P_H$.

Here, we restrict our consideration to the reduction problem. In
this case, the bundle monomorphism (\ref{510f1}) is called a {\sl
reduced $H$-structure}. \index{reduced structure}

Let $P$ (\ref{51f1}) be a principal $G$-bundle,  and let $H$,
dim$H>0$, be a closed (and, consequently, Lie) subgroup of $G$.
Then we have the composite bundle
\mar{b3223a}\beq
P\to P/H\to X, \label{b3223a}
\eeq
where
\mar{b3194}\beq
 P_\Si=P\ar^{\pi_{P\Si}} P/H \label{b3194}
\eeq
is \index{$P_\Si$} a principal bundle with a structure group $H$
and
\mar{b3193}\beq
\Si=P/H\ar^{\pi_{\Si X}} X \label{b3193}
\eeq
is a $P$-associated bundle with the typical fibre $G/H$ on which
the structure group $G$ acts on the left (see Example \ref{52e1}).

One says that a structure Lie group $G$ of a principal bundle $P$
is reduced to its closed subgroup $H$ if the following equivalent
conditions hold.

$\bullet$ A principal bundle $P$ admits a bundle atlas $\Psi_P$
(\ref{51f2}) with $H$-valued transition functions $\vr_{\al\bt}$.

$\bullet$ There exists a principal {\sl reduced subbundle}
\index{reduced subbundle} $P_H$ of $P$ with a structure group $H$.

\begin{theorem}\label{redsub} \mar{redsub}
There is one-to-one correspondence
\mar{510f2}\beq
P^h=\pi_{P\Si}^{-1}(h(X)) \label{510f2}
\eeq
between the reduced principal $H$-subbundles $i_h:P^h\to P$ of $P$
and the global sections $h$ of the quotient bundle $P/H\to X$
(\ref{b3193}).
\end{theorem}

\begin{corollary} \mar{510c1} \label{5101c}
A glance at the formula (\ref{510f2}) shows that the reduced
principal $H$-bundle $P^h$ is the restriction $h^*P_\Si$
(\ref{S10}) of the principal $H$-bundle $P_\Si$ (\ref{b3194}) to
$h(X)\subset \Si$.
\end{corollary}

In general, there is topological obstruction to reduction of a
structure group of a principal bundle to its subgroup.

\begin{theorem} \label{510a1} \mar{510a1}
In accordance with Theorem \ref{mos9}, the structure group $G$ of
a principal bundle $P$ is always reducible to its closed subgroup
$H$, if the quotient $G/H$ is diffeomorphic to a Euclidean space
$\Bbb R^m$.
\end{theorem}

In particular, this is the case of a maximal compact subgroup $H$
of a Lie group $G$. Then the following is a corollary of Theorem
\ref{510a1}.

\begin{theorem} \label{comp} \mar{comp}
A structure group $G$ of a principal bundle is always reducible to
its maximal compact subgroup $H$.
\end{theorem}

\begin{example} \label{exe} \mar{exe}
For instance, this is the case of $G=GL(n,\Bbb C)$, $H=U(n)$ and
$G=GL(n,\Bbb R)$, $H=O(n)$.
\end{example}

\begin{example} \label{exe1} \mar{exe1} Any affine bundle admits
an atlas with linear transition functions. In accordance with
Theorem \ref{510a1}, its structure group $GA(m,\Bbb R)$ is always
reducible to the linear subgroup $GL(m,\Bbb R)$ because
\be
GA(m,\Bbb R)/GL(m,\Bbb R)=\Bbb R^m.
\ee
\end{example}

Different principal $H$-subbundles $P^h$ and $P^{h'}$ of a
principal $G$-bundle $P$ are not isomorphic to each other in
general.

\begin{theorem}\label{iso1} \mar{iso1} Let a structure Lie group $G$
of a principal bundle be reducible to its closed subgroup $H$.

(i) Every vertical principal automorphism $\Phi$ of $P$  sends a
reduced principal $H$-subbundle $P^h$ of $P$ onto an isomorphic
principal $H$-subbundle $P^{h'}$.

(ii) Conversely, let two reduced subbundles $P^h$ and $P^{h'}$ of
a principal bundle $P\to X$ be isomorphic to each other, and let
$\Phi:P^h\to P^{h'}$ be their isomorphism over $X$. Then $\Phi$ is
extended to a vertical principal automorphism of $P$.
\end{theorem}

\begin{theorem}\label{iso0} \mar{iso0} If the quotient $G/H$ is homeomorphic to a
Euclidean space $\Bbb R^m$, all principal $H$-subbundles of a
principal $G$-bundle $P$ are isomorphic to each other.
\end{theorem}

There are the following properties of principal connections
compatible with a reduced structure.

\begin{theorem} \label{mos176} \mar{mos176}
Since principal connections are equivariant, every principal
connection $A_h$ on a reduced principal $H$-subbundle $P^h$ of a
principal $G$-bundle $P$ gives rise to a principal connection on
$P$.
\end{theorem}

\begin{theorem} \label{mos177} \mar{mos177}
A principal connection $A$ on a principal $G$-bundle $P$ is
reducible to a principal connection on a reduced principal
$H$-subbundle $P^h$ of $P$ iff the corresponding global section
$h$ of the $P$-associated fibre bundle $P/H\to X$ is an integral
section of the associated principal connection $A$ on $P/H\to X$.
\end{theorem}

\begin{theorem} \label{redt} \mar{redt}
Let the Lie algebra $\cG_l$ of $G$ be the direct sum
\mar{g13}\beq
\cG_l = {\got h}_l \oplus {\got m} \label{g13}
\eeq
of the Lie algebra ${\got h}_l$ of $H$ and a subspace ${\got m}$
such that ${\rm Ad}_g({\got m})\subset {\got m}$, $g\in H$ (e.g.,
$H$ is a Cartan subgroup of $G$). Let $\ol A$ be a $\cG_l$-valued
connection form (\ref{mos166}) on $P$. Then, the pull-back of the
${\got h}_l$-valued component
 of $\ol A$ onto a reduced principal $H$-subbundle $P^h$ is a
${\got h}_l$-valued connection form of a principal connection $\ol
A_h$ on $P^h$.
\end{theorem}

The following is a corollary of Theorem \ref{mos252}.

\begin{theorem} \label{mos178} \mar{mos178}
Given the composite bundle (\ref{b3223a}), let $A_\Si$ be a
principal connection on the principal $H$-bundle $P\to\Si$
(\ref{b3194}). Then, for any reduced principal $H$-bundle
$i_h:P^h\to P$, the pull-back connection $i_h^*A_\Si$
(\ref{mos83}) is a principal connection on $P^h$.
\end{theorem}

\chapter{Geometry of natural bundles}

Classical gravitation theory is formulated as field theory on
natural bundles, exemplified by tensor bundles. Therefore, we
agree to call connections on these bundles the world connections.

\section{Natural bundles}

Let $\pi:Y\to X$ be a smooth fibre bundle coordinated by $(x^\la,
y^i)$. Any automorphism $(\Phi,f)$ of $Y$, by definition, is
projected as
\be
\pi \circ \Phi= f\circ \pi
\ee
onto a diffeomorphism $f$ of its base $X$. The converse is not
true. A diffeomorphism of $X$ need not give rise to an
automorphism of $Y$, unless $Y\to X$ is a trivial bundle.

Given a one-parameter group $(\Phi_t,f_t)$ of automorphisms of
$Y$, its infinitesimal generator is a projectable vector field
\be
u=u^\la(x^\m)\dr_\la + u^i(x^\m,y^j)\dr_i
\ee
on $Y$. This vector field is projected as
\be
\tau\circ\pi= T\pi\circ u
\ee
onto a vector field $\tau=u^\la\dr_\la$ on $X$. Its flow is the
one-parameter group $(f_t)$ of diffeomorphisms of $X$ which are
projections of autmorphisms $(\Phi_t,f_t)$ of $Y$. Conversely, let
$\tau=\tau^\la\dr_\la$ be a vector field on $X$. There is a
problem of constructing its lift to a projectable vector field
\be
u= \tau^\la\dr_\la +u^i\dr_i
\ee
on $Y$ projected onto $\tau$. Such a lift always exists, but it
need not be canonical. Given a connection $\G$ on $Y$, any vector
field $\tau$ on $X$ gives rise to the horizontal vector field
$\G\tau$ (\ref{b1.85}) on $Y$. This horizontal lift
$\tau\to\G\tau$ yields a monomorphism of the $C^\infty(X)$-module
$\cT(X)$ of vector fields on $X$ to the $C^\infty(Y)$-module of
vector fields on $Y$, but this monomorphisms is not a Lie algebra
morphism, unless $\G$ is a flat connection.

Let us address the category of {\sl natural bundles}
\index{bundle!natural} $T\to X$ \index{natural bundle} which admit
the {\sl functorial lift}  \index{lift of a vector
field!functorial} $\wt\tau$ onto $T$ of any vector field $\tau$ on
$X$ such that $\tau\to\ol\tau$ is a monomorphism
\be
\cT(X)\to \cT(T), \qquad [\wt\tau,\wt\tau']=\wt{[\tau,\tau']},
\ee
of the real Lie algebra $\cT(X)$ of vector fields on $X$ to the
real Lie algebra $\cT(Y)$ of vector fields on $T$. One treats the
functorial lift $\wt\tau$ as an {\sl infinitesimal general
covariant transformation}, i.e.,  \index{general covariant
transformation!infinitesimal} an infinitesimal generator of a
local one-parameter group of general covariant transformations of
$T$.

\begin{remark} \label{gct} \mar{gct}
It should be emphasized that, in general, there exist
diffeomorphisms of $X$ which do not belong to any one-parameter
group of diffeomorphisms of $X$. In a general setting, one
therefore considers a monomorphism $f\to\wt f$ of the group of
diffeomorphisms of $X$ to the group of bundle automorphisms of a
natural bundle $T\to X$. Automorphisms $\wt f$ are called {\sl
general covariant transformations} \index{general covariant
transformation} of $T$. No vertical automorphism of $T$, unless it
is the identity morphism, is a general covariant transformation.
\end{remark}

Natural bundles are exemplified by tensor bundles (\ref{sp20}).
For instance, the tangent and cotangent bundles $TX$ and $T^*X$ of
$X$ are natural bundles. Given a vector field $\tau$ on $X$, its
functorial (or canonical) lift onto the tensor bundle $T$
(\ref{sp20}) is given by the formula (\ref{l28}). In particular,
let us recall the functorial lift (\ref{l27}) and (\ref{l27'}) of
$\tau$ onto the tangent bundle $TX$ and the cotangent bundle
$T^*X$, respectively.

\begin{remark} Any diffeomorphism $f$ of $X$ gives rise to the
tangent automorphisms $\wt f=Tf$ of $TX$ which is a general
covariant transformation of $TX$ as a natural bundle. Accordingly,
the general covariant transformation of the cotangent bundle
$T^*X$ over a diffeomorphism $f$ of its base $X$ reads
\be
\dot x'_\m=\frac{\dr x^\nu}{\dr x'^\m}\dot x_\nu.
\ee
\end{remark}

Tensor bundles over a manifold $X$ have the structure group
\mar{gl4}\beq
GL_n=GL^+(n,\Bbb R). \label{gl4}
\eeq
The \index{$GL_n$} associated principal bundle is the fibre bundle
\be
\pi_{LX}:LX\to X
\ee
of \index{$LX$} oriented linear frames in the tangent spaces to a
manifold $X$. It is called the {\sl linear frame bundle}.
\index{linear frame bundle} Its (local) sections are termed {\sl
frame fields}. \index{frame field}

Given holonomic frames $\{\dr_\m\}$ in the tangent bundle $TX$
associated with the holonomic atlas $\Psi_T$ (\ref{mos150}), every
element $\{H_a\}$ of the linear frame bundle $LX$ takes the form
$H_a=H^\m_a\dr_\m$, where $H^\m_a$ is a matrix of the natural
representation of the group $GL_n$ in $\Bbb R^n$. These matrices
constitute the bundle coordinates
\be
(x^\la, H^\m_a), \qquad H'^\m_a=\frac{\dr x'^\m}{\dr
x^\la}H^\la_a,
\ee
on $LX$ associated to its \index{holonomic!atlas!of the frame
bundle} {\sl holonomic atlas}
\mar{tty}\beq
\Psi_T=\{(U_\iota, z_\iota=\{\dr_\m\})\} \label{tty}
\eeq
given by the local frame fields $z_\iota=\{\dr_\m\}$. With respect
to these coordinates, the right action (\ref{1}) of $GL_n$ on $LX$
reads
\be
R_{gP}: H^\m_a\to H^\m_bg^b{}_a, \qquad g\in GL_n.
\ee

The frame bundle $LX$ admits the canonical $\Bbb R^n$-valued
one-form \index{$\thh_{LX}$}
\mar{b3133'}\beq
\thh_{LX} = H^a_\m dx^\m\ot t_a,\label{b3133'}
\eeq
where $\{t_a\}$ is a fixed basis for $\Bbb R^n$ and $H^a_\m$ is
the inverse matrix of $H^\m_a$.

The frame bundle $LX\to X$ belongs to the category of natural
bundles. Any diffeomorphism $f$ of $X$ gives rise to the principal
automorphism
\mar{025}\beq
\wt f: (x^\la, H^\la_a)\to (f^\la(x),\dr_\m f^\la H^\m_a)
\label{025}
\eeq
of $LX$ which is its general covariant transformation (or a {\sl
holonomic automorphism}). \index{holonomic!automorphisms} For
instance, the associated automorphism of $TX$ is the tangent
morphism $Tf$ to $f$.

Given a (local) one-parameter group of diffeomorphisms of $X$ and
its infinitesimal generator $\tau$, their lift (\ref{025}) results
in the functorial lift
\mar{tty2}\beq
\wt\tau=\tau^\m\dr_\m +\dr_\nu\tau^\al H^\nu_a\frac{\dr}{\dr
H^\al_a} \label{tty2}
\eeq
of a vector field $\tau$ on $X$ onto $LX$ defined by the condition
\be
\bL_{\wt\tau}\thh_{LX}=0.
\ee

Every $LX$-associated bundle $Y\to X$ admits a lift of any
diffeomorphism $f$ of its base to the principal automorphism $\wt
f_Y$(\ref{024}) of $Y$ associated with the principal automorphism
$\wt f$ (\ref{025}) of the liner frame bundle $LX$. Thus, all
bundles associated with the linear frame bundle $LX$ are natural
bundles. However, there are natural bundles which are not
associated with $LX$.

\begin{remark}
In a more general setting, higher order natural bundles and gauge
natural bundles are considered. Note that the linear frame bundle
$LX$ over a manifold $X$ is the set of first order jets of local
diffeomorphisms of the vector space $\Bbb R^n$ to $X$, $n=\di X$,
at the origin of $\Bbb R^n$. Accordingly, one considers $r$-order
frame bundles $L^rX$ of $r$-order jets of local diffeomorphisms of
$\Bbb R^n$ to $X$. Furthermore, given a principal bundle $P\to X$
with a structure group $G$, the $r$-order jet bundle $J^1P\to X$
of its sections fails to be a principal bundle. However, the
product
\be
W^rP=L^rX\times J^rP
\ee
is a principal bundle with the structure group $W^r_nG$ which is a
semidirect product of the group $G^r_n$ of invertible $r$-order
jets of maps $\Bbb R^n$ to itself at its origin (e.g.,
$G^1_n=GL(n,\Bbb R)$) and the group $T^r_nG$ of $r$-order jets of
morphisms $\Bbb R^n\to G$ at the origin of $\Bbb R^n$. Moreover,
if $Y\to X$ is a $P$-associated bundle, the jet bundle $J^rY\to X$
is a vector bundle associated with the principal bundle $W^rP$. It
exemplifies {\sl gauge natural bundles} \index{gauge natural
bundle} which \index{bundle!gauge natural} can be described as
fibre bundles associated with principal bundles $W^rP$. Natural
bundles are gauge natural bundles for a trivial group $G=\bb$. The
bundle of principal connections $C$ (\ref{B1}) is a first order
gauge natural bundle.
\end{remark}

\section{Linear world connections}

Since the tangent bundle $TX$ is associated with the linear frame
bundle $LX$, every world connection (\ref{B}):
\mar{B'}\beq
\G= dx^\la\otimes (\dr_\la +\G_\la{}^\m{}_\n \dot x^\n
\dot\dr_\m), \label{B'}
\eeq
on a manifold $X$ is associated with a principal connection on
$LX$. We agree to call $\G$ (\ref{B'}) the {\sl linear world
connection} \index{connection!linear!world} in \index{world
connection!linear} order to distinct it from an affine world
connection in Section 5.3.

Being principal connections on the linear frame bundle $LX$,
linear world connections  are represented by sections of the
quotient bundle
\mar{015}\beq
C_{\rm W}=J^1LX/GL_n, \label{015}
\eeq
called \index{$C_{\rm W}$} the {\sl bundle of world connections}.
\index{bundle!of world connections} With respect to the holonomic
atlas $\Psi_T$ (\ref{tty}), this bundle is provided with the
coordinates
\be
(x^\la, k_\la{}^\nu{}_\al), \qquad k'_\la{}^\nu{}_\al=
\left[\frac{\dr x'^\nu}{\dr x^\g} \frac{\dr x^\bt}{\dr x'^\al}
k_\m{}^\g{}_\bt + \frac{\dr x^\bt}{\dr x'^\al}\frac{\dr^2
x'^\nu}{\dr x^\m\dr x^\bt} \right] \frac{\dr x^\m}{\dr x'^\la},
\ee
so that, for any section $\G$ of $C_{\rm W}\to X$,
\be
k_\la{}^\nu{}_\al\circ \G=\G_\la{}^\nu{}_\al
\ee
are components of the linear world connection $\G$ (\ref{B'}).

Though the bundle of world connections $C_{\rm W}\to X$
(\ref{015}) is not $LX$-associated, it is a natural bundle. It
admits the lift
\be
\wt f_C: J^1LX/GL_n\to J^1\wt f(J^1LX)/GL_n
\ee
of any diffeomorphism $f$ of its base $X$ and, consequently, the
functorial lift
\mar{b3150}\beq
\wt\tau_C = \tau^\m\dr_\m +[\dr_\nu\tau^\al k_\m{}^\nu{}_\bt -
\dr_\bt\tau^\nu k_\m{}^\al{}_\nu - \dr_\m\tau^\nu
k_\nu{}^\al{}_\bt + \dr_{\m\bt}\tau^\al]\frac{\dr}{\dr
k_\m{}^\al{}_\bt} \label{b3150}
\eeq
of any vector field $\tau$ on $X$.

The first order jet manifold $J^1C_{\rm W}$ of the bundle of world
connections admits the canonical splitting (\ref{296}). In order
to obtain its coordinate expression, let us consider the strength
(\ref{lkp}) of the linear world connection $\G$ (\ref{B'}). It
reads
\be
F_\G=\frac12 F_{\la\m}{}^b{}_aI_b{}^a dx^\la\w dx^\m=\frac12
R_{\m\nu}{}^\al{}_\bt dx^\la\w dx^\m,
\ee
where
\be
(I_b{}^a)^\al{}_\bt=H^\al_b H^a_\bt
\ee
are generators of the group $GL_n$ (\ref{gl4}) in fibres of $TX$
with respect to the holonomic frames, and
\mar{cvv}\beq
R_{\la\m}{}^\al{}_\bt = \dr_\la \G_\m{}^\al{}_\bt - \dr_\m
\G_\la{}^\al{}_\bt + \G_\la{}^\g{}_\bt \G_\m{}^\al{}_\g -
\G_\m{}^\g{}_\bt \G_\la{}^\al{}_\g \label{cvv}
\eeq
are components if the curvature (\ref{1203}) of a linear world
connection $\G$. Accordingly, the above mentioned canonical
splitting (\ref{296}) of $J^1C_{\rm W}$ can be written in the form
\mar{0101}\ben
&&k_{\la\m}{}^\al{}_\bt=\frac12(\cR_{\la\m}{}^\al{}_\bt
+\cS_{\la\m}{}^\al{}_\bt) =
\label{0101}\\
&& \qquad \frac12(k_{\la\m}{}^\al{}_\bt - k_{\m\la}{}^\al{}_\bt +
k_\la{}^\g{}_\bt k_\m{}^\al{}_\g
-k_\m{}^\g{}_\bt k_\la{}^\al{}_\g) + \nonumber\\
&& \qquad \frac12(k_{\la\m}{}^\al{}_\bt + k_{\m\la}{}^\al{}_\bt -
k_\la{}^\g{}_\bt k_\m{}^\al{}_\g + k_\m{}^\g{}_\bt
k_\la{}^\al{}_\g). \nonumber
\een
It is readily observed that, if $\G$ is a section of $C_{\rm W}\to
X$, then
\be
\cR_{\la\m}{}^\al{}_\bt\circ J^1\G=R_{\la\m}{}^\al{}_\bt.
\ee

Because of the canonical vertical splitting (\ref{mos163}) of the
vertical tangent bundle $VTX$ of $TX$, the curvature form
(\ref{1203}) of a linear world connection $\G$ can be represented
by the tangent-valued two-form
\mar{1203a}\beq
R=\frac12R_{\la\m}{}^\al{}_\bt\dot x^\bt dx^\la\w dx^\m\ot\dr_\al
\label{1203a}
\eeq
on $TX$. Due to this representation, the \index{Ricci tensor} {\sl
Ricci tensor}
\mar{ric}\beq
R_c=\frac12R_{\la\m}{}^\la{}_\bt dx^\m\ot dx^\bt \label{ric}
\eeq
of a linear world connection $\G$ is defined.

Owing to the above mentioned vertical splitting (\ref{mos163}) of
$VTX$, the torsion form $T$ (\ref{191}) of $\G$ can be written as
the tangent-valued two-form
\mar{mos164}\ben
&& T =\frac12 T_\m{}^\n{}_\la  dx^\la\w dx^\m\ot \dr_\n,
\label{mos164}\\
&& T_\m{}^\n{}_\la  = \G_\m{}^\n{}_\la - \G_\la{}^\n{}_\m,
\nonumber
\een
on $X$. The  \index{torsion form!soldering} {\sl soldering torsion
form}
\mar{mos160}\beq
T=T_\m{}^\n{}_\la \dot x^\la dx^\m\ot\dot \dr_\nu \label{mos160}
\eeq
on $TX$ is also defined. Then one can show the following.

$\bullet$ Given a linear world connection $\G$ (\ref{B'}) and its
soldering torsion form $T$ (\ref{mos160}), the sum $\G+c T$,
$c\in\Bbb R$, is a linear world connection.

$\bullet$  Every linear world connection $\G$ defines a unique
symmetric world connection
\mar{mos172}\beq
\G'=\G -\frac12 T. \label{mos172}
\eeq

$\bullet$  If $\G$ and $\G'$ are linear world connections, then
\be
c \G+(1-c)\G'
\ee
is so for any $c\in\Bbb R$.

A manifold $X$ is said to be {\sl flat} \index{manifold!flat} if
it admits a flat linear world connection $\G$. By virtue of
Theorem \ref{gena113}, there exists an atlas of local constant
trivializations of $TX$ such that
\be
\G=dx^\la\ot\dr_\la
\ee
relative to this atlas. As a consequence, the curvature form $R$
(\ref{1203a}) of this connection equals zero. However, such an
atlas is not holonomic in general. Relative to this atlas, the
canonical soldering form (\ref{z117'}) on $TX$ reads
\be
\thh_J=H^a_\m dx^\m\dot\dr_a,
\ee
and the torsion form $T$ (\ref{191}) of $\G$ defined as the
Nijenhuis differential $d_\G\thh_J$ (\ref{1190}) need not vanish.

A manifold $X$ is called {\sl parallelizable}
\index{manifold!parallelizable} if the tangent bundle $TX\to X$ is
trivial. By virtue of Theorem \ref{gena113}, a parallelizable
manifold is flat. Conversely, a flat manifold is parallelizable if
it is simply connected.

Every linear world connection $\G$ (\ref{B'}) yields the
horizontal lift
\mar{b3180}\beq
\G\tau =\tau^\la(\dr_\la +\G_\la{}^\bt{}_\al\dot x^\al\dot\dr_\bt)
\label{b3180}
\eeq
of a vector field $\tau$ on $X$ onto the tangent bundle $TX$. A
vector field $\tau$ on $X$ is said to be {\sl parallel}
\index{vector field!parallel} relative to a connection $\G$ if it
is an integral section of $\G$. Its integral curve is called the
{\sl autoparallel} \index{autoparallel} of a world connection
$\G$.

\begin{remark} \label{symm} \mar{symm}
By virtue of Theorem \ref{ints}, any vector field on $X$ is an
integral section of some linear world connection. If $\tau(x)\neq
0$ at a point $x\in X$, there exists a coordinate system $(q^i)$
on some neighbourhood $U$ of $x$ such that $\tau^i(x)=$const. on
$U$. Then $\tau$ on $U$ is an integral section of the local
symmetric linear world connection
\mar{ffr3}\beq
\G_\tau(x)=d q^i\ot\dr_i, \qquad x\in U, \label{ffr3}
\eeq
 on $U$. The functorial lift
$\wt\tau$ (\ref{l27}) can be obtained at each point $x\in X$ as
the horizontal lift of $\tau$ by means of the local symmetric
connection (\ref{ffr3}).
\end{remark}

The horizontal lift of a vector field $\tau$ on $X$ onto the
linear frame bundle $LX$ by means of a world connection $K$ reads
\mar{mos170}\beq
\G\tau=\tau^\la\left(\dr_\la +\G_\la{}^\n{}_\al
H^\al_a\frac{\dr}{\dr H^\n_a} \right). \label{mos170}
\eeq
It is called {\sl standard} \index{vector
field!horizontal!standard} if the morphism
\be
u\rfloor\thh_{LX}: LX\to \Bbb R^n
\ee
is constant on $LX$. It is readily observed that every standard
horizontal vector field on $LX$ takes the form
\mar{mos169}\beq
u_v=H^\la_bv^b\left(\dr_\la +\G_\la{}^\n{}_\al
H^\al_a\frac{\dr}{\dr H^\n_a} \right)  \label{mos169}
\eeq
where $v=v^bt_b\in\Bbb R^n$. A glance at this expression shows
that a standard horizontal vector field is not projectable.

Since $TX$ is an $LX$-associated fibre bundle, we have the
canonical morphism
\be
LX\times\Bbb R^n\to TX,\qquad (H^\m_a, v^a) \to \dot
x^\m=H^\m_av^a.
\ee
The tangent map to this morphism sends every standard horizontal
vector field (\ref{mos169}) on $LX$ to the horizontal vector field
\mar{mos170'}\beq
u=\dot x^\la(\dr_\la +\G_\la{}^\n{}_\al \dot x^\al\dot\dr_\n)
\label{mos170'}
\eeq
on $TX$. Such a vector field on $TX$ is called {\sl holonomic}.
\index{vector field!holonomic} Given holonomic coordinates
$(x^\m,\dot x^\m, \dot{\rm x}^\m,\ddot x^\m)$ on the double
tangent bundle $TTX$, the holonomic vector field (\ref{mos170'})
defines the second order dynamic equation
\mar{mos171}\beq
\ddot x^\n=\G_\la{}^\n{}_\al\dot x^\la\dot x^\al \label{mos171}
\eeq
on $X$ which is called the {\sl geodesic equation} \index{geodesic
equation} with respect to a linear world connection $\G$.
Solutions of the geodesic equation (\ref{mos171}), called the {\sl
geodesics} \index{geodesic} of $\G$, are the projection of
integral curves of the vector field (\ref{mos170'}) in $TX$ onto
$X$. Moreover, one can show the following.

\begin{theorem}
The projection of an integral curve of any standard horizontal
vector field (\ref{mos169}) on $LX$ onto $X$ is a geodesic in $X$.
Conversely, any geodesic in $X$ is of this type.
\end{theorem}

It is readily observed that, if linear world connections $\G$ and
$\G'$ differ from each other only in the torsion, they define the
same holonomic vector field (\ref{mos170'}) and the same geodesic
equation (\ref{mos171}).

Let $\tau$ be an integral vector field of a linear world
connection $\G$, i.e., $\nabla^\G_\m\tau=0$. Consequently, it
obeys the equation $\tau^\m\nabla_\m^\G\tau=0$. Any autoparallel
of a linear world connection $\G$ is its geodesic and, conversely,
a geodesic of $\G$ is an autoparallel of its symmetric part
(\ref{mos172}).

\section{Affine world connections}

The tangent bundle $TX$ of a manifold $X$ as like as any vector
bundle possesses a natural structure of an affine bundle (see
Section 1.2). Therefore, one can consider affine connections on
$TX$, called {\sl affine world connections}. \index{world
connection!affine} Here we study them as principal connections.

Let $Y\to X$ be an affine bundle with an $k$-dimensional typical
fibre $V$. It is associated with a principal bundle $AY$ of affine
frames in $Y$, whose structure group is the general affine group
$GA(k,\Bbb R)$. Then any affine connection on $Y\to X$ can be seen
as an associated with a principal connection on $AY\to X$. These
connections are represented by global sections of the affine
bundle
\be
J^1P/GA(k,\Bbb R)\to X.
\ee
They always exist.

As was mentioned in Section 1.3.5, every affine connection $\G$
(\ref{184}) on $Y\to X$ defines a unique associated linear
connection $\ol \G$ (\ref{mos032}) on the underlying vector bundle
$\ol Y\to X$. This connection $\ol\G$ is associated with a linear
principal connection  on the principal bundle $L\ol Y$ of linear
frames in $\ol Y$ whose structure group is the general linear
group $GL(k,\Bbb R)$. We have the exact sequence of groups
\mar{mos031}\beq
0\to T_k\to GA(k,\Bbb R) \to GL(k,\Bbb R)\to \bb, \label{mos031}
\eeq
where $T_k$ is the group of translations in $\Bbb R^k$. It is
readily observed that there is the corresponding principal bundle
morphism $AY\to L\ol Y$ over $X$, and the principal connection
$\ol \G$ on $L\ol Y$ is the image of the principal connection $\G$
on $AY\to X$ under this morphism in accordance with Theorem
\ref{mos253}.

The exact sequence (\ref{mos031}) admits a splitting
\be
GL(k,\Bbb R)\to GA(k,\Bbb R),
\ee
but this splitting is not canonical. It depends on the morphism
\be
V\ni v\to v - v_0\in \ol V,
\ee
i.e., on the choice of an origin $v_0$ of the affine space $V$.
Given $v_0$, the image of the corresponding monomorphism
\be
GL(k,\Bbb R)\to GA(k,\Bbb R)
\ee
is a stabilizer
\be
G(v_0)\subset GA(k,\Bbb R)
\ee
of $v_0$. Different subgroups $G(v_0)$ and $G(v'_0)$ are related
to each other as follows:
\be
G(v'_0)=T(v'_0-v_0)G(v_0)T^{-1}(v'_0-v_0),
\ee
where $T(v'_0-v_0)$ is the translation along the vector
$(v'_0-v_0)\in\ol V$.

\begin{remark}
Accordingly, the well-known morphism of a $k$-dimensional affine
space $V$ onto a hypersurface $\ol y^{k+1}=1$ in $\Bbb R^{k+1}$
and the corresponding representation of elements of $GA(k,\Bbb R)$
by particular $(k+1)\times (k+1)$-matrices also fail to be
canonical. They depend on a point $v_0\in V$ sent to vector
$(0,\ldots,0,1)\in \Bbb R^{k+1}$.
\end{remark}

One can say something more if $Y\to X$ is a vector bundle provided
with the natural structure of an affine bundle whose origin is the
canonical zero section $\wh 0$. In this case, we have the
canonical splitting of the exact sequence (\ref{mos031}) such that
$GL(k,\Bbb R)$ is a subgroup of $GA(k,\Bbb R)$ and $GA(k,\Bbb R)$
is the semidirect product of $GL(k,\Bbb R)$ and the group
$T(k,\Bbb R)$ of translations in $\Bbb R^k$. Given a $GA(k,\Bbb
R)$-principal bundle $AY\to X$, its affine structure group
$GA(k,\Bbb R)$ is always reducible to the linear subgroup since
the quotient $GA(k,\Bbb R)/GL(k,\Bbb R)$ is a vector space $\Bbb
R^k$ provided with the natural affine structure (see Example
\ref{exe1}). The corresponding quotient bundle is isomorphic to
the vector bundle $Y\to X$. There is the canonical injection of
the linear frame bundle $LY\to AY$ onto the reduced $GL(k,\Bbb
R)$-principal subbundle of $AY$ which corresponds to the zero
section $\wh 0$ of $Y\to X$. In this case, every principal
connection on the linear frame bundle $LY$ gives rise to a
principal connection on the affine frame bundle in accordance with
Theorem \ref{mos176}. This is equivalent to the fact that any
affine connection $\G$ on a vector bundle $Y\to X$ defines a
linear connection $\ol \G$ on $Y\to X$ and that every linear
connection on $Y\to X$ can be seen as an affine one. Then any
affine connection $\G$ on the vector bundle $Y\to X$ is
represented by the sum of the associated linear connection $\ol\G$
and a basic soldering form $\si$ on $Y\to X$. Due to the vertical
splitting (\ref{12f10}), this soldering form is represented by a
global section of the tensor product $T^*X\ot Y$.

Let now $Y\to X$ be the tangent bundle $TX\to X$ considered as an
affine bundle. Then the relationship between affine and linear
world connections on $TX$ is the repetition of that we have said
in the case of an arbitrary vector bundle $Y\to X$. In particular,
any affine world connection
\mar{mos033}\beq
\G= dx^\la\ot(\dr_\la + \G_\la{}^\al{}_\m(x) \dot x^\m
+\si^\al_\la(x))\dr_\al \label{mos033}
\eeq
on $TX\to X$ is represented by the sum of the associated linear
world connection
\mar{mos034}\beq
\ol \G= \G_\la{}^\al{}_\m(x) \dot x^\m dx^\la\ot\dr_\al
\label{mos034}
\eeq
on $TX\to X$ and a basic soldering form
\mar{mos035}\beq
\si=\si^\al_\la(x) dx^\la\ot\dr_\al \label{mos035}
\eeq
 on $Y\to X$, which is the $(1,1)$-tensor field on $X$. For instance, if
$\si=\thh_X$ (\ref{b1.51}), we have the Cartan connection
(\ref{b1.97}).

It is readily observed that the soldered curvature (\ref{1186}) of
any soldering form (\ref{mos035}) equals zero. Then we obtain from
(\ref{1193a}) that the torsion (\ref{mos036}) of the affine
connection $\G$ (\ref{mos033}) with respect to $\si$
(\ref{mos035}) coincides with that of the associated linear
connection $\ol \G$ (\ref{mos034}) and reads
\mar{mos037}\ben
&& T =\frac12 T_{\la
\m}^i dx^\m\wedge dx^\la\otimes \dr_i, \nonumber  \\
&& T_\la{}^\la{}_\m =
\G_\la{}^\al{}_\nu \si_\m^\nu - \G_\m{}^\al{}_\nu \si_\la^\nu.
\label{mos037}
\een
The relation between the curvatures of an affine world connection
$\G$ (\ref{mos033}) and the associated linear connection $\ol \G$
(\ref{mos034}) is given by the general expression (\ref{1193})
where $\rho=0$ and $T$ is (\ref{mos037}).

\chapter{Geometry of graded manifolds}

In classical field theory, there are different descriptions of odd
fields on graded manifolds and supermanifolds. Both graded
manifolds and supermanifolds are phrased in terms of sheaves of
graded commutative algebras. However, graded manifolds are
characterized by sheaves on smooth manifolds, while supermanifolds
are constructed by gluing of sheaves on supervector spaces.
Treating odd fields on a manifold $X$, we follow the Serre--Swan
theorem generalized to graded manifolds (Theorem \ref{vv0}). It
states that, if a Grassmann algebra is an exterior algebra of some
projective $C^\infty(X)$-module of finite rank, it is isomorphic
to the algebra of graded functions on a graded manifold whose body
is $X$. By virtue of this theorem, odd fields on an arbitrary
manifold $X$ are described as generating elements of the structure
ring of a graded manifold whose body is $X$.

\section{Grassmann-graded algebraic calculus}

Throughout the Lectures, by the Grassmann gradation is meant $\Bbb
Z_2$-gradation. Hereafter, the symbol $\nw .$ stands for the
Grassmann parity. In the literature, a $\Bbb Z_2$-graded structure
is simply called the graded structure if there is no danger of
confusion.  Let us summarize the relevant notions of the
Grassmann-graded algebraic calculus.

An algebra $\cA$ is called {\sl graded} \index{algebra!$\Bbb
Z_2$-graded} if it is endowed with a {\sl grading automorphism}
\index{grading automorphism} $\g$ such that $\g^2=\id$. A graded
algebra falls into the direct sum $\cA=\cA_0\oplus \cA_1$ of $\Bbb
Z$-modules $\cA_0$ and $\cA_1$ of {\sl even} \index{even element}
and {\sl odd} \index{odd element} elements such that
\be
&& \g(a)=(-1)^ia, \qquad a\in\cA_i, \qquad i=0,1, \\
&& [aa']=([a]+[a']){\rm mod}\,2, \qquad a\in \cA_{[a]}, \qquad a'\in
\cA_{[a']}.
\ee
One calls $\cA_0$ and $\cA_1$ the even and odd parts of $\cA$,
respectively. The even part $\cA_0$ is a subalgebra of $\cA$ and
the odd one $\cA_1$ is an $\cA_0$-module. If $\cA$ is a {\sl
graded ring}, \index{graded!ring} then
$[\bb]=0$.\index{ring!graded}

A graded algebra $\cA$ is called {\sl graded commutative}
\index{algebra!$\Bbb Z_2$-graded!commutative} if
\be
aa'=(-1)^{[a][a']}a'a,
\ee
where $a$ and $a'$ are {\sl graded-homogeneous elements}
\index{graded-homogeneous element} of $\cA$.

Given a graded algebra $\cA$, a left {\sl graded $\cA$-module}
\index{graded!module} $Q$ \index{module!graded} is defined as a
left $\cA$-module provided with the grading automorphism $\g$ such
that
\be
&& \g(aq)=\g(a)\g(q), \qquad a\in\cA,\qquad q\in Q,\\
&& [aq]=([a]+[q]){\rm mod}\,2.
\ee
A graded module $Q$ is split into the direct sum $Q=Q_0\oplus Q_1$
of two $\cA_0$-modules $Q_0$ and $Q_1$ of even and odd elements.

If $\cK$ is a graded commutative ring, a graded $\cK$-module can
be provided with a graded {\sl $\cK$-bimodule}
\index{bimodule!graded} structure \index{graded!bimodule} by
letting
\be
qa = (-1)^{[a][q]}aq, \qquad a\in\cK, \qquad q\in Q.
\ee
A graded $\cK$-module is called {\sl free} \index{graded!module!
free} if it has a basis generated by graded-homogeneous elements.
This basis is said to be of type $(n,m)$ if it contains $n$ even
and $m$ odd elements.

In particular, by a real {\sl graded vector space} $B=B_0\oplus
B_1$ \index{graded!vector space} is \index{vector space!graded}
meant a graded $\Bbb R$-module. A real graded vector space is said
to be $(n,m)$-dimensional \index{graded!vector
space!$(n,m)$-dimensional} if $B_0=\Bbb R^n$  and $B_1=\Bbb R^m$.

Given a graded commutative ring $\cK$, the following are standard
constructions of new graded modules from old ones.

$\bullet$ The direct sum of graded modules and a graded factor
module are defined just as those of modules over a commutative
ring.

$\bullet$ The {\sl tensor product} \index{tensor product!of graded
modules} $P\ot Q$ of graded $\cK$-modules $P$ and $Q$ is an
additive group generated by elements $p\ot q$, $p\in P$, $q\in Q$,
obeying the relations
\be
&& (p+p')\ot q =p\ot q + p'\ot q, \\
&& p\ot(q+q')=p\ot q+p\ot q', \\
&&  ap\ot q=(-1)^{[p][a]}pa\ot q= (-1)^{[p][a]}p\ot aq,  \qquad a\in\cK.
\ee
In particular, the tensor algebra $\ot P$ of a graded $\cK$-module
$P$ is defined as that (\ref{spr620}) of a module over a
commutative ring. Its quotient $\w P$ with respect to the ideal
generated by elements
\be
p\ot p' + (-1)^{[p][p']}p'\ot p, \qquad p,p'\in P,
\ee
is the \index{exterior algebra!bigraded} {\sl bigraded exterior
algebra} of a \index{bigraded!exterior algebra} graded module $P$
with respect to \index{exterior product!graded} the {\sl graded
exterior product} \index{graded!exterior product}
\be
p\w p' =- (-1)^{[p][p']}p'\w p.
\ee

$\bullet$ A morphism $\Phi:P\to Q$ of graded $\cK$-modules seen as
additive groups is said to be \index{graded!morphism!even} {\sl
even graded morphism} \index{even morphism} (resp. {\sl odd graded
morphism}) \index{odd morphism} if \index{graded!morphism!odd}
$\Phi$ preserves (resp. change) the Grassmann parity of all
graded-homogeneous elements of $P$ and obeys the relations
\be
\Phi(ap)=(-1)^{[\Phi][a]}a\Phi(p), \qquad p\in P, \qquad a\in\cK.
\ee
A morphism $\Phi:P\to Q$ of graded $\cK$-modules as additive
groups is called a {\sl graded $\cK$-module morphism}
\index{graded!morphism} if it is represented by a sum of even and
odd graded morphisms. The set $\hm_\cK(P,Q)$ of graded morphisms
of a graded $\cK$-module $P$ to a graded $\cK$-module $Q$ is
naturally a graded $\cK$-module. The graded $\cK$-module
$P^*=\hm_\cK(P,\cK)$ is called the {\sl dual}
\index{graded!module!dual} of a graded $\cK$-module $P$.

A {\sl graded commutative $\cK$-ring} \index{graded!commutative
ring} $\cA$ is a graded commutative ring which also is a graded
$\cK$-module. A {\sl real graded commutative ring}
\index{graded!commutative ring!real} is said to be of rank $N$ if
it is a free algebra generated by the unit $\bb$ and $N$ odd
elements. A {\sl graded commutative Banach ring}
\index{graded!commutative ring!Banach} $\cA$ is a real graded
commutative ring which is a real Banach algebra whose norm obeys
the condition
\be
\|a_0 + a_1\|=\|a_0\| + \|a_1\|, \qquad a_0\in \cA_0, \quad a_1\in
\cA_1.
\ee

Let $V$ be a real vector space, and let $\La=\w V$ be its exterior
algebra endowed with the Grassmann gradation
\mar{+66}\beq
\La=\La_0\oplus \La_1, \qquad \La_0=\Bbb R\op\bigoplus_{k=1}
\op\w^{2k} V, \qquad \La_1=\op\bigoplus_{k=1} \op\w^{2k-1} V.
\label{+66}
\eeq
It is a real graded commutative ring, called the {\sl Grassmann
algebra}. \index{Grassmann algebra} A Grassmann algebra, seen as
an additive group, admits the decomposition
\mar{+11}\beq
\La=\Bbb R\oplus R =\Bbb R\oplus R_0\oplus R_1=\Bbb R \oplus
(\La_1)^2 \oplus \La_1, \label{+11}
\eeq
where $R$ is the {\sl ideal of nilpotents} \index{ideal!of
nilpotents} of $\La$. The corresponding projections
$\si:\La\to\Bbb R$ and $s:\La\to R$ are called the {\sl body}
\index{body map} and {\sl soul} \index{soul map} maps,
respectively.

Hereafter, we restrict our consideration to Grassmann algebras of
finite rank. Given a basis $\{c^i\}$ for the vector space $V$, the
elements of the Grassmann algebra $\La$ (\ref{+66}) take the form
\mar{z784}\beq
a=\op\sum_{k=0,1,\ldots} \op\sum_{(i_1\cdots i_k)}a_{i_1\cdots
i_k}c^{i_1}\cdots c^{i_k}, \label{z784}
\eeq
where the second sum runs through all the tuples $(i_1\cdots i_k)$
such that no two of them are permutations of each other. The
Grassmann algebra $\La$ becomes a graded commutative Banach ring
with respect to the norm
\be
\|a\|=\op\sum_{k=0} \op\sum_{(i_1\cdots i_k)}\nm{a_{i_1\cdots
i_k}}.
\ee

Let $B$ be a graded vector space. Given a Grassmann algebra $\La$,
it can be brought into a graded $\La$-module
\be
\La B=(\La B)_0\oplus (\La B)_1=(\La_0\ot B_0\oplus \La_1\ot
B_1)\oplus (\La_1\ot B_0\oplus \La_0\ot B_1),
\ee
called a {\sl superspace}. \index{superspace} The superspace
\mar{+70}\beq
B^{n|m}=[(\op\oplus^n\La_0) \oplus (\op\oplus^m\La_1)]\oplus
[(\op\oplus^n\La_1)\oplus (\op\oplus^m\La_0)] \label{+70}
\eeq
is \index{$B^{n|m}$} said to be $(n,m)$-dimensional. The graded
$\La_0$-module
\be
B^{n,m}= (\op\oplus^n\La_0) \oplus (\op\oplus^m\La_1)
\ee
is called an $(n,m)$-dimensional {\sl supervector space}.
\index{supervector space} Whenever referring to a topology on a
supervector space $B^{n,m}$, we will mean the Euclidean topology
on a $2^{N-1}[n+m]$-dimensional real vector space.

Let $\cK$ be a graded commutative ring. A graded commutative
(non-associative) $\cK$-algebra $\cG$ is called a {\sl Lie
$\cK$-superalgebra} \index{Lie superalgebra} if its product
$[.,.]$, called the {\sl Lie superbracket}, \index{Lie
superbracket} obeys the relations
\be
&& [\ve,\ve']=-(-1)^{[\ve][\ve']}[\ve',\ve],\\
&& (-1)^{[\ve][\ve'']}[\ve,[\ve',\ve'']]
+(-1)^{[\ve'][\ve]}[\ve',[\ve'',\ve]] +
(-1)^{[\ve''][\ve']}[\ve'',[\ve,\ve']] =0.
\ee
The even part $\cG_0$ of a Lie $\cK$-superalgebra $\cG$ is a Lie
$\cK_0$-algebra. A graded $\cK$-module $P$ is called a {\sl
$\cG$-module} \index{module!over a Lie superalgebra} if it is
provided with a $\cK$-bilinear map
\be
&& \cG\times P\ni (\ve,p)\to \ve p\in P, \qquad [\ve
p]=([\ve]+[p]){\rm mod}\,2,\\
&& [\ve,\ve']p=(\ve\circ\ve'-(-1)^{[\ve][\ve']}\ve'\circ\ve)p.
\ee

\section{Grassmann-graded differential calculus}

Linear differential operators on graded modules over a graded
commutative ring are defined similarly to those in commutative
geometry (Section 8.2).

Let $\cK$ be a graded commutative ring and $\cA$ a graded
commutative $\cK$-ring. Let $P$ and $Q$ be graded $\cA$-modules.
The graded $\cK$-module $\hm_\cK (P,Q)$ of graded $\cK$-module
homomorphisms $\Phi:P\to Q$ can be endowed with the two graded
$\cA$-module structures
\mar{ws11}\beq
(a\Phi)(p)= a\Phi(p),  \qquad  (\Phi\bll a)(p) = \Phi (a p),\qquad
a\in \cA, \quad p\in P, \label{ws11}
\eeq
called $\cA$- and $\cA^\bll$-module structures, respectively. Let
us put
\mar{ws12}\beq
\dl_a\Phi= a\Phi -(-1)^{[a][\Phi]}\Phi\bll a, \qquad a\in\cA.
\label{ws12}
\eeq
An element $\Delta\in\hm_\cK(P,Q)$ is said to be a $Q$-valued {\sl
graded differential operator} \index{graded!differential operator}
of \index{differential operator!graded} order $s$ on $P$ if
\be
\dl_{a_0}\circ\cdots\circ\dl_{a_s}\Delta=0
\ee
for any tuple of $s+1$ elements $a_0,\ldots,a_s$ of $\cA$. The set
$\dif_s(P,Q)$ of these operators inherits the graded module
structures (\ref{ws11}).

In particular, zero order graded differential operators obey the
condition
\be
\dl_a \Delta(p)=a\Delta(p)-(-1)^{[a][\Delta]}\Delta(ap)=0, \qquad
a\in\cA, \qquad p\in P,
\ee
i.e., they coincide with graded $\cA$-module morphisms $P\to Q$. A
first order graded differential operator $\Delta$ satisfies the
relation
\be
&& \dl_a\circ\dl_b\,\Delta(p)=
ab\Delta(p)- (-1)^{([b]+[\Delta])[a]}b\Delta(ap)-
(-1)^{[b][\Delta]}a\Delta(bp)+\\
&& \qquad (-1)^{[b][\Delta]+([\Delta]+[b])[a]}
=0, \qquad a,b\in\cA, \quad p\in P.
\ee

For instance, let $P=\cA$. Any zero order $Q$-valued graded
differential operator $\Delta$ on $\cA$ is defined by its value
$\Delta(\bb)$. Then there is a graded $\cA$-module isomorphism
\be
\dif_0(\cA,Q)=Q, \qquad Q\ni q\to \Delta_q\in \dif_0(\cA,Q),
\ee
where $\Delta_q$ is given by the equality $\Delta_q(\bb)=q$. A
first order $Q$-valued graded differential operator $\Delta$ on
$\cA$ fulfils the condition
\be
\Delta(ab)= \Delta(a)b+ (-1)^{[a][\Delta]}a\Delta(b)
-(-1)^{([b]+[a])[\Delta]} ab \Delta(\bb), \qquad  a,b\in\cA.
\ee
It is called a $Q$-valued {\sl graded derivation}
\index{graded!derivation} of \index{derivation!graded} $\cA$ if
$\Delta(\bb)=0$, i.e., the \index{Leibniz rule!Grassmann-graded}
{\sl Grassmann-graded Leibniz rule}
\mar{ws10}\beq
\Delta(ab) = \Delta(a)b + (-1)^{[a][\Delta]}a\Delta(b), \quad
a,b\in \cA, \label{ws10}
\eeq
holds. One obtains at once that any first order graded
differential operator on $\cA$ falls into the sum
\be
\Delta(a)= \Delta(\bb)a +[\Delta(a)-\Delta(\bb)a]
\ee
of a zero order graded differential operator $\Delta(\bb)a$ and a
graded derivation $\Delta(a)-\Delta(\bb)a$. If $\dr$ is a graded
derivation of $\cA$, then $a\dr$ is so for any $a\in \cA$. Hence,
graded derivations of $\cA$ constitute a graded $\cA$-module
$\gd(\cA,Q)$, called the {\sl graded derivation module}.
\index{graded!derivation module} \index{derivation module!graded}

If $Q=\cA$, the graded derivation module $\gd\cA$ also is a Lie
superalgebra over the graded commutative ring $\cK$ with respect
to the superbracket
\mar{ws14}\beq
[u,u']=u\circ u' - (-1)^{[u][u']}u'\circ u, \qquad u,u'\in \cA.
\label{ws14}
\eeq
We have the graded $\cA$-module decomposition
\mar{ws15}\beq
\dif_1(\cA) = \cA \oplus\gd\cA. \label{ws15}
\eeq

Since $\gd\cA$ is a Lie $\cK$-superalgebra, let us consider the
Chevalley--Eilenberg complex $C^*[\gd\cA;\cA]$ where the graded
commutative ring $\cA$ is a regarded as a $\gd\cA$-module. It is
the complex
\mar{ws85}\beq
0\to \cA\ar^d C^1[\gd\cA;\cA]\ar^d \cdots
C^k[\gd\cA;\cA]\ar^d\cdots \label{ws85}
\eeq
where
\be
C^k[\gd\cA;\cA]=\hm_\cK(\op\w^k \gd\cA,\cA)
\ee
are $\gd\cA$-modules of $\cK$-linear graded morphisms of the
graded exterior products $\op\w^k \gd\cA$ of the $\cK$-module
$\gd\cA$ to $\cA$. Let us bring homogeneous elements of $\op\w^k
\gd\cA$ into the form
\be
\ve_1\w\cdots\ve_r\w\e_{r+1}\w\cdots\w \e_k, \qquad
\ve_i\in\gd\cA_0, \quad \e_j\in\gd\cA_1.
\ee
Then the even coboundary operator $d$ of the complex (\ref{ws85})
is given by the expression
\mar{ws86}\ben
&& dc(\ve_1\w\cdots\w\ve_r\w\e_1\w\cdots\w\e_s)=
\label{ws86}\\
&&\op\sum_{i=1}^r (-1)^{i-1}\ve_i
c(\ve_1\w\cdots\wh\ve_i\cdots\w\ve_r\w\e_1\w\cdots\e_s)+
\nonumber \\
&& \op\sum_{j=1}^s (-1)^r\ve_i
c(\ve_1\w\cdots\w\ve_r\w\e_1\w\cdots\wh\e_j\cdots\w\e_s)
+\nonumber\\
&& \op\sum_{1\leq i<j\leq r} (-1)^{i+j}
c([\ve_i,\ve_j]\w\ve_1\w\cdots\wh\ve_i\cdots\wh\ve_j
\cdots\w\ve_r\w\e_1\w\cdots\w\e_s)+\nonumber\\
&&\op\sum_{1\leq i<j\leq s} c([\e_i,\e_j]\w\ve_1\w\cdots\w
\ve_r\w\e_1\w\cdots
\wh\e_i\cdots\wh\e_j\cdots\w\e_s)+\nonumber\\
&& \op\sum_{1\leq i<r,1\leq j\leq s} (-1)^{i+r+1}
c([\ve_i,\e_j]\w\ve_1\w\cdots\wh\ve_i\cdots\w\ve_r\w
\e_1\w\cdots\wh\e_j\cdots\w\e_s),\nonumber
\een
where the caret $\,\wh{}\,$ denotes omission. This operator is
called the {\sl graded Chevalley--Eilenberg coboundary operator}.
\index{Chevalley--Eilenberg!coboundary operator!graded}

Let us consider the extended Chevalley--Eilenberg complex
\be
0\to \cK\ar^{\rm in}C^*[\gd\cA;\cA].
\ee
This complex contains a subcomplex $\cO^*[\gd\cA]$ of $\cA$-linear
graded morphisms. The $\Bbb N$-graded module $\cO^*[\gd\cA]$ is
provided with the structure of a bigraded $\cA$-algebra with
respect to the graded exterior product
\mar{ws103'}\ben
&& \f\w\f'(u_1,...,u_{r+s})= \label{ws103'}\\
&& \qquad \op\sum_{i_1<\cdots<i_r;j_1<\cdots<j_s} {\rm
Sgn}^{i_1\cdots i_rj_1\cdots j_s}_{1\cdots r+s} \f(u_{i_1},\ldots,
u_{i_r}) \f'(u_{j_1},\ldots,u_{j_s}), \nonumber \\
&& \f\in \cO^r[\gd\cA], \qquad \f'\in \cO^s[\gd\cA], \qquad u_k\in \gd\cA,
\nonumber
\een
where $u_1,\ldots, u_{r+s}$ are graded-homogeneous elements of
$\gd\cA$ and
\be
u_1\w\cdots \w u_{r+s}= {\rm Sgn}^{i_1\cdots i_rj_1\cdots
j_s}_{1\cdots r+s} u_{i_1}\w\cdots\w u_{i_r}\w u_{j_1}\w\cdots\w
u_{j_s}.
\ee
The graded Chevalley--Eilenberg coboundary operator $d$
(\ref{ws86}) and the graded exterior product $\w$ (\ref{ws103'})
bring $\cO^*[\gd\cA]$ into a {\sl differential bigraded algebra}
\index{algebra!differential bigraded} (henceforth DBGA)
\index{DBGA} whose elements obey the relations
\mar{ws45,44}\ben
&& \f\w \f'=(-1)^{|\f||\f'|+[\f][\f']}\f'\w\f, \label{ws45} \\
&& d(\f\w\f')= d\f\w\f' +(-1)^{|\f|}\f\w d\f'. \label{ws44}
\een
It is called the {\sl graded Chevalley--Eilenberg differential
calculus} \index{Chevalley--Eilenberg!differential
calculus!Grassmann-graded} over a graded commutative $\cK$-ring
$\cA$. In particular, we have
\mar{ws47}\beq
\cO^1[\gd\cA]=\hm_\cA(\gd\cA,\cA)=\gd\cA^*. \label{ws47}
\eeq
One can extend this duality relation to the {\sl graded interior
product} \index{interior product!graded} of $u\in\gd\cA$
\index{graded!interior product} with any element $\f\in
\cO^*[\gd\cA]$ by the rules
\mar{ws46}\ben
&& u\rfloor(bda) =(-1)^{[u][b]}u(a),\qquad a,b \in\cA, \nonumber\\
&& u\rfloor(\f\w\f')=
(u\rfloor\f)\w\f'+(-1)^{|\f|+[\f][u]}\f\w(u\rfloor\f').
\label{ws46}
\een
As a consequence, any graded derivation $u\in\gd\cA$ of $\cA$
yields a derivation
\mar{+117}\ben
&& \bL_u\f= u\rfloor d\f + d(u\rfloor\f), \qquad \f\in\cO^*, \qquad
u\in\gd\cA, \label{+117} \\
&& \bL_u(\f\w\f')=\bL_u(\f)\w\f' + (-1)^{[u][\f]}\f\w\bL_u(\f'), \nonumber
\een
called the \index{Lie derivative!graded} {\sl graded Lie
derivative} of the DBGA $\cO^*[\gd\cA]$.

The minimal graded Chevalley--Eilenberg differential calculus
$\cO^*\cA\subset \cO^*[\gd\cA]$  over a graded commutative ring
$\cA$ consists of the monomials
\be
a_0da_1\w\cdots\w da_k, \qquad a_i\in\cA.
\ee
The corresponding complex
\mar{t100}\beq
0\to\cK\ar \cA\ar^d\cO^1\cA\ar^d \cdots  \cO^k\cA\ar^d \cdots
\label{t100}
\eeq
is called the {\sl bigraded de Rham complex} \index{de Rham
complex!bigraded} of \index{bigraded!de Rham complex} $\cA$.

Following the construction of a connection in commutative geometry
(see Section 8.2), one comes to the notion of a connection on
modules over a real graded commutative ring $\cA$. The following
are the straightforward counterparts of Definitions \ref{1016} and
\ref{mos088}.

\begin{definition} \label{ws33} \mar{ws33}
A {\sl connection} \index{connection!on a graded module} on a
graded $\cA$-module $P$ is a graded $\cA$-module morphism
\mar{ws34}\beq
\gd\cA\ni u\to \nabla_u\in \dif_1(P,P) \label{ws34}
\eeq
such that the first order differential operators $\nabla_u$ obey
the  \index{Leibniz rule!for a connection!Grassmann-graded} {\sl
Grassmann-graded Leibniz rule}
\mar{ws35}\beq
\nabla_u (ap)= u(a)p+ (-1)^{[a][u]}a\nabla_u(p), \quad a\in \cA,
\quad p\in P. \label{ws35}
\eeq
\end{definition}

\begin{definition} \label{ws36} \mar{ws36}
Let $P$ in Definition \ref{ws33} be a graded commutative
$\cA$-ring and $\gd P$ the derivation module of $P$ as a graded
commutative $\cK$-ring. A {\sl connection} \index{connection!on a
graded commutative ring} on a graded commutative $\cA$-ring $P$ is
a graded $\cA$-module morphism
\mar{ws37}\beq
\gd\cA\ni u\to \nabla_u\in \gd P, \label{ws37}
\eeq
which is a connection on $P$ as an $\cA$-module, i.e., it obeys
the graded Leibniz rule (\ref{ws35}).
\end{definition}

\section{Graded manifolds}

A {\sl graded manifold} \index{graded!manifold} of dimension
$(n,m)$ is defined as a local-ringed space $(Z,\gA)$ where $Z$ is
an $n$-dimensional smooth manifold $Z$ and $\gA=\gA_0\oplus\gA_1$
is a sheaf of graded commutative algebras of rank $m$ such that:

$\bullet$ there is the exact sequence of sheaves
\mar{cmp140}\beq
0\to \cR \to\gA \op\to^\si C^\infty_Z\to 0, \qquad
\cR=\gA_1+(\gA_1)^2,\label{cmp140}
\eeq
where $C^\infty_Z$ is the sheaf of smooth real functions on $Z$;

$\bullet$ $\cR/\cR^2$ is a locally free sheaf of
$C^\infty_Z$-modules of finite rank (with respect to pointwise
operations), and the sheaf $\gA$ is locally isomorphic to the
exterior product $\w_{C^\infty_Z}(\cR/\cR^2)$.

The sheaf $\gA$ is called a {\sl structure sheaf} \index{structure
sheaf!of a graded manifold} of a graded manifold $(Z,\gA)$, and a
manifold $Z$ is said to be the {\sl body} \index{body!of a graded
manifold} of $(Z,\gA)$. Sections of the sheaf $\gA$ are called
{\sl graded functions} \index{graded!function} on a graded
manifold $(Z,\gA)$. They make up a graded commutative
$C^\infty(Z)$-ring $\gA(Z)$ \index{$\gA(Z)$} called the {\sl
structure ring} of $(Z,\gA)$. \index{structure ring of a graded
manifold}

A graded manifold $(Z,\gA)$ possesses the following local
structure. Given a point $z\in Z$, there exists its open
neighborhood $U$, called a {\sl splitting domain},
\index{splitting domain} such that
\mar{+54}\beq
\gA(U)= C^\infty(U)\ot\w\Bbb R^m. \label{+54}
\eeq
This means that the restriction $\gA|_U$ of the structure sheaf
$\gA$ to $U$ is isomorphic to the sheaf $C^\infty_U\ot\w\Bbb R^m$
of sections of some exterior bundle
\be
\w E^*_U= U\times \w\Bbb R^m\to U.
\ee

The well-known {\sl Batchelor theorem} \index{Batchelor theorem}
states that such a structure of a graded manifold is global as
follows.

\begin{theorem} \label{lmp1a} \mar{lmp1a}
Let $(Z,\gA)$ be a graded manifold. There exists a vector bundle
$E\to Z$ with an $m$-dimensional typical fibre $V$ such that the
structure sheaf $\gA$ of $(Z,\gA)$ is isomorphic to the structure
sheaf $\gA_E=S_{\w E^*}$ \index{$\gA_E$} of germs of sections of
the exterior bundle $\w E^*$ (\ref{ss12f11}), whose typical fibre
is the Grassmann algebra $\w V^*$.
\end{theorem}

Note that Batchelor's isomorphism in Theorem \ref{lmp1a} fails to
be canonical. In field models, it however is fixed from the
beginning. Therefore, we restrict our consideration to graded
manifolds $(Z,\gA_E)$ whose structure sheaf is the sheaf of germs
of sections of some exterior bundle $\w E^*$. We agree to call
$(Z,\gA_E)$ a {\sl simple graded manifold} modelled over a vector
bundle $E\to Z$, \index{graded!manifold!simple} called its {\sl
characteristic vector bundle}. \index{vector
bundle!characteristic} Accordingly, the structure ring $\cA_E$
\index{$\cA_E$} of a simple graded manifold $(Z,\gA_E)$ is the
structure module
\mar{33f1}\beq
\cA_E=\gA_E(Z)=\w E^*(Z) \label{33f1}
\eeq
of sections of the exterior bundle $\w E^*$. Automorphisms of a
simple graded manifold $(Z,\gA_E)$ are restricted to those induced
by automorphisms of its characteristic vector bundles $E\to Z$
(see Remark \ref{33r1}).

Combining Batchelor Theorem \ref{lmp1a} and classical Serre--Swan
Theorem \ref{sp60}, we come to the following {\sl Serre--Swan
theorem for graded manifolds}. \index{Serre--Swan theorem!for
graded manifolds}

\begin{theorem} \label{vv0} \mar{vv0}
Let $Z$ be a smooth manifold. A graded commutative
$C^\infty(Z)$-algebra $\cA$ is isomorphic to the structure ring of
a graded manifold with a body $Z$ iff it is the exterior algebra
of some projective $C^\infty(Z)$-module of finite rank.
\end{theorem}

Given a graded manifold $(Z,\gA_E)$, every trivialization chart
$(U; z^A,y^a)$ of the vector bundle $E\to Z$ yields a splitting
domain $(U; z^A,c^a)$ of $(Z,\gA_E)$. Graded functions on such a
chart are $\La$-valued functions
\mar{z785}\beq
f=\op\sum_{k=0}^m \frac1{k!}f_{a_1\ldots a_k}(z)c^{a_1}\cdots
c^{a_k}, \label{z785}
\eeq
where $f_{a_1\cdots a_k}(z)$ are smooth functions on $U$ and
$\{c^a\}$ is the  fibre basis for $E^*$. In particular, the sheaf
epimorphism $\si$ in (\ref{cmp140}) is induced by the body map of
$\La$. One calls $\{z^A,c^a\}$ the local {\sl basis for the graded
manifold} $(Z,\gA_E)$. \index{basis!for a graded manifold}
Transition functions $y'^a=\rho^a_b(z^A)y^b$ of bundle coordinates
on $E\to Z$ induce the corresponding transformation
\mar{+6}\beq
c'^a=\rho^a_b(z^A)c^b \label{+6}
\eeq
of the associated local basis for the graded manifold $(Z,\gA_E)$
and the according coordinate transformation law of graded
functions (\ref{z785}).

\begin{remark} \mar{33r62} \label{33r62}
Strictly speaking, elements $c^a$ of the local basis for a graded
manifold are locally constant sections $c^a$ of $E^*\to X$ such
that $y_b\circ c^a=\dl^a_b$. Therefore, graded functions are
locally represented by $\La$-valued functions (\ref{z785}), but
they are not $\La$-valued functions on a manifold $Z$ because of
the transformation law (\ref{+6}).
\end{remark}

\begin{remark} \label{33r1} \mar{33r1}
In general, automorphisms of a graded manifold read
\mar{+95}\beq
c'^a=\rho^a(z^A,c^b). \label{+95}
\eeq
Considering a simple graded manifold $(Z,\gA_E)$, we restrict the
class of graded manifold transformations (\ref{+95}) to the linear
ones (\ref{+6}), compatible with given Batchelor's isomorphism.
\end{remark}

Let $E\to Z$ and $E'\to Z$ be vector bundles and $\Phi: E\to E'$
their bundle morphism over a morphism $\vf: Z\to Z'$. Then every
section $s^*$ of the dual bundle $E'^*\to Z'$ defines the
pull-back section $\Phi^*s^*$ of the dual bundle $E^*\to Z$ by the
law
\be
v_z\rfloor \Phi^*s^*(z)=\Phi(v_z)\rfloor s^*(\vf(z)), \qquad
v_z\in E_z.
\ee
It follows that the bundle morphism $(\Phi,\vf)$ yields
\index{morphism!of graded manifolds} a {\sl morphism of simple
graded manifolds}
\mar{w901}\beq
\wh\Phi: (Z,\gA_E) \to (Z',\gA_{E'}) \label{w901}
\eeq
as local-ringed spaces. This is a pair $(\vf,\vf_*\circ\Phi^*)$ of
a morphism $\vf$ of  body manifolds and the composition
$\vf_*\circ\Phi^*$ of the pull-back
\be
\cA_{E'}\ni f\to \Phi^*f\in\cA_E
\ee
of graded functions and the direct image $\vf_*$ of the sheaf
$\gA_E$ onto $Z'$. Relative to local bases $(z^A,c^a)$ and
$(z'^A,c'^a)$ for $(Z,\gA_E)$ and $(Z',\gA_{E'})$, the morphism
(\ref{w901}) of graded manifolds reads
\be
\wh\Phi(z)=\vf(z), \qquad \wh\Phi(c'^a)=\Phi^a_b(z)c^b.
\ee

Given a graded manifold $(Z,\gA)$, by the {\sl sheaf $\gd\gA$ of
graded derivations} \index{sheaf!of graded derivations}  of $\gA$
is meant a subsheaf of endomorphisms of the structure sheaf $\gA$
such that any section $u\in \gd\gA(U)$ of $\gd\gA$ over an open
subset $U\subset Z$ is a graded derivation of the real graded
commutative algebra $\gA(U)$, i.e., $u\in\gd(\gA(U))$. Conversely,
one can show that, given open sets $U'\subset U$, there is a
surjection of the graded derivation modules
\be
\gd(\gA(U))\to \gd(\gA(U')).
\ee
It follows that any graded derivation of the local graded algebra
$\gA(U)$ also is a local section over $U$ of the sheaf $\gd\gA$.
Global sections of $\gd\gA$ are  called {\sl graded vector fields}
\index{graded!vector field} on \index{vector field!graded} the
graded manifold $(Z,\gA)$. They make up the graded derivation
module $\gd\gA(Z)$ of the real graded commutative ring $\gA(Z)$.
This module is a real Lie superalgebra with the superbracket
(\ref{ws14}).

A key point is that graded vector fields $u\in\gd\cA_E$ on a
simple graded manifold $(Z,\gA_E)$ can be represented by sections
of some vector bundle as follows. Due to the canonical splitting
$VE= E\times E$, the vertical tangent bundle $VE$ of $E\to Z$ can
be provided with the fibre bases $\{\dr/\dr c^a\}$, which are the
duals of the bases $\{c^a\}$. Then graded vector fields on a
splitting domain $(U;z^A,c^a)$ of $(Z,\gA_E)$ read
\mar{hn14}\beq
u= u^A\dr_A + u^a\frac{\dr}{\dr c^a}, \label{hn14}
\eeq
where $u^\la, u^a$ are local graded functions on $U$. In
particular,
\be
\frac{\dr}{\dr c^a}\circ\frac{\dr}{\dr c^b} =-\frac{\dr}{\dr
c^b}\circ\frac{\dr}{\dr c^a}, \qquad \dr_A\circ\frac{\dr}{\dr
c^a}=\frac{\dr}{\dr c^a}\circ \dr_A.
\ee
Graded derivations (\ref{hn14}) act on graded functions $f$
(\ref{z785}) by the rule
\mar{cmp50a}\beq
u(f_{a\ldots b}c^a\cdots c^b)=u^A\dr_A(f_{a\ldots b})c^a\cdots c^b
+u^k f_{a\ldots b}\frac{\dr}{\dr c^k}\rfloor (c^a\cdots c^b).
\label{cmp50a}
\eeq
This rule implies the corresponding coordinate transformation law
\be
u'^A =u^A, \qquad u'^a=\rho^a_ju^j +u^A\dr_A(\rho^a_j)c^j
\ee
of graded vector fields. It follows that graded vector fields
(\ref{hn14}) can be represented by sections of the following
vector bundle $\cV_E\to Z$. \index{$\cV_E$} This vector bundle is
locally isomorphic to the vector bundle
\mar{+243}\beq
\cV_E|_U\approx\w E^*\op\ot_Z(E\op\oplus_Z TZ)|_U, \label{+243}
\eeq
and is characterized by an atlas of bundle coordinates
\be
(z^A,z^A_{a_1\ldots a_k},v^i_{b_1\ldots b_k}), \qquad
k=0,\ldots,m,
\ee
possessing the transition functions
\be
&& z'^A_{i_1\ldots
i_k}=\rho^{-1}{}_{i_1}^{a_1}\cdots
\rho^{-1}{}_{i_k}^{a_k} z^A_{a_1\ldots a_k}, \\
&& v'^i_{j_1\ldots j_k}=\rho^{-1}{}_{j_1}^{b_1}\cdots
\rho^{-1}{}_{j_k}^{b_k}\left[\rho^i_jv^j_{b_1\ldots b_k}+
\frac{k!}{(k-1)!} z^A_{b_1\ldots b_{k-1}}\dr_A\rho^i_{b_k}\right],
\ee
which fulfil the cocycle condition (\ref{+9}). Thus, the graded
derivation module $\gd\cA_E$ \index{$\gd\cA_E$} is isomorphic to
the structure module $\cV_E(Z)$ of global sections of the vector
bundle $\cV_E\to Z$.

There is the exact sequence
\mar{1030}\beq
0\to \w E^*\op\ot_Z E\to\cV_E\to \w E^*\op\ot_Z TZ\to 0
\label{1030}
\eeq
of vector bundles over $Z$. Its splitting
\mar{cmp70}\beq
\wt\g:\dot z^A\dr_A \to \dot z^A\left(\dr_A
+\wt\g_A^a\frac{\dr}{\dr c^a}\right) \label{cmp70}
\eeq
transforms every vector field $\tau$ on $Z$ into the graded vector
field
\mar{ijmp10}\beq
\tau=\tau^A\dr_A\to \nabla_\tau=\tau^A\left(\dr_A
+\wt\g_A^a\frac{\dr}{\dr c^a}\right), \label{ijmp10}
\eeq
which is a graded derivation of the real graded commutative ring
$\cA_E$ (\ref{33f1}) satisfying the Leibniz rule
\be
\nabla_\tau(sf)=(\tau\rfloor ds)f +s\nabla_\tau(f), \quad
f\in\cA_E, \quad s\in C^\infty(Z).
\ee
It follows that the splitting (\ref{cmp70}) of the exact sequence
(\ref{1030}) yields a connection on the graded commutative
$C^\infty(Z)$-ring $\cA_E$ in accordance with Definition
\ref{ws36}. It is called a {\sl graded connection}
\index{connection!on a graded manifold} on
\index{graded!connection} the simple graded manifold $(Z,\gA_E)$.
In particular, this connection provides the corresponding
horizontal splitting
\be
u= u^A\dr_A + u^a\frac{\dr}{\dr c^a}=u^A\left(\dr_A
+\wt\g_A^a\frac{\dr}{\dr c^a}\right) + (u^a-
u^A\wt\g_A^a)\frac{\dr}{\dr c^a}
\ee
of graded vector fields. In accordance with Theorem \ref{sp11}, a
graded connection (\ref{cmp70}) always exists.

\begin{remark} \label{+94} \mar{+94}
By virtue of the isomorphism (\ref{+54}), any connection $\wt \g$
on a graded manifold $(Z,\gA)$, restricted to a splitting domain
$U$, takes the form (\ref{cmp70}). Given two splitting domains $U$
and $U'$ of $(Z,\gA)$ with the transition functions (\ref{+95}),
the connection components $\wt\g^a_A$ obey the transformation law
\mar{+96}\beq
\wt\g'^a_A= \wt\g^b_A\frac{\dr}{\dr c^b}\rho^a +\dr_A\rho^a.
\label{+96}
\eeq
If $U$ and $U'$ are the trivialization charts of the same vector
bundle $E$ in Theorem \ref{lmp1a} together with the transition
functions (\ref{+6}), the transformation law (\ref{+96}) takes the
form
\mar{+97}\beq
\wt\g'^a_A= \rho^a_b(z)\wt\g^b_A +\dr_A\rho^a_b(z)c^b. \label{+97}
\eeq
\end{remark}

\begin{remark} \label{1031} \mar{1031}
Every linear connection
\be
\g=dz^A\ot (\dr_A +\g_A{}^a{}_by^b \dr_a)
\ee
on a vector bundle $E\to Z$ yields the graded connection
\mar{cmp73}\beq
\g_S=dz^A\ot \left(\dr_A +\g_A{}^a{}_bc^b\frac{\dr}{\dr
c^a}\right) \label{cmp73}
\eeq
on the simple graded manifold $(Z,\gA_E)$ modelled over $E$. In
view of Remark \ref{+94}, $\g_S$ also is a graded connection on
the graded manifold $(Z,\gA)\cong (Z,\gA_E)$, but its linear form
(\ref{cmp73}) is not maintained under the transformation law
(\ref{+96}).
\end{remark}

\section{Graded differential forms}

Given the structure ring $\cA_E$ of graded functions on a simple
graded manifold $(Z,\gA_E)$ and the real Lie superalgebra
$\gd\cA_E$ of its graded derivations, let us consider the graded
Chevalley--Eilenberg differential calculus
\mar{33f21}\beq
\cS^*[E;Z]=\cO^*[\gd\cA_E] \label{33f21}
\eeq
over \index{$\cS^*[E;Z]$} $\cA_E$. Since the graded derivation
module $\gd\cA_E$ is isomorphic to the structure module of
sections of the vector bundle $\cV_E\to Z$, elements of
$\cS^*[E;Z]$ are sections of the exterior bundle $\w\ol\cV_E$ of
the $\cA_E$-dual $\ol\cV_E\to Z$ of $\cV_E$. The bundle $\ol\cV_E$
is locally isomorphic to the vector bundle
\mar{+244}\beq
\ol\cV_E|_U\approx (E^*\op\oplus_Z T^*Z)|_U. \label{+244}
\eeq
With respect to the dual fibre bases $\{dz^A\}$ for $T^*Z$ and
$\{dc^b\}$ for $E^*$, sections of $\ol\cV_E$ take the coordinate
form
\be
\f=\f_A dz^A + \f_adc^a,
\ee
together with transition functions
\be
\f'_a=\rho^{-1}{}_a^b\f_b, \qquad \f'_A=\f_A
+\rho^{-1}{}_a^b\dr_A(\rho^a_j)\f_bc^j.
\ee
The duality isomorphism $\cS^1[E;Z]=\gd\cA_E^*$ (\ref{ws47}) is
given by the graded interior product
\mar{cmp65}\beq
u\rfloor \f=u^A\f_A + (-1)^{\nw{\f_a}}u^a\f_a. \label{cmp65}
\eeq
Elements of $\cS^*[E;Z]$ are called {\sl graded exterior forms}
\index{graded!exterior form} on \index{exterior form!graded} the
graded manifold $(Z,\gA_E)$.

Seen as an $\cA_E$-algebra, the DBGA $\cS^*[E;Z]$ (\ref{33f21}) on
a splitting domain $(U;z^A,c^a)$ is locally generated by the
graded one-forms $dz^A$, $dc^i$ such that
\mar{+113'}\beq
dz^A\w dc^i=-dc^i\w dz^A, \qquad dc^i\w dc^j= dc^j\w dc^i.
\label{+113'}
\eeq
Accordingly, the graded Chevalley--Eilenberg coboundary operator
$d$ (\ref{ws86}), called the {\sl graded!exterior differential},
\index{graded!exterior differential} reads
\be
d\f= dz^A \w \dr_A\f +dc^a\w \frac{\dr}{\dr c^a}\f,
\ee
where the derivatives $\dr_\la$, $\dr/\dr c^a$ act on coefficients
of graded exterior forms by the formula (\ref{cmp50a}), and they
are graded commutative with the graded forms $dz^A$ and $dc^a$.
The formulas (\ref{ws45}) -- (\ref{+117}) hold.

\begin{theorem} \label{v62} \mar{v62}
The DBGA $\cS^*[E;Z]$ (\ref{33f21}) is a minimal differential
calculus over $\cA_E$, i.e., it is generated by elements $df$,
$f\in \cA_E$.
\end{theorem}

The bigraded de Rham complex (\ref{t100}) of the minimal graded
Chevalley--Eilenberg differential calculus $\cS^*[E;Z]$ reads
\mar{+137}\beq
0\to \Bbb R\to \cA_E \ar^d \cS^1[E;Z]\ar^d\cdots
\cS^k[E;Z]\ar^d\cdots. \label{+137}
\eeq
Its cohomology $H^*(\cA_E)$  is called the {\sl de Rham cohomology
of a simple graded manifold} \index{de Rham cohomology!of a graded
manifold} $(Z,\gA_E)$.

In particular, given the DGA $\cO^*(Z)$ of exterior forms on $Z$,
there exist the canonical monomorphism
\mar{uut}\beq
\cO^*(Z)\to \cS^*[E;Z] \label{uut}
\eeq
and the body epimorphism $\cS^*[E;Z]\to \cO^*(Z)$ which are
cochain morphisms of the de Rham complexes (\ref{+137}) and
(\ref{t37}).

\begin{theorem} \label{33t3} \mar{33t3}
The de Rham cohomology of a simple graded manifold $(Z,\gA_E)$
equals the de Rham cohomology of its body $Z$.
\end{theorem}

\begin{corollary} \label{33c1} \mar{33c1}
Any closed graded exterior form is decomposed into a sum $\f=\si
+d\xi$ where $\si$ is a closed exterior form on $Z$.
\end{corollary}

\chapter{Lagrangian theory}

Lagrangian theory on fibre bundles is algebraically formulated in
terms of the variational bicomplex without appealing to the
calculus of variations. This formulation is extended to Lagrangian
theory on graded manifolds.

\section{Variational bicomplex}

Let $Y\to X$ be a fibre bundle. The DGA $\cO^*_\infty$
(\ref{5.77}), decomposed into the variational bicomplex, describes
finite order Lagrangian theories on $Y\to X$. One also considers
the variational bicomplex of the DGA $\cQ^*_\infty$ (\ref{17t4})
and different variants of the variational sequence of finite jet
order.

In order to transform the bicomplex $\cO^{*,*}_\infty$ into the
variational one, let us consider the following two operators
acting on $\cO^{*,n}_\infty$.

(i) There exists an $\Bbb R$-module endomorphism
\mar{r12}\ben
&& \vr=\op\sum_{k>0} \frac1k\ol\vr\circ h_k\circ h^n:
\cO^{*>0,n}_\infty\to \cO^{*>0,n}_\infty,
\label{r12}\\
&& \ol\vr(\f)= \op\sum_{0\leq|\La|} (-1)^{|\La|}\thh^i\w
[d_\La(\dr^\La_i\rfloor\f)], \qquad \f\in \cO^{>0,n}_\infty,
\nonumber
\een
possessing the following properties.

\begin{lemma} \label{21t1} \mar{21t1}
For any $\f\in \cO^{>0,n}_\infty$, the form $\f-\vr(\f)$ is
locally $d_H$-exact on each coordinate chart (\ref{j3}).
\end{lemma}

\begin{lemma} \label{21t2} \mar{21t2}
The operator $\vr$ obeys the relation
\mar{21f5}\beq
(\vr\circ d_H)(\psi)=0, \qquad \psi\in \cO^{>0,n-1}_\infty.
\label{21f5}
\eeq
\end{lemma}

It follows from Lemmas \ref{21t1} and \ref{21t2} that $\vr$
(\ref{r12}) is a projector.

(ii) One defines the \index{variational!operator} {\sl variational
operator}
\mar{21f1}\beq
\dl=\vr\circ d : \cO^{*,n}_\infty\to \cO^{*+1,n}_\infty,
\label{21f1}
\eeq
which is nilpotent, i.e., $\dl\circ\dl=0$, and obeys the relation
$\dl\circ\vr=\dl$.

Let us denote $\bE_k=\vr(\cO^{k,n}_\infty)$. Provided with the
operators $d_H$, $d_V$, $\vr$ and $\dl$,  the DGA $\cO^*_\infty$
is decomposed into the {\sl variational bicomplex}
\index{variational!bicomplex}
\mar{7}\beq
\begin{array}{ccccrlcrlcccrlcrl}
 & &  &  & & \vdots & & & \vdots  & & & & &
\vdots & &   & \vdots \\
& & & & _{d_V} & \put(0,-7){\vector(0,1){14}} & & _{d_V} &
\put(0,-7){\vector(0,1){14}} & &  & & _{d_V} &
\put(0,-7){\vector(0,1){14}}& & _{-\dl} & \put(0,-7){\vector(0,1){14}} \\
 &  & 0 & \to & &\cO^{1,0}_\infty &\op\to^{d_H} & &
\cO^{1,1}_\infty & \op\to^{d_H} & \cdots & & &
\cO^{1,n}_\infty &\op\to^\vr &  & \bE_1\to  0\\
& & & & _{d_V} &\put(0,-7){\vector(0,1){14}} & & _{d_V} &
\put(0,-7){\vector(0,1){14}} & & &  & _{d_V} &
\put(0,-7){\vector(0,1){14}}
 & & _{-\dl} & \put(0,-7){\vector(0,1){14}} \\
0 & \to & \Bbb R & \to & & \cO^0_\infty &\op\to^{d_H} & &
\cO^{0,1}_\infty & \op\to^{d_H} & \cdots & & &
\cO^{0,n}_\infty & \equiv &  & \cO^{0,n}_\infty \\
& & & & & \put(0,-7){\vector(0,1){14}} & &  &
\put(0,-7){\vector(0,1){14}} & & & &  &
\put(0,-7){\vector(0,1){14}} & &  & \\
0 & \to & \Bbb R & \to & & \cO^0(X) &\op\to^d & & \cO^1(X) &
\op\to^d & \cdots & & &
\cO^n(X) & \op\to^d & 0 &  \\
& & & & &\put(0,-5){\vector(0,1){10}} & & &
\put(0,-5){\vector(0,1){10}} & &  &  & &
\put(0,-5){\vector(0,1){10}} & &  & \\
& & & & &0 & &  & 0 & & & & &  0 & &  &
\end{array}
\label{7}
\eeq

\noindent It possesses the following cohomology.

\begin{theorem} \label{g90} \mar{g90}
The second row from the bottom and the last column of the
variational bicomplex (\ref{7}) make up the
\index{variational!complex} {\sl variational complex}
\index{complex!variational}
\mar{b317}\beq
0\to\Bbb R\to \cO^0_\infty \ar^{d_H}\cO^{0,1}_\infty\cdots
\op\longrightarrow^{d_H} \cO^{0,n}_\infty  \op\longrightarrow^\dl
\bE_1 \op\longrightarrow^\dl \bE_2 \ar \cdots\,. \label{b317}
\eeq
Its cohomology is isomorphic to the de Rham cohomology of a fibre
bundle $Y$, namely,
\mar{j20}\beq
H^{k<n}(d_H;\cO^*_\infty)=H^{k<n}_{\rm DR}(Y), \qquad H^{k\geq
n}(\dl; \cO^*_\infty)=H^{k\geq n}_{\rm DR}(Y). \label{j20}
\eeq
\end{theorem}

\begin{theorem} \label{21t7} \mar{21t7}
The rows of contact forms of the variational bicomplex (\ref{7})
are exact sequences.
\end{theorem}

The cohomology isomorphism (\ref{j20}) gives something more. Due
to the relations $d_H\circ h_0= h_0\circ d$ and $\dl\circ\vr=\dl$,
we have the cochain morphism of the de Rham complex (\ref{5.13})
of the DGA $\cO^*_\infty$ to its variational complex (\ref{b317}).
By virtue of Theorems \ref{j4} and \ref{g90}, the corresponding
homomorphism of their cohomology groups is an isomorphism. Then
the splitting of a closed form $\f\in\cO^*_\infty$ in Corollary
\ref{j21} leads to the following decompositions.

\begin{theorem} \label{t41} \mar{t41}
Any $d_H$-closed form $\f\in\cO^{0,m}$, $m< n$, is represented by
a sum
\mar{t60'}\beq
\f=h_0\si+ d_H \xi, \qquad \xi\in \cO^{m-1}_\infty, \label{t60'}
\eeq
where $\si$ is a closed $m$-form on $Y$. Any $\dl$-closed form
$\f\in\cO^{k,n}$ is split into
\mar{t42a-c}\ben
&& \f=h_0\si +
d_H\xi, \qquad k=0, \qquad \xi\in \cO^{0,n-1}_\infty,
\label{t42a}\\
&& \f=\vr(\si) +\dl(\xi), \qquad k=1, \qquad \xi\in
\cO^{0,n}_\infty,
\label{t42b}\\
&& \f=\vr(\si) +\dl(\xi), \qquad k>1, \qquad \xi\in \bE_{k-1},
\label{t42c}
\een
where $\si$ is a closed $(n+k)$-form on $Y$.
\end{theorem}

\section{Lagrangian theory on fibre bundles}

In Lagrangian formalism on fibre bundles, a finite order {\sl
Lagrangian} \index{Lagrangian} and its {\sl Euler--Lagrange
operator} \index{Euler--Lagrange operator} are defined as elements
\index{$\cE_L$}
\mar{21f10,12}\ben
&& L=\cL\om\in \cO^{0,n}_\infty, \label{21f10}\\
&& \dl L=\cE_L=\cE_i\thh^i\w\om \in \bE_1, \label{21f11}\\
&& \cE_i=\op\sum_{0\leq|\La|}(-1)^{|\La|}d_\La(\dr^\La_i \cL),
\label{21f12}
\een
of the variational complex (\ref{b317}) (see the notation
(\ref{gm141})). Components $\cE_i$ (\ref{21f12}) \index{$\cE_i$}
of the Euler--Lagrange operator (\ref{21f11}) are called the {\sl
variational derivatives}. \index{variational!derivative} Elements
of $\bE_1$ are called the {\sl Euler--Lagrange-type operators}.
\index{Euler--Lagrange-type operator}

Hereafter, we call a pair $(\cO^*_\infty, L)$ the {\sl Lagrangian
system}. \index{Lagrangian system} The following are corollaries
of Theorem \ref{t41}.

\begin{corollary} \label{lmp112'} \mar{lmp112'}
A finite order Lagrangian $L$ (\ref{21f10}) is {\sl variationally
trivial}, \index{Lagrangian!variationally trivial} i.e.,
$\dl(L)=0$ iff
\mar{tams3}\beq
L=h_0\si + d_H \xi, \qquad \xi\in \cO^{0,n-1}_\infty,
\label{tams3}
\eeq
where $\si$ is a closed $n$-form on $Y$.
\end{corollary}

\begin{corollary} \label{21t12} \mar{21t12}
A finite order Euler--Lagrange-type operator $\cE\in \bE_1$
satisfies the {\sl Helmholtz condition} \index{Heimholtz
condition} $\dl(\cE)=0$ iff
\be
\cE=\dl L + \vr(\si), \qquad L\in\cO^{0,n}_\infty,
\ee
where $\si$ is a closed $(n+1)$-form on $Y$.
\end{corollary}

A glance at the expression (\ref{21f11}) shows that, if a
Lagrangian $L$ (\ref{21f10}) is of $r$-order, its Euler--Lagrange
operator $\cE_L$ is of $2r$-order. Its kernel is called the {\sl
Euler--Lagrange equation}. Euler--Lagrange equations traditionally
came from the {\sl variational formula}
\index{variational!formula}
\mar{+421}\beq
dL=\dl L-d_H\Xi_L \label{+421}
\eeq
of the calculus of variations. In formalism of the variational
bicomplex, this formula is a corollary of Theorem \ref{21t7}.

\begin{corollary} \label{g93} \mar{g93}
The exactness of the row of one-contact forms of the variational
bicomplex (\ref{7}) at the term $\cO^{1,n}_\infty$ relative to the
projector $\vr$ provides the $\Bbb R$-module decomposition
\be
\cO^{1,n}_\infty=\bE_1\oplus d_H(\cO^{1,n-1}_\infty).
\ee
In particular, any Lagrangian $L$ admits the decomposition
(\ref{+421}).
\end{corollary}

Defined up to a $d_H$-closed term, a form $\Xi_L\in \cO^n_\infty$
in the variational formula (\ref{+421}) reads
\mar{21f30}\ben
&&\Xi_L=L+[(\dr^\la_i\cL-d_\m F^{\m\la}_i)\thh^i
+\op\sum_{s=1}F^{\la\nu_s\ldots\nu_1}_i
\thh^i_{\nu_s\ldots\nu_1}]\w\om_\la, \label{21f30}\\
&& F_i^{\nu_k\ldots\nu_1}= \dr_i^{\nu_k\ldots\nu_1}\cL-d_\m
F_i^{\m\nu_k\ldots\nu_1}+ \si_i^{\nu_k\ldots\nu_1}, \qquad
k=2,3,\ldots,\nonumber
\een
where $\si_i^{\nu_k\ldots\nu_1}$ are local functions such that
\be
\si_i^{(\nu_k\nu_{k-1})\ldots\nu_1}=0.
\ee
The form $\Xi_L$ (\ref{21f30}) possesses the following properties:

$\bullet$ $h_0(\Xi_L)=L$,

$\bullet$  $h_0(\vt\rfloor d\Xi_L)=\vt^i\cE_i\om$ for any
derivation $\vt$ (\ref{g3}).

\noindent Consequently, $\Xi_L$ is a Lepage equivalent of a
Lagrangian $L$.

A special interest is concerned with Lagrangian theories on an
affine bundle $Y\to X$. Since $X$ is a strong deformation retract
of an affine bundle $Y$, the de Rham cohomology of $Y$ equals that
of $X$. In this case, the cohomology (\ref{j20}) of the
variational complex (\ref{b317}) equals the de Rham cohomology of
$X$, namely,
\mar{21f33}\ben
&& H^{<n}(d_H;\cO^*_\infty)= H^{<n}_{\rm DR}(X), \nonumber\\
&& H^n(\dl;\cO^*_\infty)=H^n_{\rm DR}(X), \label{21f33}\\
&& H^{>n}(\dl;\cO^*_\infty)=0. \nonumber
\een
It follows that every $d_H$-closed form $\f\in \cO^{0,m<n}_\infty$
is represented by the sum
\mar{t2}\beq
\f=\si + d_H\xi, \qquad \xi\in \cO^{0,m-1}_\infty, \label{t2}
\eeq
where $\si$ is a closed $m$-form on $X$. Similarly, any
variationally trivial Lagrangian takes the form
\mar{t2'}\beq
L=\si + d_H\xi, \qquad \xi\in \cO^{0,n-1}_\infty, \label{t2'}
\eeq
where $\si$ is an $n$-form on $X$.

In view of the cohomology isomorphism (\ref{21f33}), if $Y\to X$
is an affine bundle, let us restrict our consideration to the
\index{variational!complex!short} {\sl short variational complex}
\mar{j25}\beq
0\to\Bbb R\to \cO^0_\infty \ar^{d_H}\cO^{0,1}_\infty\cdots
\op\longrightarrow^{d_H} \cO^{0,n}_\infty  \op\longrightarrow^\dl
\bE_1, \label{j25}
\eeq
whose non-trivial cohomology equals that of the variational
complex (\ref{b317}). Let us consider a DGA $\cP^*_\infty\subset
\cO^*_\infty$ \index{$\cP^*_\infty$} of exterior forms whose
coefficients are polynomials in jet coordinates $y^i_\La$, $0\leq
|\La|$, of the continuous bundle $J^\infty Y\to X$.

\begin{theorem} \label{21t40} \mar{21t40}
The cohomology of the short variational complex
\mar{21f34}\beq
0\to\Bbb R\to \cP^0_\infty \ar^{d_H}\cP^{0,1}_\infty\cdots
\op\longrightarrow^{d_H} \cP^{0,n}_\infty  \op\longrightarrow^\dl
0 \label{21f34}
\eeq
of the polynomial algebra $\cP^*_\infty$ equals that of the
complex (\ref{j25}), i.e., the de Rham cohomology of $X$.
\end{theorem}

Given a Lagrangian system $(\cO^*_\infty, L)$, its {\sl
infinitesimal transformations} \index{infinitesimal
transformation!of a Lagrangian system} are defined to be contact
derivations of the ring $\cO^0_\infty$.

A derivation $\vt\in \gd \cO^0_\infty$ (\ref{g3}) is called {\sl
contact} \index{contact derivation} \index{derivation!contact} if
the Lie derivative $\bL_\up$ preserves the ideal of contact forms
of the DGA $\cO^*_\infty$, i.e., the Lie derivative $\bL_\up$ of a
contact form is a contact form.

\begin{lemma}
A derivation $\vt$ (\ref{g3}) is contact iff it takes the form
\mar{g4}\beq
\vt=\up^\la\dr_\la +\up^i\dr_i +\op\sum_{0<|\La|}
[d_\La(\up^i-y^i_\m\up^\m)+y^i_{\m+\La}\up^\m]\dr^\La_i.
\label{g4}
\eeq
\end{lemma}

The expression (\ref{55.5}) enables one to regard a contact
derivation $\vt$ (\ref{g4}) as an infinite order jet prolongation
$\vt=J^\infty\up$ of its restriction
\mar{j15}\beq
\up=\up^\la\dr_\la +\up^i\dr_i \label{j15}
\eeq
to the ring $C^\infty(Y)$. Since coefficients $\up^\la$ and
$\up^i$ of $\up$ (\ref{j15}) depend generally on jet coordinates
$y^i_\La$, $0<|\La|$, one calls $\up$ (\ref{j15}) a {\sl
generalized vector field}. \index{vector field!generalized} It
\index{generalized vector field} can be represented as a section
of some pull-back bundle
\be
J^rY\op\times_Y TY \to J^rY.
\ee
A contact derivation $\vt$ (\ref{g4}) is called {\sl projectable},
\index{contact derivation!projectable} if
\index{derivation!contact!projectable} the generalized vector
field $\up$ (\ref{j15}) projects onto a vector field
$\up^\la\dr_\la$ on $X$.

Any contact derivation $\vt$ (\ref{g4}) admits the horizontal
splitting
\mar{g5,'}\ben
&&\vt=\vt_H +\vt_V=\up^\la d_\la + [\up_V^i\dr_i +
\op\sum_{0<|\La|}
d_\La\up_V^i\dr_i^\La], \label{g5}\\
&& \up=\up_H+ \up_V= \up^\la d_\la +(\up^i-y^i_\m\up^\m)\dr_i,
\label{g5'}
\een
relative to the canonical connection $\nabla$ (\ref{17f20}) on the
$C^\infty(X)$-ring $\cO^0_\infty$.

\begin{lemma}
Any \index{contact derivation!vertical} {\sl vertical contact
derivation} \index{derivation!contact!vertical}
\mar{gg6}\beq
\vt=\up^i\dr_i +\op\sum_{0<|\La|} d_\La \up^i\dr_i^\La \label{gg6}
\eeq
obeys the relations
\mar{g6,'}\ben
&& \vt\rfloor d_H\f=-d_H(\vt\rfloor\f), \label{g6}\\
&& \bL_\vt(d_H\f)=d_H(\bL_\vt\f), \qquad \f\in\cO^*_\infty.
\label{g6'}
\een
\end{lemma}

The global decomposition (\ref{+421}) leads to the following first
variational formula (Theorem \ref{g75}) and the first Noether
theorem (Theorem \ref{j22}).

\begin{theorem}  \label{g75} \mar{g75}
Given a  Lagrangian $L\in\cO^{0,n}_\infty$, its Lie derivative
$\bL_\up L$ along a contact derivation $\up$ (\ref{g5}) fulfils
the \index{first variational formula} {\sl first variational
formula}
\mar{g8}\beq
\bL_\vt L= \up_V\rfloor\dl L +d_H(h_0(\vt\rfloor\Xi_L)) +\cL d_V
(\up_H\rfloor\om), \label{g8}
\eeq
where $\Xi_L$ is the Lepage equivalent (\ref{21f30}) of $L$.
\end{theorem}

A contact derivation $\vt$ (\ref{g4}) is called a {\sl variational
symmetry} \index{variational!symmetry} of
\index{symmetry!variational} a Lagrangian $L$ if the Lie
derivative $\bL_\vt L$ is $d_H$-exact, i.e.,
\mar{22f1}\beq
\bL_\vt L=d_H\si. \label{22f1}
\eeq

\begin{lemma} \label{22l10} \mar{22l10}
A glance at the expression (\ref{g8}) shows the following.

(i)  A contact derivation $\vt$ is a variational symmetry only if
it is projectable.

(ii) Any projectable contact derivation is a variational symmetry
of a variationally trivial Lagrangian. It follows that, if $\vt$
is a variational symmetry of a Lagrangian $L$, it also is a
variational symmetry of a Lagrangian $L+L_0$, where $L_0$ is a
variationally trivial Lagrangian.

(iii) A projectable contact derivations $\vt$ is a variational
symmetry iff its vertical part $\up_V$ (\ref{g5}) is well.

(iv) A projectable contact derivations $\vt$ is a variational
symmetry iff the density $\up_V\rfloor \dl L$ is $d_H$-exact.
\end{lemma}

It is readily observed that variational symmetries of a Lagrangian
$L$ constitute a real vector subspace $\ccG_L$ of the derivation
module $\gd \cO^0_\infty$. By virtue of item (ii) of Lemma
\ref{22l10}, the Lie bracket
\be
\bL_{[\vt,\vt']}= [\bL_\vt,\bL_{\vt'}]
\ee
of variational symmetries is a variational symmetry and,
therefore, their vector space $\ccG_L$ is a real Lie algebra. The
following is the {\sl first Noether theorem}. \index{first Noether
theorem} \index{Noether theorem!first}

\begin{theorem} \label{j22} \mar{j22} If a contact derivation $\vt$
(\ref{g4}) is a variational symmetry (\ref{22f1}) of a Lagrangian
$L$, the first variational formula (\ref{g8}) restricted to the
kernel of the Euler--Lagrange operator $\Ker\cE_L$ leads to
\index{conservation law!weak} the \index{weak conservation law}
{\sl weak conservation law}
\mar{22f2}\beq
0\ap d_H(h_0(\vt\rfloor\Xi_L)-\si) \label{22f2}
\eeq
{\sl on the shell} \index{on-shell} $\dl L=0$.
\end{theorem}

A variational symmetry $\vt$ of a Lagrangian $L$ is called its
{\sl exact symmetry} \index{symmetry!exact} or, simply, a {\sl
symmetry} \index{symmetry} if
\mar{22f0}\beq
\bL_\vt L=0. \label{22f0}
\eeq
Symmetries of a Lagrangian $L$ constitute a real vector space,
which is a real Lie algebra. Its vertical symmetries $\up$
(\ref{gg6}) obey the relation
\be
\bL_\up L=\up\rfloor dL
\ee
and, therefore, make up a $\cO^0_\infty$-module which is a Lie
$C^\infty(X)$-algebra.

If $\vt$ is an exact symmetry of a Lagrangian $L$, the weak
conservation law (\ref{22f2}) takes the form
\mar{22f5}\beq
0\ap d_H(h_0(\vt\rfloor\Xi_L))=-d_H\cJ_\up, \label{22f5}
\eeq
where \index{$\cJ_\up$}
\mar{22f6}\beq
\cJ_\up=\cJ^\m_\up\om_\m= -h_0(\vt\rfloor\Xi_L) \label{22f6}
\eeq
is called the {\sl symmetry current}. \index{symmetry current} Of
course, the symmetry current (\ref{22f6}) is defined with the
accuracy of a $d_H$-closed term.

Let $\vt$ be an exact symmetry of a Lagrangian $L$. Whenever $L_0$
is a variationally trivial Lagrangian, $\vt$ is a variational
symmetry of the Lagrangian $L+L_0$ such that the weak conservation
law (\ref{22f2}) for this Lagrangian is reduced to the weak
conservation law (\ref{22f5}) for a Lagrangian $L$ as follows:
\be
\bL_\vt(L+L_0)=d_H\si\ap d_H\si -d_H\cJ_\up.
\ee

\begin{remark} In accordance with the standard terminology,
variational and exact symmetries generated by generalized vector
fields (\ref{j15}) are called {\sl generalized symmetries}
\index{symmetry!generalized} because they depend on derivatives of
variables. Accordingly, by variational symmetries and symmetries
one means only those generated by vector fields $u$ on $Y$. We
agree to \index{symmetry!classical}  call them
\index{variational!symmetry!classical} {\sl classical symmetries}.
\end{remark}

Let $\vt$ be a classical variational symmetry of a Lagrangian $L$,
i.e.,  $\vt$ (\ref{g4}) is the jet prolongation of a vector field
$u$ on $Y$. Then the relation
\mar{22f9}\beq
\bL_\vt\cE_L=\dl(\bL_\vt L) \label{22f9}
\eeq
holds. It follows that $\vt$ also is a symmetry of the
Euler--Lagrange operator $\cE_L$ of $L$, i.e., $\bL_\vt\cE_L=0$.
However, the equality (\ref{22f9}) fails to be true in the case of
generalized symmetries.

\begin{definition} \mar{s7} \label{s7}
Let $E\to X$ be a vector bundle and $E(X)$ the $C^\infty(X)$
module $E(X)$ of sections of $E\to X$. Let $\zeta$ be a linear
differential operator on $E(X)$ taking values into the vector
space $\ccG_L$ of variational symmetries of a Lagrangian $L$ (see
Definition \ref{ws131}). Elements
\mar{gg1}\beq
u_\xi=\zeta(\xi) \label{gg1}
\eeq
of $\im\zeta$ are called the {\sl gauge symmetry} \index{gauge
symmetry} of \index{symmetry!gauge} a Lagrangian $L$ parameterized
by sections $\xi$ of $E\to X$. They are called the {\sl gauge
parameters}. \index{gauge!parameters}
\end{definition}

\begin{remark} \label{22r100} \mar{22r100}
The differential operator $\zeta$ in Definition \ref{s7} takes its
values into the vector space $\ccG_L$  as a subspace of the
$C^\infty(X)$-module $\gd \cO^0_\infty$, but it sends the
$C^\infty(X)$-module $E(X)$ into the real vector space
$\ccG_L\subset \gd \cO^0_\infty$. The differential operator
$\zeta$ is assumed to be at least of first order.
\end{remark}

Equivalently, the gauge symmetry (\ref{gg1}) is given by a section
$\wt\zeta$ of the fibre bundle
\be
(J^rY\op\times_Y J^mE)\op\times_Y TY\to J^rY\op\times_Y J^mE
\ee
(see Definition \ref{ch538}) such that
\be
u_\xi=\zeta(\xi)=\wt\zeta\circ\xi
\ee
for any section $\xi$ of $E\to X$. Hence, it is a generalized
vector field $u_\zeta$ on the product $Y\times E$  represented by
a section of the pull-back bundle
\be
J^k(Y\op\times_X E)\op\times_Y T(Y\op\times_X E)\to
J^k(Y\op\times_X E), \qquad k={\rm max}(r,m),
\ee
which lives in $TY\subset T(Y\times E)$. This generalized vector
field yields a contact derivation $J^\infty u_\zeta$ (\ref{g4}) of
the real ring $\cO^0_\infty[Y\times E]$ which obeys the following
condition.

\begin{condition} \label{22c1} \mar{22c1} Given a Lagrangian
\be
L\in \cO^{0,n}_\infty E\subset \cO^{0,n}_\infty[Y\times E],
\ee
let us consider its Lie derivative
\mar{gg5}\beq
\bL_{J^\infty u_\zeta} L=J^\infty u_\zeta\rfloor dL + d(J^\infty
u_\zeta\rfloor \label{gg5}L)
\eeq
where $d$ is the exterior differential of $\cO^0_\infty[Y\times
E]$. Then, for any section $\xi$ of $E\to X$, the pull-back $\xi^*
\bL_\vt$ is $d_H$-exact.
\end{condition}

It follows from the first variational formula (\ref{g8}) for the
Lie derivative (\ref{gg5}) that Condition \ref{22c1} holds only if
$u_\zeta$ projects onto a generalized vector field on $E$ and, in
this case, iff the density $(u_\zeta)_V\rfloor \cE$ is
$d_H$-exact. Thus, we come to the following equivalent definition
of gauge symmetries.

\begin{definition} \label{s7'} \mar{s7'} Let $E\to X$ be a vector
bundle. A gauge symmetry of a Lagrangian $L$ parameterized by
sections $\xi$ of $E\to X$ is defined as a contact derivation
$\vt=J^\infty u$ of the real ring $\cO^0_\infty[Y\times E]$ such
that:

(i) it vanishes on the subring $\cO^0_\infty E$,

(ii) the generalized vector field $u$ is linear in coordinates
$\chi^a_\La$ on $J^\infty E$, and it projects onto a generalized
vector field on $E$, i.e., it takes the form
\mar{gg2}\beq
u=\left(\op\sum_{0\leq|\La|\leq m}
u^{\la\La}_a(x^\m)\chi^a_\La\right)\dr_\la +
\left(\op\sum_{0\leq|\La|\leq m}
u^{i\La}_a(x^\m,y^j_\Si)\chi^a_\La\right)\dr_i, \label{gg2}
\eeq

(iii) the vertical part of $u$ (\ref{gg2}) obeys the equality
\mar{gg10}\beq
u_V\rfloor \cE=d_H\si. \label{gg10}
\eeq
\end{definition}

For the sake of convenience, we also call a generalized vector
field (\ref{gg2}) the gauge symmetry. By virtue of item (iii) of
Definition \ref{s7'}, $u$ (\ref{gg2}) is a gauge symmetry iff its
vertical part is so.

Gauge symmetries possess the following particular properties.

(i) Let $E'\to X$ be a vector bundle and $\zeta'$ a linear
$E(X)$-valued differential operator on the $C^\infty(X)$-module
$E'(X)$ of sections of $E'\to X$. Then
\be
u_{\zeta'(\xi')} =(\zeta\circ\zeta')(\xi')
\ee
also is a gauge symmetry of $L$ parameterized by sections $\xi'$
of $E'\to X$. It factorizes through the gauge symmetries $u_\f$
(\ref{gg1}).

(ii) If a gauge symmetry is an exact Lagrangian symmetry, the
corresponding conserved symmetry current $\cJ_u$ (\ref{22f6}) is
reduced to a superpotential (see Theorem \ref{supp'}).

(iii) The {\sl direct second Noether theorem} \index{Noether
theorem!second!direct}  associates to a gauge symmetry of a
Lagrangian $L$ the Noether identities of its Euler--Lagrange
operator $\dl L$.

\begin{theorem} \label{ggg3} \mar{ggg3}
Let $u$ (\ref{gg2}) be a gauge symmetry of a Lagrangian $L$, then
its Euler--Lagrange operator $\dl L$ obeys the Noether identities
\mar{gg11}\ben
&&\cE_a=\op\sum_{0\leq |\La|}(-1)^{|\La|}d_\La[(u^{i\La}_a-y^i_\la
u^{\la\La}_a)\cE_i]= \label{gg11}\\
&&\qquad \op\sum_{0\leq |\La|}\eta(u^i_a-y^i_\la
u^\la_a)^\La d_\La\cE_i=0 \nonumber
\een
(see Notation \ref{42n10}).
\end{theorem}

It follows from direct second Noether Theorem \ref{ggg3} that
gauge symmetries of Lagrangian field theory characterize its
degeneracy. A problem is that any Lagrangian possesses gauge
symmetries and, therefore, one must separate them into the trivial
and non-trivial ones. Moreover, gauge symmetries can be {\sl
reducible}, \index{gauge symmetry!reducible} i.e., $\Ker\zeta\neq
0$. To solve these problems, we follow a different definition of
gauge symmetries as those associated to non-trivial Noether
identities by means of inverse second Noether Theorem \ref{w35}.

\section{Grassmann-graded Lagrangian theory}

We start with the following definition of jets of odd variables.
Let us consider a vector bundle $F\to X$ and the simple graded
manifolds $(X,\cA_{J^rF})$ modelled over the vector bundles
$J^rF\to X$. There is the direct system of the corresponding DBGA
\be
\cS^*[F;X]\ar \cS^*[J^1F;X]\ar\cdots \cS^*[J^rF;X]\ar\cdots
\ee
of graded exterior forms on graded manifolds $(X,\cA_{J^rF})$. Its
direct limit $\cS^*_\infty[F;X]$ \index{$\cS^*_\infty[F;X]$} is
the Grassmann-graded counterpart of the DGA $\cP^*_\infty$.

In order to describe Lagrangian theories both of even and odd
variables, let us consider a composite bundle
\mar{cvc}\beq
F\to Y\to X \label{cvc}
\eeq
where $F\to Y$ is a vector bundle provided with bundle coordinates
$(x^\la, y^i, q^a)$. We call the simple graded manifold
$(Y,\gA_F)$ modelled over $F\to Y$ the {\sl composite graded
manifold}. \index{graded!manifold!composite} Let us associate to
this graded manifold the following DBGA $\cS^*_\infty[F;Y]$.

It is readily observed that the jet manifold $J^rF$ of $F\to X$ is
a vector bundle $J^rF\to J^rY$ coordinated by $(x^\la, y^i_\La,
q^a_\La)$, $0\leq |\La|\leq r$. Let $(J^rY,\gA_r)$ be a simple
graded manifold modelled over this vector bundle. Its local basis
is $(x^\la, y^i_\La, c^a_\La)$, $0\leq|\La|\leq r$. Let
\mar{34f2}\beq
\cS^*_r[F;Y]=\cS^*_r[J^rF;J^rY] \label{43f2}
\eeq
denote \index{$\cS^*_r[F;Y]$} the DBGA of graded exterior forms on
the simple graded manifold $(J^rY,\gA_r)$. In particular, there is
a cochain monomorphism
\mar{34f3}\beq
\cO^*_r=\cO^*(J^rY)\to \cS^*_r[F;Y]. \label{34f3}
\eeq

The surjection
\be
\pi^{r+1}_r:J^{r+1}Y\to J^rY
\ee
yields an epimorphism of graded manifolds
\be
(\pi^{r+1}_r,\wh \pi^{r+1}_r):(J^{r+1}Y,\gA_{r+1}) \to
(J^rY,\gA_r),
\ee
including the sheaf monomorphism
\be
\wh \pi^{r+1}_r:\pi_r^{r+1*}\gA_r\to \gA_{r+1},
\ee
where $\pi_r^{r+1*}\gA_r$ is the pull-back onto $J^{r+1}Y$ of the
continuous fibre bundle $\gA_r\to J^rY$. This sheaf monomorphism
induces the monomorphism of the canonical presheaves $\ol \gA_r\to
\ol \gA_{r+1}$, which associates to each open subset $U\subset
J^{r+1}Y$ the ring of sections of $\gA_r$ over $\pi^{r+1}_r(U)$.
Accordingly, there is a monomorphism of the structure rings
\mar{34f1}\beq
\pi_r^{r+1*}:\cS^0_r[F;Y]\to \cS^0_{r+1}[F;Y] \label{34f1}
\eeq
of graded functions on graded manifolds $(J^rY,\gA_r)$ and
$(J^{r+1}Y,\gA_{r+1})$. By virtue of Lemma \ref{v62}, the
differential calculus $\cS^*_r[F;Y]$ and $\cS^*_{r+1}[F;Y]$ are
minimal. Therefore, the monomorphism (\ref{34f1}) yields that of
the DBGA
\mar{v4'}\beq
\pi_r^{r+1*}:\cS^*_r[F;Y]\to \cS^*_{r+1}[F;Y]. \label{v4'}
\eeq
As a consequence, we have the direct system of DBGAs
\mar{j2}\beq
\cS^*[F;Y]\ar^{\pi^*} \cS^*_1[F;Y]\ar\cdots \cS^*_{r-1}[F;Y]
\op\ar^{\pi^{r*}_{r-1}}\cS^*_r[F;Y]\ar\cdots. \label{j2}
\eeq
The DBGA \index{$\cS^*_\infty[F;Y]$} $\cS^*_\infty[F;Y]$
\index{algebra!$\cS^*_\infty[F;Y]$} that we associate to the
composite graded manifold $(Y,\gA_F)$ is defined as the direct
limit
\mar{5.77a}\beq
\cS^*_\infty [F;Y]=\op\lim^\to \cS^*_r[F;Y] \label{5.77a}
\eeq
of the direct system (\ref{j2}). It consists of all graded
exterior forms $\f\in \cS^*[F_r;J^rY]$ on graded manifolds
$(J^rY,\gA_r)$ modulo the monomorphisms (\ref{v4'}). Its elements
obey the relations (\ref{ws45}) -- (\ref{ws44}).

The cochain monomorphisms $\cO^*_r\to \cS^*_r[F;Y]$ (\ref{34f3})
provide a monomorphism of the direct system (\ref{5.7}) to the
direct system (\ref{j2}) and, consequently, the monomorphism
\mar{v7}\beq
\cO^*_\infty\to \cS^*_\infty[F;Y] \label{v7}
\eeq
of their direct limits. In particular, $\cS^*_\infty[F;Y]$ is an
$\cO^0_\infty$-algebra. Accordingly, the body epimorphisms
$\cS^*_r[F;Y]\to \cO^*_r$ yield the epimorphism of
$\cO^0_\infty$-algebras
\mar{v7'}\beq
\cS^*_\infty[F;Y]\to \cO^*_\infty.  \label{v7'}
\eeq
It is readily observed that the morphisms (\ref{v7}) and
(\ref{v7'}) are cochain morphisms between the de Rham complex
(\ref{5.13}) of the DGA $\cO^*_\infty$ (\ref{5.77}) and the de
Rham complex
\mar{g110}\beq
0\to\Bbb R\ar \cS^0_\infty[F;Y]\ar^d \cS^1_\infty[F;Y]\cdots
\ar^d\cS^k_\infty[F;Y] \ar\cdots \label{g110}
\eeq
of the DBGA $\cS^0_\infty[F;Y]$. Moreover, the corresponding
homomorphisms of cohomology groups of these complexes are
isomorphisms as follows.

\begin{theorem} \label{v9} \mar{v9} There is an isomorphism
\mar{v10'}\beq
H^*(\cS^*_\infty[F;Y])= H^*_{DR}(Y) \label{v10'}
\eeq
of the cohomology $H^*(\cS^*_\infty[F;Y])$ of the de Rham complex
(\ref{g110}) to the de Rham cohomology  $H^*_{DR}(Y)$ of $Y$.
\end{theorem}

\begin{corollary} \mar{34c1} \label{34c1}
Any closed graded form $\f\in \cS^*_\infty[F;Y]$ is decomposed
into the sum $\f=\si +d\xi$ where $\si$ is a closed exterior form
on $Y$.
\end{corollary}

Similarly to the DGA $\cO^*_\infty$ (\ref{5.77}), one can think of
elements of $\cS^*_\infty[F;Y]$ as being {\sl graded differential
forms} \index{differential form!graded} on
\index{graded!differential form} the infinite order jet manifold
$J^\infty Y$. We can restrict $\cS^*_\infty[F;Y]$ to the
coordinate chart (\ref{j3}) of $J^\infty Y$ and say that
$\cS^*_\infty[F;Y]$ as an $\cO^0_\infty$-algebra is locally
generated by  the elements
\be
(c^a_\La,
dx^\la,\thh^a_\La=dc^a_\La-c^a_{\la+\La}dx^\la,\thh^i_\La=
dy^i_\La-y^i_{\la+\La}dx^\la), \qquad 0\leq |\La|,
\ee
where $c^a_\La$, $\thh^a_\La$ are odd and $dx^\la$, $\thh^i_\La$
are even. We agree to call $(y^i,c^a)$ the local {\sl generating
basis} \index{basis!generating} for \index{generating basis}
$\cS^*_\infty[F;Y]$. Let the collective symbol $s^A$ stand for its
elements. Accordingly, the notation $s^A_\La$ and
\be
\thh^A_\La=ds^A_\La- s^A_{\la+\La}dx^\la
\ee
is introduced. For the sake of simplicity, we further denote
$[A]=[s^A]$. \index{$[A]$}

The DBGA $\cS^*_\infty[F;Y]$ is split into
$\cS^0_\infty[F;Y]$-modules $\cS^{k,r}_\infty[F;Y]$ of
\index{horizontal!form!graded} {\sl $k$-contact and $r$-horizontal
graded forms} \index{contact form!graded} together with the
corresponding projections
\be
h_k:\cS^*_\infty[F;Y]\to \cS^{k,*}_\infty[F;Y], \qquad
h^m:\cS^*_\infty[F;Y]\to \cS^{*,m}_\infty[F;Y].
\ee
Accordingly, the graded exterior differential $d$ on
$\cS^*_\infty[F;Y]$ falls into the sum $d=d_V+d_H$ of
\index{differential!vertical!graded} the {\sl vertical graded
differential}
\be
d_V \circ h^m=h^m\circ d\circ h^m, \qquad d_V(\f)=\thh^A_\La \w
\dr^\La_A\f, \qquad \f\in\cS^*_\infty[F;Y],
\ee
and the \index{differential!total!graded} {\sl total graded
differential}
\be
d_H\circ h_k=h_k\circ d\circ h_k, \qquad d_H\circ h_0=h_0\circ d,
\qquad d_H(\f)=dx^\la\w d_\la(\f),
\ee
where
\be
d_\la = \dr_\la + \op\sum_{0\leq|\La|} s^A_{\la+\La}\dr_A^\La
\ee
are the {\sl graded total derivatives}. \index{total
derivative!graded} These differentials obey the nilpotent
relations (\ref{mbm}).

Similarly to the DGA $\cO^*_\infty$, the DBGA $\cS^*_\infty[F;Y]$
is provided with the graded projection endomorphism
\be
&& \vr=\op\sum_{k>0} \frac1k\ol\vr\circ h_k\circ h^n:
\cS^{*>0,n}_\infty[F;Y]\to \cS^{*>0,n}_\infty[F;Y], \\
&& \ol\vr(\f)= \op\sum_{0\leq|\La|} (-1)^{\nm\La}\thh^A\w
[d_\La(\dr^\La_A\rfloor\f)], \qquad \f\in \cS^{>0,n}_\infty[F;Y],
\ee
such that $\vr\circ d_H=0$, and with the nilpotent graded
variational operator \index{variational!operator!graded}
\mar{34f10}\beq
\dl=\vr\circ d \cS^{*,n}_\infty[F;Y]\to
\cS^{*+1,n}_\infty[F;Y].\label{34f10}
\eeq
With these operators the DBGA $\cS^{*,}_\infty[F;Y]$ is decomposed
into the {\sl Grassmann-graded variational bicomplex}.
\index{variational!bicomplex!graded} We restrict our consideration
to its short \index{variational!complex!graded} {\sl variational
subcomplex}
\mar{g111}\ben
&& 0\to \Bbb R\to \cS^0_\infty[F;Y]\ar^{d_H}\cS^{0,1}_\infty[F;Y]
\cdots \ar^{d_H} \label{g111}\\
&& \qquad \cS^{0,n}_\infty[F;Y]\ar^\dl \bE_1,\qquad
\bE_1=\vr(\cS^{1,n}_\infty[F;Y]), \nonumber
\een
and the subcomplex of one-contact graded forms
\mar{g112}\beq
 0\to \cS^{1,0}_\infty[F;Y]\ar^{d_H} \cS^{1,1}_\infty[F;Y]\cdots
\ar^{d_H}\cS^{1,n}_\infty[F;Y]\ar^\vr \bE_1\to 0. \label{g112}
\eeq

\begin{theorem} \label{v11} \mar{v11}
Cohomology of the complex (\ref{g111}) equals the de Rham
cohomology $H^*_{DR}(Y)$ of $Y$.
\end{theorem}

\begin{theorem} \label{v11'} \mar{v11'}
The complex (\ref{g112}) is exact.
\end{theorem}

Decomposed into the variational bicomplex, the DBGA
$\cS^*_\infty[F;Y]$ describes Grassmann-graded Lagrangian theory
on the composite graded manifold $(Y,\gA_F)$. Its {\sl graded
Lagrangian} \index{Lagrangian!graded} is defined as an element
\mar{0709}\beq
L=\cL\om\in \cS^{0,n}_\infty[F;Y] \label{0709}
\eeq
of the graded variational complex (\ref{g111}), while the graded
exterior form
\mar{0709'}\beq
\dl L= \thh^A\w \cE_A\om=\op\sum_{0\leq|\La|}
 (-1)^{|\La|}\thh^A\w d_\La (\dr^\La_A L)\om\in \bE_1 \label{0709'}
\eeq
is said to be its {\sl graded Euler--Lagrange operator}. We agree
to call a pair $(\cS^{0,n}_\infty[F;Y],L)$ the {\sl
Grassmann-graded Lagrangian system}. \index{Lagrangian
system!Grassmann-graded}

The following is a corollary of Theorem \ref{v11}.

\begin{theorem} \label{cmp26} \mar{cmp26}
Every $d_H$-closed graded form $\f\in\cS^{0,m<n}_\infty[F;Y]$
falls into the sum
\mar{g214}\beq
\f=h_0\si + d_H\xi, \qquad \xi\in \cS^{0,m-1}_\infty[F;Y],
\label{g214}
\eeq
where $\si$ is a closed $m$-form on $Y$. Any $\dl$-closed (i.e.,
variationally trivial) Grassmann-graded Lagrangian $L\in
\cS^{0,n}_\infty[F;Y]$ is the sum
\mar{g215}\beq
L=h_0\si + d_H\xi, \qquad \xi\in \cS^{0,n-1}_\infty[F;Y],
\label{g215}
\eeq
where $\si$ is a closed $n$-form on $Y$.
\end{theorem}

\begin{corollary} \mar{34c5} \label{34c5}
Any variationally trivial odd Lagrangian is $d_H$-exact.
\end{corollary}

The exactness of the complex (\ref{g112}) at the term
$\cS^{1,n}_\infty[F;Y]$ results in the following.

\begin{theorem} \label{g103} \mar{g103}
Given a graded Lagrangian $L$, there is the decomposition
\mar{g99,'}\ben
&& dL=\dl L - d_H\Xi_L,
\qquad \Xi\in \cS^{n-1}_\infty[F;Y], \label{g99}\\
&& \Xi_L=L+\op\sum_{s=0} \thh^A_{\nu_s\ldots\nu_1}\w
F^{\la\nu_s\ldots\nu_1}_A\om_\la, \label{g99'}\\
&& F_A^{\nu_k\ldots\nu_1}= \dr_A^{\nu_k\ldots\nu_1}\cL-d_\la
F_A^{\la\nu_k\ldots\nu_1} +\si_A^{\nu_k\ldots\nu_1},\qquad
k=1,2,\ldots,\nonumber
\een
where local graded functions $\si$ obey the relations
\be
\si^\nu_A=0,\qquad \si_A^{(\nu_k\nu_{k-1})\ldots\nu_1}=0.
\ee
\end{theorem}

The form $\Xi_L$ (\ref{g99'}) provides a global Lepage equivalent
of a graded Lagrangian $L$. \index{Lepage equivalent!of a graded
Lagrangian}

Given a Grassmann-graded Lagrangian system $(\cS^*_\infty[F;Y],
L)$, by its {\sl infinitesimal transformations}
\index{infinitesimal transformation!of a Lagrangian
system!Grassmann-graded} are meant contact graded derivations of
the real graded commutative ring $\cS^0_\infty[F;Y]$. They
constitute a $\cS^0_\infty[F;Y]$-module $\gd \cS^0_\infty[F;Y]$
which is a real Lie superalgebra with the Lie superbracket
(\ref{ws14}).

\begin{theorem} \label{35t1} \mar{35t1}
The derivation module $\gd\cS^0_\infty[F;Y]$ is isomorphic to the
$\cS^0_\infty[F;Y]$-dual $(\cS^1_\infty[F;Y])^*$ of the module of
graded one-forms $\cS^1_\infty[F;Y]$. It follows that the DBGA
$\cS^*_\infty[F;Y]$ is minimal differential calculus over the real
graded commutative ring $\cS^0_\infty[F;Y]$.
\end{theorem}

Let $\vt\rfloor\f$, $\vt\in \gd\cS^0_\infty[F;Y]$, $\f\in
\cS^1_\infty[F;Y]$, denote the corresponding interior product.
Extended to the DBGA $\cS^*_\infty[F;Y]$, it obeys the rule
\be
\vt\rfloor(\f\w\si)=(\vt\rfloor \f)\w\si
+(-1)^{|\f|+[\f][\vt]}\f\w(\vt\rfloor\si), \qquad \f,\si\in
\cS^*_\infty[F;Y].
\ee

Restricted to a coordinate chart (\ref{j3}) of $J^\infty Y$, the
algebra $\cS^*_\infty[F;Y]$ is a free $\cS^0_\infty[F;Y]$-module
generated by one-forms $dx^\la$, $\thh^A_\La$. Due to the
isomorphism stated in Theorem \ref{35t1}, any graded derivation
$\vt\in\gd\cS^0_\infty[F;Y]$ takes the local form
\mar{gg3}\ben
&& \vt=\vt^\la \dr_\la + \vt^A\dr_A + \op\sum_{0<|\La|}\vt^A_\La
\dr^\La_A, \label{gg3}\\
&& \dr^\La_A\rfloor dy_\Si^B=\dl_A^B\dl^\La_\Si.
\een

Every graded derivation $\vt$ (\ref{gg3}) yields the graded Lie
derivative
\be
&& \bL_\vt\f=\vt\rfloor d\f+ d(\vt\rfloor\f), \qquad \f\in
\cS^*_\infty[F;Y],\\
&& \bL_\vt(\f\w\si)=\bL_\vt(\f)\w\si
+(-1)^{[\vt][\f]}\f\w\bL_\vt(\si),
\ee
of the DBGA $\cS^*_\infty[F;Y]$. A graded derivation $\vt$
(\ref{gg3}) is called {\sl contact}
\index{derivation!graded!contact} if \index{contact
derivation!graded} the Lie derivative $\bL_\vt$ preserves the
ideal of contact graded forms.

\begin{lemma}
With respect to the local generating basis $(s^A)$ for the DBGA
$\cS^*_\infty[F;Y]$, any its contact graded derivation takes the
form
\mar{g105}\beq
\vt=\up_H+\up_V=\up^\la d_\la + [\up^A\dr_A +\op\sum_{|\La|>0}
d_\La(\up^A-s^A_\m\up^\m)\dr_A^\La], \label{g105}
\eeq
where $\up_H$ and $\up_V$ denotes the horizontal and vertical
parts of $\vt$.
\end{lemma}

A glance at the expression (\ref{g105}) shows that a contact
graded derivation $\vt$ is an infinite order jet prolongation of
its restriction
\mar{jj15}\beq
\up=\up^\la\dr_\la +\up^A\dr_A \label{jj15}
\eeq
to the graded commutative ring $S^0[F;Y]$. We call $\up$
(\ref{jj15}) \index{generalized vector field!graded} the
\index{graded!vector field!generalized} {\sl generalized graded
vector field}. It is readily justified the following (see Lemma
\ref{gg6}).

\begin{lemma} Any vertical contact graded derivation
\mar{j40}\beq
\vt= \up^A\dr_A +\op\sum_{|\La|>0} d_\La\up^A\dr_A^\La \label{j40}
\eeq
satisfies the relations
\mar{g233,'}\ben
&& \vt\rfloor d_H\f=-d_H(\vt\rfloor\f), \label{g233}\\
 && \bL_\vt(d_H\f)=d_H(\bL_\vt\f)
\label{g233'}
\een
for all $\f\in\cS^*_\infty[F;Y]$.
\end{lemma}

Then the forthcoming assertions are the straightforward
generalizations of Theorem \ref{g75}, Lemma \ref{22l10} and
Theorem \ref{j22}.

A corollary of the decomposition (\ref{g99}) is the {\sl first
variational formula} \index{first variational formula!for a graded
Lagrangian} for a graded Lagrangian.

\begin{theorem} \label{j44} \mar{j44}
The Lie derivative of a graded Lagrangian along any contact graded
derivation (\ref{g105}) obeys the first variational formula
\mar{g107}\beq
\bL_\vt L= \up_V\rfloor\dl L +d_H(h_0(\vt\rfloor \Xi_L)) + d_V
(\up_H\rfloor\om)\cL, \label{g107}
\eeq
where $\Xi_L$ is the Lepage equivalent (\ref{g99'}) of $L$.
\end{theorem}

A contact graded derivation $\vt$ (\ref{g105}) is called a {\sl
variational symmetry} \index{variational!symmetry!of a graded
Lagrangian} (strictly speaking, a variational {\sl supersymmetry})
\index{supersymmetry} of a graded Lagrangian  $L$ if the Lie
derivative $\bL_\vt L$ is $d_H$-exact, i.e.,
\mar{35f1}\beq
\bL_\vt L=d_H\si. \label{35f1}
\eeq

\begin{lemma} \mar{35l10} \label{35l10}
A glance at the expression (\ref{g107}) shows the following.

(i) A contact graded derivation $\vt$ is a variational symmetry
only if it is projected onto $X$.

(ii) Any projectable contact graded derivation is a variational
symmetry of a variationally trivial graded Lagrangian. It follows
that, if $\vt$ is a variational symmetry of a graded Lagrangian
$L$, it also is a variational symmetry of a Lagrangian $L+L_0$,
where $L_0$ is a variationally trivial graded Lagrangian.

(iii) A contact graded derivations $\vt$ is a variational symmetry
iff its vertical part $\up_V$ (\ref{g105}) is well.

(iv) It is a variational symmetry iff the graded density
$\up_V\rfloor \dl L$ is $d_H$-exact.
\end{lemma}

Variational symmetries of a graded Lagrangian $L$ constitute a
real vector subspace $\ccG_L$ of the graded derivation module
$\gd\cS^0_\infty[F;Y]$. By virtue of item (ii) of Lemma
\ref{35l10}, the Lie superbracket
\be
\bL_{[\vt,\vt']}= [\bL_\vt,\bL_{\vt'}]
\ee
of variational symmetries is a variational symmetry and,
therefore, their vector space $\ccG_L$ \index{$\ccG_L$} is a real
Lie superalgebra.

A corollary of the first variational formula (\ref{g107}) is the
{\sl first Noether theorem} for graded Lagrangians. \index{first
Noether theorem!for a graded Lagrangian}

\begin{theorem} \label{j45} \mar{j45} If a contact graded derivation $\vt$
(\ref{g105}) is a variational symmetry (\ref{35f1}) of a graded
Lagrangian $L$, the first variational formula (\ref{g107})
restricted to Ker$\,\dl L$ leads to the weak conservation law
\mar{35f2}\beq
0\ap d_H(h_0(\vt\rfloor\Xi_L)-\si). \label{35f2}
\eeq
\end{theorem}

A vertical contact graded derivation $\vt$ (\ref{j40}) is said to
be \index{nilpotent derivation} {\sl nilpotent}
\index{derivation!nilpotent} if
\mar{g133}\ben
&&\bL_\vt(\bL_\vt\f)= \op\sum_{0\leq|\Si|,0\leq|\La|}
(\up^B_\Si\dr^\Si_B(\up^A_\La)\dr^\La_A + \label{g133}\\
&& \qquad (-1)^{[s^B][\up^A]}\up^B_\Si\up^A_\La\dr^\Si_B \dr^\La_A)\f=0
\nonumber
\een
for any horizontal graded form $\f\in S^{0,*}_\infty$.

\begin{lemma} \label{041} \mar{041} A vertical contact graded
derivation (\ref{j40}) is  nilpotent only if it is odd and iff the
equality
\be
\bL_\vt(\up^A)=\op\sum_{0\leq|\Si|} \up^B_\Si\dr^\Si_B(\up^A)=0
\ee
holds for all $\up^A$.
\end{lemma}

For the sake of brevity, the common symbol $\up$ further stands
for a generalized graded vector field $\up$, the contact graded
derivation $\vt$  determined by $\up$,  and the Lie derivative
$\bL_\vt$. We agree to call all these operators, simply, a {\sl
graded derivation of a field system algebra}.
\index{graded!derivation!of a field system algebra}

\begin{remark} \label{rr35} \mar{rr5}
For the sake of convenience,  \index{right derivation} {\sl right
derivations} \index{derivation!right}
\mar{rgh}\beq
\op\up^\lto =\rdr_A\up^A \label{rgh}
\eeq
also are \index{$\op\up^\lto$} considered. They act on graded
functions and differential forms $\f$ on the right by the rules
\be
&& \op\up^\lto(\f)=d\f\lfloor\op\up^\lto +d(\f\lfloor\op\up^\lto), \\
&&\op\up^\lto(\f\w\f')=(-1)^{[\f']}\op\up^\lto(\f)\w\f'+
\f\w\op\up^\lto(\f'),\\
&& \thh_{\La A}\lfloor\rdr^{\Si B}=\dl^A_B\dl^\Si_\La.
\ee
One associates to any graded right derivation $\op\up^\lto$
(\ref{rgh}) the left one
\mar{rgh2}\ben
&& \up^l=(-1)^{[\up][A]}\up^A\dr_A, \label{rgh2}\\
&&\up^l(f)=(-1)^{[\up][f]}\up(f), \qquad f\in \cS^0_\infty[F;Y].
\nonumber
\een
\end{remark}

\section{Noether identities}

The degeneracy of Lagrangian theory is characterized by a set of
non-trivial reducible Noether identities. Any Euler--Lagrange
operator satisfies Noether identities (henceforth NI) \index{NI}
which therefore must be separated into the trivial and non-trivial
ones. These NI can obey first-stage NI, which in turn are subject
to the second-stage ones, and so on. Thus, there is a hierarchy of
higher-stage NI which also are separated into the trivial and
non-trivial ones. If certain conditions hold, one can associate to
a Grassmann-graded Lagrangian system the exact Koszul--Tate
complex possessing the boundary operator whose nilpotentness is
equivalent to all non-trivial NI and higher-stage NI. The inverse
second Noether theorem formulated in homology terms associates to
this Koszul--Tate complex the cochain sequence of ghosts with the
ascent operator, called the gauge operator, whose components are
non-trivial gauge and higher-stage gauge symmetries of Lagrangian
theory.

Let $(\cS^*_\infty[F;Y],L)$ be a Grassmann-graded Lagrangian
system. Without a lose of generality, let a Lagrangian $L$ be
even. Its Euler--Lagrange operator $\dl L$ (\ref{0709'}) is
assumed to be at least of order 1 in order to guarantee that
transition functions of $Y$ do not vanish on-shell. This
Euler--Lagrange operator $\dl L\in \bE_1$ takes its values into
the graded vector bundle
\mar{41f33}\beq
\ol{VF}=V^*F\op\ot_F\op\w^n T^*X\to F, \label{41f33}
\eeq
where $V^*F$ is the vertical cotangent bundle of $F\to X$. It
however is not a vector bundle over $Y$. Therefore, we restrict
our consideration to the case of a pull-back composite bundle $F$
(\ref{cvc}), that is,
\mar{41f1}\beq
F=Y\op\times_X F^1\to Y\to X, \label{41f1}
\eeq
where $F^1\to X$ is a vector bundle. Let us introduce the
following notation.

\begin{notation}
Given the vertical tangent bundle $VE$ of a fibre bundle $E\to X$,
by its {\sl density-dual bundle} \index{bundle!density-dual} is
meant \index{density-dual bundle} the fibre bundle
\mar{41f2}\beq
\ol{VE}=V^*E\op\ot_E \op\w^n T^*X. \label{41f2}
\eeq
If $E\to X$ is a vector bundle, we have
\mar{41f3}\beq
\ol{VE}=\ol E\op\times_X E, \qquad \ol E=E^*\op\ot_X\op\w^n T^*X,
\label{41f3}
\eeq
where $\ol E$ \index{$\ol E$} is called the {\sl density-dual}
\index{density-dual vector bundle} of $E$. Let
\be
E=E^0\op\oplus_X E^1
\ee
be a {\sl graded vector bundle} over $X$. \index{vector
bundle!graded} Its {\sl graded density-dual} \index{density-dual
vector bundle!graded} is defined to be
\be
\ol E=\ol E^1\op\oplus_X \ol E^0.
\ee
In these terms, we treat the composite bundle $F$ (\ref{cvc}) as a
graded vector bundle over $Y$ possessing only odd part. The
density-dual $\ol{VF}$ (\ref{41f2}) of the vertical tangent bundle
$VF$ of $F\to X$ is $\ol{VF}$ (\ref{41f33}). If $F$ (\ref{cvc}) is
the pull-back bundle (\ref{41f1}), then
\mar{41f4}\beq
\ol{VF}=((\ol F^1\op\oplus_Y V^*Y)\op\ot_Y\op\w^n T^*X)\op\oplus_Y
F^1 \label{41f4}
\eeq
is a graded vector bundle over $Y$. Given a graded vector bundle
\be
E=E^0\op\oplus_Y E^1\to Y,
\ee
we consider the composite bundle $E\to E^0\to X$ and the
\index{$\cP^*_\infty[E;Y]$} DBGA
\index{algebra!$\cP^*_\infty[E;Y]$} (\ref{5.77a}):
\mar{41f5}\beq
\cP^*_\infty[E;Y]=\cS^*_\infty[E;E^0]. \label{41f5}
\eeq
\end{notation}

Let us consider the density-dual $\ol{VF}$ (\ref{41f4}) of the
vertical tangent bundle $VF\to F$, and let us enlarge the DBGA
$\cS^*_\infty[F;Y]$ to the DBGA $\cP^*_\infty[\ol{VF};Y]$
(\ref{41f5}) with the local generating basis $(s^A, \ol s_A),
\qquad [\ol s_A]=([A]+1){\rm mod}\,2$. Following the physical
terminology, we agree to call its elements $\ol s_A$ the {\sl
antifields} \index{antifield} of antifield number Ant$[\ol s_A]=
1$. The DBGA $\cP^*_\infty[\ol{VF};Y]$ is endowed with the
nilpotent right graded \index{$\ol\dl$} derivation $\ol\dl=\rdr^A
\cE_A$, where $\cE_A$ are the variational derivatives
(\ref{0709'}). Then we have the chain complex
\mar{v042}\beq
0\lto \im\ol\dl \llr^{\ol\dl} \cP^{0,n}_\infty[\ol{VF};Y]_1
\llr^{\ol\dl} \cP^{0,n}_\infty[\ol{VF};Y]_2 \label{v042}
\eeq
of graded densities of antifield number $\leq 2$. Its
one-boundaries $\ol\dl\Phi$, $\Phi\in
\cP^{0,n}_\infty[\ol{VF};Y]_2$, by very definition, vanish
on-shell.

\begin{lemma} \label{41l1} \mar{41l1} One can associate to any
Grassmann-graded Lagrangian system $(\cS^*_\infty[F;Y],L)$ the
chain complex (\ref{v042}).
\end{lemma}

Any one-cycle
\mar{0712}\beq
\Phi= \op\sum_{0\leq|\La|} \Phi^{A,\La}\ol s_{\La A} \om \in
\cP^{0,n}_\infty[\ol{VF};Y]_1\label{0712}
\eeq
of the complex (\ref{v042}) is a differential operator on the
bundle $\ol{VF}$ such that it is linear on fibres of $\ol{VF}\to
F$ and its kernel contains the graded Euler--Lagrange operator
$\dl L$ (\ref{0709'}), i.e.,
\mar{0713}\beq
\ol\dl\Phi=0, \qquad \op\sum_{0\leq|\La|} \Phi^{A,\La}d_\La
\cE_A\om=0. \label{0713}
\eeq
Thus, the one-cycles (\ref{0712}) define the NI (\ref{0713}) of
the Euler--Lagrange operator $\dl L$, which we call \index{Noether
identities} {\sl Noether identities} (NI) of the Grassmann-graded
Lagrangian system $(\cS^*_\infty[F;Y],L)$.

In particular, one-chains $\Phi$ (\ref{0712}) are necessarily NI
if they are boundaries. Accordingly, {\sl non-trivial NI}
\index{Noether identities!non-trivial} modulo the trivial ones are
associated to elements of the first homology $H_1(\ol\dl)$ of the
complex (\ref{v042}). A Lagrangian $L$ is called {\sl degenerate}
\index{Lagrangian!degenerate} if there are non-trivial NI.

Non-trivial NI obey first-stage NI. To describe them, let us
assume that the module $H_1(\ol \dl)$ is finitely generated.
Namely, there exists a graded projective $C^\infty(X)$-module
$\cC_{(0)}\subset H_1(\ol \dl)$ of finite rank with a local basis
$\{\Delta_r\om\}$:
\mar{41f7}\beq
\Delta_r\om=\op\sum_{0\leq|\La|} \Delta_r^{A,\La}\ol s_{\La
A}\om,\qquad \Delta_r^{A,\La}\in \cS^0_\infty[F;Y], \label{41f7}
\eeq
such that any element $\Phi\in H_1(\ol \dl)$ factorizes as
\mar{xx2}\beq
\Phi= \op\sum_{0\leq|\Xi|} \Phi^{r,\Xi} d_\Xi \Delta_r \om, \qquad
\Phi^{r,\Xi}\in \cS^0_\infty[F;Y], \label{xx2}
\eeq
through elements (\ref{41f7}) of $\cC_{(0)}$. Thus, all
non-trivial NI (\ref{0713}) result from the NI
\mar{v64}\beq
\ol\dl\Delta_r= \op\sum_{0\leq|\La|} \Delta_r^{A,\La} d_\La
\cE_A=0, \label{v64}
\eeq
called the {\sl complete NI}. \index{Noether identities!complete}
Clearly, the factorization (\ref{xx2}) is independent of
specification of a local basis $\{\Delta_r\om\}$.

A Lagrangian system whose non-trivial NI are finitely generated is
called {\sl finitely degenerate}. \index{Lagrangian
system!finitely degenerate} Hereafter, {\sl degenerate Lagrangian
systems} \index{Lagrangian system!finitely degenerate} only of
this type are considered.

By virtue of Serre--Swan Theorem \ref{vv0}, the graded module
$\cC_{(0)}$ is isomorphic to a module of sections of the
density-dual $\ol E_0$ of some graded vector bundle $E_0\to X$.
Let us enlarge $\cP^*_\infty[\ol{VF};Y]$ to the DBGA
\mar{41f14}\beq
\ol\cP^*_\infty\{0\}=\cP^*_\infty[\ol{VF}\op\oplus_Y \ol E_0;Y]
\label{41f14}
\eeq
possessing the local generating basis $(s^A,\ol s_A, \ol c_r)$
where $\ol c_r$ are {\sl Noether antifields}
\index{antifield!Noether} of Grassmann parity $[\ol
c_r]=([\Delta_r]+1){\rm mod}\,2$ and antifield number ${\rm
Ant}[\ol c_r]=2$. The DBGA (\ref{41f14}) is provided with the odd
right graded derivation $\dl_0=\ol\dl + \rdr^r\Delta_r$ which is
nilpotent iff the complete NI (\ref{v64}) hold. Then $\dl_0$ is a
boundary operator of the chain complex
\mar{v66}\beq
0\lto \im\ol\dl \op\lto^{\ol\dl}
\cP^{0,n}_\infty[\ol{VF};Y]_1\op\lto^{\dl_0}
\ol\cP^{0,n}_\infty\{0\}_2 \op\lto^{\dl_0}
\ol\cP^{0,n}_\infty\{0\}_3 \label{v66}
\eeq
of graded densities of antifield number $\leq 3$. Let $H_*(\dl_0)$
denote its homology. We have
\be
H_0(\dl_0)=H_0(\ol\dl)=0.
\ee
Furthermore, any one-cycle $\Phi$ up to a boundary takes the form
(\ref{xx2}) and, therefore, it is a $\dl_0$-boundary. Hence,
$H_1(\dl_0)=0$, i.e., the complex (\ref{v66}) is one-exact.

\begin{lemma} \label{41l2} \mar{41l2}
If the homology $H_1(\ol\dl)$ of the complex (\ref{v042}) is
finitely generated in the above mentioned sense, this complex can
be extended to the one-exact chain complex (\ref{v66}) with a
boundary operator whose nilpotency conditions are equivalent to
the complete NI (\ref{v64}).
\end{lemma}

Let us consider the second homology $H_2(\dl_0)$ of the complex
(\ref{v66}). Its two-chains  read
\mar{41f9}\beq
\Phi= G + H= \op\sum_{0\leq|\La|} G^{r,\La}\ol c_{\La r}\om +
\op\sum_{0\leq|\La|,|\Si|} H^{(A,\La)(B,\Si)}\ol s_{\La A}\ol
s_{\Si B}\om. \label{41f9}
\eeq
Its two-cycles define the \index{Noether identities!first stage}
{\sl first-stage NI}
\mar{v79}\beq
\dl_0 \Phi=0, \qquad   \op\sum_{0\leq|\La|} G^{r,\La}d_\La\Delta_r
\om =-\ol\dl H. \label{v79}
\eeq

The first-stage NI (\ref{v79}) are {\sl trivial} \index{Noether
identities!first stage!trivial} either if a two-cycle $\Phi$
(\ref{41f9}) is a $\dl_0$-boundary or its summand $G$ vanishes
on-shell. Therefore, non-trivial first-stage NI fails to exhaust
the second homology $H_2(\dl_0)$ the complex (\ref{v66}) in
general.

\begin{lemma} \label{v134'} \mar{v134'}
{\sl Non-trivial first-stage NI} \index{Noether identities!first
stage!non-trivial} modulo the trivial ones are identified with
elements of the homology $H_2(\dl_0)$ iff any $\ol\dl$-cycle
$\f\in \ol\cP^{0,n}_\infty\{0\}_2$ is a $\dl_0$-boundary.
\end{lemma}

A degenerate Lagrangian system is called {\sl reducible}
\index{Lagrangian system!reducible} (resp. {\sl irreducible})
\index{Lagrangian system!irreducible} if it admits (resp. does not
admit) non-trivial first stage NI.

If the condition of Lemma \ref{v134'} is satisfied, let us assume
that non-trivial first-stage NI are finitely generated as follows.
There exists a graded projective $C^\infty(X)$-module
$\cC_{(1)}\subset H_2(\dl_0)$ of finite rank with a local basis
$\{\Delta_{r_1}\om\}$:
\mar{41f13}\beq
\Delta_{r_1}\om=\op\sum_{0\leq|\La|} \Delta_{r_1}^{r,\La}\ol
c_{\La r}\om + h_{r_1}\om,   \label{41f13}
\eeq
such that any element $\Phi\in H_2(\dl_0)$ factorizes as
\mar{v80'}\beq
\Phi= \op\sum_{0\leq|\Xi|} \Phi^{r_1,\Xi} d_\Xi \Delta_{r_1}\om,
\qquad \Phi^{r_1,\Xi}\in \cS^0_\infty[F;Y], \label{v80'}
\eeq
through elements (\ref{41f13}) of $\cC_{(1)}$. Thus, all
non-trivial first-stage NI (\ref{v79}) result from the equalities
\mar{v82'}\beq
 \op\sum_{0\leq|\La|} \Delta_{r_1}^{r,\La} d_\La \Delta_r +\ol\dl
h_{r_1} =0, \label{v82'}
\eeq
called the {\sl complete first-stage NI}. \index{Noether
identities!first-stage!complete}

The complete first-stage NI obey second-stage NI, and so on.
Iterating the arguments, one comes to the following.

A degenerate Grassmann-graded Lagrangian system
$(\cS^*_\infty[F;Y],L)$ is called {\sl $N$-stage reducible}
\index{Lagrangian system!$N$-stage reducible} if it admits
finitely generated non-trivial $N$-stage NI, but no non-trivial
$(N+1)$-stage ones. It is characterized as follows.

$\bullet$ There are graded vector bundles $E_0,\ldots, E_N$ over
$X$, and the DBGA $\cP^*_\infty[\ol{VF};Y]$ is enlarged to the
DBGA
\mar{v91}\beq
\ol\cP^*_\infty\{N\}=\cP^*_\infty[\ol{VF}\op\oplus_Y \ol
E_0\op\oplus_Y\cdots\op\oplus_Y \ol E_N;Y] \label{v91}
\eeq
with the local generating basis
\be
(s^A,\ol s_A, \ol c_r, \ol c_{r_1}, \ldots, \ol c_{r_N})
\ee
where $\ol c_{r_k}$ are {\sl Noether $k$-stage antifields}
\index{antifield!$k$-stage} of antifield number Ant$[\ol
c_{r_k}]=k+2$.

$\bullet$ The DBGA (\ref{v91}) admits with the nilpotent right
graded derivation \index{$\dl_{\rm KT}$}
\mar{v92,'}\ben
&&\dl_{\rm KT}=\dl_N=\ol\dl +
\op\sum_{0\leq|\La|}\rdr^r\Delta_r^{A,\La}\ol s_{\La A} +
\op\sum_{1\leq k\leq N}\rdr^{r_k} \Delta_{r_k},
\label{v92}\\
&& \Delta_{r_k}\om= \op\sum_{0\leq|\La|}
\Delta_{r_k}^{r_{k-1},\La}\ol c_{\La r_{k-1}}\om +
\label{v92'}\\
&& \qquad \op\sum_{0\leq |\Si|, |\Xi|}(h_{r_k}^{(r_{k-2},\Si)(A,\Xi)}\ol
c_{\Si r_{k-2}}\ol s_{\Xi A}+...)\om \in
\ol\cP^{0,n}_\infty\{k-1\}_{k+1}, \nonumber
\een
of \index{$\dl_N$} antifield number -1. The index $k=-1$ here
stands for $\ol s_A$. The nilpotent derivation $\dl_{\rm KT}$
(\ref{v92}) is called the {\sl Koszul--Tate operator}.
\index{Koszul--Tate operator}

$\bullet$ With this graded derivation, the module
$\ol\cP^{0,n}_\infty\{N\}_{\leq N+3}$ of densities of antifield
number $\leq (N+3)$ is decomposed into the exact {\sl Koszul--Tate
chain complex} \index{Koszul--Tate complex}
\mar{v94}\ben
&& 0\lto \im \ol\dl \llr^{\ol\dl}
\cP^{0,n}_\infty[\ol{VF};Y]_1\llr^{\dl_0}
\ol\cP^{0,n}_\infty\{0\}_2\llr^{\dl_1}
\ol\cP^{0,n}_\infty\{1\}_3\cdots
\label{v94}\\
&& \qquad
 \llr^{\dl_{N-1}} \ol\cP^{0,n}_\infty\{N-1\}_{N+1}
\llr^{\dl_{\rm KT}} \ol\cP^{0,n}_\infty\{N\}_{N+2}\llr^{\dl_{\rm
KT}} \ol\cP^{0,n}_\infty\{N\}_{N+3} \nonumber
\een
which satisfies the following {\sl homology regularity condition}.
\index{homology regularity condition}

\begin{condition} \label{v155} \mar{v155} Any $\dl_{k<N}$-cycle
\be
\f\in \ol\cP_\infty^{0,n}\{k\}_{k+3}\subset
\ol\cP_\infty^{0,n}\{k+1\}_{k+3}
\ee
is a $\dl_{k+1}$-boundary.
\end{condition}

$\bullet$ The nilpotentness $\dl_{\rm KT}^2=0$ of the Koszul--Tate
operator (\ref{v92}) is equivalent to the complete non-trivial NI
(\ref{v64}) and the \index{Noether identities!higher-stage} {\sl
complete non-trivial $(k\leq N)$-stage NI}
\mar{v93}\ben
&& \op\sum_{0\leq|\La|} \Delta_{r_k}^{r_{k-1},\La}d_\La
\left(\op\sum_{0\leq|\Si|} \Delta_{r_{k-1}}^{r_{k-2},\Si}\ol
c_{\Si
r_{k-2}}\right) = \label{v93}\\
&& \qquad -  \ol\dl\left(\op\sum_{0\leq |\Si|,
|\Xi|}h_{r_k}^{(r_{k-2},\Si)(A,\Xi)}\ol c_{\Si r_{k-2}}\ol s_{\Xi
A}\right). \nonumber
\een
This item means the following.

\begin{theorem} Any $\dl_k$-cocycle $\Phi\in
\cP^{0,n}_\infty\{k\}_{k+2}$ is a $k$-stage NI, and {\it vice
versa}.
\end{theorem}

\begin{theorem}
Any trivial $k$-stage NI is a $\dl_k$-boundary $\Phi\in
\cP^{0,n}_\infty\{k\}_{k+2}$.
\end{theorem}

\begin{theorem} All non-trivial $k$-stage NI, by
assumption, factorize as
\be
\Phi= \op\sum_{0\leq|\Xi|} \Phi^{r_k,\Xi} d_\Xi \Delta_{r_k}\om,
\qquad \Phi^{r_1,\Xi}\in \cS^0_\infty[F;Y],
\ee
through the complete ones (\ref{v93}).
\end{theorem}

It may happen that a Grassmann-graded Lagrangian field system
possesses non-trivial NI of any stage. However, we restrict our
consideration to $N$-reducible Lagrangian systems for a finite
integer $N$.

\section{Gauge symmetries}

Different variants of the second Noether theorem have been
suggested in order to relate reducible NI and gauge symmetries.
The {\sl inverse second Noether theorem} \index{Noether
theorem!second!inverse}(Theorem \ref{w35}), that we formulate in
homology terms, associates to the Koszul--Tate complex (\ref{v94})
of non-trivial NI the cochain sequence (\ref{w108}) with the
ascent operator $\bu$ (\ref{w108'}) whose components are
non-trivial gauge and higher-stage gauge symmetries of Lagrangian
system. Let us start with the following notation.

\begin{notation} \label{42n1} \mar{42n1}
Given the DBGA $\ol\cP^*_\infty\{N\}$ (\ref{v91}), we consider the
DBGA
\mar{w5}\beq
\cP^*_\infty\{N\}=\cP^*_\infty[F\op\oplus_Y E_0\op\oplus_Y\cdots
\op\oplus_Y E_N;Y], \label{w5}
\eeq
possessing the local generating basis
\be
(s^A, c^r, c^{r_1}, \ldots, c^{r_N}), \qquad [c^{r_k}]=([\ol
c_{r_k}]+1){\rm mod}\,2,
\ee
and the DBGA
\mar{w6}\beq
P^*_\infty\{N\}=\cP^*_\infty[\ol{VF}\op\oplus_Y E_0\oplus\cdots
\op\oplus_Y E_N \op\oplus_Y \ol E_0\op\oplus_Y\cdots\op\oplus_Y
\ol E_N;Y] \label{w6}
\eeq
with the local generating basis
\be
(s^A, \ol s_A, c^r, c^{r_1}, \ldots, c^{r_N},\ol c_r, \ol c_{r_1},
\ldots, \ol c_{r_N}).
\ee
Their elements $c^{r_k}$ are called $k$-stage {\sl ghosts}
\index{ghost} of ghost number gh$[c^{r_k}]=k+1$ and antifield
number
\be
{\rm Ant}[c^{r_k}]=-(k+1).
\ee
The $C^\infty(X)$-module $\cC^{(k)}$ of $k$-stage ghosts is the
density-dual of the module $\cC_{(k+1})$ of $(k+1)$-stage
antifields. The DBGAs $\ol\cP^*_\infty\{N\}$ (\ref{v91}) and
$\cP^*_\infty\{N\}$ (\ref{w5}) are subalgebras of
$P^*_\infty\{N\}$ (\ref{w6}). The Koszul--Tate operator $\dl_{\rm
KT}$ (\ref{v92}) is naturally extended to a graded derivation of
the DBGA $P^*_\infty\{N\}$.
\end{notation}

\begin{notation} \label{42n10} \mar{42n10} Any
graded differential form $\f\in \cS^*_\infty[F;Y]$ and any finite
tuple $(f^\La)$, $0\leq |\La|\leq k$, of local graded functions
$f^\La\in \cS^0_\infty[F;Y]$ obey the following relations:
\mar{qq1}\ben
&& \op\sum_{0\leq |\La|\leq k} f^\La d_\La \f\w \om= \op\sum_{0\leq
|\La|}(-1)^{|\La|}d_\La (f^\La)\f\w \om +d_H\si,
\label{qq1a}\\
&& \op\sum_{0\leq |\La|\leq k} (-1)^{|\La|}d_\La(f^\La \f)=
\op\sum_{0\leq |\La|\leq k} \eta (f)^\La d_\La \f, \label{qq1b}\\
&& \eta (f)^\La = \op\sum_{0\leq|\Si|\leq k-|\La|}(-1)^{|\Si+\La|}
\frac{(|\Si+\La|)!}{|\Si|!|\La|!} d_\Si f^{\Si+\La},
\label{qq1c}\\
&& \eta(\eta(f))^\La=f^\La. \label{qq1d}
\een
\end{notation}

\begin{theorem} \label{w35} \mar{w35} Given the Koszul--Tate complex (\ref{v94}),
the module of graded densities $\cP_\infty^{0,n}\{N\}$ is
decomposed into the cochain sequence
\mar{w108,'}\ben
&& 0\to \cS^{0,n}_\infty[F;Y]\ar^{\bu}
\cP^{0,n}_\infty\{N\}^1\ar^{\bu}
\cP^{0,n}_\infty\{N\}^2\ar^{\bu}\cdots, \label{w108}\\
&& \bu=u + u^{(1)}+\cdots + u^{(N)}=\label{w108'}\\
&& \qquad u^A\frac{\dr}{\dr s^A} +
u^r\frac{\dr}{\dr c^r} +\cdots  + u^{r_{N-1}}\frac{\dr}{\dr
c^{r_{N-1}}}, \nonumber
\een
graded \index{$\bu$} in ghost number. Its ascent operator $\bu$
(\ref{w108'}) is an odd graded derivation of ghost number 1 where
\mar{w33}\beq
u= u^A\frac{\dr}{\dr s^A}, \qquad u^A =\op\sum_{0\leq|\La|}
c^r_\La\eta(\Delta^A_r)^\La, \label{w33}
\eeq
is a variational symmetry of a graded Lagrangian $L$ and the
graded derivations
\mar{w38}\beq
u^{(k)}= u^{r_{k-1}}\frac{\dr}{\dr
c^{r_{k-1}}}=\op\sum_{0\leq|\La|}
c^{r_k}_\La\eta(\Delta^{r_{k-1}}_{r_k})^\La\frac{\dr}{\dr
c^{r_{k-1}}}, \quad k=1,\ldots,N, \label{w38}
\eeq
obey the relations
\mar{w34}\ben
&&\op\sum_{0\leq|\Si|} d_\Si u^{r_{k-1}}\frac{\dr}{\dr
c^{r_{k-1}}_\Si} u^{r_{k-2}} =\ol\dl(\al^{r_{k-2}}),\label{w34}\\
&& \al^{r_{k-2}} = -\op\sum_{0\leq|\Si|}
\eta(h_{r_k}^{(r_{k-2})(A,\Xi)})^\Si d_\Si(c^{r_k} \ol s_{\Xi A}).
\nonumber
\een
\end{theorem}

A glance at the expression (\ref{w33}) shows that the variational
symmetry $u$ is a linear differential operator on the
$C^\infty(X)$-module $\cC^{(0)}$ of ghosts with values into the
real space $\cG_L$ of variational symmetries. Following Definition
\ref{s7} extended to Lagrangian theories of odd variables, we call
$u$ (\ref{w33}) the {\sl gauge symmetry of a graded Lagrangian}
\index{gauge symmetry!of a graded Lagrangian} $L$ which is
associated to the complete NI (\ref{v64}).

\begin{remark} \label{r10r} \mar{r10r}
In contrast with Definitions \ref{s7} and \ref{s7'}, gauge
symmetries $u$ (\ref{w33}) are parameterized by ghosts, but not
gauge parameters. Given a gauge symmetry $u$ (\ref{gg2}) defined
as a derivation of the real ring $\cO^0_\infty[Y\times E]$, one
can associate to it the gauge symmetry
\mar{gg2'}\beq
u=\left(\op\sum_{0\leq|\La|\leq m}
u^{\la\La}_a(x^\m)c^a_\La\right)\dr_\la +
\left(\op\sum_{0\leq|\La|\leq m}
u^{i\La}_a(x^\m,y^j_\Si)c^a_\La\right)\dr_i, \label{gg2'}
\eeq
which is an odd graded derivation of the real ring
$\cS^0_\infty[E;Y]$, and {\it vice versa}.
\end{remark}

Turn now to the relation (\ref{w34}). For $k=1$, it takes the form
\be
\op\sum_{0\leq|\Si|} d_\Si u^r\frac{\dr}{\dr c^r_\Si} u^A =\ol
\dl(\al^A)
\ee
of a first-stage gauge symmetry condition on-shell which the
non-trivial gauge symmetry $u$ (\ref{w33}) satisfies. Therefore,
one can treat the odd graded derivation
\be
u^{(1)}= u^r\frac{\dr}{\dr c^r}, \qquad u^r=\op\sum_{0\leq|\La|}
c^{r_1}_\La\eta(\Delta^r_{r_1})^\La,
\ee
as a {\sl first-stage gauge symmetry} \index{gauge
symmetry!first-stage} associated to the complete first-stage NI
\be
 \op\sum_{0\leq|\La|} \Delta_{r_1}^{r,\La}d_\La
\left(\op\sum_{0\leq|\Si|} \Delta_r^{A,\Si}\ol s_{\Si A}\right) =
- \ol\dl\left(\op\sum_{0\leq |\Si|,
|\Xi|}h_{r_1}^{(B,\Si)(A,\Xi)}\ol s_{\Si B}\ol s_{\Xi A}\right).
\ee

Iterating the arguments, one comes to the relation (\ref{w34})
which provides a $k$-stage gauge symmetry condition which is
associated to the complete $k$-stage NI (\ref{v93}). The odd
graded derivation $u_{(k)}$ (\ref{w38}) is called the {\sl
$k$-stage gauge symmetry}. \index{gauge symmetry!$k$-stage}

In accordance with Theorem \ref{w35}, components of the ascent
operator $\bu$ (\ref{w108'}) are complete non-trivial gauge and
higher-stage gauge symmetries. Therefore, we agree to call this
operator the {\sl gauge operator}. \index{gauge!operator}

Being a variational symmetry, a gauge symmetry $u$ (\ref{w33})
defines the weak conservation law (\ref{35f2}). Let $u$ be an
exact Lagrangian symmetry. In this case, the associated symmetry
current
\mar{42f50}\beq
\cJ_u= -h_0(u\rfloor\Xi_L) \label{42f50}
\eeq
is conserved. The peculiarity of gauge conservation laws always is
that the symmetry current (\ref{42f50}) is reduced to a
superpotential as follows.

\begin{theorem} \label{supp'} \mar{supp'}
If $u$ (\ref{w33}) is an exact gauge symmetry of a graded
Lagrangian $L$, the corresponding conserved symmetry current
$\cJ_u$ (\ref{42f50}) takes the form
\mar{b381'}\beq
\cJ_u= W+ d_HU=(W^\m +d_\nu U^{\nu\m})\om_\m, \label{b381'}
\eeq
where the term $W$ vanishes on-shell, and $U$ is a horizontal
$(n-2)$-form.
\end{theorem}

\chapter{Topics on commutative geometry}

Several relevant topics on commutative geometry and algebraic
topology are compiled in this Lecture.

\section{Commutative algebra}

An {\sl algebra} $\cA$ \index{algebra} is an additive group which
is additionally provided with distributive multiplication. All
algebras  throughout the Lectures are associative, unless they are
Lie algebras.  A {\sl ring} \index{ring} is defined to be a {\sl
unital} algebra, \index{algebra!unital} i.e., it contains a unit
element $\bb\neq 0$.

A subset $\cI$ of an algebra $\cA$ is called a left (resp. right)
{\sl ideal} \index{ideal} if it is a subgroup of the additive
group $\cA$ and $ab\in \cI$ (resp. $ba\in\cI$) for all $a\in \cA$,
$b\in \cI$. If $\cI$ is both a left and right ideal, it is called
a two-sided ideal. An ideal is a subalgebra, but a {\sl proper
ideal} \index{ideal!proper} (i.e., $\cI\neq \cA$) of a ring is not
a subring because it does not contain the unit element.

Let $\cA$ be a commutative ring. Of course, its ideals are
two-sided. Its proper ideal is said to be {\sl maximal}
\index{ideal!maximal} if it does not belong to another proper
ideal. A commutative ring $\cA$ is called {\sl local} \index{local
ring} \index{ring!local} if it has a unique maximal ideal. This
ideal consists of all non-invertible elements of $\cA$.

Given an ideal $\cI\subset \cA$, the additive factor group
$\cA/\cI$ is an algebra, called the {\sl factor algebra}.
\index{factor!algebra} If $\cA$ is a ring, then $\cA/\cI$ is so.
If $\cI$ is a maximal ideal, the factor ring $\cA/\cI$ is a field.

Given an algebra $\cA$, an additive group $P$ is said to be a left
(resp. right) {\sl $\cA$-module} \index{module} if it is provided
with distributive multiplication $\cA\times P\to P$ by elements of
$\cA$ such that $(ab)p=a(bp)$ (resp. $(ab)p=b(ap)$) for all
$a,b\in\cA$ and $p\in P$.  If $\cA$ is a ring, one additionally
assumes that $\bb p=p=p\bb$ for all $p\in P$. Left and right
module structures are usually written by means of left and right
multiplications $(a,p)\to ap$ and $(a,p)\to pa$, respectively. If
$P$ is both a left module over an algebra $\cA$ and a right module
over an algebra $\cA'$, it is called an $(\cA-\cA')$-bimodule (an
$\cA$-bimodule if $\cA=\cA'$). \index{bimodule} If $\cA$ is a
commutative algebra, an $\cA$-bimodule $P$ is said to be {\sl
commutative} \index{bimodule!commutative} if $ap=pa$ for all $a\in
\cA$ and $p\in P$. Any left or right module over a commutative
algebra $\cA$ can be brought into a commutative bimodule.
Therefore, unless otherwise stated, any module over a commutative
algebra $\cA$ is called an $\cA$-module. A module over a field is
called a {\sl vector space}. \index{vector space}

If an algebra $\cA$ is a module over a commutative ring $\cK$, it
is said to be a $\cK$-algebra.

Hereafter, all associative algebras are assumed to be commutative.

The following are standard constructions of new modules from old
ones.

$\bullet$ The {\sl direct sum} \index{direct sum!of modules}
$P_1\oplus P_2$ of $\cA$-modules $P_1$ and $P_2$ is the additive
group $P_1\times P_2$ provided with the $\cA$-module structure
\be
a(p_1,p_2)=(ap_1,ap_2), \qquad p_{1,2}\in P_{1,2}, \qquad a\in\cA.
\ee
Let $\{P_i\}_{i\in I}$ be a set of modules. Their direct sum
$\oplus P_i$ consists of elements $(\ldots, p_i,\ldots)$ of the
Cartesian product $\prod P_i$ such that $p_i\neq 0$ at most for a
finite number of indices $i\in I$.

$\bullet$ Given a submodule $Q$ of an $\cA$-module $P$, the
quotient $P/Q$ of the additive group $P$ with respect to its
subgroup $Q$ also is provided with an $\cA$-module structure. It
is called a {\sl factor module}. \index{factor!module}

$\bullet$ The set $\hm_\cA(P,Q)$ of $\cA$-linear morphisms of an
$\cA$-module $P$ to an $\cA$-module $Q$ is naturally an
$\cA$-module. The $\cA$-module
\be
P^*=\hm_\cA(P,\cA)
\ee
is called the {\sl dual} \index{module!dual} \index{dual module}
of an $\cA$-module $P$. There is a monomorphism  $P\to P^{**}$.

$\bullet$ The {\sl tensor product} \index{tensor product!of
modules} $P\ot Q$ of $\cA$-modules $P$ and $Q$ is an additive
group which is generated by elements $p\ot q$, $p\in P$, $q\in Q$,
obeying the relations
\be
&& (p+p')\ot q =p\ot q + p'\ot q, \\
&& p\ot(q+q')=p\ot q+p\ot q', \\
&&  pa\ot q= p\ot aq, \qquad p\in P, \qquad q\in Q, \qquad
a\in\cA.
\ee
It is provided with the $\cA$-module structure
\be
a(p\ot q)=(ap)\ot q=p\ot (qa)=(p\ot q)a.
\ee
In particular, we have the following.

(i) If a ring $\cA$ is treated as an $\cA$-module, the tensor
product $\cA\ot_\cA Q$ is canonically isomorphic to $Q$ via the
assignment
\be
\cA\ot_\cA Q\ni a\ot q \leftrightarrow aq\in Q.
\ee

(ii) The {\sl tensor product of Abelian groups} $G$ and $G'$
\index{tensor product!of Abelian groups} is defined as their
tensor product $G\ot G'$ as $\Bbb Z$-modules.

(iii) The {\sl tensor product of commutative algebras} $\cA$ and
$\cA'$ \index{tensor product!of commutative algebras} is defined
as their tensor product $\cA\ot\cA'$ as modules provided with the
multiplication
\be
(a\ot a')(b\ot b')= (aa')\ot bb'.
\ee

An $\cA$-module $P$ is called {\sl free} \index{module!free} if it
has a {\sl basis}, \index{basis!for a module} i.e., a linearly
independent subset $I\subset P$ spanning $P$ such that each
element of $P$ has a unique expression as a linear combination of
elements of $I$ with a finite number of non-zero coefficients from
an algebra $\cA$. Any vector space is free. Any module is
isomorphic to a quotient of a free module. A module is said to be
{\sl finitely generated} \index{module!finitely generated} (or of
{\sl finite rank}) \index{module!of finite rank} if it is a
quotient of a free module with a finite basis.

One says that a module $P$ is {\sl projective}
\index{module!projective} if it is a direct summand of a free
module, i.e., there exists a module $Q$ such that $P\oplus Q$ is a
free module. A module $P$ is projective iff $P=\bp S$ where $S$ is
a free module and $\bp$ is a projector of $S$, i.e., $\bp^2=\bp$.

\begin{theorem}
Any projective module over a local ring is free.
\end{theorem}

Now we focus on exact sequences, direct and inverse limits of
modules. A composition of module morphisms
\be
P\ar^i Q\ar^j T
\ee
is said to be {\sl exact} \index{exact sequence!of modules} at $Q$
if $\Ker j=\im i$. A composition of module morphisms
\mar{spr13}\beq
0\to P\ar^i Q\ar^j T\to 0 \label{spr13}
\eeq
is called a {\sl short exact sequence} \index{exact sequence!of
modules!short} if it is exact at all the terms $P$, $Q$, and $T$.
This condition implies that: (i) $i$ is a monomorphism,  (ii)
$\Ker j=\im i$, and (iii) $j$ is an epimorphism onto the quotient
$T=Q/P$.

\begin{theorem} \label{spr183} \mar{spr183}
Given an exact sequence of modules (\ref{spr13}) and another
$\cA$-module $R$, the sequence of modules
\be
0\to\hm_\cA(T,R)\ar^{j^*} \hm_\cA(Q,R)\ar^{i^*} \hm(P,R)
\ee
is exact at the first and second terms, i.e., $j^*$ is a
monomorphism, but $i^*$ need not be an epimorphism.
\end{theorem}

One says that the exact sequence (\ref{spr13}) is {\sl split}
\index{exact sequence!of modules!split} if there exists a
monomorphism $s:T\to Q$ such that $j\circ s=\id T$ or,
equivalently,
\be
Q=i(P)\oplus s(T) \cong P\oplus T.
\ee

\begin{theorem} \label{spr183a} \mar{spr183a}
The exact sequence (\ref{spr13}) is always split if $T$ is a
projective module.
\end{theorem}

A {\sl directed set} \index{directed set} $I$ is a set with an
order relation $<$ which satisfies the following three conditions:
(i) $i<i$, for all $i\in I$; (ii) if $i<j$ and $j< k$, then $i<k$;
(iii) for any $i,j\in I$, there exists $k\in I$ such that $i<k$
and $j<k$. It may happen that $i\neq j$, but $i<j$ and $j<i$
simultaneously.

A family of modules $\{P_i\}_{i\in I}$ (over the same algebra),
indexed by a directed set $I$, is called a {\sl direct system}
\index{direct system of modules} if, for any pair $i<j$, there
exists a morphism $r^i_j:P_i\to P_j$ such that
\be
r^i_i=\id P_i, \qquad r^i_j\circ r^j_k=r^i_k, \qquad i<j<k.
\ee
A direct system of modules admits a {\sl direct limit}.
\index{direct limit} This is a module $P_\infty$ together with
morphisms $r^i_\infty: P_i\to P_\infty$ such that
$r^i_\infty=r^j_\infty\circ r^i_j$ for all $i<j$. The module
$P_\infty$ consists of elements of the direct sum $\oplus_I P_i$
modulo the identification of elements of $P_i$ with their images
in $P_j$ for all $i<j$. An example of a direct system is a {\sl
direct sequence} \index{direct sequence}
\mar{spr1}\beq
P_0\ar P_1\ar \cdots P_i\ar^{r^i_{i+1}}\cdots, \qquad I=\Bbb N.
\label{spr1}
\eeq
Note that direct limits also exist in the categories of
commutative and graded commutative algebras and rings, but not in
categories containing non-Abelian groups.

\begin{theorem} \label{spr170} \mar{spr170}
Direct limits commute with direct sums and tensor products of
modules. Namely, let $\{P_i\}$ and $\{Q_i\}$ be two direct systems
of modules over the same algebra which are indexed by the same
directed set $I$, and let $P_\infty$ and $Q_\infty$ be their
direct limits. Then the direct limits of the direct systems
$\{P_i\oplus Q_i\}$ and $\{P_i\ot Q_i\}$ are $P_\infty\oplus
Q_\infty$ and $P_\infty\ot Q_\infty$, respectively.
\end{theorem}

\begin{theorem} \label{spr170'} \mar{spr170'}
A morphism of a direct system $\{P_i, r^i_j\}_I$ to a direct
system $\{Q_{i'}, \rho^{i'}_{j'}\}_{I'}$ consists of an order
preserving map $f:I\to I'$ and morphisms $F_i:P_i\to Q_{f(i)}$
which obey the compatibility conditions
\be
\rho^{f(i)}_{f(j)}\circ F_i=F_j\circ r^i_j.
\ee
If $P_\infty$ and $Q_\infty$ are limits of these direct systems,
there exists a unique morphism $F_\infty: P_\infty\to Q_\infty$
such that
\be
\rho^{f(i)}_\infty\circ F_i=F_\infty\circ r^i_\infty.
\ee
\end{theorem}

\begin{theorem} \label{dlim1} \mar{dlim1}
Direct limits preserve monomorphisms and epimorphisms, i.e., if
all $F_i:P_i\to Q_{f(i)}$ are monomorphisms or epimorphisms, so is
$\Phi_\infty:P_\infty\to Q_\infty$. Let short exact sequences
\mar{spr186}\beq
0\to P_i\ar^{F_i} Q_i\ar^{\Phi_i} T_i\to 0 \label{spr186}
\eeq
for all $i\in I$ define a short exact sequence of direct systems
of modules $\{P_i\}_I$, $\{Q_i\}_I$, and $\{T_i\}_I$ which are
indexed by the same directed set $I$. Then their direct limits
form a short exact sequence
\mar{spr187}\beq
0\to P_\infty\ar^{F_\infty} Q_\infty\ar^{\Phi_\infty} T_\infty\to
0. \label{spr187}
\eeq
\end{theorem}

In particular, the direct limit of factor modules $Q_i/P_i$ is the
factor module $Q_\infty/P_\infty$. By virtue of Theorem
\ref{spr170}, if all the exact sequences (\ref{spr186}) are split,
the exact sequence (\ref{spr187}) is well.

\begin{remark} \label{ws40} \mar{ws40} Let $P$ be an $\cA$-module.
We denote \index{$P^{\ot k}$}
\be
P^{\ot k}=\op\ot^kP.
\ee
Let us consider the direct system of $\cA$-modules
\be
\cA\ar (\cA\oplus P)\ar \cdots  (\cA\oplus P\oplus\cdots\oplus
P^{\ot k})\ar \cdots.
\ee
Its direct limit \index{$\ot P$}
\mar{spr620}\beq
\ot P=\cA\oplus P\oplus\cdots\oplus P^{\ot k}\oplus\cdots
\label{spr620}
\eeq
is an $\Bbb N$-graded $\cA$-algebra with respect to the tensor
product $\ot$. It is called the {\sl tensor algebra} \index{tensor
algebra} of a module $P$. Its quotient with respect to the ideal
generated by elements
\be
p\ot p'+p'\ot p, \qquad p,p'\in P,
\ee
is an $\Bbb N$-graded commutative algebra, called the {\sl
exterior algebra} \index{exterior algebra} of $P$.
\end{remark}

Given an {\sl inverse sequences} \index{inverse sequence} of
modules
\mar{spr2}\beq
P^0\lla P^1\lla \cdots P^i\op\lla^{\pi^{i+1}_i}\cdots,
\label{spr2}
\eeq
its {\sl inductive limit} \index{inductive limit} is a module
$P^\infty$ together with morphisms $\pi^\infty_i: P^\infty\to P^i$
such that $\pi^\infty_i=\pi^j_i\circ \pi^\infty_j$ for all $i<j$.
It consists of elements $(\ldots,p^i,\ldots)$, $p^i\in P^i$, of
the Cartesian product $\prod P^i$ such that $p^i=\pi^j_i(p^j)$ for
all $i<j$.

\begin{theorem} \label{spr3} \mar{spr3}
Inductive limits preserve monomorphisms, but not epimorphisms. Let
exact sequences
\be
0\to P^i\ar^{F^i} Q^i\ar^{\Phi^i} T^i, \qquad i\in\Bbb N,
\ee
for all $i\in\Bbb N$ define an exact sequence of inverse systems
of modules $\{P^i\}$, $\{Q^i\}$ and $\{T^i\}$. Then their
inductive limits form an exact sequence
\be
0\to P^\infty\ar^{F^\infty} Q^\infty\ar^{\Phi^\infty} T^\infty.
\ee
\end{theorem}

In contrast with direct limits, the inductive ones exist in the
category of groups which are not necessarily commutative.

\section{Differential operators on modules}

This Section addresses the notion of a linear differential
operator on a module over a commutative ring.

Let $\cK$ be a commutative ring and $\cA$ a commutative
$\cK$-ring. Let $P$ and $Q$ be $\cA$-modules. The $\cK$-module
$\hm_\cK (P,Q)$ of $\cK$-module homomorphisms $\Phi:P\to Q$ can be
endowed with the two different $\cA$-module structures
\mar{5.29}\beq
(a\Phi)(p)= a\Phi(p),  \qquad  (\Phi\bll a)(p) = \Phi (a p),\qquad
a\in \cA, \quad p\in P. \label{5.29}
\eeq
For the sake of convenience, we refer to the second one as the
$\cA^\bll$-module structure. Let us put
\mar{spr172}\beq
\dl_a\Phi= a\Phi -\Phi\bll a, \qquad a\in\cA. \label{spr172}
\eeq

\begin{definition} \label{ws131} \mar{ws131}
An element $\Delta\in\hm_\cK(P,Q)$ is called a $Q$-valued {\sl
differential operator} \index{differential operator!on a module}
of order $s$ on $P$ if
\be
\dl_{a_0}\circ\cdots\circ\dl_{a_s}\Delta=0
\ee
for any tuple of $s+1$ elements $a_0,\ldots,a_s$ of $\cA$. The set
$\dif_s(P,Q)$ of these operators inherits the $\cA$- and
$\cA^\bll$-module structures (\ref{5.29}).
\end{definition}

In particular, zero order differential operators obey the
condition
\be
\dl_a \Delta(p)=a\Delta(p)-\Delta(ap)=0, \qquad a\in\cA, \qquad
p\in P,
\ee
and, consequently, they coincide with $\cA$-module morphisms $P\to
Q$. A first order differential operator $\Delta$ satisfies the
condition
\mar{ws106}\beq
\dl_b\circ\dl_a\,\Delta(p)= ba\Delta(p) -b\Delta(ap)
-a\Delta(bp)+\Delta(abp) =0, \quad a,b\in\cA. \label{ws106}
\eeq

The following fact reduces the study of $Q$-valued differential
operators on an $\cA$-module $P$ to that of $Q$-valued
differential operators on the ring $\cA$.

\begin{theorem} \label{ws109} \mar{ws109}
Let us consider the $\cA$-module morphism
\mar{n2}\beq
h_s: \dif_s(\cA,Q)\to Q, \qquad h_s(\Delta)=\Delta(\bb).
\label{n2}
\eeq
Any $Q$-valued $s$-order differential operator $\Delta\in
\dif_s(P,Q)$ on $P$ uniquely factorizes as
\mar{n13}\beq
\Delta:P\ar^{\gf_\Delta} \dif_s(\cA,Q)\ar^{h_s} Q \label{n13}
\eeq
through the morphism $h_s$ (\ref{n2}) and some homomorphism
\mar{n0}\beq
\gf_\Delta: P\to \dif_s(\cA,Q), \qquad (\gf_\Delta
p)(a)=\Delta(ap), \qquad a\in \cA, \label{n0}
\eeq
of the $\cA$-module $P$ to the $\cA^\bll$-module $\dif_s(\cA,Q)$.
The assignment $\Delta\to\gf_\Delta$ defines the isomorphism
\mar{n1}\beq
\dif_s(P,Q)=\hm_{\cA-\cA^\bll}(P,\dif_s(\cA,Q)). \label{n1}
\eeq
\end{theorem}

Let $P=\cA$. Any zero order $Q$-valued differential operator
$\Delta$ on $\cA$ is defined by its value $\Delta(\bb)$. Then
there is an isomorphism
\be
\dif_0(\cA,Q)=Q
\ee
via the association
\be
Q\ni q\to \Delta_q\in \dif_0(\cA,Q),
\ee
where $\Delta_q$ is given by the equality $\Delta_q(\bb)=q$. A
first order $Q$-valued differential operator $\Delta$ on $\cA$
fulfils the condition
\be
\Delta(ab)=b\Delta(a)+ a\Delta(b) -ba \Delta(\bb), \qquad
a,b\in\cA.
\ee
It is called a $Q$-valued {\sl derivation} \index{derivation} of
$\cA$ if $\Delta(\bb)=0$, i.e., the \index{Leibniz rule} {\sl
Leibniz rule}
\mar{+a20}\beq
\Delta(ab) = \Delta(a)b + a\Delta(b), \qquad  a,b\in \cA,
\label{+a20}
\eeq
holds. One obtains at once that any first order differential
operator on $\cA$ falls into the sum
\be
\Delta(a)= a\Delta(\bb) +[\Delta(a)-a\Delta(\bb)]
\ee
of the zero order differential operator $a\Delta(\bb)$ and the
derivation $\Delta(a)-a\Delta(\bb)$. If $\dr$ is a $Q$-valued
derivation of $\cA$, then $a\dr$ is well for any $a\in \cA$.
Hence, $Q$-valued derivations of $\cA$ constitute an $\cA$-module
$\gd(\cA,Q)$, called the {\sl derivation module}.
\index{derivation module} There is the $\cA$-module decomposition
\mar{spr156'}\beq
\dif_1(\cA,Q) = Q \oplus\gd(\cA,Q). \label{spr156'}
\eeq

If $P=Q=\cA$, the derivation module $\gd\cA$ \index{$\gd\cA$} of
$\cA$ also is a Lie $\cK$-algebra with respect to the Lie bracket
\mar{+860}\beq
[u,u']=u\circ u' - u'\circ u, \qquad u,u'\in\cA. \label{+860}
\eeq
Accordingly, the decomposition (\ref{spr156'}) takes the form
\mar{spr156}\beq
\dif_1(\cA) = \cA \oplus\gd\cA. \label{spr156}
\eeq

\begin{definition} \label{1016} \mar{1016}
A {\sl connection} \index{connection!on a module} on an
$\cA$-module $P$ is an $\cA$-module morphism
\mar{1017}\beq
\gd\cA\ni u\to \nabla_u\in \dif_1(P,P) \label{1017}
\eeq
such that the first order differential operators $\nabla_u$ obey
the {\sl Leibniz rule} \index{Leibniz rule!for a connection}
\mar{1018}\beq
\nabla_u (ap)= u(a)p+ a\nabla_u(p), \quad a\in \cA, \quad p\in P.
\label{1018}
\eeq
\end{definition}

Though $\nabla_u$ (\ref{1017}) is called a connection, it in fact
is a {\sl covariant differential} \index{covariant differential!on
a module} on a module $P$.

Let $P$ be a commutative $\cA$-ring and $\gd P$ the derivation
module of $P$ as a $\cK$-ring. The $\gd P$ is both a $P$- and
$\cA$-module. Then Definition \ref{1016} is modified as follows.

\begin{definition} \label{mos088} \mar{mos088}
A {\sl connection} \index{connection!on a ring} on an $\cA$-ring
$P$ is an $\cA$-module morphism
\mar{mos090}\beq
\gd\cA\ni u\to \nabla_u\in \gd P\subset \dif_1(P,P),
\label{mos090}
\eeq
which is a connection on $P$ as an $\cA$-module.
\end{definition}

\section{Homology and cohomology of complexes}

This Section summarizes the relevant basics on complexes of
modules over a commutative ring.

Let $\cK$ be a commutative ring. A sequence
\mar{b3255}\beq
0\lto B_0 \lla^{\dr_1} B_1 \lla^{\dr_2}\cdots B_p\lla^{\dr_{p+1}}
\cdots \label{b3255}
\eeq
of $\cK$-modules $B_p$ and homomorphisms $\dr_p$ is said to be a
{\sl chain complex} \index{complex!chain} if
\be
\dr_p\circ \dr_{p+1} =0, \qquad  p\in \Bbb N,
\ee
i.e., $\im \dr_{p+1}\subset \Ker \dr_p$. Homomorphisms $\dr_p$ are
called {\sl boundary operators}. \index{boundary operator}
Elements of a module $B_p$, its submodules $\Ker \dr_p\subset B_p$
and $\im \dr_{p+1}\subset\Ker \dr_p$ are called {\sl $p$-chains},
\index{chain} {\sl $p$-cycles} \index{cycle} and {\sl
$p$-boundaries}, \index{boundary} respectively. The {\sl $p$-th
homology group} \index{homology} of the chain complex $B_*$
(\ref{b3255}) is the factor module
\be
H_p(B_*)= \Ker \dr_p/\im \dr_{p+1}.
\ee
It is a $\cK$-module. In particular, we have
$H_0(B_*)=B_0/\im\dr_1$. The chain complex (\ref{b3255}) is exact
at a term $B_p$ iff $H_p(B_*)=0$. This complex is said to be {\sl
$k$-exact} \index{complex!$k$-exact} if its homology groups
$H_{p\leq k}(B_*)$ are trivial. It is called {\sl exact}
\index{complex!exact} if all its homology groups are trivial,
i.e., it is an exact sequence.

A sequence
\mar{b3256}\beq
0\to B^0 \ar^{\dl^0} B^1 \ar^{\dl^1}\cdots B^p\ar^{\dl^p}\cdots
\label{b3256}
\eeq
of modules $B^p$ and their homomorphisms $\dl^p$ is said to be a
{\sl cochain complex} \index{complex!cochain} (or, simply, a {\sl
complex}) \index{complex} if
\be
\dl^{p+1}\circ \dl^p =0, \qquad  p\in \Bbb N,
\ee
i.e., $\im \dl^p\subset \Ker \dl^{p+1}$. The homomorphisms $\dl^p$
are called {\sl coboundary operators}. \index{coboundary operator}
Elements of a module $B^p$, its submodules $\Ker \dl^p\subset B^p$
and  $\im \dl^{p-1}$ are called {\sl $p$-cochains},
\index{cochain} {\sl $p$-cocycles} \index{cocycle} and {\sl
$p$-coboundaries}, \index{coboundary} respectively. The {\sl
$p$-th cohomology group} \index{cohomology} of the complex $B^*$
(\ref{b3256}) is the factor module
\be
H^p(B^*)= \Ker \dl^p/\im \dl^{p-1}.
\ee
It is a $\cK$-module. In particular, $H^0(B^*)=\Ker \dl^0$. The
complex (\ref{b3256}) is exact at a term $B^p$ iff $H^p(B^*)=0$.
This complex is an exact sequence if all its cohomology groups are
trivial.

A complex $(B^*,\dl^*)$ is called {\sl acyclic}
\index{complex!acyclic} if its cohomology groups $H^{0<p}(B^*)$
are trivial. A complex $(B^*,\dl^*)$ is said to be a {\sl
resolution} \index{resolution} of a module $B$ if it is acyclic
and $H^0(B^*)=\Ker\dl^0=B$.

The following are the standard constructions of new complexes from
old ones.

$\bullet$ Given complexes $(B^*_1,\dl^*_1)$ and $(B^*_2,\dl^*_2)$,
their {\sl direct sum} \index{direct sum!of complexes}
$B^*_1\oplus B^*_2$ is a complex of modules
\be
(B^*_1\oplus B^*_2)^p=B^p_1\oplus B^p_2
\ee
with respect to the coboundary operators
\be
\dl^p_\oplus(b^p_1 + b^p_2)=\dl^p_1b^p_1 +\dl^p_2 b^p_2.
\ee

$\bullet$ Given a subcomplex $(C^*,\dl^*)$ of a complex
$(B^*,\dl^*)$, the {\sl factor complex} \index{factor!complex}
$B^*/C^*$ is defined as a complex of factor modules $B^p/C^p$
provided with the coboundary operators $\dl^p[b^p]=[\dl^p b^p]$,
where $[b^p]\in B^p/C^p$ denotes the coset of an element $b^p$.

$\bullet$ Given complexes $(B^*_1,\dl^*_1)$ and $(B^*_2,\dl^*_2)$,
their {\sl tensor product} \index{tensor product!of complexes}
$B^*_1\ot B^*_2$ is a complex of modules
\be
(B^*_1\ot B^*_2)^p=\op\oplus_{k+r=p} B^k_1\ot B^r_2
\ee
with respect to the coboundary operators
\be
\dl^p_\ot(b^k_1\ot b^r_2)=(\dl_1^kb^k_1)\ot b^r_2 +
(-1)^kb^k_1\ot(\dl_2^rb^r_2).
\ee

A {\sl cochain morphism} \index{cochain morphism} of complexes
\mar{spr32'}\beq
\g:B^*_1\to B^*_2 \label{spr32'}
\eeq
is defined as a family of degree-preserving homomorphisms
\be
\g^p: B^p_1\to B^p_2, \qquad p\in\Bbb N,
\ee
such that
\be
\dl^p_2\circ\g^p=\g^{p+1}\circ\dl^p_1, \qquad p\in\Bbb N.
\ee
It follows that if $b^p\in B^p_1$ is a cocycle or a coboundary,
then $\g^p(b^p)\in B^p_2$ is so. Therefore, the cochain morphism
of complexes (\ref{spr32'}) yields an induced homomorphism of
their cohomology groups
\be
[\g]^*: H^*(B^*_1) \to H^*(B^*_2).
\ee

Let short exact sequences
\be
0\to C^p\ar^{\g_p} B^p \ar^{\zeta_p} F^p\to 0
\ee
for all $p\in\Bbb N$ define a short exact sequence of complexes
\mar{spr34'}\beq
0\to C^*\ar^\g B^* \ar^\zeta F^*\to 0, \label{spr34'}
\eeq
where $\g$ is a cochain monomorphism and $\zeta$ is a cochain
epimorphism onto the quotient $F^*=B^*/C^*$.

\begin{theorem} \label{spr36'} \mar{spr36'}
The short exact sequence  of complexes (\ref{spr34'}) yields the
long exact sequence of their cohomology groups
\mar{spr35'}\ben
&& 0\to H^0(C^*)\ar^{[\g]^0} H^0(B^*)\ar^{[\zeta]^0} H^0(F^*)\ar^{\tau^0}
H^1(C^*)\ar\cdots \label{spr35'}\\
&& \qquad \ar H^p(C^*)\ar^{[\g]^p} H^p(B^*)\ar^{[\zeta]^p}
H^p(F^*)\ar^{\tau^p} H^{p+1}(C^*)\ar\cdots.\nonumber
\een
\end{theorem}

\begin{theorem} \label{spr37'} \mar{spr37'}
A direct sequence of complexes
\mar{spr55}\beq
B^*_0\ar B^*_1\ar\cdots B^*_k\ar^{\g^k_{k+1}} B^*_{k+1}\ar \cdots
\label{spr55}
\eeq
admits a direct limit $B^*_\infty$ which is a complex whose
cohomology $H^*(B^*_\infty)$ is a direct limit of the direct
sequence of cohomology groups
\be
H^*(B^*_0)\ar H^*(B^*_1)\ar \cdots H^*(B^*_k)\ar^{[\g^k_{k+1}]}
H^*(B^*_{k+1})\ar \cdots.
\ee
\end{theorem}

\section{Differential calculus over a commutative ring}

Let $\cG$ be a Lie algebra over a commutative ring $\cK$. Let
$\cG$ act on a $\cK$-module $P$ on the left such that
\be
[\ve,\ve']p=(\ve\circ \ve'-\ve'\circ \ve)p, \qquad \ve,\ve'\in\cG.
\ee
Then one calls $P$ the Lie algebra {\sl $\cG$-module}.
\index{module!over a Lie algebra} Let us consider
$\cK$-multilinear skew-symmetric maps
\be
c^k:\op\times^k\cG\to P.
\ee
They form a $\cG$-module $C^k[\cG;P]$. Let us put $C^0[\cG;P]=P$.
We obtain the cochain complex
\mar{spr997}\beq
0\to P\ar^{\dl^0} C^1[\cG;P]\ar^{\dl^1} \cdots C^k[\cG;P]
\ar^{\dl^k} \cdots \label{spr997}
\eeq
with respect \index{Chevalley--Eilenberg!coboundary operator} to
the {\sl Chevalley--Eilenberg coboundary operators}
\index{coboundary operator!Chevalley--Eilenberg}
\mar{spr132}\ben
&& \dl^kc^k (\ve_0,\ldots,\ve_k)=\op\sum_{i=0}^k(-1)^i\ve_ic^k(\ve_0,\ldots,
\wh\ve_i, \ldots, \ve_k)+ \label{spr132}\\
&& \qquad \op\sum_{1\leq i<j\leq k}
(-1)^{i+j}c^k([\ve_i,\ve_j], \ve_0,\ldots, \wh\ve_i, \ldots,
\wh\ve_j,\ldots, \ve_k), \nonumber
\een
where the caret $\,\wh{}\,$ denotes omission. For instance, we
have
\mar{spr133,4}\ben
&& \dl^0p(\ve_0)=\ve_0p, \label{spr133}\\
&& \dl^1c^1(\ve_0,\ve_1)=\ve_0c^1(\ve_1)-\ve_1c^1(\ve_0) -
c^1([\ve_0,\ve_1]). \label{spr134}
\een
The complex (\ref{spr997}) is called the {\sl Chevalley--Eilenberg
complex}, \index{complex!Chevalley--Eilenberg} and
\index{Chevalley--Eilenberg!complex} its cohomology $H^*(\cG,P)$
is the {\sl Chevalley--Eilenberg cohomology} of a
\index{cohomology!Chevalley--Eilenberg} Lie algebra $\cG$ with
coefficients in $P$. \index{Chevalley--Eilenberg!cohomology}

Let $\cA$ be a commutative $\cK$-ring. Since the derivation module
$\gd\cA$ of $\cA$ is a Lie $\cK$-algebra, one can associate to
$\cA$ the Chevalley--Eilenberg complex $C^*[\gd\cA;\cA]$. Its
subcomplex of $\cA$-multilinear maps is a DGA, also called the
differential calculus over $\cA$. By a gradation throughout this
Section is meant the $\Bbb N$-gradation.

A {\sl graded algebra} \index{algebra!graded}
\index{graded!algebra}$\Om^*$ over a commutative ring $\cK$ is
defined as a direct sum $\Om^*= \op\oplus_k \Om^k$ of
$\cK$-modules $\Om^k$, provided with an associative multiplication
law $\al\cdot\bt$, $\al,\bt\in \Om^*$, such that $\al\cdot\bt\in
\Om^{|\al|+|\bt|}$, where $|\al|$ denotes the degree of an element
$\al\in \Om^{|\al|}$. In particular, it follows that $\Om^0$ is a
(non-commutative) $\cK$-algebra $\cA$, while $\Om^{k>0}$ are
$\cA$-bimodules and $\Om^*$ is an $(\cA-\cA)$-algebra. A graded
algebra  is said to be {\sl graded commutative}
\index{algebra!graded!commutative} \index{graded!commutative
algebra} if
\be
\al\cdot\bt=(-1)^{|\al||\bt|}\bt\cdot \al, \qquad \al,\bt\in
\Om^*.
\ee

A graded algebra $\Om^*$ is called the {\sl differential graded
algebra} \index{algebra!differential graded} (DGA) \index{DGA} or
the {\sl differential calculus} \index{differential calculus} over
$\cA$ if it is a cochain complex of $\cK$-modules
\mar{spr260}\beq
0\to \cK\ar\cA\ar^\dl\Om^1\ar^\dl\cdots\Om^k\ar^\dl\cdots
\label{spr260}
\eeq
relative to a coboundary operator $\dl$ \index{Leibniz
rule!graded} which obeys the \index{graded!Leibniz rule} {\sl
graded Leibniz rule}
\mar{1006}\beq
\dl(\al\cdot\bt)=\dl\al\cdot\bt +(-1)^{|\al|}\al\cdot \dl\bt.
\label{1006}
\eeq
In particular, $\dl:\cA\to \Om^1$ is a $\Om^1$-valued derivation
of a $\cK$-algebra $\cA$. The cochain complex (\ref{spr260}) is
said to be the {\sl abstract de Rham complex} \index{de Rham
complex!abstract} of \index{complex!de Rham!abstract} the DGA
$(\Om^*,\dl)$. Cohomology $H^*(\Om^*)$ of the complex
(\ref{spr260}) is called the \index{de Rham cohomology!abstract}
{\sl abstract de Rham cohomology}. \index{cohomology!de
Rham!abstract}

A morphism $\g$ between two DGAs $(\Om^*,\dl)$ and $(\Om'^*,\dl')$
is defined as a cochain morphism, i.e., $\g\circ\dl=\g\circ \dl'$.
It yields the corresponding morphism of the abstract de Rham
cohomology groups of these algebras.

One considers the minimal differential graded subalgebra
$\Om^*\cA$ of the DGA $\Om^*$ which contains $\cA$. Seen as an
$(\cA-\cA)$-algebra, it is generated by the elements $\dl a$,
$a\in \cA$, and consists of monomials
\be
\al=a_0\dl a_1\cdots \dl a_k, \qquad a_i\in \cA,
\ee
whose product obeys the \index{juxtaposition rule} {\sl
juxtaposition rule}
\be
(a_0\dl a_1)\cdot (b_0\dl b_1)=a_0\dl (a_1b_0)\cdot \dl b_1-
a_0a_1\dl b_0\cdot \dl b_1
\ee
in accordance with the equality (\ref{1006}). The DGA
$(\Om^*\cA,\dl)$ is called the {\sl minimal differential calculus}
\index{differential calculus!minimal} over $\cA$.

Let now $\cA$ be a commutative $\cK$-ring possessing a non-trivial
Lie algebra $\gd\cA$ of derivations. We consider the extended
Chevalley--Eilenberg complex
\be
0\to \cK\ar^{\rm in}C^*[\gd\cA;\cA]
\ee
of the Lie algebra $\gd\cA$ with coefficients in the ring $\cA$,
regarded as a $\gd\cA$-module. It is easily justified that this
complex contains a subcomplex $\cO^*[\gd\cA]$
\index{$\cO^*[\gd\cA]$} of $\cA$-multilinear skew-symmetric maps
\mar{+840'}\beq
\f^k:\op\times^k \gd\cA\to \cA \label{+840'}
\eeq
with respect to the Chevalley--Eilenberg coboundary operator
\mar{+840}\ben
&& d\f(u_0,\ldots,u_k)=\op\sum^k_{i=0}(-1)^iu_i
(\f(u_0,\ldots,\wh{u_i},\ldots,u_k)) +\label{+840}\\
&& \qquad \op\sum_{i<j} (-1)^{i+j}
\f([u_i,u_j],u_0,\ldots, \wh u_i, \ldots, \wh u_j,\ldots,u_k).
\nonumber
\een
In particular, we have
\be
&& (d a)(u)=u(a), \qquad a\in\cA, \qquad u\in\gd\cA, \\
&&(d\f)(u_0,u_1)= u_0(\f(u_1)) -u_1(\f(u_0)) -\f([u_0,u_1]), \quad \f\in
\cO^1[\gd\cA], \\
&& \cO^0[\gd\cA]=\cA, \\
&&\cO^1[\gd\cA]=\hm_\cA(\gd\cA,\cA)=\gd\cA^*.
\ee
It follows that $d(\bb)=0$ and $d$ is a $\cO^1[\gd\cA]$-valued
derivation of $\cA$.

The graded module $\cO^*[\gd\cA]$ is provided with the structure
of a graded $\cA$-algebra with respect to the exterior product
\mar{ws103}\ben
&& \f\w\f'(u_1,...,u_{r+s})= \label{ws103}\\
&& \qquad \op\sum_{i_1<\cdots<i_r;j_1<\cdots<j_s} {\rm
sgn}^{i_1\cdots i_rj_1\cdots j_s}_{1\cdots r+s} \f(u_{i_1},\ldots,
u_{i_r}) \f'(u_{j_1},\ldots,u_{j_s}), \nonumber \\
&& \f\in \cO^r[\gd\cA], \qquad \f'\in \cO^s[\gd\cA], \qquad u_k\in \gd\cA, \nonumber
\een
where sgn$^{...}_{...}$ is the sign of a permutation. This product
obeys the relations
\mar{ws99}\ben
&& d(\f\w\f')=d(\f)\w\f' +(-1)^{|\f|}\f\w d(\f'),
\quad \f,\f'\in \cO^*[\gd\cA], \nonumber\\
&& \f\w \f' =(-1)^{|\f||\f'|}\f'\w \f. \label{ws99}
\een
By virtue of the first one, $\cO^*[\gd\cA]$ is a differential
graded $\cK$-algebra, called the {\sl Chevalley--Eilenberg
differential calculus} \index{Chevalley--Eilenberg!differential
calculus} over \index{differential calculus!Chevalley--Eilenberg}
a $\cK$-ring $\cA$. The relation (\ref{ws99}) shows that
$\cO^*[\gd\cA]$ is a graded commutative algebra.

The {\sl minimal Chevalley--Eilenberg differential calculus}
$\cO^*\cA$  \index{$\cO^*\cA$} over a ring
\index{Chevalley--Eilenberg!differential calculus!minimal} $\cA$
\index{differential calculus!Chevalley--Eilenberg!minimal}
consists of the monomials
\be
a_0da_1\w\cdots\w da_k, \qquad a_i\in\cA.
\ee
Its complex
\mar{t10}\beq
0\to\cK\ar \cA\ar^d\cO^1\cA\ar^d \cdots  \cO^k\cA\ar^d \cdots
\label{t10}
\eeq
is said to be the {\sl de Rham complex} \index{de Rham complex!of
a ring} of a $\cK$-ring $\cA$, and its cohomology $H^*(\cA)$ is
called the {\sl de Rham cohomology} \index{de Rham cohomology!of a
ring} of $\cA$.

\section{Sheaf cohomology}

A {\sl sheaf} \index{sheaf} on a topological space $X$ is a
continuous fibre bundle $\pi:S\to X$ in modules over a commutative
ring $\cK$, where the surjection $\pi$ is a local homeomorphism
and fibres $S_x$, $x\in X$, called the {\sl stalks}, \index{stalk}
are provided with the discrete topology. Global sections of a
sheaf $S$ make up a $\cK$-module $S(X)$, called the {\sl structure
module} \index{structure module! of a sheaf} of $S$.

Any sheaf is generated by a presheaf. A {\sl presheaf}
\index{presheaf} $S_\sU$ on a topological space $X$ is defined if
a module $S_U$ over a commutative ring $\cK$ is assigned to every
open subset $U\subset X$ $(S_\emptyset=0)$ and if, for any pair of
open subsets $V\subset U$, there exists the restriction morphism
$r_V^U:S_U\rightarrow S_V$ such that
\be
r_U^U=\id S_U,\qquad  r_W^U=r_W^Vr_V^U, \qquad W\subset V\subset
U.
\ee

Every presheaf $S_\sU$ on a topological space $X$ yields a sheaf
on $X$ whose stalk $S_x$ at a point $x\in X$ is the direct limit
of the modules $S_U,\,x\in U$, with respect to the restriction
morphisms $r_V^U$. It means that, for each open neighborhood $U$
of a point $x$, every element $s\in S_U$ determines an element
$s_x\in S_x$, called the {\sl germ} of $s$ at $x$. Two elements
$s\in S_U$ and $s'\in S_V$ belong to the same germ at $x$ iff
there exists an open neighborhood $W\subset U\cap V$ of $x$ such
that $r_W^Us=r_W^Vs'$.

\begin{example} \label{spr7} \mar{spr7}
Let $C^0_\sU$ be the presheaf of continuous real functions on a
topological space $X$. Two such functions $s$ and $s'$ define the
same germ $s_x$ if they coincide on an open neighborhood  of $x$.
Hence, we obtain the {\sl sheaf $C^0_X$ of continuous functions}
\index{sheaf!of continuous functions} on $X$. Similarly, the {\sl
sheaf $C^\infty_X$ \index{$C^\infty_X$} of smooth functions}
\index{sheaf!of smooth functions} on a smooth manifold $X$ is
defined. Let us also mention the presheaf of real functions which
are constant on connected open subsets of $X$. It generates the
{\sl constant sheaf} on $X$ \index{sheaf!constant} denoted by
$\Bbb R$.
\end{example}

Different presheaves may generate the same sheaf. Conversely,
every sheaf $S$ defines a presheaf $S(\sU)$ of modules $S(U)$ of
its local sections. It is called the {\sl canonical presheaf}
\index{presheaf!canonical} of the sheaf $S$. If a sheaf $S$ is
constructed from a presheaf $S_\sU$, there are natural module
morphisms
\be
S_U\ni s\to s(U)\in S(U), \qquad s(x)= s_x, \quad x\in U,
\ee
which are neither monomorphisms nor epimorphisms in general. For
instance, it may happen that a non-zero presheaf defines a zero
sheaf. The sheaf generated by the canonical presheaf of a sheaf
$S$ coincides with $S$.

A direct sum and a tensor product of presheaves (as families of
modules)  and sheaves (as fibre bundles in modules) are naturally
defined. By virtue of Theorem \ref{spr170}, a direct sum (resp. a
tensor product) of presheaves generates a direct sum (resp. a
tensor product) of the corresponding sheaves.

\begin{remark} \label{spr190'} \mar{spr190'}
In a different terminology, a sheaf is introduced as a presheaf
which satisfies the following additional axioms.

(S1) Suppose that $U\subset X$ is an open subset and $\{U_\al\}$
is its open cover. If $s,s'\in S_U$ obey the condition
\be
r^U_{U_\al}(s)=r^U_{U_\al}(s')
\ee
for all $U_\al$, then $s=s'$.

(S2) Let $U$ and $\{U_\al\}$ be as in previous item. Suppose that
we are given a family of presheaf elements $\{s_\al\in
S_{U_\al}\}$ such that
\be
r^{U_\al}_{U_\al\cap U_\la}(s_\al)=r^{U_\la}_{U_\al\cap
U_\la}(s_\la)
\ee
for all $U_\al$, $U_\la$. Then there exists a presheaf element
$s\in S_U$ such that $s_\al=r^U_{U_\al}(s)$.

\noindent Canonical presheaves are in one-to-one correspondence
with presheaves obeying these axioms. For instance, presheaves of
continuous, smooth and locally constant functions in Example
\ref{spr7} satisfy the axioms (S1) -- (S2).
\end{remark}

\begin{remark} \mar{rrr1} \label{rrr1}
The notion of a sheaf can be extended to sets, but not to
non-commutative groups. One can consider a presheaf of such
groups, but it generates a sheaf of sets because a direct limit of
non-commutative groups need not be a group. The first (but not
higher) cohomology of $X$ with coefficients in this sheaf is
defined.
\end{remark}

A {\sl morphism of a presheaf} \index{morphism!of presheaves}
$S_\sU$ to a presheaf $S'_\sU$ on the same topological space $X$
is defined as a set of module morphisms $\g_U:S_U\to S'_U$ which
commute with restriction morphisms. A morphism of presheaves
yields a {\sl morphism of sheaves} \index{morphism!of sheaves}
generated by these presheaves. This is a bundle morphism over $X$
such that $\g_x: S_x\to S'_x$ is the direct limit of morphisms
$\g_U$, $x\in U$. Conversely, any morphism of sheaves $S\to S'$ on
a topological space $X$ yields a morphism of canonical presheaves
of local sections of these sheaves. Let $\hm(S|_U,S'|_U)$ be the
commutative group of sheaf morphisms $S|_U\to S'|_U$ for any open
subset $U\subset X$. These groups are assembled into a presheaf,
and define the sheaf $\hm(S,S')$ on $X$. There is a monomorphism
\mar{+212}\beq
\hm(S,S')(U)\to \hm(S(U),S'(U)), \label{+212}
\eeq
which need not be an isomorphism.

By virtue of Theorem \ref{dlim1}, if a presheaf morphism is a
monomorphism or an epimorphism, so is the corresponding sheaf
morphism. Furthermore, the following holds.

\begin{theorem} \label{spr29} \mar{spr29}
A short exact sequence
\mar{spr208}\beq
0\to S'_\sU\to S_\sU\to S''_\sU\to 0 \label{spr208}
\eeq
of presheaves on the same topological space yields the short exact
sequence of sheaves generated by these presheaves
\mar{ms0102}\beq
0\to S'\to S\to S''\to 0, \label{ms0102}
\eeq
where the {\sl factor sheaf} \index{factor!sheaf} $S''=S/S'$ is
isomorphic to that generated by the factor presheaf
$S''_\sU=S_\sU/S'_\sU$. If the exact sequence of presheaves
(\ref{spr208}) is split, i.e.,
\be
S_\sU\cong S'_\sU\oplus S''_\sU,
\ee
the corresponding splitting
\be
S\cong S'\oplus S''
\ee
of the exact sequence of sheaves (\ref{ms0102}) holds.
\end{theorem}

The converse is more intricate. A sheaf morphism induces a
morphism of the corresponding canonical presheaves. If $S\to S'$
is a monomorphism,
\be
S(\sU)\to S'(\sU)
\ee
also is a monomorphism. However, if $S\to S'$ is an epimorphism,
\be
S(\sU)\to S'(\sU)
\ee
need not be so. Therefore, the short exact sequence (\ref{ms0102})
of sheaves yields the exact sequence of the canonical presheaves
\mar{ms0103'}\beq
0\to S'(\sU)\to S(\sU)\to S''(\sU), \label{ms0103'}
\eeq
where $S(\sU)\to S''(\sU)$ is not necessarily an epimorphism. At
the same time, there is the short exact sequence of presheaves
\mar{ms0103}\beq
0\to S'(\sU)\to S(\sU)\to S''_\sU \to 0, \label{ms0103}
\eeq
where the factor presheaf
\be
S''_\sU=S(\sU)/S'(\sU)
\ee
generates the factor sheaf $S''=S/S'$, but need not be its
canonical presheaf.

Let us turn now to sheaf cohomology. Note that only proper covers
are considered.

Let $S_\sU$ be a presheaf of modules on a topological space $X$,
and let $\gU=\{U_i\}_{i\in I}$ be an open cover of $X$. One
constructs a cochain complex where a $p$-cochain is defined as a
function $s^p$ which associates an element
\mar{rrr5}\beq
s^p(i_0,\ldots,i_p)\in S_{U_{i_0}\cap\cdots\cap U_{i_p}}
\label{rrr5}
\eeq
to each $(p+1)$-tuple $(i_0,\ldots,i_p)$ of indices in $I$. These
$p$-cochains are assembled into a module $C^p(\gU,S_\sU)$. Let us
introduce the coboundary operator
\mar{spr180}\ben
&& \delta^p:C^p(\gU,S_\sU)\to C^{p+1}(\gU,S_\sU), \nonumber\\
&& \dl^ps^p(i_0,\ldots,i_{p+1})=\op\sum_{k=0}^{p+1}(-1)^kr_W^
{W_k}s^p(i_0,\ldots,\wh i_k,\ldots,i_{p+1}), \label{spr180}\\
&&
W=U_{i_0}\cap\ldots\cap U_{i_{p+1}},\qquad
W_k=U_{i_0}\cap\cdots\cap\wh U_{i_k}\cap\cdots\cap
U_{i_{p+1}}.\nonumber
\een
One can easily check that $\delta^{p+1}\circ\delta^p=0$. Thus, we
obtain the cochain complex of modules
\mar{spr181}\beq
0\to C^0(\gU,S_\sU)\ar^{\dl^0}\cdots C^p(\gU,S_\sU)\ar^{\dl^p}
C^{p+1}(\gU,S_\sU)\ar\cdots. \label{spr181}
\eeq
Its cohomology groups
\be
H^p(\gU;S_\sU)=\Ker\dl^p/\im\dl^{p-1}
\ee
are modules. Of course, they depend on an open cover $\gU$ of $X$.

Let $\gU'$ be a refinement of the cover $\gU$. Then there is a
morphism of cohomology groups
\mar{rrr3}\beq
H^*(\gU;S_\sU)\rightarrow H^*(\gU';S_\sU). \label{rrr3}
\eeq
Let us take the direct limit of cohomology groups $H^*(\gU;S_\sU)$
relative to these morphisms, where $\gU$ runs through all open
covers of $X$. This limit $H^*(X;S_\sU)$ is called the cohomology
of $X$ with coefficients in the presheaf $S_\sU$.

Let $S$ be a sheaf on a topological space $X$. {\sl Cohomology of
$X$ with coefficients in $S$} \index{cohomology!with coefficients
in a sheaf} or, simply, {\sl sheaf cohomology} of $X$ \index{sheaf
cohomology} is defined as cohomology
\be
H^*(X;S)=H^*(X;S(\sU))
\ee
with coefficients in the canonical presheaf $S(\sU)$ of the sheaf
$S$.

In this case, a $p$-cochain $s^p\in C^p(\gU,S(\sU))$ is a
collection
\be
s^p=\{s^p(i_0,\ldots,i_p)\}
\ee
of local sections $s^p(i_0,\ldots,i_p)$ of the sheaf $S$ over
$U_{i_0}\cap\cdots\cap U_{i_p}$ for each $(p+1)$-tuple
$(U_{i_0},\ldots,U_{i_p})$ of elements of the cover $\gU$. The
coboundary operator (\ref{spr180}) reads
\be
\dl^ps^p(i_0,\ldots,i_{p+1})=\op\sum_{k=0}^{p+1}(-1)^k
s^p(i_0,\ldots,\wh i_k,\ldots,i_{p+1})|_{U_{i_0}\cap\cdots\cap
U_{i_{p+1}}}.
\ee
For instance, we have
\mar{spr188,9}\ben
&& \dl^0s^0(i,j)=[s^0(j) -s^0(i)]|_{U_i\cap U_j},
\label{spr188}\\
&& \dl^1s^1(i,j,k)=[s^1(j,k)-s^1(i,k)
   +s^1(i,j)]|_{U_i\cap U_j\cap U_k}. \label{spr189}
\een
A glance at the expression (\ref{spr188}) shows that a
zero-cocycle is a collection $s=\{s(i)\}_I$ of local sections of
the sheaf $S$ over $U_i\in\gU$ such that $s(i)=s(j)$ on $U_i\cap
U_j$. It follows from the axiom (S2) in Remark \ref{spr190'} that
$s$ is a global section of the sheaf $S$, while each $s(i)$ is its
restriction $s|_{U_i}$ to $U_i$. Consequently, the cohomology
group $H^0(\gU;S(\sU))$ is isomorphic to the structure module
$S(X)$ of global sections of the sheaf $S$. A one-cocycle is a
collection $\{s(i,j)\}$ of local sections of the sheaf $S$ over
overlaps $U_i\cap U_j$ which satisfy the \index{cocycle
condition!for a sheaf} {\sl cocycle condition}
\mar{spr192'}\beq
[s(j,k)-s(i,k) +s(i,j)]|_{U_i\cap U_j\cap U_k}=0. \label{spr192'}
\eeq

If $X$ is a paracompact space, the study of its sheaf cohomology
is essentially simplified due to the following fact.

\begin{theorem} \label{spr225} \mar{spr225}
Cohomology of a paracompact space $X$ with coefficients in a sheaf
$S$ coincides with cohomology of $X$ with coefficients in any
presheaf generating the sheaf $S$.
\end{theorem}

\begin{remark} \label{spr200} \mar{spr200}
We follow the definition of a {\sl paracompact topological space}
\index{paracompact space} as a Hausdorff space such that any its
open cover admits a {\sl locally finite} \index{locally finite
cover} open refinement, i.e., any point has an open neighborhood
which intersects only a finite number of elements of this
refinement. A topological space $X$ is paracompact iff any cover
$\{U_\xi\}$ of $X$ admits a subordinate {\sl partition of unity}
\index{partition of unity} $\{f_\xi\}$, i.e.:

(i) $f_\xi$ are real positive continuous functions on $X$;

(ii) supp$\,f_\xi\subset U_\xi$;

(iii) each point $x\in X$ has an open neighborhood which
intersects only a finite number of the sets supp$\,f_\xi$;

(iv) $\op\sum_\xi f_\xi(x)=1$ for all $x\in X$.
\end{remark}

The key point of the analysis of sheaf cohomology is that short
exact sequences of sheaves yield long exact sequences of their
cohomology groups.

Let $S_\sU$ and $S'_\sU$ be presheaves on the same topological
space $X$. It is readily observed that, given an open cover $\gU$
of $X$, any morphism $S_\sU\to S'_\sU$  yields a cochain morphism
of complexes
\be
C^*(\gU,S_\sU)\to C^*(\gU,S'_\sU)
\ee
and the corresponding morphism
\be
H^*(\gU;S_\sU)\to H^*(\gU;S'_\sU)
\ee
of cohomology groups of these complexes. Passing to the direct
limit through all refinements of $\gU$, we come to a morphism of
cohomology groups
\be
H^*(X;S_\sU)\to H^*(X;S'_\sU)
\ee
of $X$ with coefficients in the presheaves $S_\sU$ and $S'_\sU$.
In particular, any sheaf morphism $S\to S'$ yields a morphism of
canonical presheaves
\be
S(\{U\})\to S'(\{U\})
\ee
and the corresponding cohomology morphism
\be
H^*(X;S)\to H^*(X;S').
\ee

By virtue of Theorems \ref{spr36'} and \ref{spr37'}, every short
exact sequence
\mar{spr220}\beq
0\to S'_\sU\ar S_\sU\ar S''_\sU\to 0 \label{spr220}
\eeq
of presheaves on the same topological space $X$ and the
corresponding exact sequence of complexes (\ref{spr181}) yield the
long exact sequence
\mar{spr221}\ben
&& 0\to H^0(X;S'_\sU)\ar H^0(X;S_\sU)\ar H^0(X;S''_\sU)\ar \label{spr221}\\
&& \qquad
H^1(X;S'_\sU) \ar\cdots  H^p(X;S'_\sU)\ar H^p(X;S_\sU)\ar \nonumber\\
&& \qquad  H^p(X;S''_\sU)\ar
H^{p+1}(X;S'_\sU) \ar\cdots \nonumber
\een
of the cohomology groups of $X$ with coefficients in these
presheaves. This result is extended to the exact sequence of
sheaves. Let
\mar{spr226}\beq
0\to S'\ar S\ar S'' \to 0 \label{spr226}
\eeq
be a short  exact sequence of sheaves on $X$. It yields the short
exact sequence of presheaves (\ref{ms0103}) where the presheaf
$S''_\sU$ generates the sheaf $S''$. If $X$ is paracompact,
\be
H^*(X;S''_\sU)=H^*(X;S'')
\ee
in accordance with Theorem \ref{spr225}, and we have the exact
sequence of sheaf cohomology
\mar{spr227}\ben
&& 0\to H^0(X;S')\ar H^0(X;S)\ar H^0(X;S'')\ar
\label{spr227}\\
&& \qquad H^1(X;S') \ar\cdots   H^p(X;S')\ar H^p(X;S)\ar \nonumber
\\
&& \qquad H^p(X;S'')\ar H^{p+1}(X;S') \ar\cdots\,. \nonumber
\een

Let us turn now to the abstract de Rham theorem which provides a
powerful tool of studying algebraic systems on paracompact spaces.

Let us consider an exact sequence of sheaves
\mar{spr228}\beq
0\to S\ar^h S_0\ar^{h^0} S_1\ar^{h^1}\cdots S_p\ar^{h^p}\cdots.
\label{spr228}
\eeq
It is said to be a {\sl resolution of the sheaf}
\index{resolution!of a sheaf} $S$ if each sheaf $S_{p\geq 0}$ is
{\sl acyclic}, \index{sheaf!acyclic} i.e., its cohomology groups
$H^{k>0}(X;S_p)$ vanish.

Any exact sequence of sheaves (\ref{spr228}) yields the sequence
of their structure modules
\mar{spr229}\beq
0\to S(X)\ar^{h_*} S_0(X)\ar^{h^0_*} S_1(X)\ar^{h^1_*}\cdots
S_p(X)\ar^{h^p_*}\cdots \label{spr229}
\eeq
which is always exact at terms $S(X)$ and $S_0(X)$ (see the exact
sequence (\ref{ms0103'})). The sequence (\ref{spr229}) is a
cochain complex because
\be
h^{p+1}_*\circ h^p_*=0.
\ee
If $X$ is a paracompact space and the exact sequence
(\ref{spr228}) is a resolution of $S$, the forthcoming {\sl
abstract de Rham theorem} \index{de Rham theorem!abstract}
establishes an isomorphism of cohomology of the complex
(\ref{spr229}) to cohomology of $X$ with coefficients in the sheaf
$S$.

\begin{theorem} \label{spr230} \mar{spr230}
Given a resolution (\ref{spr228}) of a sheaf $S$ on a paracompact
topological  space $X$ and the induced complex (\ref{spr229}),
there are isomorphisms
\mar{spr231}\beq
H^0(X;S)=\Ker h^0_*, \qquad H^q(X;S)=\Ker h^q_*/\im h^{q-1}_*,
\qquad q>0. \label{spr231}
\eeq
\end{theorem}

A sheaf $S$  on a paracompact space $X$ is called {\sl fine}
\index{sheaf!fine} if, for each locally finite open cover $\gU
=\{U_i\}_{i\in I}$ of $X$,  there exists a system $\{h_i\}$ of
endomorphisms $h_i:S\to S$ such that:

(i) there is a closed subset $V_i\subset U_i$ and $h_i(S_x)=0$ if
$x\not\in V_i$,

(ii) $\op\sum_{i\in I}h_i$ is the identity map of $S$.

\begin{theorem}  \label{107t1} \mar{107t1}
A fine sheaf on a paracompact space is acyclic.
\end{theorem}

There is the following important example of fine sheaves.

\begin{theorem}  \label{spr256} \mar{spr256}
Let $X$ be a paracompact topological  space which admits a
partition of unity performed by elements of the structure module
$\gA(X)$ of some sheaf $\gA$ of real functions on $X$. Then any
sheaf $S$ of $\gA$-modules on $X$, including $\gA$ itself, is
fine.
\end{theorem}

In particular, the sheaf $C^0_X$ of continuous functions on a
paracompact topological space is fine, and so is any sheaf of
$C^0_X$-modules.

\section{Local-ringed spaces}

Local-ringed spaces are sheafs of local rings. For instance,
smooth manifolds, represented by sheaves of real smooth functions,
make up a subcategory of the category of local-ringed spaces.

A sheaf $\gR$ on a topological space $X$ is said to be a {\sl
ringed space} \index{ringed space} if its stalk $\gR_x$ at each
point $x\in X$ is a real commutative ring. A ringed space is often
denoted by a pair $(X,\gR)$ of a topological space $X$ and a sheaf
$\gR$ of rings on $X$. They are called the {\sl body}
\index{body!of a ringed space} and the {\sl structure sheaf}
\index{structure sheaf!of a ringed space} of a ringed space,
respectively.

A ringed space is said to be a {\sl local-ringed space}
\index{local-ringed space} (a {\sl geometric space})
\index{geometric space} if it is a sheaf of local rings.

For instance, the sheaf $C^0_X$ of continuous real functions on a
topological space $X$ is a local-ringed space. Its stalk $C^0_x$,
$x\in X$, contains the unique maximal ideal of germs of functions
vanishing at $x$.

Morphisms of local-ringed spaces are defined as those of sheaves
on different topological spaces as follows.

Let $\vf:X\to X'$ be a continuous map. Given a sheaf $S$ on $X$,
its {\sl direct image} \index{image of a sheaf!direct} $\vf_*S$
\index{direct image of a sheaf} on $X'$ is generated by the
presheaf of assignments
\be
X'\supset U'\to S(\vf^{-1}(U'))
\ee
for any open subset $U'\subset X'$. Conversely, given a sheaf $S'$
on $X'$, its {\sl inverse image} \index{image of a sheaf!inverse}
$\vf^*S'$ \index{inverse image of a sheaf} on $X$ is defined as
the pull-back onto $X$ of the continuous fibre bundle $S'$ over
$X'$, i.e., $\vf^*S'_x=S_{\vf(x)}$. This sheaf is generated by the
presheaf which associates to any open $V\subset X$ the direct
limit of modules $S'(U)$ over all open subsets $U\subset X'$ such
that $V\subset f^{-1}(U)$.

\begin{remark} \label{spr201} \mar{spr201}
Let $i:X\to X'$ be a closed subspace of $X'$. Then $i_*S$ is a
unique sheaf on $X'$ such that
\be
i_*S|_X=S, \qquad i_*S|_{X'\setminus X}=0.
\ee
Indeed, if $x'\in X\subset X'$, then
\be
i_*S(U')= S(U'\cap X)
\ee
for any open neighborhood $U$ of this point. If $x'\not\in X$,
there exists its neighborhood $U'$ such that $U'\cap X$ is empty,
i.e., $i_*S(U')=0$. The sheaf $i_*S$ is called the {\sl trivial
extension} \index{trivial extension of a sheaf} of the sheaf $S$.
\end{remark}

By a  \index{morphism!of ringed spaces} {\sl morphism of ringed
spaces}
\be
(X,\gR)\to (X',\gR')
\ee
is meant a pair $(\vf,\wh\vf)$ of a continuous map $\vf:X\to X'$
and a sheaf morphism $\wh\vf:\gR'\to \vf_*\gR$ or, equivalently, a
sheaf morphisms $\vf^*\gR'\to \gR$. Restricted to each stalk, a
sheaf morphism $\Phi$ is assumed to be a ring homomorphism. A
morphism of ringed spaces is said to be:

$\bullet$ a monomorphism if $\vf$ is an injection and $\Phi$ is an
epimorphism,

$\bullet$ an epimorphism if $\vf$ is a surjection, while $\Phi$ is
a monomorphism.

Let $(X,\gR)$ be a local-ringed space. By a {\sl sheaf $\gd \gR$
of derivations} \index{sheaf!of derivations} of the sheaf $\gR$ is
meant a subsheaf of endomorphisms of $\gR$ such that any section
$u$ of $\gd \gR$ over an open subset $U\subset X$ is a derivation
of the real ring $\gR(U)$. It should be emphasized that, since the
monomorphism (\ref{+212}) is not necessarily an isomorphism, a
derivation of the ring $\gR(U)$ need not be a section of the sheaf
$\gd \gR|_U$. Namely, it may happen that, given open sets
$U'\subset U$, there is no restriction morphism
\be
\gd (\gR(U)) \to\gd (\gR(U')).
\ee

Given a local-ringed space $(X,\gR)$, a sheaf $P$ on $X$ is called
a {\sl sheaf of $\gR$-modules} \index{sheaf!of modules} if every
stalk $P_x$, $x\in X$, is an $\gR_x$-module or, equivalently, if
$P(U)$ is an $\gR(U)$-module for any open subset $U\subset X$. A
sheaf of $\gR$-modules $P$ is said to be {\sl locally free}
\index{sheaf!locally free} if there exists an open neighborhood
$U$ of every point $x\in X$ such that $P(U)$ is a free
$\gR(U)$-module. If all these free modules are of finite rank
(resp. of the same finite rank), one says that $P$ is of {\sl
finite type} \index{sheaf!locally free!of finite type} (resp. of
constant rank). \index{sheaf!locally free!of constant rank} The
structure module of a locally free sheaf is called a {\sl locally
free module}. \index{module!locally free}

The following is a generalization of Theorem \ref{spr256}.

\begin{theorem} \label{spr256'} \mar{spr256'} Let $X$ be a paracompact
space which admits a partition of unity by elements of the
structure module $S(X)$ of some sheaf $S$ of real functions on
$X$. Let $P$ be a sheaf of $S$-modules. Then $P$ is fine and,
consequently, acyclic.
\end{theorem}

Assumed to be paracompact, a smooth manifold $X$ admits a
partition of unity performed by smooth real functions. It follows
that the sheaf $C^\infty_X$ of smooth real functions on $X$ is
fine, and so is any sheaf of $C^\infty_X$-modules, e.g., the
sheaves of sections of smooth vector bundles over $X$.

Similarly to the sheaf $C^0_X$ of continuous functions, the sheaf
$C^\infty_X$ of smooth real functions on a smooth manifold $X$ is
a local-ringed spaces. Its stalk $C^\infty_x$ at a point $x\in X$
has a unique maximal ideal $\m_x$ of germs of smooth functions
vanishing at $x$. Though the sheaf $C^\infty_X$ is defined on a
topological space $X$, it fixes a unique smooth manifold structure
on $X$ as follows.

\begin{theorem} \label{+26} \mar{+26}
Let $X$ be a paracompact topological space and $(X,\gR)$ a
local-ringed space. Let $X$ admit an open cover $\{U_i\}$ such
that the sheaf $\gR$ restricted to each $U_i$ is isomorphic to the
local-ringed space  $(\Bbb R^n, C^\infty_{R^n})$. Then $X$ is an
$n$-dimensional smooth manifold together with a natural
isomorphism of local-ringed spaces $(X,\gR)$ and $(X,C^\infty_X)$.
\end{theorem}

One can think of this result as being an alternative definition of
smooth real manifolds in terms of local-ringed spaces. In
particular, there is one-to-one correspondence between smooth
manifold morphisms $X\to X'$ and the $\Bbb R$-ring morphisms
$C^\infty(X')\to C^\infty(X)$.

For instance, let $Y\to X$ be a smooth vector bundle. The germs of
its sections make up a sheaf of $C^\infty_X$-modules, called the
{\sl structure sheaf} $S_Y$ \index{structure sheaf!of a vector
bundle} of a vector bundle $Y\to X$. The sheaf $S_Y$ is fine. The
structure module of this sheaf coincides with the structure module
$Y(X)$ of global sections of a vector bundle $Y\to X$. The
following {\sl Serre--Swan theorem} \index{Serre--Swan theorem}
shows that these modules exhaust all projective modules of finite
rank over $C^\infty(X)$. Originally proved for bundles over a
compact base $X$, this theorem has been extended to an arbitrary
$X$.

\begin{theorem} \label{sp60} \mar{sp60}
Let $X$ be a smooth manifold. A $C^\infty(X)$-module $P$ is
isomorphic to the structure module of a smooth vector bundle over
$X$ iff it is a projective module of finite rank.
\end{theorem}

This theorem states the categorial equivalence between the vector
bundles over a smooth manifold $X$ and projective modules of
finite rank over the ring $C^\infty(X)$ of smooth real functions
on $X$. The following are corollaries  of this equivalence

$\bullet$ The structure module $Y^*(X)$ of the dual $Y^*\to X$ of
a vector bundle $Y\to X$ is the $C^\infty(X)$-dual $Y(X)^*$ of the
structure module $Y(X)$ of $Y\to X$.

$\bullet$ Any exact sequence of vector bundles
\mar{t51}\beq
0\to Y \ar Y'\ar Y''\to 0 \label{t51}
\eeq
over the same base $X$ yields the exact sequence
\mar{t52}\beq
0\to Y(X) \ar Y'(X)\ar Y''(X)\to 0 \label{t52}
\eeq
of their structure modules, and {\it vice versa}. In accordance
with Theorem \ref{sp11}, the exact sequence (\ref{t51}) is always
split. Every its splitting defines that of the exact sequence
(\ref{t52}), and {\it vice versa}.

$\bullet$ The derivation module of the real ring $C^\infty(X)$
coincides with the $C^\infty(X)$-module $\cT(X)$ of vector fields
on $X$, i.e., with the structure module of the tangent bundle $TX$
of $X$. Hence, it is a projective $C^\infty(X)$-module of finite
rank. It is the $C^\infty(X)$-dual $\cT(X)=\cO^1(X)^*$ of the
structure module $\cO^1(X)$ of the cotangent bundle $T^*X$ of $X$
which is the module of differential one-forms on $X$ and,
conversely,
\be
\cO^1(X)=\cT(X)^*.
\ee

$\bullet$ Therefore, if $P$ is a $C^\infty(X)$-module, one can
reformulate Definition \ref{1016} of a connection on $P$ as
follows. A connection on a $C^\infty(X)$-module $P$ is a
$C^\infty(X)$-module morphism
\mar{t56}\beq
\nabla: P\to \cO^1(X)\ot P, \label{t56}
\eeq
which satisfies the Leibniz rule
\be
\nabla(fp)=df\ot p +f\nabla(p), \qquad f\in C^\infty(X), \qquad
p\in P.
\ee
It associates to any vector field $\tau\in\cT(X)$ on $X$ a first
order differential operator $\nabla_\tau$ on $P$ which obeys the
Leibniz rule
\mar{t58}\beq
\nabla_\tau(fp)=(\tau\rfloor df)p +f\nabla_\tau p. \label{t58}
\eeq
In particular, let $Y\to X$ be a vector bundle and $Y(X)$ its
structure module. The notion of a connection on the structure
module $Y(X)$ is equivalent to the standard geometric notion of a
connection on a vector bundle $Y\to X$.

Since the derivation module of the real ring $C^\infty(X)$ is the
$C^\infty(X)$-module $\cT(X)$ of vector fields on $X$ and
\be
\cO^1(X)=\cT(X)^*,
\ee
the Chevalley--Eilenberg differential calculus over the real ring
$C^\infty(X)$ is exactly the DGA $(\cO^*(X),d)$ of exterior forms
on $X$, where the Chevalley--Eilenberg coboundary operator $d$
(\ref{+840}) coincides with the exterior differential. Moreover,
one can show that $(\cO^*(X),d)$ is a minimal differential
calculus, i.e., the $C^\infty(X)$-module $\cO^1(X)$ is generated
by elements $df$, $f\in C^\infty(X)$. Therefore, the de Rham
complex (\ref{t10}) of the real ring $C^\infty(X)$ is the {\sl de
Rham complex} \index{de Rham complex!of exterior forms}
\mar{t37}\beq
0\to \Bbb R\ar C^\infty(X)\ar^d \cO^1(X)\ar^d\cdots \cO^k(X)\ar^d
\cdots \label{t37}
\eeq
of exterior forms on a manifold $X$.

The de Rham cohomology of the complex (\ref{t37}) is called the
{\sl de Rham cohomology} \index{de Rham cohomology!of a manifold}
\index{cohomology!de Rham!of a manifold} $H^*_{\rm DR}(X)$ of $X$.
To describe them, let us consider the  \index{de Rham complex!of
sheaves} {\sl de Rham complex}
\mar{t67}\beq
0\to \Bbb R\ar C^\infty_X\ar^d \cO^1_X\ar^d\cdots \cO^k_X\ar^d
\cdots \label{t67}
\eeq
of sheaves $\cO^k_X$, $k\in\Bbb N_+$, of germs of exterior forms
on $X$. These sheaves are fine. Due to the {\sl Poincar\'e lemma},
\index{Poincar\'e lemma} the complex (\ref{t67}) is exact and,
thereby, is a fine resolution of the constant sheaf $\Bbb R$ on a
manifold $X$. Then a corollary of Theorem \ref{spr230} is the
classical {\sl de Rham theorem}. \index{de Rham theorem}

\begin{theorem} \label{t60} \mar{t60} There is an isomorphism
\mar{t61}\beq
H^k_{\rm DR}(X)=H^k(X;\Bbb R) \label{t61}
\eeq
of the de Rham cohomology $H^*_{\rm DR}(X)$ of a manifold $X$ to
cohomology of $X$ with coefficients in the constant sheaf $\Bbb
R$.
\end{theorem}

\bibliographystyle{alpha}
\bibliographystyle{plain}
\addcontentsline{toc}{chapter}{Bibliography}
\bibliography{conn99}



\addcontentsline{toc}{chapter}{Index}

\begin{theindex}

  \item $B^{n, \m}${86}
  \item $C^\infty_X$, 133
  \item $C_{\rm W}$, 76
  \item $D_\G$, 41
  \item $G$-bundle, 55
    \subitem principal, 56
    \subitem smooth, 56
  \item $GL_n$, 74
  \item $HY$, 37
  \item $H^1(X;G^0_X)$, 55
  \item $J^1Y$, 23
  \item $J^1\Phi$, 24
  \item $J^1_\Si Y$, 47
  \item $J^1s$, 24
  \item $J^1u$, 24
  \item $J^\infty Y$, 32
  \item $J^k u$, 27
  \item $J^rY$, 25
  \item $J^r\Phi$, 27
  \item $J^rs$, 27
  \item $LX$, 74
  \item $L_G$, 52
  \item $L_g$, 52
  \item $P^G$, 65
  \item $P^{\ot k}$, 124
  \item $P_\Si$, 71
  \item $R_G$, 52
  \item $R_g$, 52
  \item $R_{GP}$, 57
  \item $R_{gP}$, 57
  \item $TZ$, 3
  \item $T^*Z$, 3
  \item $T_GP$, 58
  \item $Tf$, 3
  \item $VY$, 11
  \item $V\G$, 49
  \item $V\Phi$, 11
  \item $V^*Y$, 12
  \item $V^*\G$, 50
  \item $V_GP$, 58
  \item $V_\Si Y$, 48
  \item $V_\Si^*Y$, 48
  \item $Y(X)$, 9
  \item $Y^h$, 9
  \item $[A]$, 108
  \item $\G\tau$, 38
  \item $\G^*$, 44
  \item $\bL_u$, 18
  \item $\bb$, 51
  \item $\bu$, 119
  \item $\cA_E$, 91
  \item $\cE_L$, 99
  \item $\cE_i$, 99
  \item $\cF$, 64
  \item $\cG_l$, 52
  \item $\cG_r$, 52
  \item $\cJ_\up$, 103
  \item $\cO^*(Z)$, 16
  \item $\cO^*[\gd\cA]$, 132
  \item $\cO^*\cA$, 133
  \item $\cO^*_\infty Y$, 33
  \item $\cO^*_\infty$, 33
  \item $\cO^{k,m}_\infty$, 34
  \item $\cO_k^*$, 27
  \item $\cP^*_\infty$, 101
  \item $\cP^*_\infty[E;Y]$, 114
  \item $\cQ^*_\infty$, 33
  \item $\cS^*[E;Z]$, 95
  \item $\cS^*_\infty[F;X]$, 106
  \item $\cS^*_\infty[F;Y]$, 107
  \item $\cS^*_r[F;Y]$, 106
  \item $\cT(Z)$, 14
  \item $\cV_E$, 93
  \item $\ccG(X)$, 66
  \item $\ccG_L$, 111
  \item $\dl_N$, 117
  \item $\dl_{\rm KT}$, 117
  \item $\dot\dr_\la$, 14
  \item $\e_m$, 52
  \item $\gA(Z)$, 90
  \item $\gA_E$, 90
  \item $\gd\cA$, 127
  \item $\gd\cA_E$, 93
  \item $\nabla^\G$, 41
  \item $\ol E$, 113
  \item $\ol dy^i$, 12
  \item $\ol\dl$, 114
  \item $\om$, 17
  \item $\om_\la$, 17
  \item $\op\up^\lto$, 112
  \item $\ot P$, 125
  \item $\pi^1$, 23
  \item $\pi^1_0$, 23
  \item $\pi^\infty_r$, 32
  \item $\pi_k^r$, 26
  \item $\pi_{Y\Si}$, 8
  \item $\pi_{\Si X}$, 8
  \item $\psi_\xi$, 5
  \item $\thh^i$, 24
  \item $\thh^i_\La$, 29
  \item $\thh_Z$, 19
  \item $\thh_{LX}$, 75
  \item $\ve_m$, 52
  \item $\vr_{\xi\zeta}$, 5
  \item $\w Y$, 10
  \item $\wh 0$, 9
  \item $\wt D$, 49
  \item $\wt\tau$, 14
  \item $\{_\la{}^\nu{}_\m\}$, 45
  \item $d_H$, 34
  \item $d_V$, 34
  \item $d_\la$, 24
  \item $f^*Y$, 8
  \item $f^*\G$, 39
  \item $f^*\f$, 17
  \item $h_0$, 29
  \item $u\rfloor\f$, 18
  \item $u_\xi$, 66
  \item ${\rm Ad}_g$, 53

  \indexspace

  \item action of a group, 51
    \subitem effective, 51
    \subitem free, 51
    \subitem on the left, 51
    \subitem on the right, 51
    \subitem transitive, 51
  \item action of a structure group
    \subitem on $J^1P$, 60
    \subitem on $P$, 57
    \subitem on $TP$, 58
  \item adjoint representation
    \subitem of a Lie algebra, 53
    \subitem of a Lie group, 53
  \item affine bundle, 12
    \subitem morphism, 13
  \item algebra, 121
    \subitem $\Bbb Z_2$-graded, 83
      \subsubitem commutative, 83
    \subitem $\cO^*_\infty Y$, 33
    \subitem $\cO^*_\infty$, 33
    \subitem $\cP^*_\infty[E;Y]$, 114
    \subitem $\cS^*_\infty[F;Y]$, 107
    \subitem differential bigraded, 89
    \subitem differential graded, 131
    \subitem graded, 131
      \subsubitem commutative, 131
    \subitem unital, 121
  \item annihilator of a distribution, 15
  \item antifield, 114
    \subitem $k$-stage, 117
    \subitem Noether, 115
  \item associated bundles, 55
  \item automorphism
    \subitem associated, 69
    \subitem principal, 65
  \item autoparallel, 79

  \indexspace

  \item base of a fibred manifold, 5
  \item basic form, 17
  \item basis
    \subitem for a graded manifold, 91
    \subitem for a module, 123
    \subitem generating, 108
  \item Batchelor theorem, 90
  \item Bianchi identity
    \subitem first, 43
    \subitem second, 42
  \item bigraded
    \subitem de Rham complex, 89
    \subitem exterior algebra, 84
  \item bimodule, 121
    \subitem commutative, 121
    \subitem graded, 84
  \item body
    \subitem of a graded manifold, 90
    \subitem of a ringed space, 139
  \item body map, 85
  \item boundary, 128
  \item boundary operator, 128
  \item bundle
    \subitem $P$-associated, 68
    \subitem affine, 12
    \subitem associated, 70
      \subsubitem canonically, 70
    \subitem atlas, 6
      \subsubitem associated, 68
      \subsubitem holonomic, 11
      \subsubitem of constant local trivializations, 47
    \subitem automorphism, 7
      \subsubitem vertical, 7
    \subitem composite, 9
    \subitem continuous, 6
      \subsubitem locally trivial, 6
    \subitem coordinates, 6
      \subsubitem affine, 13
      \subsubitem linear, 9
    \subitem cotangent, 3
    \subitem density-dual, 113
    \subitem epimorphism, 7
    \subitem exterior, 10
    \subitem gauge natural, 76
    \subitem isomorphism, 7
    \subitem lift, 70
    \subitem locally trivial, 6
    \subitem monomorphism, 7
    \subitem morphism, 7
      \subsubitem affine, 13
      \subsubitem linear, 9
      \subsubitem of principal bundles, 58
    \subitem natural, 74
    \subitem of principal connections, 61
    \subitem of world connections, 76
    \subitem principal, 56
    \subitem product, 8
    \subitem smooth, 6
    \subitem tangent, 3
      \subsubitem affine, 13
      \subsubitem vertical, 11
    \subitem with a structure group, 55
      \subsubitem principal, 56
      \subsubitem smooth, 56

  \indexspace

  \item canonical principal connection, 64
  \item Cartan connection, 46
  \item chain, 128
  \item Chevalley--Eilenberg
    \subitem coboundary operator, 130
      \subsubitem graded, 88
    \subitem cohomology, 131
    \subitem complex, 131
    \subitem differential calculus, 132
      \subsubitem Grassmann-graded, 89
      \subsubitem minimal, 133
  \item Christoffel symbols, 45
  \item classical solution, 30
  \item closed map, 5
  \item coboundary, 128
  \item coboundary operator, 128
    \subitem Chevalley--Eilenberg, 130
  \item cochain, 128
  \item cochain morphism, 129
  \item cocycle, 128
  \item cocycle condition, 5
    \subitem for a sheaf, 137
  \item codistribution, 15
  \item coframe, 3
  \item cohomology, 128
    \subitem Chevalley--Eilenberg, 131
    \subitem de Rham
      \subsubitem abstract, 131
      \subsubitem of a manifold, 143
    \subitem with coefficients in a sheaf, 136
  \item complex, 128
    \subitem $k$-exact, 128
    \subitem acyclic, 128
    \subitem chain, 128
    \subitem Chevalley--Eilenberg, 131
    \subitem cochain, 128
    \subitem de Rham
      \subsubitem abstract, 131
    \subitem exact, 128
    \subitem variational, 98
  \item component of a connection, 37
  \item composite bundle, 9
  \item composite connection, 48
  \item connection, 37
    \subitem affine, 45
    \subitem composite, 48
    \subitem covertical, 50
    \subitem dual, 43
    \subitem flat, 46
    \subitem linear, 43
      \subsubitem world, 76
    \subitem on a graded commutative ring, 90
    \subitem on a graded manifold, 94
    \subitem on a graded module, 89
    \subitem on a module, 127
    \subitem on a ring, 127
    \subitem principal, 61
      \subsubitem associated, 69
      \subsubitem canonical, 64
    \subitem projectable, 48
    \subitem reducible, 41
    \subitem vertical, 50
  \item connection form, 38
    \subitem of a principal connection, 62
      \subsubitem local, 62
    \subitem vertical, 39
  \item conservation law
    \subitem weak, 102
  \item contact derivation, 101
    \subitem graded, 110
    \subitem projectable, 101
    \subitem vertical, 101
  \item contact form, 30
    \subitem graded, 108
    \subitem local, 24
      \subsubitem of higher jet order, 29
  \item contraction, 18
  \item cotangent bundle, 3
    \subitem vertical, 12
  \item covariant derivative, 41
  \item covariant differential, 41
    \subitem on a module, 127
    \subitem vertical, 49
  \item curvature, 42
    \subitem of a principal connection, 63
    \subitem of a world connection, 44
    \subitem of an associated principal connection, 69
    \subitem soldered, 42
  \item curvature-free connection, 46
  \item curve integral, 14
  \item cycle, 128

  \indexspace

  \item DBGA, 89
  \item de Rham cohomology
    \subitem abstract, 131
    \subitem of a graded manifold, 95
    \subitem of a manifold, 143
    \subitem of a ring, 133
  \item de Rham complex
    \subitem abstract, 131
    \subitem bigraded, 89
    \subitem of a ring, 133
    \subitem of exterior forms, 143
    \subitem of sheaves, 143
  \item de Rham theorem, 143
    \subitem abstract, 139
  \item density, 17
  \item density-dual bundle, 113
  \item density-dual vector bundle, 113
    \subitem graded, 114
  \item derivation, 127
    \subitem contact, 101
      \subsubitem projectable, 101
      \subsubitem vertical, 101
    \subitem graded, 87
      \subsubitem contact, 110
    \subitem nilpotent, 112
    \subitem right, 112
  \item derivation module, 127
    \subitem graded, 87
  \item DGA, 131
  \item differential
    \subitem covariant, 41
      \subsubitem vertical, 49
    \subitem exterior, 17
    \subitem total, 34
      \subsubitem graded, 108
    \subitem vertical, 34
      \subsubitem graded, 108
  \item differential calculus, 131
    \subitem Chevalley--Eilenberg, 132
      \subsubitem minimal, 133
    \subitem minimal, 132
  \item differential equation, 30
    \subitem associated to a differential operator, 31
  \item differential form, 33
    \subitem graded, 107
  \item differential ideal, 15
  \item differential operator
    \subitem as a morphism, 31
    \subitem as a section, 30
    \subitem graded, 86
    \subitem of standard form, 31
    \subitem on a module, 126
  \item direct image of a sheaf, 140
  \item direct limit, 124
  \item direct sequence, 124
  \item direct sum
    \subitem of complexes, 129
    \subitem of modules, 122
  \item direct system of modules, 123
  \item directed set, 123
  \item distribution, 15
    \subitem horizontal, 37
    \subitem involutive, 15
  \item domain, 7
  \item dual module, 122
  \item dual vector bundle, 10

  \indexspace

  \item Ehresmann connection, 39
  \item equivalent $G$-bundle atlases, 55
  \item equivalent $G$-bundles, 55
  \item equivalent bundle atlases, 6
  \item equivariant
    \subitem automorphism, 65
    \subitem connection, 61
    \subitem function, 65
  \item Euler--Lagrange operator, 99
  \item Euler--Lagrange-type operator, 99
  \item even element, 83
  \item even morphism, 84
  \item exact sequence
    \subitem of modules, 123
      \subsubitem short, 123
      \subsubitem split, 123
    \subitem of vector bundles, 10
      \subsubitem short, 10
      \subsubitem split, 11
  \item exterior algebra, 125
    \subitem bigraded, 84
  \item exterior bundle, 10
  \item exterior differential, 17
  \item exterior form, 16
    \subitem basic, 17
    \subitem graded, 95
    \subitem horizontal, 17
  \item exterior product, 16
    \subitem graded, 84
    \subitem of vector bundles, 10

  \indexspace

  \item factor
    \subitem algebra, 121
    \subitem bundle, 11
    \subitem complex, 129
    \subitem module, 122
    \subitem sheaf, 135
  \item fibration, 5
  \item fibre, 5
  \item fibre bundle, 5
  \item fibre coordinates, 6
  \item fibred coordinates, 5
  \item fibred manifold, 4
  \item fibrewise morphism, 7
  \item first Noether theorem, 102
    \subitem for a graded Lagrangian, 111
  \item first variational formula, 102
    \subitem for a graded Lagrangian, 111
  \item flow, 14
  \item foliated manifold, 16
  \item foliation, 15
    \subitem horizontal, 46
    \subitem simple, 16
  \item Fr\"olicher--Nijenhuis bracket, 19
  \item frame, 9
    \subitem holonomic, 3
  \item frame field, 74

  \indexspace

  \item gauge
    \subitem algebra, 59
    \subitem algebra bundle, 59
    \subitem group, 66
    \subitem operator, 120
    \subitem parameters, 104
    \subitem transformation, 65
      \subsubitem infinitesimal, 66
  \item gauge natural bundle, 76
  \item gauge symmetry, 104
    \subitem $k$-stage, 120
    \subitem first-stage, 120
    \subitem of a graded Lagrangian, 120
    \subitem reducible, 105
  \item general covariant transformation, 74
    \subitem infinitesimal, 74
  \item generalized vector field, 101
    \subitem graded, 110
  \item generating basis, 108
  \item geodesic, 80
  \item geodesic equation, 80
  \item geometric space, 139
  \item ghost, 118
  \item graded
    \subitem algebra, 131
    \subitem bimodule, 84
    \subitem commutative algebra, 131
    \subitem commutative ring, 85
      \subsubitem Banach, 85
      \subsubitem real, 85
    \subitem connection, 94
    \subitem derivation, 87
      \subsubitem of a field system algebra, 112
    \subitem derivation module, 87
    \subitem differential form, 107
    \subitem differential operator, 86
    \subitem exterior differential, 95
    \subitem exterior form, 95
    \subitem exterior product, 84
    \subitem function, 90
    \subitem interior product, 89
    \subitem Leibniz rule, 131
    \subitem manifold, 90
      \subsubitem composite, 106
      \subsubitem simple, 91
    \subitem module, 84
      \subsubitem  free, 84
      \subsubitem dual, 85
    \subitem morphism, 85
      \subsubitem even, 84
      \subsubitem odd, 84
    \subitem ring, 83
    \subitem vector field, 92
      \subsubitem generalized, 110
    \subitem vector space, 84
      \subsubitem $(n,m)$-dimensional, 84
  \item graded-homogeneous element, 84
  \item grading automorphism, 83
  \item Grassmann algebra, 85
  \item group bundle, 65

  \indexspace

  \item Heimholtz condition, 99
  \item holonomic
    \subitem atlas, 11
      \subsubitem of the frame bundle, 75
    \subitem automorphisms, 75
    \subitem coframe, 3
    \subitem coordinates, 3
    \subitem frame, 3
  \item homogeneous space, 52
  \item homology, 128
  \item homology regularity condition, 117
  \item horizontal
    \subitem distribution, 37
    \subitem foliation, 46
    \subitem form, 17
      \subsubitem graded, 108
    \subitem lift
      \subsubitem of a path, 39
      \subsubitem of a vector field, 38
    \subitem projection, 29
    \subitem splitting, 38
      \subsubitem canonical, 40
    \subitem vector field, 38

  \indexspace

  \item ideal, 121
    \subitem maximal, 121
    \subitem of nilpotents, 85
    \subitem proper, 121
  \item image of a sheaf
    \subitem direct, 140
    \subitem inverse, 140
  \item imbedded submanifold, 4
  \item imbedding, 4
  \item immersion, 4
  \item induced coordinates, 11
  \item inductive limit, 125
  \item infinitesimal generator, 14
  \item infinitesimal transformation
    \subitem of a Lagrangian system, 101
      \subsubitem Grassmann-graded, 110
  \item integral curve, 14
  \item integral manifold, 15
    \subitem maximal, 15
  \item integral section of a connection, 41
  \item interior product, 18
    \subitem graded, 89
    \subitem of vector bundles, 10
  \item inverse image of a sheaf, 140
  \item inverse sequence, 125

  \indexspace

  \item jet
    \subitem first order, 23
    \subitem higher order, 25
    \subitem infinite order, 32
  \item jet bundle, 24
    \subitem affine, 24
  \item jet coordinates, 23
  \item jet manifold, 23
    \subitem higher order, 25
    \subitem infinite order, 32
  \item jet prolongation
    \subitem functor, 26
    \subitem of a morphism, 24
      \subsubitem higher order, 27
    \subitem of a section, 24
      \subsubitem higher order, 27
    \subitem of a structure group action, 60
    \subitem of a vector field, 24
      \subsubitem higher order, 27
  \item juxtaposition rule, 131

  \indexspace

  \item kernel
    \subitem of a bundle morphism, 7
    \subitem of a differential operator, 31
    \subitem of a vector bundle morphism, 10
  \item Koszul--Tate complex, 117
  \item Koszul--Tate operator, 117

  \indexspace

  \item Lagrangian, 99
    \subitem degenerate, 115
    \subitem graded, 109
    \subitem variationally trivial, 99
  \item Lagrangian system, 99
    \subitem $N$-stage reducible, 117
    \subitem finitely degenerate, 115
    \subitem Grassmann-graded, 109
    \subitem irreducible, 116
    \subitem reducible, 116
  \item leaf, 15
  \item left-invariant form, 54
    \subitem canonical, 54
  \item Leibniz rule, 127
    \subitem for a connection, 127
      \subsubitem Grassmann-graded, 89
    \subitem graded, 131
    \subitem Grassmann-graded, 87
  \item Lepage equivalent
    \subitem of a graded Lagrangian, 109
  \item Levi--Civita connection, 45
  \item Levi--Civita symbol, 17
  \item Lie algebra
    \subitem left, 52
    \subitem right, 52
  \item Lie algebra bundle, 59
  \item Lie bracket, 14
  \item Lie derivative
    \subitem graded, 89
    \subitem of a tangent-valued form, 20
    \subitem of an exterior form, 18
  \item Lie superalgebra, 86
  \item Lie superbracket, 86
  \item lift of a bundle, 70
  \item lift of a vector field
    \subitem canonical, 14
    \subitem functorial, 74
    \subitem horizontal, 38
  \item linear derivative of an affine morphism, 13
  \item linear frame bundle, 74
  \item local diffeomerphism, 4
  \item local ring, 121
  \item local-ringed space, 139
  \item locally finite cover, 137

  \indexspace

  \item manifold, 3
    \subitem fibred, 4
    \subitem flat, 78
    \subitem parallelizable, 78
  \item matrix group, 54
  \item module, 121
    \subitem dual, 122
    \subitem finitely generated, 123
    \subitem free, 123
    \subitem graded, 84
    \subitem locally free, 141
    \subitem of finite rank, 123
    \subitem over a Lie algebra, 130
    \subitem over a Lie superalgebra, 86
    \subitem projective, 123
  \item morphism
    \subitem of fibre bundles, 7
    \subitem of graded manifolds, 92
    \subitem of presheaves, 134
    \subitem of ringed spaces, 140
    \subitem of sheaves, 134
  \item Mourer--Cartan equation, 54
  \item multi-index, 25

  \indexspace

  \item natural bundle, 74
  \item NI, 113
  \item Nijenhuis differential, 19
  \item nilpotent derivation, 112
  \item Noether identities, 115
    \subitem complete, 115
    \subitem first stage, 116
      \subsubitem non-trivial, 116
      \subsubitem trivial, 116
    \subitem first-stage
      \subsubitem complete, 116
    \subitem higher-stage, 117
    \subitem non-trivial, 115
  \item Noether theorem
    \subitem first, 102
    \subitem second
      \subsubitem direct, 105
      \subsubitem inverse, 118

  \indexspace

  \item odd element, 83
  \item odd morphism, 84
  \item on-shell, 102
  \item open map, 4

  \indexspace

  \item paracompact space, 137
  \item partition of unity, 137
  \item path, 39
  \item Poincar\'e lemma, 143
  \item presheaf, 133
    \subitem canonical, 133
  \item principal
    \subitem automorphism, 65
      \subsubitem of a connection bundle, 66
      \subsubitem of an associated bundle, 69
    \subitem bundle, 56
      \subsubitem continuous, 57
    \subitem connection, 61
      \subsubitem associated, 69
      \subsubitem canonical, 64
      \subsubitem conjugate, 63
    \subitem vector field, 66
      \subsubitem vertical, 66
  \item product connection, 41
  \item proper cover, 6
  \item proper map, 4
  \item pull-back
    \subitem bundle, 8
    \subitem connection, 39
    \subitem form, 17
    \subitem section, 8
    \subitem vertical-valued form, 21

  \indexspace

  \item rank of a morphism, 4
  \item reduced structure, 70
  \item reduced subbundle, 71
  \item reducible connection, 41
  \item representation of a Lie algebra, 53
  \item resolution, 128
    \subitem of a sheaf, 138
  \item restriction of a bundle, 8
  \item Ricci tensor, 78
  \item right derivation, 112
  \item right structure constants, 52
  \item right-invariant form, 54
  \item ring, 121
    \subitem graded, 83
    \subitem local, 121
  \item ringed space, 139

  \indexspace

  \item section, 5
    \subitem global, 5
    \subitem integral, 41
    \subitem local, 5
    \subitem of a jet bundle, 24
      \subsubitem integrable, 24
    \subitem zero-valued, 9
  \item Serre--Swan theorem, 141
    \subitem for graded manifolds, 91
  \item sheaf, 133
    \subitem acyclic, 138
    \subitem constant, 133
    \subitem fine, 139
    \subitem locally free, 141
      \subsubitem of constant rank, 141
      \subsubitem of finite type, 141
    \subitem of continuous functions, 133
    \subitem of derivations, 140
    \subitem of graded derivations, 92
    \subitem of modules, 141
    \subitem of smooth functions, 133
  \item sheaf cohomology, 136
  \item smooth manifold, 3
  \item soldered curvature, 42
  \item soldering form, 20
    \subitem basic, 20
  \item soul map, 85
  \item splitting domain, 90
  \item stalk, 133
  \item strength, 63
    \subitem canonical, 64
    \subitem form, 64
    \subitem of a linear connection, 70
  \item structure group, 55
    \subitem action, 57
    \subitem reduction, 70
  \item structure module
    \subitem  of a sheaf, 133
    \subitem of a vector bundle, 9
  \item structure ring of a graded manifold, 90
  \item structure sheaf
    \subitem of a graded manifold, 90
    \subitem of a ringed space, 139
    \subitem of a vector bundle, 141
  \item subbundle, 7
  \item submanifold, 4
  \item submersion, 4
    \subitem continuous, 6
  \item superspace, 86
  \item supersymmetry, 111
  \item supervector space, 86
  \item symmetry, 103
    \subitem classical, 103
    \subitem exact, 103
    \subitem gauge, 104
    \subitem generalized, 103
    \subitem variational, 102
  \item symmetry current, 103

  \indexspace

  \item tangent bundle, 3
    \subitem affine, 13
    \subitem vertical, 11
  \item tangent morphism, 3
    \subitem vertical, 11
  \item tangent prolongation
    \subitem of a group action, 53
    \subitem of a structure group action, 58
  \item tangent-valued form, 18
    \subitem canonical, 18
    \subitem horizontal, 20
      \subsubitem projectable, 20
  \item tensor algebra, 125
  \item tensor bundle, 11
  \item tensor product
    \subitem of Abelian groups, 122
    \subitem of commutative algebras, 122
    \subitem of complexes, 129
    \subitem of graded modules, 84
    \subitem of modules, 122
    \subitem of vector bundles, 10
  \item tensor product connection, 44
  \item torsion form, 42
    \subitem of a world connection, 44
    \subitem soldering, 78
  \item total derivative, 24
    \subitem graded, 108
    \subitem higher order, 25
    \subitem infinite order, 35
  \item total space, 4
  \item transition functions, 5
    \subitem $G$-valued, 55
  \item trivial extension of a sheaf, 140
  \item trivialization chart, 6
  \item trivialization morphism, 6
  \item tubular neighborhood, 16
  \item typical fibre, 5

  \indexspace

  \item variational
    \subitem bicomplex, 98
      \subsubitem graded, 108
    \subitem complex, 98
      \subsubitem graded, 108
      \subsubitem short, 100
    \subitem derivative, 99
    \subitem formula, 99
    \subitem operator, 98
      \subsubitem graded, 108
    \subitem symmetry, 102
      \subsubitem classical, 103
      \subsubitem of a graded Lagrangian, 111
  \item vector bundle, 9
    \subitem characteristic, 91
    \subitem dual, 10
    \subitem graded, 114
  \item vector field, 14
    \subitem complete, 14
    \subitem fundamental, 58
    \subitem generalized, 101
    \subitem graded, 92
    \subitem holonomic, 79
    \subitem horizontal, 38
      \subsubitem standard, 79
    \subitem integrable, 28
    \subitem left-invariant, 52
    \subitem parallel, 79
    \subitem principal, 66
    \subitem projectable, 14
      \subsubitem on a jet manifold, 28
    \subitem right-invariant, 52
    \subitem subordinate to a distribution, 15
    \subitem vertical, 14
  \item vector space, 122
    \subitem graded, 84
  \item vector-valued form, 21
  \item vertical automorphism, 7
  \item vertical splitting, 12
    \subitem of a vector bundle, 12
    \subitem of an affine bundle, 13
  \item vertical-valued form, 20

  \indexspace

  \item weak conservation law, 102
  \item Whitney sum
    \subitem of vector bundles, 10
  \item world connection, 44
    \subitem affine, 80
    \subitem linear, 76
    \subitem on a tensor bundle, 44
    \subitem on the cotangent bundle, 44
    \subitem symmetric, 44
  \item world metric, 45

\end{theindex}

\end{document}